\documentclass[aps,prb,onecolumn,twoside,floatfix,showpacs]{revtex4-1}
\usepackage{graphicx}
\usepackage{amsmath}
\usepackage{longtable}
\usepackage{color}      
\usepackage[normalem]{ulem}

\begin{document}
  \date{\today}


\preprint{}

\title{Phase stability of ternary fcc and bcc Fe-Cr-Ni alloys}


\author{Jan S. Wr\'obel}
\email[]{jan.wrobel@ccfe.ac.uk, jan.wrobel@inmat.pw.edu.pl}

\author{Duc Nguyen-Manh}
\author{Mikhail Yu. Lavrentiev}
\author{Marek Muzyk}
\author{Sergei L. Dudarev}
\affiliation{CCFE, Culham Science Centre, Abingdon, Oxon OX14 3DB, UK}

\begin{abstract}
The phase stability of fcc and bcc magnetic binary Fe-Cr, Fe-Ni and Cr-Ni alloys, and ternary Fe-Cr-Ni alloys is investigated using a combination of Density Functional Theory (DFT), Cluster Expansion (CE) and Magnetic Cluster Expansion (MCE) approaches. Energies, magnetic moments, and volumes of more than 500 alloy structures have been evaluated using DFT, and the predicted most stable configurations are compared with experimental observations. Deviations from the Vegard law in fcc Fe-Cr-Ni alloys, resulting from the non-linear variation of atomic magnetic moments as functions of alloy composition, are observed. Accuracy of the CE model is assessed against the DFT data, where for ternary Fe-Cr-Ni alloys the cross-validation error is found to be less than 12 meV/atom. A set of cluster interaction parameters is defined for each alloy, where it is used for predicting new ordered alloy structures. Fcc Fe$_2$CrNi phase with Cu$_2$NiZn-like crystal structure is predicted to be the global ground state of ternary Fe-Cr-Ni alloys, with the lowest chemical ordering temperature of 650K. DFT-based Monte Carlo (MC) simulations are applied to the investigation of order-disorder transitions in Fe-Cr-Ni alloys. Enthalpies of formation of ternary alloys predicted by MC simulations at 1600K, combined with magnetic correction derived from MCE, are in excellent agreement with experimental values measured at 1565K. The relative stability of fcc and bcc phases is assessed by comparing the free energies of alloy formation. Evaluation of the free energies involved the application of a dedicated algorithm for computing configurational entropies of the alloys. Chemical order is analyzed, as a function of temperature and composition, in terms of the Warren-Cowley Short-Range Order (SRO) parameters and effective chemical pairwise interactions. In addition to compositions close to binary intermetallic phases CrNi$_2$, FeNi, FeNi$_3$ and FeNi$_8$, pronounced chemical order is found in fcc alloys near the centre of the ternary alloy composition triangle. The calculated SRO parameters compare favourably with experimental data on binary and ternary alloys. Finite temperature {\it magnetic} properties of fcc Fe-Cr-Ni alloys are investigated using an MCE Hamiltonian parameterized using a DFT database of energies and magnetic moments computed for a large number of alloy configurations. MCE simulations show that the ordered ternary Fe$_2$CrNi alloy phase remains magnetic up to 850-900 K due to strong anti-ferromagnetic coupling between (Fe,Ni) and Cr atoms in the ternary Fe-Cr-Ni matrix.
\end{abstract}


\maketitle


\section{Introduction\protect\\}

Fe-Cr-Ni alloys are one of the most studied ternary alloy systems. Their significance stems from the fact that they form the basis for many types of austenitic, ferritic and martensitic steels. Ternary Fe-Cr-Ni and binary Fe-Cr, Fe-Ni and Ni-Cr alloys exhibit diverse magnetic, thermodynamic and mechanical properties, which make them suitable for a variety of applications. This alloy family includes several outstanding examples, like Invar\cite{Guillaume1897} and Permalloy\cite{Arnold1923}. Fe-Cr-Ni based steels, including austenitic 304 and 316 steels, are widely used as structural materials for light water and fast breeder fission reactors\cite{Klueh,Toyama2012}. Inconel alloys X-750 and 718 are used in reactor core components\cite{Rowcliffe2009}. Fe-Cr-based steels F82H and Eurofer are among candidate structural materials for tritium breeding blankets of fusion reactors \cite{Stork2014}. Since the stability of materials in extreme conditions is affected by many factors, extensive and accurate knowledge of how materials respond to temperature and irradiation over extended periods of time is required. The selection of optimal alloy compositions is therefore one of the objectives of fission and fusion materials research. For example, there is a perception that bcc alloys like V-Cr-Ti alloys or ferritic steels exhibit better resistance to radiation swelling in comparison with fcc alloys \cite{Boutard2008}. However, it has been shown by Satoh \textit{et al.} \cite{Satoh2007} that in the fcc Fe$_{55}$Cr$_{15}$Ni$_{30}$ alloy irradiated up to 6 dpa swelling is also significantly reduced when temperature is above 350$^{\circ}$C.

Because of the broad range of applications of Fe-Cr-Ni alloys, their phase diagram has been extensively assessed from the thermodynamic perspective. Microstructure of Fe-Cr-Ni steels is well described by the Schaeffler diagram\cite{Ferry2006}. The phase composition of steels can be controlled by varying Cr and Ni content, since chromium is a ferrite (bcc phase) stabilizer and nickel is an austenite (fcc phase) stabilizer. A thermodynamic model for Fe-Cr-Ni alloys employing CALPHAD method has been developed using interpolation of elevated temperature experimental data\cite{Rees1949,Hattersley1966,Cook1952}. Due to the relatively slow kinetics of relaxation towards equilibrium at low temperatures, the amount of experimental information about the low temperature part of the phase diagram is limited.  This information can instead be derived from {\it ab initio} DFT simulations\cite{Koermann2014}, as was recently demonstrated for binary Fe-Ni alloys in Ref. \onlinecite{Cacciamani2010}. A recent revision of the Fe-Cr-Ni CALPHAD phase diagram is given in Ref. \onlinecite{Franke2011}, where both magnetic and chemical ordering temperatures of binary Fe-Ni alloys were extrapolated to ternary alloys.

There have been only a few DFT studies of Fe-Cr-Ni ternary alloys. Properties of the alloys in the dilute Cr and Ni limit were analyzed in Refs. \onlinecite{Klaver2012,Hepburn2013}. The Coherent Potential Approximation (CPA) was used by the authors of Refs. \onlinecite{Vitos2002,Vitos2006,Delczeg2012}. Recently\cite{Piochaud2014}, Special Quasi-random Structures (SQS) \cite{Zunger1990} were used for investigating point defects in fcc Fe$_{70}$Cr$_{20}$Ni$_{10}$ alloys. In all these studies, Fe-Cr-Ni alloys were assumed to be fully chemically disordered. This assumption is not realistic, since there is direct experimental evidence showing that many Fe-Cr-Ni alloys exhibit short-range order\cite{Dimitrov1986,Cenedese1984,Menshikov1997}. Whilst chemical SRO is naturally expected for ternary alloy compositions close to the known binary intermetallic phases like FeNi$_3$, FeNi and CrNi$_2$, SRO in FeNi$_3$ alloyed with Cr is found to decrease rapidly as a function of Cr content \cite{Marwick1987}. Unexpectedly, a significant degree of chemical order is observed in alloys with compositions very different from that of binary intermetallic phases, for example in Fe$_{56}$Cr$_{21}$Ni$_{23}$ \cite{Cenedese1984}, Fe$_{64}$Cr$_{16}$Ni$_{20}$, Fe$_{59}$Cr$_{16}$Ni$_{25}$ \cite{Dimitrov1986} and Fe$_{34}$Ni$_{46}$Cr$_{20}$ \cite{Menshikov1997}.

Chemical order in alloys, and various properties of ordered alloys, can be analyzed using a combination of first-principles calculations and statistical mechanics simulations based on a generalization of the Ising alloy model. In the CE model, the energy of an alloy is represented by a series in cluster functions, where the resulting expression for the energy has the form of a generalized Ising Hamiltonian containing several coupling parameters known as Effective Cluster Interactions (ECIs)\cite{Sanchez1984}. Various methods have been developed to compute ECIs from first principles. The most often used is the Structure Inversion Method (SIM), based on the Connolly-Williams approximation \cite{Connolly1983}, and the coherent potential approximation used in combination with the Generalized Perturbation Method (CPA-GPM). In the CPA-GPM scheme, a random alloy is constructed by considering average occupancies of lattice sites by atoms of alloy components, where coupling parameters are computed using a perturbation approach \cite{Ruban2008}. In SIM, energies of ordered structures are computed using DFT, and then ECIs are obtained through least-squares fitting. Both techniques have been successfully applied to binary alloy sub-systems of Fe-Cr-Ni \cite{Klaver2006,Nguyen-Manh2007,Nguyen-Manh2012,Lavrentiev2007,Ruban2008,Barabash2009,Ekholm2010,Rahaman2014}. However, ternary Fe-Cr-Ni alloys have not received attention.

In this study we use SIM, since the accuracy of ECIs is primarily controlled by the approximations involved in {\it ab initio} calculations of energies of input structures, and by the cross-validation error between DFT and CE. The last but not least critical issue to consider here is the broad variety of magnetic configurations characterizing fcc and bcc Fe-Cr-Ni alloys. For example, fcc Fe$_{80-x}$Ni$_x$Cr$_{20}$ alloys ($10<x<30$) exhibit ferromagnetic, anti-ferromagnetic, or spin-glass type magnetic order, or a mixture of all of them \cite{Majumdar1984}. To find the most stable atomic structures needed for parameterizing the CE model, many magnetic configurations were computed and their energies compared. Variation of magnetic properties as functions of alloy composition was investigated, including the occurrence of magneto-volume effects in Fe-Cr-Ni alloys.

Effective cluster interaction parameters, obtained by mapping DFT energies of stable collinear magnetic configurations to CE, are used in quasi-canonical MC simulations. Here we investigate the phase stability and chemical order of fcc and bcc Fe-Cr-Ni alloys at finite temperatures and generate representative alloy structures for future DFT analysis of radiation defects in alloys. We also analyze magnetic properties of Fe-Cr-Ni alloys at low and high temperatures using MCE-based Monte Carlo simulations.

The paper is structured as follows. In Section II, we describe the CE formalism for multi-component alloys, focusing on the ternary alloy systems, and derive formulae for short-range order parameters expressed in terms of cluster functions. In Section III we analyze the phase stability and magnetic properties of alloy structures predicted by DFT at 0 K. Finite temperature phase stability and chemical order are investigated using quasi-canonical MC simulations in Section IV. Finite-temperature magnetic properties are explored by MCE simulations in Sections V. Conclusions are given in Section VI.

\section{Computational methodology\protect\\}

\subsection{Cluster expansion formalism for ternary alloys}

The stability of ternary alloy phases can be investigated using a combination of quantum-mechanical DFT calculations and lattice statistical mechanics simulations. The enthalpy of mixing of an alloy, which can be evaluated using DFT, is defined as
\begin{eqnarray}
\Delta H^{lat}_{DFT} (\vec{\sigma})&=&E^{lat}_{tot}(A_{c_B}B_{c_B}C_{c_C},\vec{\sigma})-c_AE^{lat}_{tot}(A) \nonumber \\
&-&c_BE^{lat}_{tot}(B)-c_CE^{lat}_{tot}(C),
\label{eq:Mixing_DFT}
\end{eqnarray}
where $c_A$, $c_B$ and $c_C$ are the average concentrations of alloy components A, B and C. $E^{lat}_{tot}$ are the total energies of relevant structures defined assuming a certain crystal lattice. Superscript $lat$ denotes the chosen lattice type: face-centred cubic (fcc) or body-centred cubic (bcc). An atomic alloy configuration is specified by a vector of configurational variables $\vec{\sigma}$.

In cluster expansion, the configurational enthalpy of mixing of a ternary alloy is defined as \cite{Walle2009}
\begin{equation}
\Delta H_{CE}(\vec{\sigma}) = \sum_{\omega}m_{\omega}J_{\omega}\left\langle \Gamma_{\omega'}(\vec{\sigma})\right\rangle_\omega ,
\label{eq:CE_1}
\end{equation}
where summation is performed over all the clusters $\omega$ that are distinct under group symmetry operations of the underlying lattice, $m^{lat}_\omega$ are multiplicity factors indicating the number of clusters equivalent to $\omega$ by symmetry (divided by the number of lattice sites), $\left\langle \Gamma_{\omega'}(\vec{\sigma})\right\rangle$ are the cluster functions defined as products of \textit{functions} of occupation variables on a specific cluster $\omega$ averaged over all the clusters $\omega'$ that are equivalent by symmetry to cluster $\omega$. $J_{\omega}$ are the concentration-independent Effective Cluster Interaction (ECI) parameters, derived from a set of \textit{ab-initio} calculations using the structure inversion method \cite{Connolly1983}.

A cluster $\omega$ is defined by its size (number of lattice points) $|\omega|$, and the relative positions of points. Coordinates of points in each cluster considered here for fcc and bcc lattices are listed in Table \ref{tab:ECI_def_ternary}. For clarity, each cluster $\omega$ is described by two parameters $(|\omega|,n)$, where $|\omega|$ is the cluster size and $n$ is a label, defined in Table \ref{tab:ECI_def_ternary}.

In binary alloys, lattice site occupation variables are usually defined as $\sigma_i=\pm1$, where $\sigma$ indicates whether site \textit{i} is occupied by an atom of type A ($\sigma_i=+1$) or B ($\sigma_i=-1$). In this case the cluster function is defined as a product of occupation variables over all the sites included in cluster $\omega$
\begin{equation}
\Gamma_{\omega,n}(\vec{\sigma}) = \sigma_1\sigma_2\ldots\sigma_{|\omega|}.
\label{eq:CE_2}
\end{equation}
In a $K$-component system, a cluster function is not a simple product of occupation variables. Instead, it is defined as a product of orthogonal point functions $\gamma_{j_i,K}(\sigma_i)$,
\begin{equation}
\Gamma_{\omega,n}^{(s)}(\vec{\sigma}) = \gamma_{j_1,K}(\sigma_1)\gamma_{j_2,K}(\sigma_2)\ldots\gamma_{j_{|\omega|},K}(\sigma_{|\omega|}),
\label{eq:CE_3}
\end{equation}
where sequence $(s) =(j_{1} j_{2} \ldots\ j_{|\omega|})$ is the {\it decoration} \cite{Sandberg2007} of cluster by point functions. All the decorations of clusters, which are not symmetry-equivalent for fcc and/or bcc ternary alloys, are given in Table \ref{tab:ECI_def_ternary} together with their multiplicities $m_{|\omega|,n}^{(s)}$ and effective cluster interactions $J_{|\omega|,n}^{(s)}$.

The number of possible decorations of clusters by non-zero point functions is a permutation with repetitions, $\left.(K-1)^{|\omega|}\right.$. Effective cluster interactions for those clusters are given in Table \ref{tab:ECI_def_ternary} only once, together with the corresponding multiplicity factor $m_{|\omega|,n}$.   In ternary alloys, occupation variables and point functions can be defined in various ways. For example, in Ref. \onlinecite{Sanchez1984,Wolverton1994} occupation variables are defined as $\sigma_i=-1,0,+1$ and point functions as: $\gamma_{0,3}=1$ (for the zero cluster), $\gamma_{1,3}(\sigma_i)=\sqrt{\frac{3}{2}}\sigma_i$, and $\gamma_{2,3}(\sigma_i)=\sqrt{2}(1-\frac{3}{2}\sigma_i^2)$.

We define occupation variables and point functions following Ref. \onlinecite{Walle2009}. This allows us to apply the same formulae as for a $K$-component system
\begin{equation}
\gamma_{j,K}\left(\sigma_i\right)=\begin{cases}
 1 & \textrm{ if }j=0\textrm{ }, \\
 -\cos\left(2\pi\lceil\frac{j}{2}\rceil\frac{\sigma_i}{K}\right) & \textrm{ if }j>0\textrm{ and odd},  \\
 -\sin\left(2\pi\lceil\frac{j}{2}\rceil\frac{\sigma_i}{K}\right) & \textrm{ if }j>0\textrm{ and even},
\end{cases}
\label{eq:point_functions}
\end{equation}
where $\sigma_i = 0,1,2,\ldots,\left(K-1\right)$, $j$ is the index of point functions ($j=0,1,2,\ldots,(K-1)$), and where $\lceil \frac{j}{2} \rceil$ denotes an operation where we take the integer plus one value of a non-integer number, for example $\lceil 2.5 \rceil=3$. In ternary alloys, index $K$ equals 3. In what follows we will drop it to simplify notations. Occupation variables are now defined as $\sigma=0,1,2$, referring to the constituent components of the alloy \textit{A}, \textit{B} and \textit{C}, which here correspond to Fe, Cr, and Ni, respectively.

The enthalpy of mixing (Eq. \ref{eq:CE_1}) of a ternary alloy on a lattice can now be written as

\begin{widetext}

\begin{eqnarray}
\Delta H_{CE}(\vec{\sigma}) &=& \sum_{|\omega|,n,s}m_{|\omega|,n}^{(s)}J_{|\omega|,n}^{(s)}\left\langle \Gamma_{|\omega'|,n'}^{(s')}(\vec{\sigma})\right\rangle_{|\omega|,n,s} \nonumber \\
 &=& J_{1,1}^{(0)}\left\langle \Gamma_{1,1}^{(0)}\right\rangle+J_{1,1}^{(1)}\left\langle \Gamma_{1,1}^{(1)}\right\rangle + J_{1,1}^{(2)}\left\langle \Gamma_{1,1}^{(2)}\right\rangle+\sum_{n=1}^{pairs} \left(m_{2,n}^{(11)}J_{2,n}^{(11)}\left\langle \Gamma_{2,n}^{(11)}\right\rangle + m_{2,n}^{(12)}J_{2,n}^{(12)}\left\langle \Gamma_{2,n}^{(12)}\right\rangle \right. \nonumber \\
&+& \left. m_{2,n}^{(22)}J_{2,n}^{(22)}\left\langle \Gamma_{2,n}^{(22)}\right\rangle\right) + \sum_{n=1}^{multibody} \ldots
\label{eq:CE_expanded_1}
\end{eqnarray}
Expressions for fcc and bcc alloys differ because of their different multiplicity factors, $m_{|\omega|,n}^{(s)}$, given in Table \ref{tab:ECI_def_ternary}.

\begin{center}
\begin{longtable}{ccccccccc}
\multicolumn{9}{c}{\parbox{12cm}{TABLE I. Size $|\omega|$, label $n$, decoration $(s)$, multiplicity $m_{|\omega|,n}^{(s)}$ and coordinates of points in the relevant clusters on fcc and bcc lattices. $J_{|\omega|,n}^{(s)}$ (in meV) are the effective cluster interaction parameters for fcc and bcc ternary Fe-Cr-Ni alloys. Index $(s)$ is the same as the sequence of points in the relevant cluster.}}
\label{tab:ECI_def_ternary} \\

\hline	
\hline	
 \multicolumn{3}{c}{ } &  \multicolumn{3}{c}{fcc} &  \multicolumn{3}{c}{bcc} \\
    $|\omega|$ & $n$ & ($s$) & Coordinates & $m_{|\omega|,n}^{(s)}$ & $J_{|\omega|,n}^{(s)}$ & Coordinates & $m_{|\omega|,n}^{(s)}$     & $J_{|\omega|,n}^{(s)}$ \\
\hline
\endfirsthead

\multicolumn{9}{c}{ TABLE I. (\textit{Continued})} \\

\hline	
\hline	
 \multicolumn{3}{c}{ } &  \multicolumn{3}{c}{fcc} &  \multicolumn{3}{c}{bcc} \\
    $|\omega|$ & $n$ & ($s$) & Coordinates & $m_{|\omega|,n}^{(s)}$ & $J_{|\omega|,n}^{(s)}$ & Coordinates & $m_{|\omega|,n}^{(s)}$     & $J_{|\omega|,n}^{(s)}$ \\
\hline
\endhead

\hline \hline
\endfoot

\hline \hline
\endlastfoot

    1     & 1     & (0)   & (0,0,0) & 1     & -77.281 & (0,0,0)  & 1     & 132.945 \\
          &       & (1)   &       & 1     & -60.747 &     & 1     & 47.929 \\
          &       & (2)   &       & 1     & 2.847 &       & 1     & -168.929 \\
    2     & 1     & (1,1) & (0,0,0; $\frac{1}{2}$,$\frac{1}{2}$,0) & 6     & 4.329 & (0,0,0; $\frac{1}{2}$,$\frac{1}{2}$,$\frac{1}{2}$) & 4     & -54.656 \\*
          &       & (1,2) &       & 12    & -2.057 &       & 8     & -4.140 \\*
          &       & (2,2) &       & 6     & -2.039 &       & 4     & -64.784 \\
    2     & 2     & (1,1) & (0,0,0; 1,0,0) & 3     & -9.596 & (0,0,0; 1,0,0) & 3     & -19.159 \\*
          &       & (1,2) &       & 6     & 7.284 &       & 6     & 7.332 \\*
          &       & (2,2) &       & 3     & -31.827 &       & 3     & -19.253 \\*
    2     & 3     & (1,1) & (0,0,0; 1,$\frac{1}{2}$,$\frac{1}{2}$) & 12    & 3.345 & (0,0,0; 1,0,1) & 6     & -1.547 \\*
          &       & (1,2) &       & 24    & -0.702 &       & 12    & 11.871 \\*
          &       & (2,2) &       & 12    & 4.224 &       & 6     & 8.392 \\
    2     & 4     & (1,1) & (0,0,0; 1,1,0) & 6     & -1.990 & (0,0,0; 1$\frac{1}{2}$,$\frac{1}{2}$,$\frac{1}{2}$) & 12    & 2.466 \\*
          &       & (1,2) &       & 12    & 1.192 &       & 24    & 0.564 \\*
          &       & (2,2) &       & 6     & 6.662 &       & 12    & -2.660 \\
    2     & 5     & (1,1) & (0,0,0; 1$\frac{1}{2}$,$\frac{1}{2}$,$\frac{1}{2}$) & 6     & -2.034 & (0,0,0; 1,1,1) & 4     & 1.602 \\*
          &       & (1,2) &       & 12    & 0.724 &       & 8     & -1.368 \\*
          &       & (2,2) &       & 6     & 2.036 &       & 4     & 3.031 \\
    3     & 1     & (1,1,1) & (0,0,0; $\frac{1}{2}$,0,$\frac{1}{2}$; & 8     & -9.015 & (1,0,0; $\frac{1}{2}$,$\frac{1}{2}$,$\frac{1}{2}$; & 12    & -6.961 \\*
          &       & (2,1,1) & 0,$\frac{1}{2}$,$\frac{1}{2}$) & 24    & 3.847 & 0,0,0) & 24    & 8.827 \\*
          &       & (1,2,1) &       &       &       &       & 12    & 1.620 \\*
          &       & (2,2,1) &       & 24    & -6.544 &       & 24    & -1.954 \\*
          &       & (2,1,2) &       &       &       &       & 12    & 22.895 \\*
          &       & (2,2,2) &       & 8     & 12.492 &       & 12    & 2.934 \\
    3     & 2     & (1,1,1) & (1,0,0; $\frac{1}{2}$,-$\frac{1}{2}$,0; & 12    & -3.019 & ($\frac{1}{2}$,-$\frac{1}{2}$,-$\frac{1}{2}$; 0,0,0; & 12    & -6.255 \\*
          &       & (2,1,1) & 0,0,0) & 24    & -0.470 &  -$\frac{1}{2}$,-$\frac{1}{2}$,$\frac{1}{2}$) & 24    & 2.510 \\*
          &       & (1,2,1) &       & 12    & -1.778 &       & 12    & -1.292 \\*
          &       & (2,2,1) &       & 24    & 5.371 &       & 24    & 6.122 \\*
          &       & (2,1,2) &       & 12    & 6.310 &       & 12    & 6.580 \\*
          &       & (2,2,2) &       & 12    & -0.126 &       & 12    & 4.334 \\
    3     & 3     & (1,1,1) & ($\frac{1}{2}$,$\frac{1}{2}$,0; 0,0,0; & 24    & 0.821 &       &       &  \\*
          &       & (2,1,1) &  -$\frac{1}{2}$,0,$\frac{1}{2}$) & 48    & -0.017 &       &       &  \\*
          &       & (1,2,1) &       & 24    & 0.931 &       &       &  \\*
          &       & (2,2,1) &       & 48    & 0.369 &       &       &  \\*
          &       & (2,1,2) &       & 24    & 2.657 &       &       &  \\*
          &       & (2,2,2) &       & 24    & -3.945 &       &       &  \\
    4     & 1     & (1,1,1,1) & (0,0,0; $\frac{1}{2}$,$\frac{1}{2}$,0; & 2     & -12.978 & (1,0,0; $\frac{1}{2}$,-$\frac{1}{2}$,$\frac{1}{2}$; & 6     & -12.095 \\*
          &       & (2,1,1,1) & $\frac{1}{2}$,0,$\frac{1}{2}$; 0,$\frac{1}{2}$,$\frac{1}{2}$) & 8     & -1.931 & $\frac{1}{2}$,$\frac{1}{2}$,$\frac{1}{2}$; 0,0,0) & 24    & -13.020 \\*
          &       & (2,2,1,1) &       & 12    & 4.987 &       & 24    & 0.000 \\*
          &       & (1,2,2,1) &       &       &       &       & 12    & 0.000 \\*
          &       & (2,2,2,1) &       & 8     & -1.140 &       & 24    & 0.000 \\*
          &       & (2,2,2,2) &       & 2     & 0.824 &       & 6     & 0.007 \\
    4     & 2     & (1,1,1,1) & (1,0,0; $\frac{1}{2}$,0,$\frac{1}{2}$; & 12    & -1.452 &       &       &  \\*
          &       & (2,1,1,1) & $\frac{1}{2}$,-$\frac{1}{2}$,0; 0,0,0) & 24    & 1.076 &       &       &  \\*
          &       & (1,2,1,1) &       & 24    & -1.775 &       &       &  \\*
          &       & (2,2,1,1) &       & 48    & 1.114 &       &       &  \\*
          &       & (1,2,2,1) &       & 12    & -0.581 &       &       &  \\*
          &       & (2,2,2,1) &       & 24    & -5.109 &       &       &  \\*
          &       & (2,1,1,2) &       & 12    & 4.130 &       &       &  \\*
          &       & (2,2,1,2) &       & 24    & 2.549 &       &       &  \\*
          &       & (2,2,2,2) &       & 12    & 6.127 &       &       &  \\
    5     & 1     & (1,1,1,1,1) & (1,0,0; $\frac{1}{2}$,0,-$\frac{1}{2}$; & 6     & 4.219 & (1,0,0; $\frac{1}{2}$,-$\frac{1}{2}$,$\frac{1}{2}$; & 12    & -6.356 \\*
          &       & (2,1,1,1,1) & $\frac{1}{2}$,0,$\frac{1}{2}$; $\frac{1}{2}$,-$\frac{1}{2}$,0; & 24    & -1.263 & $\frac{1}{2}$,$\frac{1}{2}$,$\frac{1}{2}$; 0,0,0; & 24    & 7.696 \\*
          &       & (1,2,1,1,1) & 0,0,0) &       &       & 0,0,1) & 24    & -15.998 \\*
          &       & (2,2,1,1,1) &       & 24    & 0.626 &       & 48    & 15.385 \\*
          &       & (1,2,2,1,1) &       & 12    & 1.676 &       & 12    & -20.341 \\*
          &       & (2,2,2,1,1) &       & 24    & -0.360 &       & 24    & 14.846 \\*
          &       & (1,1,1,2,1) &       & 6     & -6.115 &       & 12    & -5.003 \\*
          &       & (2,1,1,2,1) &       & 24    & -1.565 &       & 24    & -3.067 \\*
          &       & (1,2,1,2,1) &       &       &       &       & 24    & -3.221 \\*
          &       & (2,2,1,2,1) &       & 24    & 3.258 &       & 48    & 1.070 \\*
          &       & (1,2,2,2,1) &       & 12    & 2.284 &       & 12    & -3.473 \\*
          &       & (2,2,2,2,1) &       & 24    & -1.400 &       & 24    & -1.255 \\*
          &       & (2,1,1,1,2) &       &       &       &       & 12    & 11.683 \\*
          &       & (2,2,1,1,2) &       &       &       &       & 24    & -1.192 \\*
          &       & (2,2,2,1,2) &       & 6     & 1.565 &       & 12    & -2.460 \\*
          &       & (2,1,1,2,2) &       &       &       &       & 12    & -6.855 \\*
          &       & (2,2,1,2,2) &       &       &       &       & 24    & -7.050 \\*
          &       & (2,2,2,2,2) &       & 6     & -6.793 &       & 12    & -5.397 \\*
\end{longtable}%
\end{center}

\end{widetext}

Configuration averages $\left\langle \Gamma_{|\omega|,n}^{(s)}(\vec{\sigma})\right\rangle$ in  Eq. \ref{eq:CE_expanded_1} can be expressed in terms of point, pair and multi-body probabilities. An average point correlation function can be calculated using the equation
\begin{equation}
\left\langle \Gamma_{1,1}^{(s)}\right\rangle=\langle \gamma_j\rangle=\sum_{k=1}^3T_{jk}\times\langle p^{(k)}\rangle=\sum_{k=1}^3T_{jk}c_k ,
\label{eq:point_corr_function}
\end{equation}
where $k=0,1,2$, $p^{(k)}$ are the site-occupation operators counting the number of sites occupied by the same atom type \cite{Ducastelle1991}. Average values of site-occupation operators $\left\langle p^{(k)}\right\rangle=c_k$ are concentrations $c_A$, $c_B$ and $c_C$, and $T_{ij}$ are elements of the point probability matrix given, through Eq. \ref{eq:point_functions}, by
\begin{equation}
 \left[\begin{array}{c}
 \left\langle \gamma_0 \right\rangle \\
 \left\langle \gamma_1 \right\rangle \\
 \left\langle \gamma_2 \right\rangle \\
 \end{array} \right]=
 \left[ \begin{array}{ccc}
 1 & 1 & 1 \\
 -1 & \frac{1}{2} & \frac{1}{2} \\
 0 & -\frac{\sqrt{3}}{2} & \frac{\sqrt{3}}{2} \\
 \end{array} \right]
\left[\begin{array}{c}
 \left\langle p^{(0)} \right\rangle \\
 \left\langle p^{(1)} \right\rangle \\
 \left\langle p^{(2)} \right\rangle \\
 \end{array} \right].
\label{eq:CE_M_Matrix}
\end{equation}

The three average point functions are therefore
\begin{eqnarray}
\left\langle \Gamma_{1,1}^{(0)}\right\rangle&=&\langle\gamma_0\rangle = \sum_ic_i\gamma_0(\sigma_i)= 1 \nonumber \\
\left\langle \Gamma_{1,1}^{(1)}\right\rangle&=&\langle\gamma_1\rangle =\sum_ic_i\gamma_1(\sigma_i)=\frac{1}{2}\left(-2c_A+c_B+c_C\right)=\frac{1}{2}\left(1-3c_A\right) \nonumber \\
\left\langle \Gamma_{1,1}^{(2)}\right\rangle&=&\langle\gamma_2\rangle = \sum_ic_i\gamma_2(\sigma_i)=\frac{\sqrt{3}}{2}\left(c_C-c_B\right) .
\label{eq:Gamma_point_clust}
\end{eqnarray}

Similarly to Eq. \ref{eq:point_corr_function}, the average cluster functions for pairwise clusters ($n$-th nearest neighbours) are linear functions of the average pairwise probabilities. They are given by
\begin{eqnarray}
\left\langle \Gamma_{2,n}^{(ij)}\right\rangle&=&\langle \gamma_i,\gamma_j\rangle_{n}=\sum_{h=1}^3\sum_{k=1}^3T_{ih}T_{jk}\times\langle p^{(h)}p^{(k)}\rangle_{n}
\nonumber \\
&=&\sum_{h=1}^3\sum_{k=1}^3\gamma_i(\sigma_h)\gamma_j(\sigma_k)y_n^{hk},
\label{eq:pair_cluster_funct}
\end{eqnarray}
where $T_{ih}$ and $T_{jk}$ are elements of the point probability matrix (Eq. \ref{eq:CE_M_Matrix}), and $y_n^{hk}$ is the temperature-dependent probability of finding atom $h$ near atom $k$ in the $n$-th nearest neighbour coordination shell, given by \cite{Ducastelle1991}
\begin{equation}
y_n^{hk}=\langle p^{(h)}p^{(k)}\rangle_{n}=\langle p^{(h)}\rangle\langle p^{(k)}\rangle\left(1-\alpha_n^{hk}\right)=c_hc_k\left(1-\alpha_n^{hk}\right).
\label{eq:pair_probability}
\end{equation}
Here $\alpha_n^{hk}$ is the Warren-Cowley short-range parameter for atoms $h$ and $k$ in the $n$-th neighbour shell, defined as the deviation from entirely random distribution of atoms in the alloy. Average cluster functions for the three pairs of non-equivalent atoms are therefore
\begin{eqnarray}
\left\langle \Gamma_{2,n}^{(11)}\right\rangle&=&\langle\gamma_1,\gamma_1\rangle_{n}=
\nonumber \\
&=&\frac{1}{4}\left(1+3y_n^{AA}-6y_n^{AB}-6y_n^{AC}\right) \nonumber \\
\left\langle \Gamma_{2,n}^{(12)}\right\rangle&=&\langle\gamma_1,\gamma_2\rangle_{n}=
\nonumber \\
&=&\frac{\sqrt{3}}{4}\left(-y_n^{BB}+y_n^{CC}+2y_n^{AB}-2y_n^{AC}\right) \nonumber \\
\left\langle\Gamma_{2,n}^{(22)}\right\rangle&=&\langle\gamma_2,\gamma_2\rangle_{n}=\frac{3}{4}\left(y_n^{BB}+y_n^{CC}-2y_n^{BC}\right).
\label{eq:Gamma_pairs}
\end{eqnarray}

Rewriting Eq. \ref{eq:CE_expanded_1} in terms of average point and pair functions given by Eqs. \ref{eq:Gamma_point_clust} and \ref{eq:Gamma_pairs}, we find that the configurational enthalpy of mixing for a ternary alloy can be expressed as a function of concentrations $c_i$ and average pair probabilities $y_n^{ij}$ via
\begin{widetext}
\begin{eqnarray}
\Delta H_{CE}(\vec{\sigma}) &=& J_1^{(0)}+J_1^{(1)}\left(1-3c_A\right) + J_1^{(2)}\frac{\sqrt{3}}{2}\left(c_C-c_B\right) \nonumber \\
&+&\sum_{n}^{pairs} \left[\frac{1}{4}m_{2,n}^{(11)}J_{2,n}^{(11)}\left(1+3y_n^{AA}-6y_n^{AB}-6y_n^{AC}\right) + \frac{\sqrt{3}}{4}m_{2,n}^{(12)}J_{2,n}^{(12)}\left(-y_n^{BB}+y_n^{CC}+2y_n^{AB}-2y_n^{AC}\right) \right. \nonumber \\
&+& \left.\frac{3}{4}m_{2,n}^{(22)}J_{2,n}^{(22)}\left(y_n^{BB}+y_n^{CC}-2y_n^{BC}\right) \right] + \sum_{n}^{multibody} \ldots
\label{eq:CE_expanded_2}
\end{eqnarray}
\end{widetext}
Detailed expressions, with analytic formulae, for the average cluster functions of 3-body clusters as well as for the enthalpy of mixing represented as a function of average triple probabilities $y_n^{ijk}$, are given in Appendix A.

\subsection{Chemical short-range order parameters}

Short-range order in ternary alloys can be investigated by analyzing chemical pairwise interactions between unlike atoms. These pairwise interactions are related to $J_{|\omega|,n}^{(s)}$, where $|\omega|=2$ and $J_{|\omega|,n}^{(s)}$ are given by an inner product of the cluster function $\Gamma_{2,n}^{(s)}$ and the corresponding energy \cite{Asta1991,Wolverton1994}, namely
\begin{equation}
J_{2,n}^{(s)}=\langle \Gamma_{2,n}^{(s)}(\vec{\sigma}), E(\vec{\sigma}) \rangle= \rho_0^{(s)}\sum_{\left\{\vec{\sigma}\right\}}\Gamma_{2,n}^{(s)}(\vec{\sigma})E(\vec{\sigma}).
\label{eq:ECI_def}
\end{equation}
Summation in the above equation is performed over all possible configurations and $\rho_0^{(s)}$ is a normalization constant chosen to satisfy the orthonormality criterion for cluster functions $\Gamma_{2,n}^{(s)}$. Effective cluster interactions in ternary alloys for pairs of non-zero point functions with indices (11),(12),(21) and (22) can now be written as
\begin{equation}
J_{2,n}^{(ij)}=\frac{4}{9}\sum_{h,k} E_n^{hk}\gamma_i(\sigma_h)\gamma_j(\sigma_k).
\label{eq:ECI}
\end{equation}
where $E_n^{hk}$ is the average energy of configurations with atom $h$ being in the $n$-th nearest neighbour shell of atom $k$. From Eq. \ref{eq:ECI}, ECI for pairs with indices (11),(12),(21) and (22) are
\begin{eqnarray}
J_{2,n}^{(11)}&=&\frac{1}{9}\left(4E_n^{AA}+E_n^{BB}+E_n^{CC}-2E_n^{AB}-2E_n^{BA}\right.
\nonumber \\
&-&\left.2E_n^{AC}-2E_n^{CA}+E_n^{BC}+E_n^{CB}\right),
\nonumber \\
J_{2,n}^{(12)}&=&\frac{1}{2}\left(J_{2,n}^{(12)}+J_{2,n}^{(21)}\right)=\frac{\sqrt{3}}{9}\left(-E_n^{BB}+E_n^{CC}\right.
\nonumber \\
&+&\left.E_n^{AB}+E_n^{BA}-E_n^{AC}-E_n^{CA}\right) \nonumber \\
J_{2,n}^{(22)}&=&\frac{1}{3}\left(E_n^{BB}+E_n^{CC}-E_n^{BC}-E_n^{CB}\right).
\label{eq:ECI_pairs}
\end{eqnarray}

A chemical pairwise interaction between atoms $i$ and $j$ in the $n$-th neighbour shell in a ternary alloy is defined as the effective cluster interaction between pairwise clusters in binary alloys \cite{Asta1991,Wolverton1994}
\begin{equation}
V_n^{ij}=\frac{1}{4}\left(E_n^{ii}+E_n^{jj}-E_n^{ij}-E_n^{ji}\right),
\label{eq:ECI_binary}
\end{equation}
where energies $E_n^{ii}$, $E_n^{jj}$, $E_n^{ij}$ and $E_n^{ji}$ are averaged over all the ternary alloy configurations. From Eqs. \ref{eq:ECI_pairs} and \ref{eq:ECI_binary}, a relation between chemical pairwise interactions involving unlike atoms, and effective cluster interactions of pairwise clusters in a ternary alloy, can be written in matrix form as
\begin{equation}
 \left[\begin{array}{c}
 V_n^{AB} \\
 V_n^{AC} \\
 V_n^{BC} \\
 \end{array} \right]=
 \left[ \begin{array}{ccc}
 \frac{9}{16} & \frac{-3\sqrt{3}}{8} & \frac{3}{16} \\
 \frac{9}{16} & \frac{3\sqrt{3}}{8} & \frac{3}{16} \\
 0 & 0 & \frac{3}{4} \\
 \end{array} \right]
\left[\begin{array}{c}
 J_{2,n}^{(11)} \\
 J_{2,n}^{(12)} \\
 J_{2,n}^{(22)} \\
 \end{array} \right].
\label{eq:VvsJ_Matrix}
\end{equation}

As for the binary alloy case, chemical pairwise interactions $V_n^{ij}$ have a simple meaning: $V_n^{ij}>0$ corresponds to attraction and $V_n^{ij}<0$ to repulsion between atoms $i$ and $j$. These interactions will be used in the analysis of SRO in Fe-Cr-Ni ternary alloys in Section IV.C. With Eq. \ref{eq:CE_expanded_2} expressed in terms of chemical pairwise interactions, the configurational enthalpy of mixing of a ternary alloy is given by
\begin{widetext}
\begin{eqnarray}
\Delta H_{CE}(\vec{\sigma}) &=& J_1^{(0)}+J_1^{(1)}\left(1-3c_A\right) + J_1^{(2)}\frac{\sqrt{3}}{2}\left(c_C-c_B\right) \nonumber \\
&-&4\sum_{n}^{pairs} \left(V_n^{AB}y_n^{AB} + V_n^{AC}y_n^{AC}+V_n^{BC}y_n^{BC}\right) + \sum_{n}^{multibody} \ldots ,
\label{eq:CE_vs_V}
\end{eqnarray}
\end{widetext}
SRO involving atoms $i$ and $j$ in the $n$-th nearest neighbour shell in either binary or ternary alloys can be described using the Warren-Cowley parameters $\alpha_n^{ij}$
\begin{equation}
\alpha_n^{ij}=1-\frac{\left\langle p^{(i)},p^{(j)} \right\rangle_{n} }{\left\langle p^{(i)}\right\rangle \left\langle p^{(j)}\right\rangle} =1-\frac{y_n^{ij}}{c_ic_j}=1-\frac{P_n^{i-j}}{c_j}.
\label{eq:SRO_definition}
\end{equation}
Here $n$ is a coordination sphere index, $c_i$ and $c_j$ are the concentrations of $i$'s and $j$'s atoms, and $P_n^{i-j}=y_n^{ij}/c_i$ is the conditional probability of finding atom $i$ in the $n$-th coordination sphere of atom $j$, see for example Ref. \onlinecite{DeFontaine1971}. As in the binary alloy case, $\alpha_n^{ij}$ vanishes if $P_n^{i-j}=c_j$, meaning that there is no (positive or negative) preference for a given atom to be surrounded by atoms of any other type. Segregation gives rise to positive $\alpha_n^{ij}$, whereas a negative value of $\alpha_n^{ij}$ indicates ordering. If at low concentration of atoms $j$, each atom $j$ is surrounded only by atoms $i$, i.e. ($P_n^{i-j}=1$), then $\alpha_n^{ij}$ acquires the lowest possible value $\alpha_{n,min}^{ij}=-(1-c_j)/c_j$.

SRO parameters can be expressed in terms of average point and pair correlation functions. Inverting Eqs. \ref{eq:CE_M_Matrix}, \ref{eq:pair_cluster_funct} and \ref{eq:SRO_definition}, analytical formulae for SRO parameters in a ternary alloy become

\begin{widetext}
\begin{eqnarray}
\alpha_n^{AB}&=&1-\frac{2-2\langle\gamma_1\rangle-2\sqrt{3}\langle\gamma_2\rangle-4\langle\gamma_1,\gamma_1\rangle_n+4\sqrt{3}\langle\gamma_1,\gamma_2\rangle_n}{2(1-2\langle\gamma_1\rangle)(1+\langle\gamma_1\rangle-\sqrt{3}\langle\gamma_2\rangle)} \nonumber \\
\alpha_n^{BC}&=&1-\frac{2+4\langle\gamma_1\rangle+2\langle\gamma_1,\gamma_1\rangle_n -6\langle\gamma_2,\gamma_2\rangle_n}{2(1+\langle\gamma_1\rangle-\sqrt{3}\langle\gamma_2\rangle)(1+\langle\gamma_1\rangle+\sqrt{3}\langle\gamma_2\rangle)} \nonumber \\
\alpha_n^{AC}&=&1-\frac{2-2\langle\gamma_1\rangle+2\sqrt{3}\langle\gamma_2\rangle-4\langle\gamma_1,\gamma_1\rangle_n - 4\sqrt{3}\langle\gamma_1,\gamma_2\rangle_n} {2(1-2\langle\gamma_1\rangle)(1+\langle\gamma_1\rangle+\sqrt{3}\langle\gamma_2\rangle)}.
\label{eq:SRO_pairs}
\end{eqnarray}
\end{widetext}

Since both point and pair correlation functions are generated by the ATAT package\cite{Walle2002} used in the present study, the SRO parameters of ternary alloys are going to be calculated using Eq. \ref{eq:SRO_pairs}.

\subsection{Magnetic Cluster Expansion}

Magnetic Cluster Expansion has been successfully applied to a number of binary systems, including bcc and fcc Fe-Cr \cite{Lavrentiev2010,Lavrentiev2011a} and fcc Fe-Ni \cite{Lavrentiev2014}. In MCE \cite{Lavrentiev2009,Lavrentiev2011}, each alloy configuration is defined by its chemical ($\sigma_i$) {\it and} magnetic ($\mathbf{M}_i$) degrees of freedom. MCE parameters are derived from DFT data on 30 ordered ternary Fe-Cr-Ni structures (see Supplementary Material), spanning the entire alloy composition range, together with DFT data on pure elements. Parametrization also used 29 binary fcc Fe-Ni configurations analysed in a recent application of MCE to fcc Fe-Ni alloys \cite{Lavrentiev2014}. We note that deriving exchange coupling parameters for non-collinear Hamiltonians from collinear \textit{ab initio} calculations is a known approach that provided a number of significant results for a broad variety of magnetic systems. This includes recent studies of MnSi by Hortamani {\it et al.} \cite{Hortamani2009}, and Fe$_{65}$Ni$_{35}$ by Liot and Abrikosov \cite{Liot2009}. Our own work on Fe and Fe/Cr interfaces \cite{Lavrentiev2010,Lavrentiev2011a}, which followed the same approach,  agrees well with experiment and non-collinear {\it ab initio} calculations, thus further validating the above approach to the parametrization of Magnetic Cluster Expansion. To simplify applications of MCE to ternary alloys and reduce the number of fitting parameters, we use an MCE Hamiltonian that includes only pairwise interactions. In this approximation, the energy of an arbitrary alloy configuration $\left(\left\{\sigma_i\right\}, \left\{\mathbf{M}_i\right\} \right)$ is written in the Heisenberg-Landau form as
\begin{widetext}
\begin{eqnarray}
H_{MCE}\left(\left\{\sigma_i\right\},\left\{\mathbf{M}_i\right\}\right)&=& \sum_i\mathcal{I}_{\sigma_i}^{(1)}+\sum_{ij\in 1NN}\mathcal{I}_{\sigma_i\sigma_j}^{(1NN)}+\sum_{ij\in 2NN}\mathcal{I}_{\sigma_i\sigma_j}^{(2NN)}+\ldots \nonumber \\
&+&\sum_iA_{\sigma_i}\mathbf{M}_i^2+\sum_iB_{\sigma_i}\mathbf{M}_i^4+\ldots \nonumber \\
&+&\sum_{ij\in 1NN}\mathcal{J}_{\sigma_i\sigma_j}^{(1NN)}\mathbf{M}_i\cdot \mathbf{M}_j+\sum_{ij\in 2NN}\mathcal{J}_{\sigma_i\sigma_j}^{(2NN)}\mathbf{M}_i\cdot \mathbf{M}_j+\ldots,
\label{eq:MCE_Hamiltonian}
\end{eqnarray}
\end{widetext}
where $\sigma_i, \sigma_j =$ Fe, Cr and Ni, and the non-magnetic and Heisenberg magnetic interaction parameters $\mathcal{I}_{ij}$ and $\mathcal{J}_{ij}$ for each coordination shell are represented by 3$\times$3 matrices. We take into account interactions that extend up to the fourth nearest neighbour coordination shell. Together, there are 24 independent non-magnetic and 24 independent magnetic interaction parameters. At the first stage of fitting, the on-site magnetic terms $A$, $B$, $C$,... were fitted using the energy versus magnetic moment curves computed for pure ferromagnetic Fe, Ni, and Cr. For chromium, only quadratic and quartic Landau expansion terms were used, while for iron and nickel the Landau expansion was extended to the 8$^{th}$-order in magnetic moment\cite{Lavrentiev2014}. The dependence of the on-site terms on atomic environment was neglected in order to reduce the number of parameters in the Hamiltonian. Following the methodology described in Ref. \cite{Lavrentiev2014}, the interaction terms $\mathcal{I}$ and $\mathcal{J}$ were fitted to DFT data on total energies and magnetic moments on each site in the simulation cell. Most of the alloy structures used for parameterizing the Fe-Cr-Ni MCE Hamiltonian (Eq. \ref{eq:MCE_Hamiltonian}) belong to the Fe-rich corner of the ternary alloy composition triangle. Hence we expect that MCE predictions are going to be most reliable for alloys where Fe content exceeds 50 at.\%.

\subsection{Computational details}

DFT calculations were performed using the Projector Augmented Wave (PAW) method implemented in VASP\cite{Kresse1996, Kresse1996a}. Exchange and correlation were treated in the generalized gradient approximation GGA-PBE \cite{Perdew1996}. To accelerate DFT calculations, we used PAW potentials without semi-core $p$ electron contribution. The core configurations of Fe, Cr and Ni in PAW potentials were [Ar]3d$^7$4s$^1$, [Ar]3d$^5$4s$^1$ and [Ar]3d$^9$4s$^1$, respectively.

Total energies were calculated using the Monkhorst-Pack mesh\cite{Monkhorst1976} of $k$-points in the Brillouin zone, with $k$-mesh spacing of 0.2 $\AA^{-1}$. This corresponds to 14$\times$14$\times$14 or 12$\times$12$\times$12 $k$-point meshes for a two-atom bcc cubic cell or a four-atom fcc cubic cell, respectively. The plane wave cut-off energy used in the calculations was 400 eV. The total energy convergence criterion was set to 10$^{-6}$ eV/cell, and force components were relaxed to 10$^{-3}$ eV/$\AA$.

Mapping DFT energies to CE was performed using the ATAT package\cite{Walle2002}. In order to find CE parameters for binary fcc alloys we used a database of 28 structures from Table I of Ref. \onlinecite{Barabash2006}.  For binary bcc alloys we used the 58 structures from Table I of Ref. \onlinecite{Nguyen-Manh2007}. For ternary fcc alloys we used the 98 structures from Fig. 2 of Ref. \onlinecite{Garbulsky1994}. To our knowledge, there is no database of structures of ternary bcc alloys available at present. We constructed the input ternary bcc structures using binary structures of Ref. \onlinecite{Nguyen-Manh2007} as a starting point. The symmetry and the number of non-equivalent positions (NEPs) in each structure was checked, and structures for which the number of NEPs was greater than two were included in the ternary bcc  structure database. The resulting input database for bcc ternary alloys consists of 94 structures. These structures are described in detail in Appendix B.

Most of the collinear spin-polarized DFT calculations were performed assuming that the initial magnetic moments of Fe, Cr and Ni atoms were +3, -1 and +1 $\mu_B$, respectively. Since magnetic properties of Fe-Cr-Ni alloys are very complex in comparison with binary alloys, full relaxations starting from various initial magnetic configurations were performed in order to find the most stable magnetic order characterizing a given structure. Such an investigation was especially critical for fcc Fe-rich structures, where the energies of competing magnetic configurations are very close.

Initial values of ECIs, derived by mapping to CE the DFT energies computed for the most stable magnetic configurations of input structures, provide a starting point for further refinement of CE parameters, which is performed by generating new structures. The complexity of magnetic properties of Fe-Cr-Ni alloys made it impossible to perform this refinement fully automatically, as is possible in the case of non-magnetic alloys. For example, the above choice of initial values of magnetic moments did not always lead to the most stable magnetic configurations. Hence results had to be filtered following an approach proposed in Ref. \onlinecite{Barabash2009}. For Fe-Cr-Ni alloys this meant that some of the structures had to be recalculated assuming an alternative initial magnetic configuration or, in a few extreme cases, the less stable structures were eliminated if their energies proved difficult to fit to a consistent set of ECIs.

Despite the fact that performing fully automatic refinement of CE parameters was not possible, reasonable values of cross-validation error between DFT and CE formation enthalpies were achieved, proving that the final set of ECI describes interatomic interactions in Fe-Cr-Ni system fairly well. A detailed description of ECIs, the number of structures used in the fitting, and the cross-validation error between DFT and CE data is given in Section III.

Quasi-canonical MC simulations were performed using the ATAT package \cite{Walle2002}. Most of the simulations were performed using a cell containing 8000 atoms in the form of 20$\times$20$\times$20 {\it primitive} fcc or bcc unit cells. For each composition, simulations were performed starting from a disordered high-temperature state (usually $T$ = 2500 K). The alloy was then cooled down with the temperature step of $\Delta T$ = 100K, with 5000 MC steps per atom at both thermalization and accumulation stages. Test simulations were also performed with 2000 MC steps at each of these stages. Since the results were not significantly different, there was no need to test with more than 5000 MC steps.

A database of enthalpies of mixing and magnetic moments of ternary fcc Fe-Cr-Ni structures derived from DFT and used for fitting the MCE Hamiltonian (see Section V) is given in Supplementary Material.

\section{Phase stability and magnetic properties at 0 K}

\subsection{Pure Elements}

\begin{table*}
\caption{Volume per atom $V$, energy with respect to the energy of the ground state, $E-E_{GS}$, and magnetic moment per atom $|m_{tot}|$, computed for various structures of pure elements, compared to available experimental data.
         \label{tab:Fe}}
\begin{ruledtabular}
    \begin{tabular}{cccccc}
    Struct. Name & $V$ (\AA$^3$/atom) & $V^{Expt.}$ (\AA$^3$/atom) & $E-E_{GS}$ (eV) & $|m_{tot}|$ ($\mu_B$) & $|m_{tot}^{Expt.}|$ ($\mu_B$) \\
\hline
    bcc-Fe (FM) -GS & 11.35 & 11.70\cite{Acet1994} & 0.000 & 2.199 & 2.22\cite{Crangle1963} \\
    bcc-Fe (NM) & 10.46 & & 0.475 & 0.000 & \\
    bcc-Fe (AFMSL) & 10.87 & & 0.444 & 1.290 & \\
    bcc-Fe (AFMDL) & 11.34 & & 0.163 & 2.104 & \\
    bcc-Fe (AFMTL) & 11.35 & & 0.112 & 4$\times$2.087; & \\
		 &  &  & & 2$\times$2.351 & \\
    fcc-Fe (NM) & 10.22 & & 0.167 & 0.000 & \\
    fcc-Fe (FM-HS) & 11.97 & 12.12\cite{Acet1994} & 0.153 & 2.572 & \\
    fcc-Fe (FM-LS) & 10.52 & & 0.162 & 1.033 & \\
    fcc-Fe (AFMSL) & 10.76 & 11.37\cite{Acet1994} & 0.100 & 1.574 & 0.75\cite{Abrahams1962} \\
    fcc-Fe (AFMDL) & 11.20 & & 0.082 & 2.062 & \\
    fcc-Fe (AFMTL) & 11.45 & & 0.082 & 8$\times$2.155; & \\
     &  &  & & 4$\times$2.429 & \\
    \\
\hline
    bcc-Cr (AFMSL) -GS & 11.63 & 11.94\cite{Kittel1971}& 0.000 & 1.070 & \\
    bcc-Cr (NM) & 11.41 & & 0.011 & 0.000 & \\
    fcc-Cr (NM) & 11.75 & & 0.405 & 0.000 & \\
\\
\hline
    fcc-Ni (FM) -GS & 10.91 & 10.90 \cite{Kittel1971} & 0.000 & 0.641 & 0.60\cite{Crangle1963}\\
    fcc-Ni (NM) & 10.84 & & 0.056 & 0.000 & \\
    bcc-Ni (FM) & 11.00 & & 0.092 & 0.569 & \\
    bcc-Ni (NM) & 10.90 & & 0.107 & 0.000 & \\

    \end{tabular}%
\end{ruledtabular}
 \end{table*}

Magnetism of Fe-Cr-Ni alloys gives rise to several structural and magnetic instabilities. This effect is well known in pure iron. \textit{Ab initio} analysis of structural and magnetic phase stability of iron was performed in Refs. \onlinecite{Herper1999,Moruzzi1989}. Our calculations confirm that the most stable Fe phase at 0 K is the ferromagnetic (FM) bcc phase. Anti-ferromagnetic single layer (AFMSL) and anti-ferromagnetic double layer (AFMDL) fcc structures are more stable than the high-spin (HS) and low-spin (LS) ferromagnetic configurations. We have extended analysis of anti-ferromagnetism in iron to anti-ferromagnetic triple layer (AFMTL) fcc and bcc structures. We have found that fcc-Fe AFMTL has the same energy per atom as fcc-Fe AFMDL but they have significantly different volumes, see Table \ref{tab:Fe}. The bcc Fe AFMTL structure of iron is more stable than bcc Fe AFMSL and bcc Fe AFMDL, but it is still less stable than bcc Fe FM.

DFT calculations confirm that the most stable collinear magnetic Cr and Ni phases at 0 K are anti-ferromagnetic bcc and ferromagnetic fcc. Ferromagnetic bcc Ni and non-magnetic fcc Cr are 0.096 eV/atom and 0.405 eV/atom less stable than fcc Ni and bcc Cr, respectively.

Since the ground states of Fe, Cr and Ni belong to different crystal lattices, the phase stability of Fe-Cr-Ni alloys and binary sub-systems is analyzed in terms of their enthalpies of formation, defined as the energy of the alloy, calculated at zero pressure, with respect to the energies of ferromagnetic bcc-Fe, ferromagnetic fcc-Ni, and anti-ferromagnetic bcc-Cr. To investigate properties of alloys on fcc and bcc crystal lattices, stabilities of fcc and bcc alloys have also been analyzed in terms of their enthalpy of mixing, defined as the energy of an alloy with respect to the energies of fcc or bcc structures of pure elements, where the choice of bcc or fcc depends on the choice of the crystal structure of the alloy under consideration.

\subsection{Fe-Ni binary alloys}

\begin{figure*}
  \includegraphics[width=\linewidth]{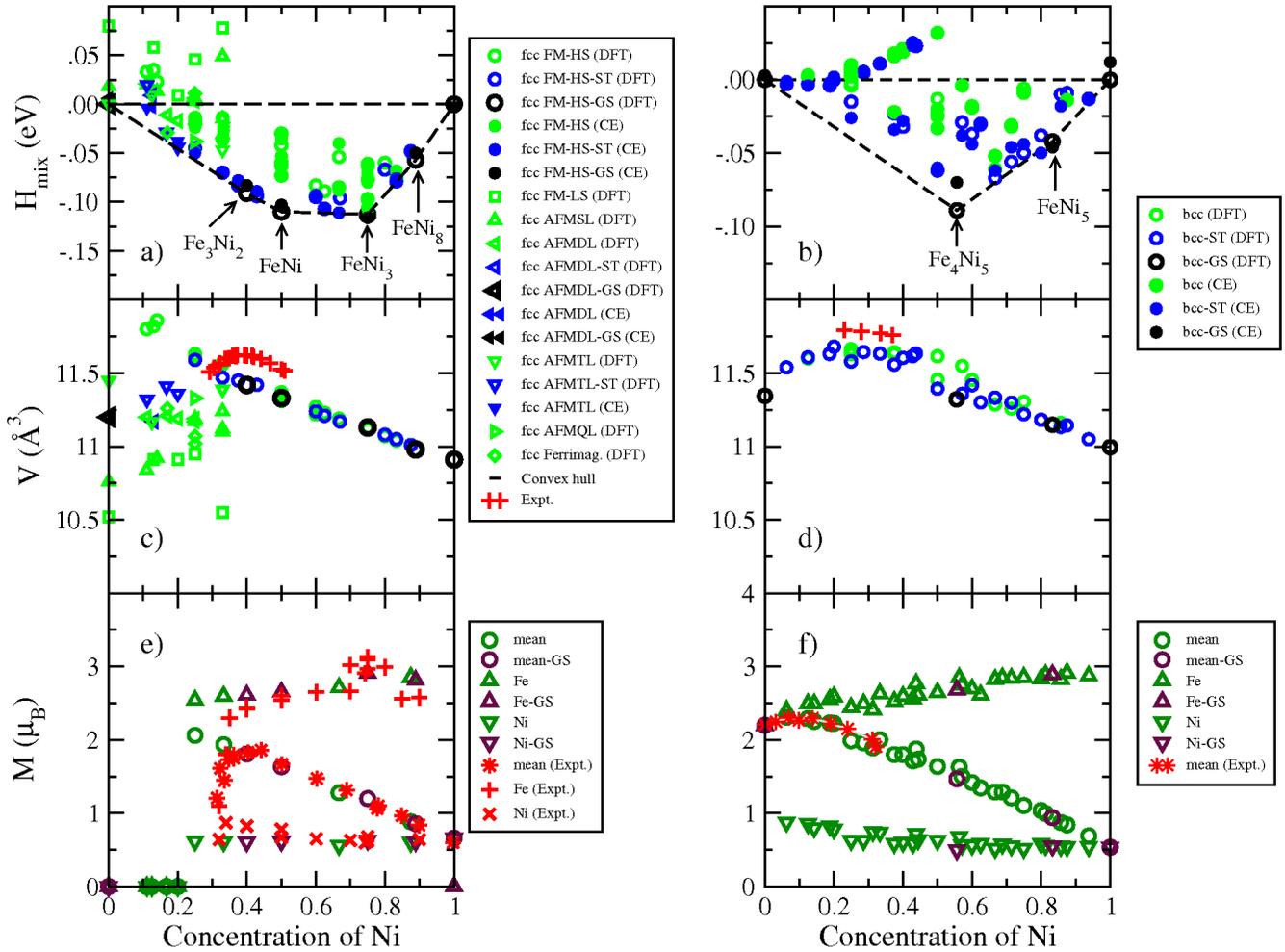}	
			\caption{
		(Color online) Enthalpies of mixing (a,b), volumes per atom (c,d) and magnetic moments (e,f) of Fe-Ni structures on fcc (a,c,e) and bcc (b,d,f) lattices, calculated using DFT. Experimental data are taken from Refs. \onlinecite{Landolt,Crangle1963,Chamberod1979}. GS refers to the ground state on fcc (a,c) or bcc (b,d) crystal lattices; ST is the most stable structure and magnetic configuration for the corresponding alloy composition.}
		\label{fig:FeNi_results}
		\end{figure*}

There is extensive literature on models for Fe-Ni alloys, see for example Refs. \onlinecite{Barabash2009,Mohri2004,Tucker2008,Abrikosov2007,Crisan2002,Ekholm2010,Ruban2005,Ruban2007}. Recently \cite{Lavrentiev2014} we used a DFT database to parameterize the Magnetic Cluster Expansion and to investigate magnetic properties of Fe-Ni alloys. In this sub-section, we compare our DFT results with previous experimental and theoretical studies, focusing on the stability of magnetic configurations and on equilibrium volumes of alloy structures.

Our results agree with an assertion, derived from simulations \cite{Barabash2009,Mohri2004,Tucker2008} and experiments \cite{Massalski1990,Reuter1989}, that fcc FeNi (L1$_0$), FeNi$_3$ (L1$_2$) and FeNi$_8$ (Pt$_8$Ti-like\cite{Barabash2009}) compounds are the global (on both fcc {\it and} bcc lattices) alloy ground states for the relevant compositions, see Fig. \ref{fig:FeNi_results}(a). Our results agree with Ref. \onlinecite{Barabash2009} in that the fcc ferromagnetic Z1(100) phase of Fe$_3$Ni (see Fig. 3 in Ref. \onlinecite{Barabash2009}) is more stable than L1$_2$, contrary to what was previously assumed according to Refs. \onlinecite{Massalski1990,Crisan2002,Mohri2004,Mohri2009}.

In Ref. \onlinecite{Barabash2009} the AFMDL configuration of fcc Fe-Ni alloys was not investigated, despite the fact that AFMDL represents the most stable magnetic configuration of fcc-Fe, see our Table \ref{tab:Fe} and Refs. \onlinecite{Herper1999,Moruzzi1989,Klaver2012}. In relation to the AFMDL structure of fcc Fe-Ni, the Z1 Fe$_3$Ni structure\cite{Lu1991} does not represent the ground state, and instead an alternative fcc ground state, Fe$_3$Ni$_2$ with $I4/mmm$ symmetry, is predicted by CE, see Fig. \ref{fig:FeNi_results}(a). None of the AFM fcc structures is the actual ground state, however the energies of fcc Fe$_5$Ni AFMTL, ferri-magnetic fcc Fe$_5$Ni, and fcc Fe$_4$Ni AFMTL, are fairly close to the bottom of the zero temperature phase stability curve. The existence of these magnetic structures may affect finite temperature stability of fcc alloys.

Our CE calculations also predict two bcc ground states, Fe$_4$Ni$_5$ (VZn-like\cite{Nguyen-Manh2007}) and FeNi$_5$ (of $Cmmm$ symmetry) that are still less stable than fcc structures of similar compositions, see Figs. \ref{fig:FeNi_results}(b) and \ref{fig:formation_binaries}. Fe$_4$Ni$_5$ (VZn-like) bcc structure is predicted as the lowest energy alloy configuration by both DFT and CE simulations.

Enthalpies of mixing of fcc and bcc Fe-Ni structures calculated using DFT and CE are compared in Fig. \ref{fig:FeNi_results}(a,b). To remain consistent with the treatment of binary alloys Fe-Cr and Ni-Cr, we used the same sets of cluster interaction parameters, namely five two-body, three three-body, two four-body, one five-body clusters, for fcc binary alloys, and five two-body, two three-body, one four-body, one five-body clusters for the corresponding bcc alloys. A set of ECIs obtained by mapping energies of structures from DFT to CE is given in Fig. \ref{fig:ECI_binaries} and Table \ref{tab:ECI_binary} in Appendix C. The cross-validation errors between DFT and CE are 8.1 and 10.9 meV/atom for fcc and bcc Fe-Ni alloys, respectively.

The magnitude and sign of ECIs explain the behaviour of fcc and bcc Fe-Ni alloys found in simulations. In fcc alloys the first and third  nearest neighbour (1NN and 3NN) pair interactions are positive, whereas the second nearest neighbour (2NN) interaction is negative. In binary alloys, from Eqs. \ref{eq:CE_1} and \ref{eq:CE_2}, this favours having the unlike atoms occupying the first and the third neighbour coordination shell, and the like atoms occupying the second neighbour shell. For the fcc lattice this favours the formation of L1$_2$ intermetallic phase, which is the ground state of fcc Fe-Ni alloy.  In bcc alloys the 1NN Fe-Ni pair interaction is negative, corresponding to repulsive interaction between the unlike atoms in the first neighbour shell. The 2NN pair interaction is positive and similar in its magnitude to the 1st ECI. As a result, bcc Fe-Ni alloys exhibit several intermetallic phases with negative enthalpies of mixing.

The atomic volumes of fcc and bcc alloys shown in Fig. \ref{fig:FeNi_results}(c,d) are not linear functions of Ni content. This non-linearity stems from the difference between atomic sizes of Fe and Ni {\it and} magnetism, see Fig. \ref{fig:FeNi_results}(e,f). Bcc alloys with low Ni content have larger volume per atom than pure Fe, despite the fact that Ni atoms have smaller size. This is correlated with the fact that the Fe$_{15}$Ni structure has the largest average atomic magnetic moment, 2.31 $\mu_B$. In fcc Fe-Ni alloy the non-linearity of atomic volume as a function of Ni content is even more pronounced, since alloys with Ni content lower than 25\% exhibit anti-ferromagnetic interaction between Fe and Ni, resulting in higher atomic density than ferromagnetically ordered alloys. Experimental measurements \cite{Chamberod1979} show that the average atomic volume is maximum for Fe-Ni alloys with $\sim 37$ at. \% Ni. This is correlated with the fact that the Fe$_3$Ni$_2$ intermetallic phase has the largest volume per atom, see Fig.\ref{fig:FeNi_results}(c). There are several structures with smaller Ni content that are ferromagnetically ordered at 0K and have larger volumes per atom than Fe$_3$Ni$_2$. Those structures are metastable, and alloys with Ni concentration below 40 at. \% Ni are mixtures of ferromagnetic Fe$_3$Ni$_2$, anti-ferromagnetic Fe, and metastable ferromagnetic and anti-ferromagnetic alloy phases. Near 25 at. \% Ni concentration the most stable magnetic configurations are ferromagnetic, however the energy difference between them and anti-ferromagnetic phases, characterized by smaller volumes, is fairly small. In particular, the most stable structure corresponding to 33 at. \% Ni is a ferromagnetic $\beta$-phase\cite{Barabash2009} where the enthalpy of mixing is -0.070 eV/atom and the atomic volume is 11.47 $\AA^3$ per atom. The AFMTL structure is 0.023 eV/atom less stable, and has the atomic volume of 11.39 $\AA^3$, whereas AFMSL is 0.039 eV/atom less stable than FM and has the volume of 11.24  $\AA^3$ per atom.  The coexistence of structures with different magnetic order and different atomic volumes but similar energies is the origin of the Invar effect \cite{Entel1993}.

\subsection{Fe-Cr binary alloys}

\begin{figure*}
 \includegraphics[width=\linewidth]{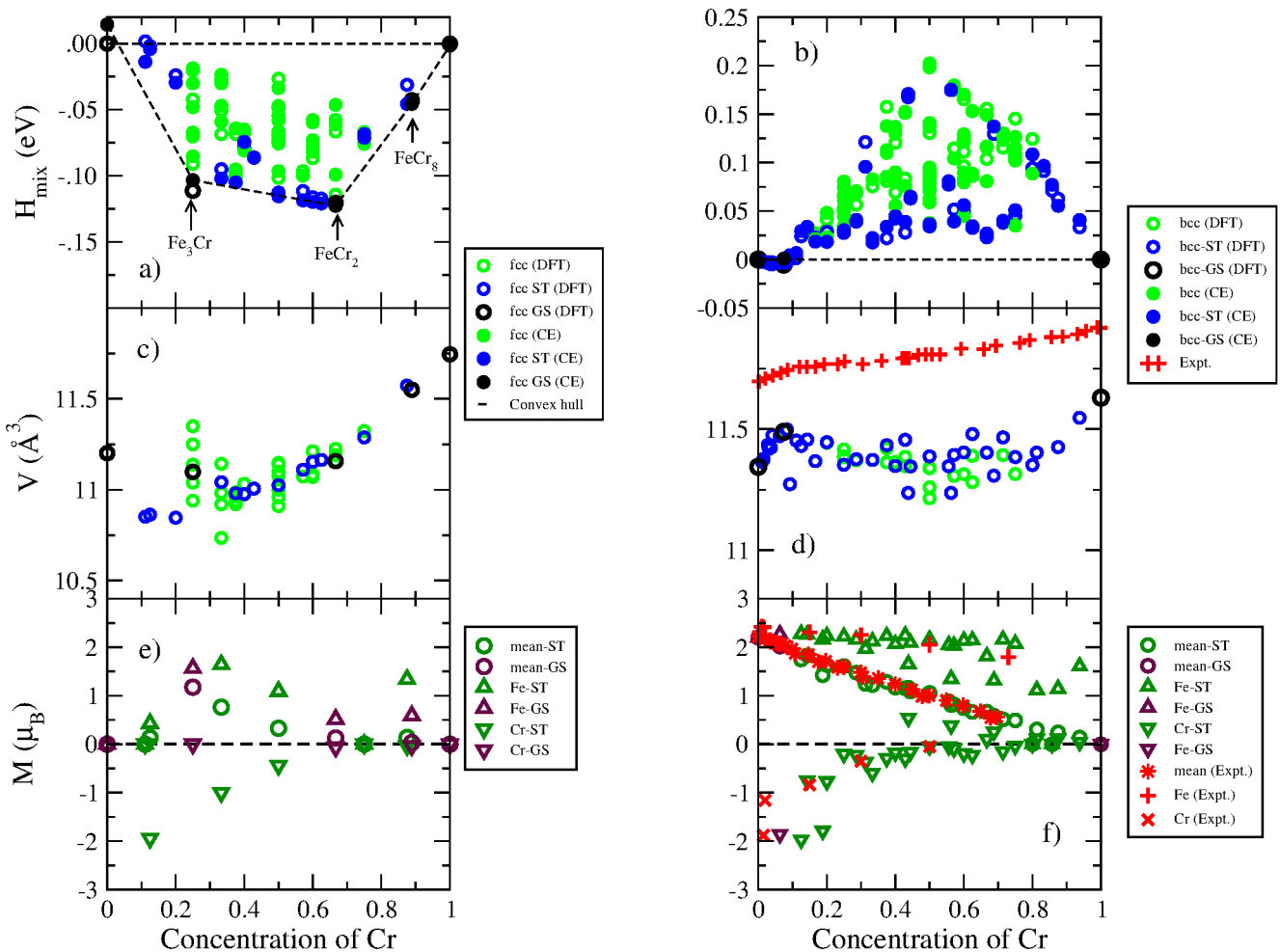}	
			\caption{
(Color online) Enthalpies of mixing (a,b), volumes per atom (c,d) and magnetic moments (e,f) predicted by DFT for fcc (a,c,e) and bcc (b,d,f) Fe-Cr alloys. Experimental data are taken from Refs. \onlinecite{Aldred1976,Aldred1976a,Kajzar1980}. GS - ground states of alloys on fcc (a,c) or bcc (b,d) crystal lattices, ST - the most stable structure for a given composition.}		
\label{fig:FeCr_results}
		\end{figure*}

Extensive theoretical \cite{Nguyen-Manh2007,Nguyen-Manh2008,Nguyen-Manh2009,Nguyen-Manh2012,Lavrentiev2007,Lavrentiev2010,Olsson2003,Olsson2006,Zhang2009,Erhart2008} and experimental\cite{Mirebeau1984} investigations show that low Cr bcc Fe-Cr alloys form intermetallic phases where the most stable structures contain between 6.25  and 7.41 at. \% Cr \cite{Nguyen-Manh2007,Erhart2008}. Results of calculations shown in Fig. \ref{fig:FeCr_results}(b) confirm those findings. For fcc Fe-Cr alloys, we predict three new ground states: Fe$_3$Cr(L1$_2$), FeCr$_2$($\beta2$(100)\cite{Barabash2009}) and FeCr$_8$ (Pt$_8$Ti-like) that are all significantly less stable than bcc structures, see Figs. \ref{fig:FeCr_results}(a) and \ref{fig:formation_binaries}. Enthalpies of mixing of ordered Fe$_3$Cr and FeCr$_2$ structures are -0.111 and -0.120 eV/atom, and are approximately 0.05 eV/atom lower than those calculated for fcc Fe-Cr random alloys.

Comparison between enthalpies of mixing of fcc and bcc Fe-Cr alloys calculated using DFT and CE is shown in Figs. \ref{fig:FeCr_results}(a) and \ref{fig:FeCr_results}(b). A full set of ECIs derived by mapping DFT energies to CE is given in Fig. \ref{fig:ECI_binaries}(c,d) and Table \ref{tab:ECI_binary} in Appendix C. The cross-validation error between DFT and CE is 11.3 and 10.6 meV/atom for fcc and bcc Fe-Cr alloys, respectively. Similarly to fcc Fe-Ni alloys, the first and the third nearest neighbour (1NN and 3NN) pair interactions are positive and the second nearest neighbour (2NN) interaction is negative, favouring the L1$_2$ intermetallic phase, which is also the ground state of fcc Fe-Cr alloy. The 1NN pair interaction in bcc Fe-Cr alloys is negative, as in bcc Fe-Ni alloys, implying repulsive interaction between the unlike atoms in the first nearest neighbour coordination shell. ECIs of bcc Fe-Cr alloys were previously analyzed in Ref. \onlinecite{Lavrentiev2007}. Despite the fact that our DFT calculations use a different set of clusters, our results are in agreement with Ref. \onlinecite{Lavrentiev2007} in that the dominant negative 1NN pair interaction and positive fifth nearest neighbour pair interaction together give rise to the formation of Fe - 6.25 at.\% Cr $\alpha$-phase.

Atomic volumes of bcc Fe-Cr alloys remain nearly constant over a broad range of alloy compositions, exhibiting small variation in the interval of 0.3 $\AA{}^3$ per atom, see Fig.\ref{fig:FeCr_results}(d). There are two exceptions to this rule. The volume per atom in Cr-rich alloys decreases as a function of Fe content. This can be explained by the fact that Fe impurities interfere with anti-ferromagnetic ordering of magnetic moments in pure Cr, reducing the magnitude of moments and the strength of magnetic interactions, see Fig. \ref{fig:FeCr_results}(f). This also affects the average atomic volume. In Fe-rich alloys, atomic volume increases linearly with Cr content, reaching a maximum of 11.50 $\AA^3$ per atom at 8.33 at. \% Cr. This confirms previous theoretical predictions derived using CPA and SQS methods \cite{Olsson2006,Zhang2009}, which show a local maximum of atomic volume (lattice parameter) in random bcc Fe-Cr alloys at approximately 10 at. \% Cr. These theoretical predictions are in agreement with experimental data\cite{Pearson1958}, where the observed deviation from Vegard's law is largest at $\sim$ 10 at. \% Cr . This effect probably results from magneto-volume coupling and strong anti-ferromagnetic interaction between Fe and Cr atoms. At low density magnetic moments are larger and the energy of atomic structure is lower, hence Cr impurities in Fe tend to increase volume per atom in the $\alpha$-phase. The increase is almost linear in Cr content until a critical concentration is reached and Cr starts segregating.

At variance with DFT analysis of ordered structures performed here, and earlier studies of random alloys \cite{Olsson2006,Zhang2009}, the experimentally measured atomic volume in alloys with Cr concentration higher than 10 \% continues to increase linearly towards the limit of pure Cr. The likely reason for the lack of agreement between DFT and experiment is that neither the ordered structures treated here nor the random alloys investigated in Refs. \onlinecite{Olsson2006,Zhang2009} are representative of real bcc Fe-Cr alloys, where alloy microstructure is a mixture of $\alpha$-phase and Cr clusters, as shown in Figs. \ref{fig:FeCr_results}(b,d) by black circles\cite{Lavrentiev2007}.

The composition dependence of atomic volume in fcc Fe-Cr alloys differs significantly from what is found in fcc Fe-Ni alloys. Due to strong anti-ferromagnetic interaction between Fe and Cr atoms, anti-ferromagnetic or ferri-magnetic order dominates in the entire range of alloy compositions, see Fig. \ref{fig:FeCr_results}(e). Volume decrease caused by anti-ferromagnetic ordering in Fe-rich fcc Fe-Ni alloys is also present in the entire range of alloy compositions. Volume decrease as a function of Cr concentration is particularly strongly pronounced in Cr-rich fcc Fe-Cr alloys.

Magnetic moments of Fe and Cr atoms as well as the average magnetic moment of ordered bcc Fe-Cr structures are similar to those predicted for random alloys in Refs. \onlinecite{Olsson2006,Klaver2006}. They agree well with the available experimental data \cite{Aldred1976,Aldred1976a,Kajzar1980}.

\subsection{Cr-Ni binary system}

\begin{figure*}
 \includegraphics[width=\linewidth]{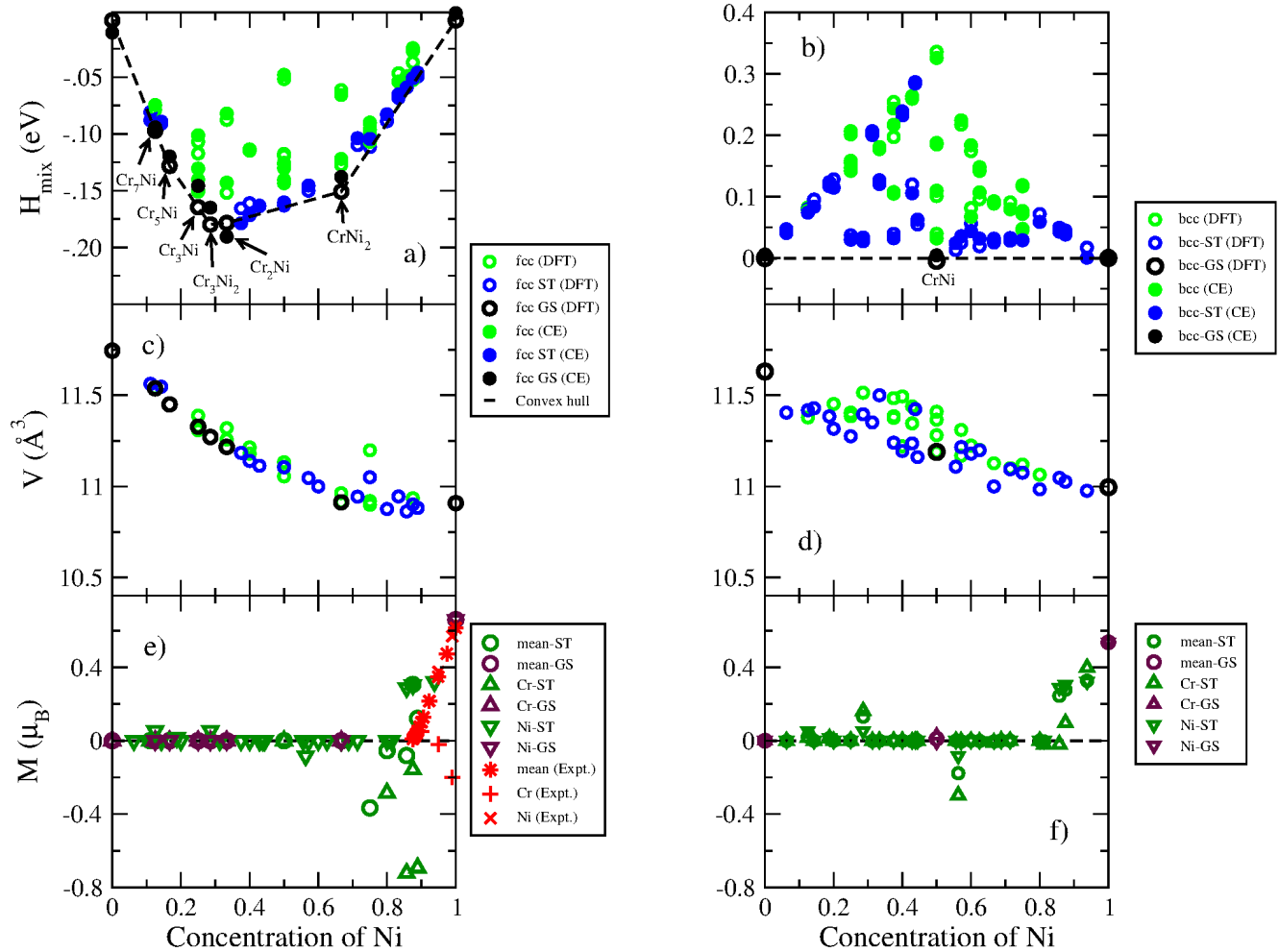}	
			\caption{
(Color online) Enthalpies of mixing (a,b), volumes per atom (c,d) and magnetic moments (e,f) calculated using DFT for Cr-Ni alloys on fcc (a,c,e) and bcc (b,d,f) lattices. Experimental data are taken from Ref. \onlinecite{Landolt}. GS - ground states of alloys on fcc (a,c) or bcc (b,d) lattices, ST - the most stable structure found for a given alloy composition.}
		\label{fig:CrNi_results}
\end{figure*}
		
DFT and CE simulations of fcc Cr-Ni alloys were performed in Ref. \onlinecite{Tucker2008}. Our analysis confirms the conclusion, derived from simulations and experiment, that there is only one globally stable ground state of the alloy, realized on the CrNi$_2$ (MoPt$_2$-like) ordered structure. We find a further five fcc ground states: Cr$_7$Ni (of $Cmmm$ symmetry, predicted by CE), Cr$_5$Ni (also predicted by CE, with $Cmmm$ symmetry), Cr$_3$Ni-Z1(100), Cr$_5$Ni$_2$ (of $I4/mmm$ symmetry, also predicted by CE), and Cr$_2$Ni-$\beta1$(100). The last of these is characterized by a large positive value of the enthalpy of formation, and is less stable than bcc alloys with the same composition, see Figs. \ref{fig:CrNi_results}(a) and \ref{fig:formation_binaries}. We find only one alloy configuration on a bcc lattice that has small negative enthalpy of mixing, CrNi (predicted by CE, with $Cmmn$ symmetry and $H_{mix}=-4$ meV/atom).

Comparison of enthalpies of mixing of fcc and bcc Cr-Ni alloys calculated using DFT and CE is given in Fig. \ref{fig:CrNi_results}(a,b). A full set of ECIs found by mapping the energies of structures from DFT to CE is given in Fig. \ref{fig:ECI_binaries} and Table \ref{tab:ECI_binary} in Appendix C. Cross-validation errors between DFT and CE are 14.2 and 12.8 meV/atom for fcc and bcc Cr-Ni alloys, respectively. Similarly to fcc Fe-Ni and Fe-Cr alloys, the first and third nearest neighbour (1NN and 3NN) pair interactions in fcc Cr-Ni alloys are positive and the second nearest neighbour (2NN) interaction is negative. Unlike the other two binary systems, the ground state of fcc Cr-Ni alloys is MoPt$_2$-like phase. The ECI parameters derived from our DFT calculations and the cross-validation error between DFT and CE are in agreement with those of Ref. \onlinecite{Tucker2008}. The negative 1NN pair interaction in bcc Cr-Ni system is the largest of all the binary alloys. Because of that, there is only one bcc intermetallic phase, CrNi, of \textit{Cmmn} symmetry, which has small negative enthalpy of mixing ($-4$ meV/atom).

Variation of atomic volume as a function of Ni content in both fcc and bcc alloys is more linear than in Fe-Ni and Fe-Cr alloys because magnetic interactions are weaker, see Figs. \ref{fig:CrNi_results}(c-f). Similarly to Fe-Cr alloys, the difference between atomic volumes of alloys with low and high concentration of Cr is more significant in fcc than bcc alloys.

\begin{figure*}
			\centering
			\begin{minipage}{.50\textwidth}
			  	\centering
			  	a) \includegraphics[width=.9\linewidth]{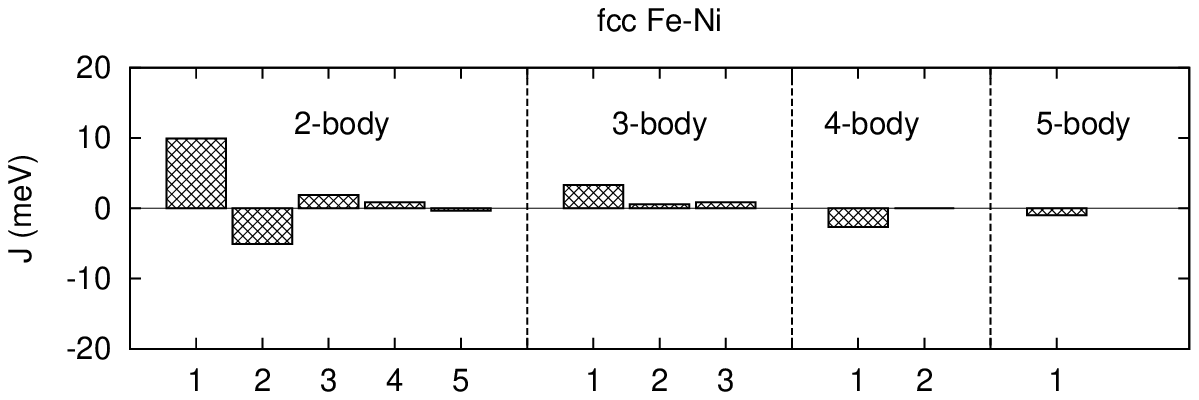}			  	
			\end{minipage}%
			\begin{minipage}{.50\textwidth}
			  	\centering
			  	b) \includegraphics[width=.9\linewidth]{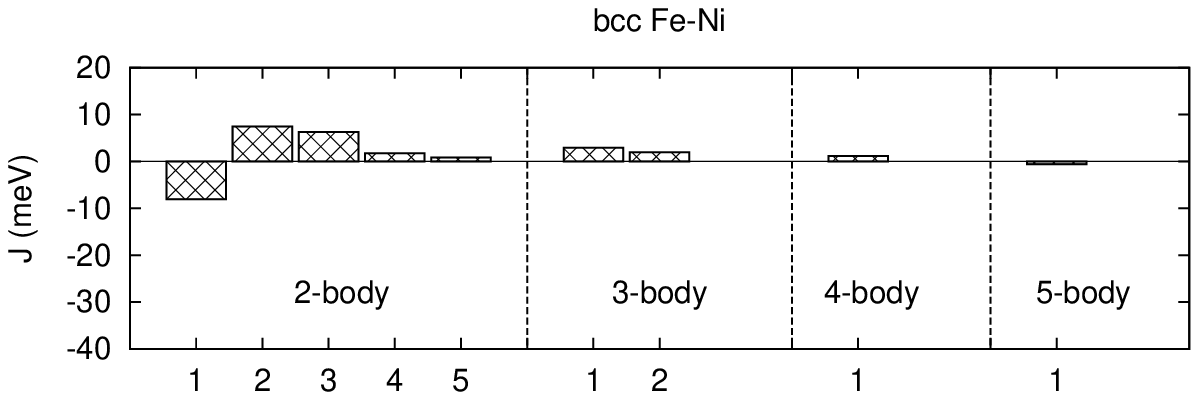}
			\end{minipage}
			\newline
			\begin{minipage}{.50\textwidth}
			  	\centering
			  	c) \includegraphics[width=.9\linewidth]{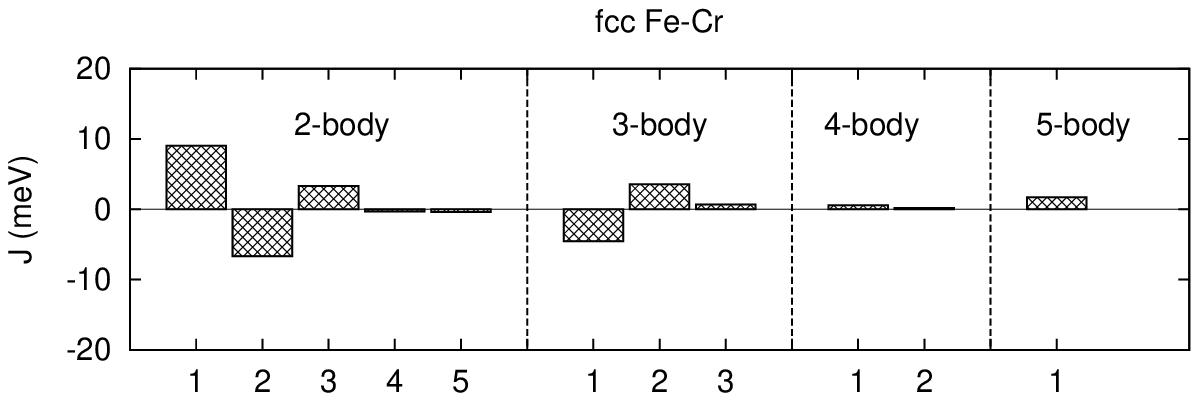}			  	
			\end{minipage}%
			\begin{minipage}{.50\textwidth}
			  	\centering
			  	d) \includegraphics[width=.9\linewidth]{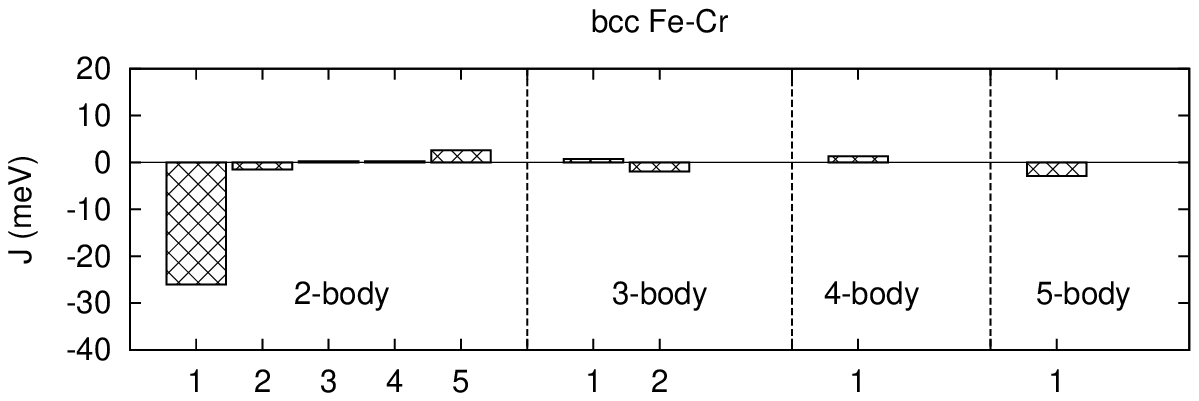}
			\end{minipage}
			\newline
			\begin{minipage}{.50\textwidth}
			  	\centering
			  	e) \includegraphics[width=.9\linewidth]{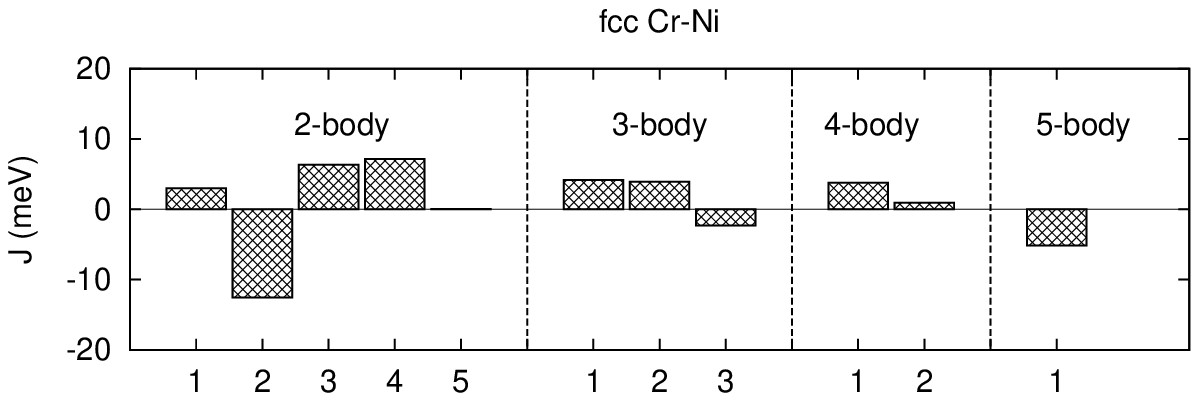}			  	
			\end{minipage}%
			\begin{minipage}{.50\textwidth}
			  	\centering
			  	f) \includegraphics[width=.9\linewidth]{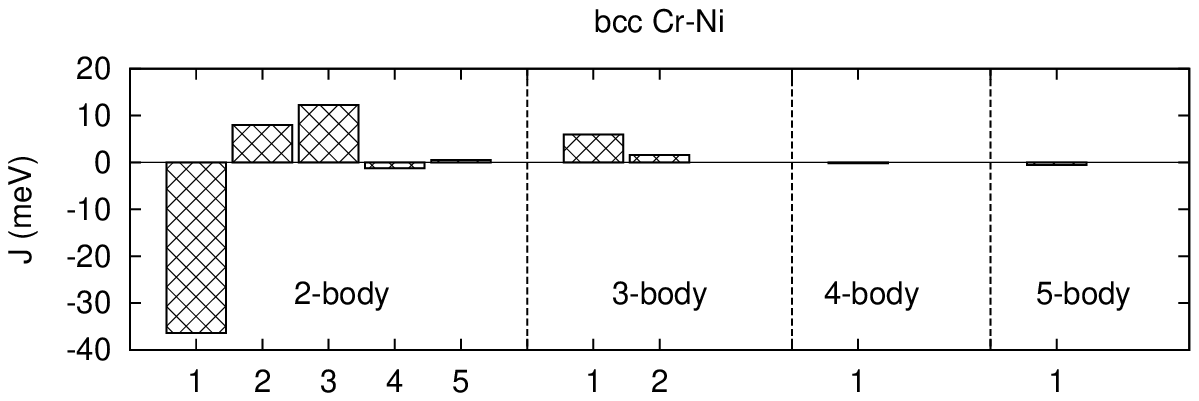}
			\end{minipage}
			\caption{ Effective cluster interactions (ECIs) derived using CE method for fcc Fe-Ni (a), bcc Fe-Ni (b), fcc Fe-Cr (c), bcc Fe-Cr (d),fcc Cr-Ni (e), bcc Cr-Ni (f) alloys.
		}
		\label{fig:ECI_binaries}
\end{figure*}

\subsection{Fe-Cr-Ni ternary system}

\begin{figure*}
			\centering
			\begin{minipage}{.50\textwidth}
			  	\centering
			  	a)\includegraphics[width=.9\linewidth]{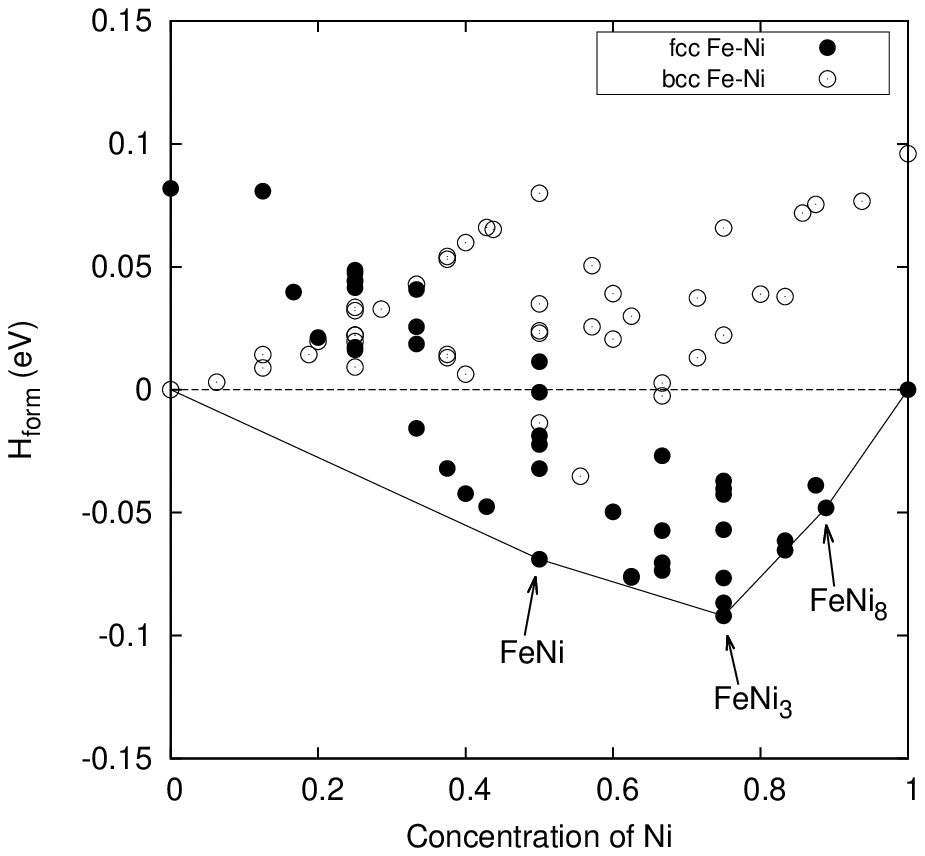}			  	
			\end{minipage}%
			\begin{minipage}{.50\textwidth}
			  	\centering
			  	b)\includegraphics[width=.9\linewidth]{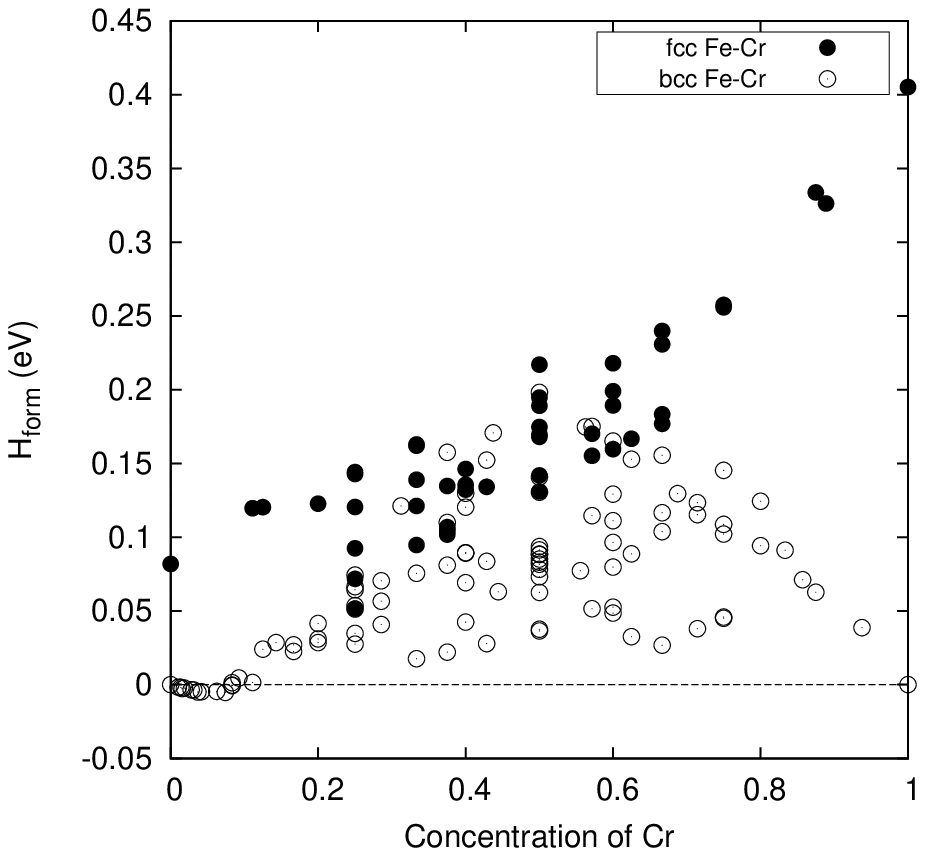}
			\end{minipage}
			\newline
			\begin{minipage}{.50\textwidth}
			  	\centering
			  	c)\includegraphics[width=.9\linewidth]{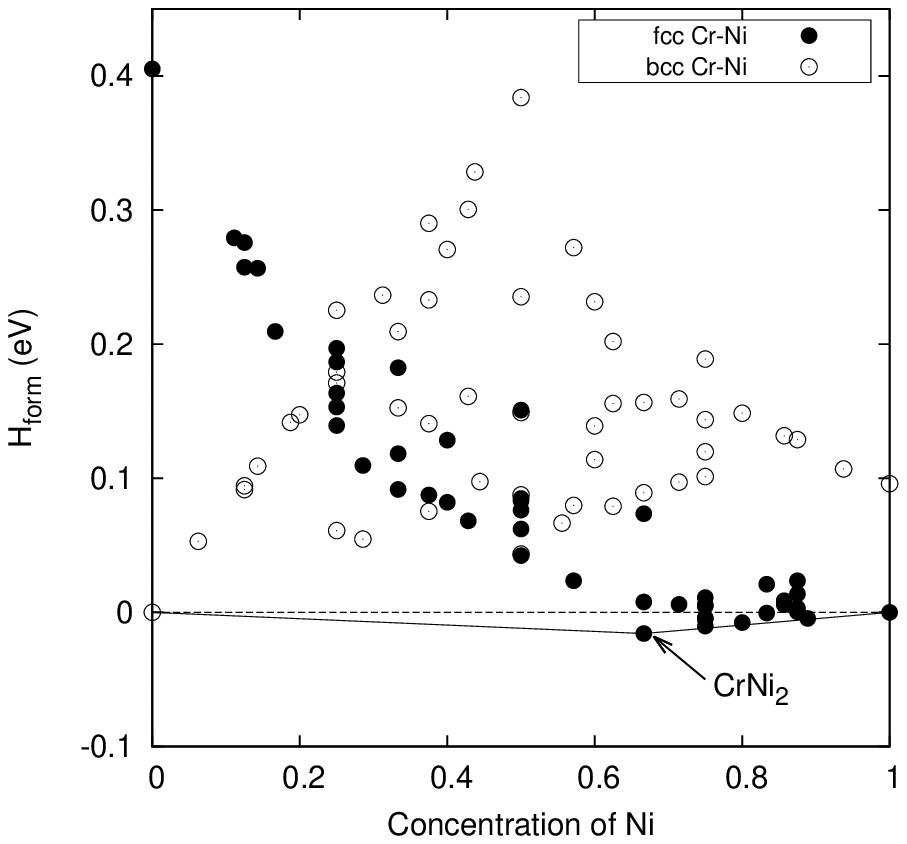}
			\end{minipage}
			\caption{
		Enthalpies of formation of Fe-Ni (a), Fe-Cr (b), and Cr-Ni (c) binary structures.}
		\label{fig:formation_binaries}
		\end{figure*}

The stability of fcc and bcc phases of ternary Fe-Cr-Ni alloys, and the corresponding binary alloys, is defined with respect to bcc Fe, bcc Cr and fcc Ni, as mentioned previously. Enthalpies of formation of Fe-Ni, Fe-Cr and Cr-Ni alloys are shown in Figs. \ref{fig:formation_binaries}(a), \ref{fig:formation_binaries}(b) and \ref{fig:formation_binaries}(c), respectively. The Ni-rich fcc Fe-Ni and Cr-Ni alloys are usually more stable than bcc alloys of similar composition, whereas alloys with smaller Ni content tend to adopt bcc structure. In Fe-Cr alloys, energies of fcc phases are always higher than the energies of bcc structures. Even so, meta-stable fcc Fe-Cr structures and interactions between the unlike atoms in fcc Fe-Cr alloys prove critical to understanding chemical ordering in Fe-Cr-Ni alloy system. From the list of ground states associated with each lattice type shown in Figs. \ref{fig:FeNi_results}, \ref{fig:FeCr_results} and \ref{fig:CrNi_results}, we conclude that there are only four binary fcc phases: FeNi, FeNi$_3$, FeNi$_8$ and CrNi$_2$, and only one binary bcc Fe-Cr phase, namely the $\alpha$-phase, which are the global ground states of the alloys. Enthalpies of formation, volumes and magnetic moments per atom, and space groups of the relevant alloy structures are given in Table \ref{tab:Enthalpies_of_GS}.

\begin{figure}
\includegraphics[width=\columnwidth]{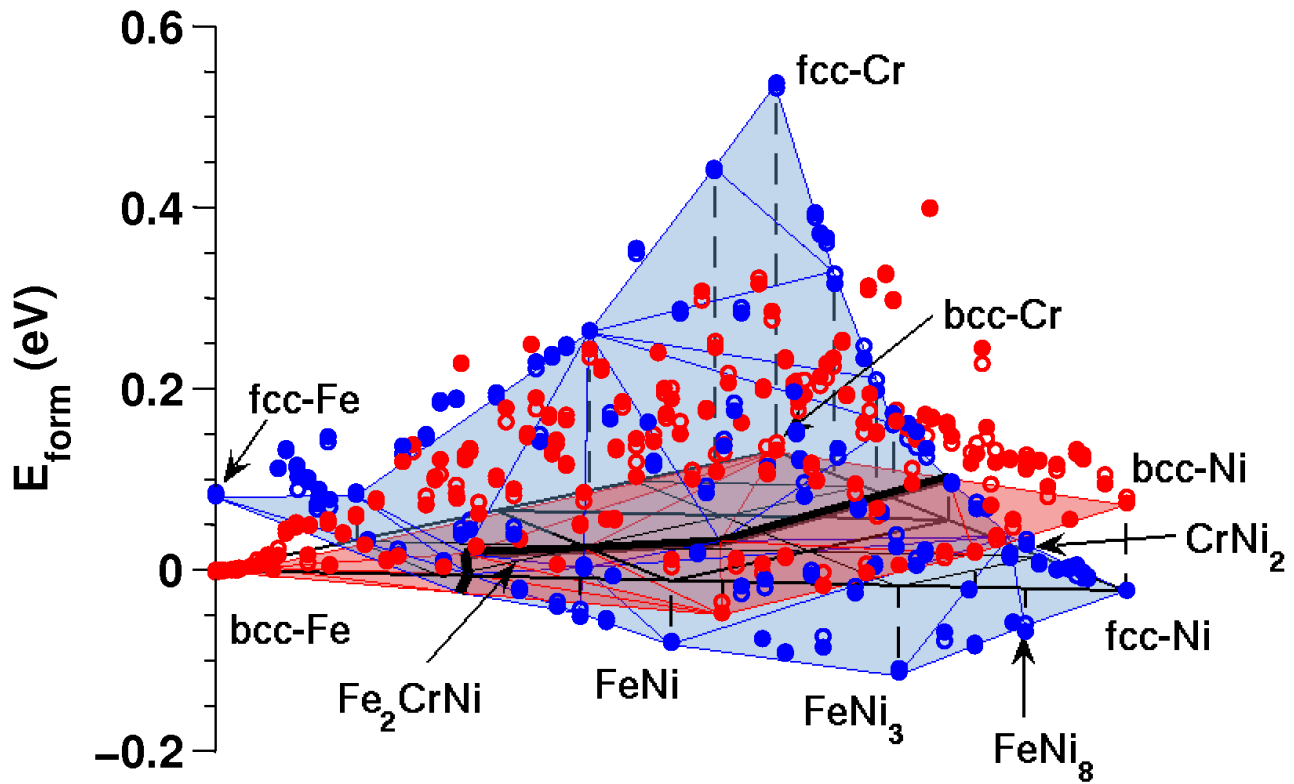}
\caption{(Color online) Enthalpies of formation predicted by DFT (filled circles) and CE (open circles) for ternary Fe-Cr-Ni alloys at 0K. Only the most stable structures for each composition are shown. Blue and red circles show computed formation enthalpies of fcc and bcc Fe-Cr-Ni ternary alloys. Blue and red surfaces show convex hulls for fcc and bcc crystal structures, respectively. Black solid line corresponds to the intersection between fcc and bcc convex hulls. Cross-validation errors between DFT and CE are 10.2 and 11.2 meV/atom for fcc and bcc ternary alloys, respectively.
\label{fig:phase_stability_FeCrNi}}
\end{figure}

Enthalpies of formation of fcc and bcc Fe-Cr-Ni alloys derived from DFT and CE are compared in Fig. \ref{fig:phase_stability_FeCrNi}. The most stable fcc and bcc structures form convex hulls, shown in Fig. \ref{fig:phase_stability_FeCrNi} by blue and red surfaces, respectively. The line of intersection between these two surfaces corresponds to the zero Kelvin fcc-bcc phase transition, which occurs if the enthalpies of formation of bcc an fcc alloys are equal. There is no Fe-Cr-Ni ternary alloy configuration on bcc lattice that has negative enthalpy of formation. Fcc alloy structures have negative enthalpy of formation in the Ni-rich limit of alloy compositions. This region of negative enthalpy of formation is elongated along the Fe-Ni edge of the alloy composition triangle. The L1$_2$-based fcc Fe$_2$CrNi phase, similar to Cu$_2$NiZn alloy phase, is the global ground state of Fe-Cr-Ni alloys. The enthalpy of formation, volume per atom, magnetic moments of each atom, as well as the space group of Fe$_2$CrNi structure, are given in Table \ref{tab:Enthalpies_of_GS}.

\begin{figure*}
			\centering
			\begin{minipage}{\textwidth}
			  	\centering
			  	a)\includegraphics[width=0.9\textwidth]{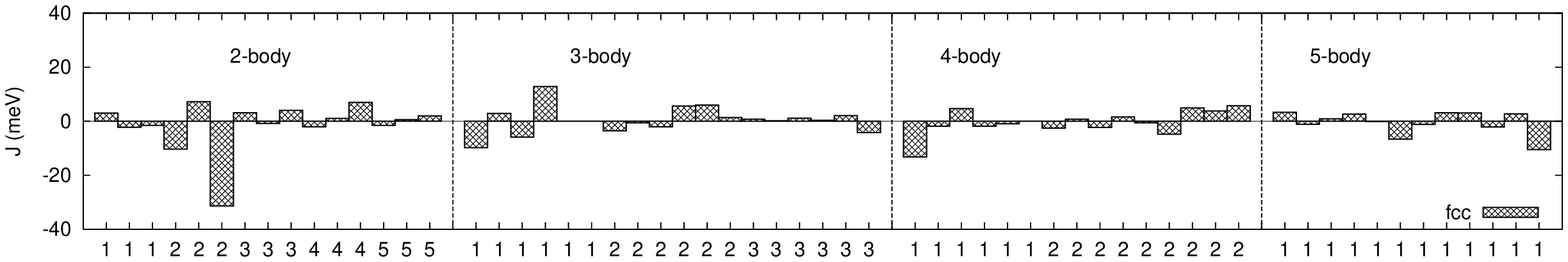}			  	
			\end{minipage}%
						\newline
			\begin{minipage}{\textwidth}
			  	\centering
			  	b)\includegraphics[width=0.9\textwidth]{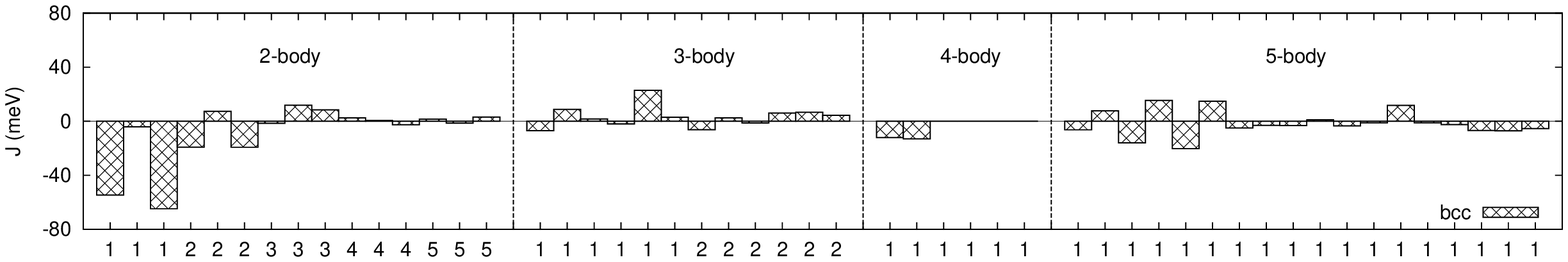}			  	
			\end{minipage}%
		\caption{
	Effective cluster interactions obtained using the CE method for fcc (a) and bcc (b) Fe-Cr-Ni ternary alloys.
        \label{fig:ECI_FeCrNi_ternary}}
\end{figure*}

ECIs of ternary fcc and bcc alloys are derived by mapping DFT energies onto CE for 248 fcc and 246 bcc structures, respectively. In CE simulations we used the same set of clusters as in fcc (five two-body, three three-body, two four-body, one five-body clusters) and bcc (five two-body, two three-body, one four-body, one five-body clusters) binary alloys. Since in ternary alloys each cluster can be decorated by point functions in various ways (see Section II.A and Table I), the number of ECIs is much larger than the number of clusters taken into consideration. Namely, we have 15 two-body, 16 three-body, 14 four-body, 12 five-body clusters for fcc alloys and 15 two-body, 12 three-body, 6 four-body, 18 five-body clusters for bcc alloys. Values of all the optimized ECIs for ternary alloys are given in Fig. \ref{fig:ECI_FeCrNi_ternary} and Table \ref{tab:ECI_def_ternary}. Cross-validation errors between DFT and CE are 10.2 and 11.2 meV/atom for fcc and bcc ternary alloys, respectively.

\begin{table*}
\caption{Enthalpies of formation of the lowest energy intermetallic phases of fcc Fe-Cr-Ni ternary alloys.
        \label{tab:Enthalpies_of_GS}}
\begin{ruledtabular}
    \begin{tabular}{cccccccc}
    Structure & Space & Wyckoff    & Mag. space & (Mag.) Wyckoff    & $V$     & $H_{form}$ & $M$   \\
              & group& positions    & group     & positions    & (eV)     & $H_{form}$ &  ($\mu_B$)   \\
     \hline
    FeNi  & $P4/mmm$ & Fe$_1$ 2$e$ & $P4/mm'm'$ & Fe$_1$ 2$e$ & 11.33 & -0.069 &        2.66 \\
    (L1$_0$) &       & Ni$_1$ 1$a$ &       & Ni$_1$ 1$a$ &       &       &  0.63 \\
          &       & Ni$_2$ 1$c$ &       & Ni$_2$ 1$c$ &       &       &  0.63 \\
    FeNi$_3$ & $Pm-3m$ & Fe$_1$ 1$a$ & $Pm'm'm$ & Fe$_1$ 1$a$ & 11.13 & -0.091 &   2.91 \\
    (L1$_2$) &       & Ni$_1$ 3$c$ &       & Ni$_1$ 1$f$ &       &       &        0.59 \\
          &       &       &       & Ni$_2$ 1$d$ &       &       &        0.58 \\
          &       &       &       & Ni$_3$ 1$g$ &       &       &        0.72 \\
    FeNi$_8$ & $I4/mmm$ & Fe$_1$ 2$a$ & $P-1$   & Fe$_1$ 1$a$ & 10.98 & -0.051 &        2.81 \\
    (NbNi$_8$) &       & Ni$_1$ 8$h$ &       & Ni$_1 $2$i$ &       &       &        0.60 \\
          &       & Ni$_2$ 8$i$ &       & Ni$_2$ 2$i$ &       &       &        0.63 \\
          &       &       &       & Ni$_3$ 2$i$ &       &       &        0.61 \\
          &       &       &       & Ni$_4$ 2$i$ &       &       &        0.61 \\
    CrNi$_2$ & $Immm$  & Cr$_1$ 2$a$ & $Immm1'$ & Cr$_1$ 2$a$ & 10.91 & -0.016 &        0.00 \\
    (MoPt$_2$) &       & Ni$_1$ 4$e$ &       & Ni$_1$ 4$e$ &       &       &        0.00 \\
    Fe$_2$CrNi & $P4/mmm$ & Cr$_1$ 1$c$ & $Pm'm'm$ & Cr$_1$ 1$f$ & 11.37 & -0.026 &        -2.44 \\
    (Cu$_2$NiZn) &       & Fe$_1$ 2$e$ &       & Fe$_1$ 1$d$ &       &       &        2.05 \\
          &       & Ni$_1$ 1$a$ &       & Fe$_2$ 1$g$ &       &       &        2.12 \\
          &       &       &       & Ni$_1$ 1$a$ &       &       &        0.15 \\
    \end{tabular}%
\end{ruledtabular}
\end{table*}

\begin{figure*}
			\centering
			\begin{minipage}{.50\textwidth}
			  	\centering
			  	a)\includegraphics[width=.9\linewidth]{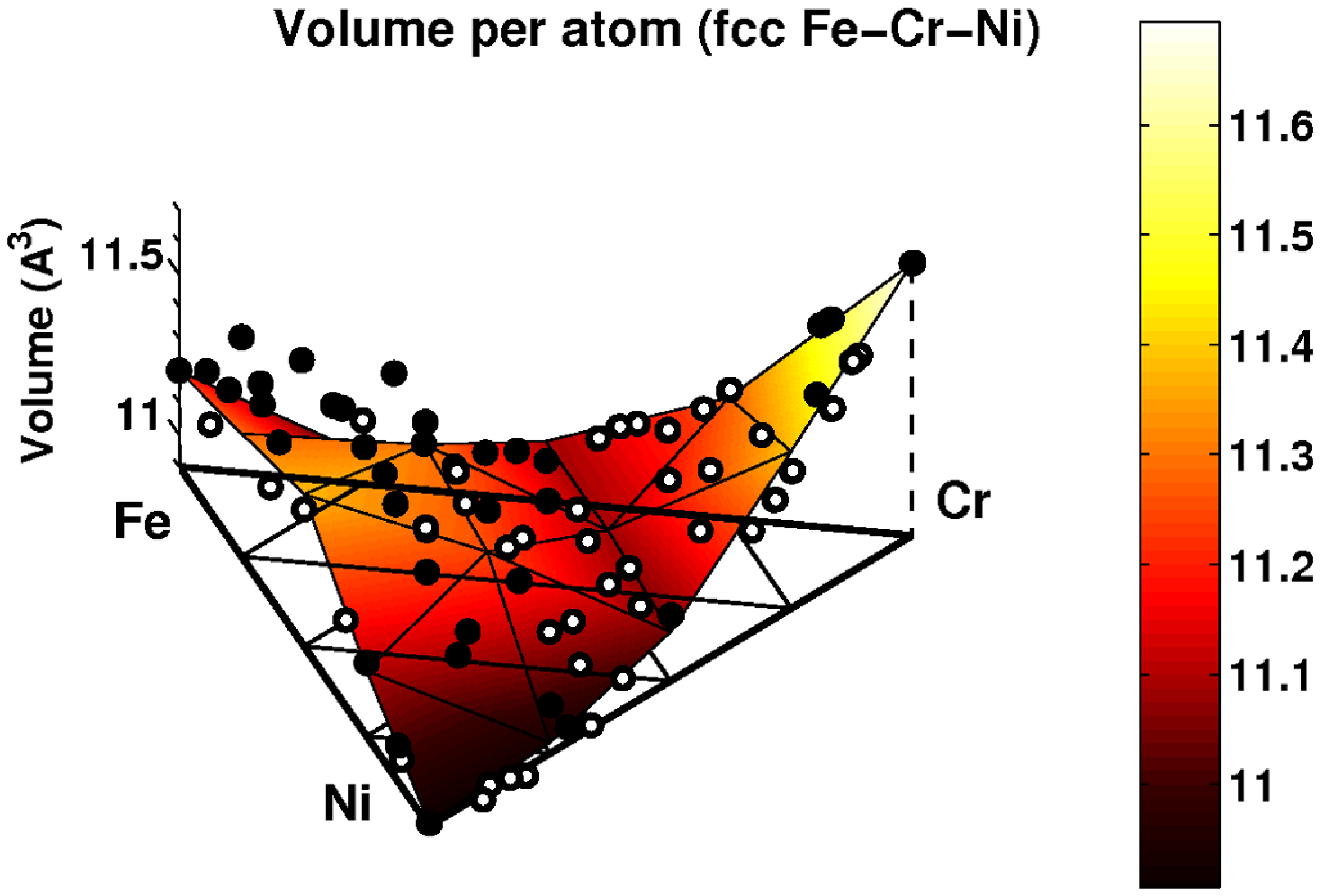}			  	
			\end{minipage}%
			\begin{minipage}{.50\textwidth}
			  	\centering
			  	b)\includegraphics[width=.9\linewidth]{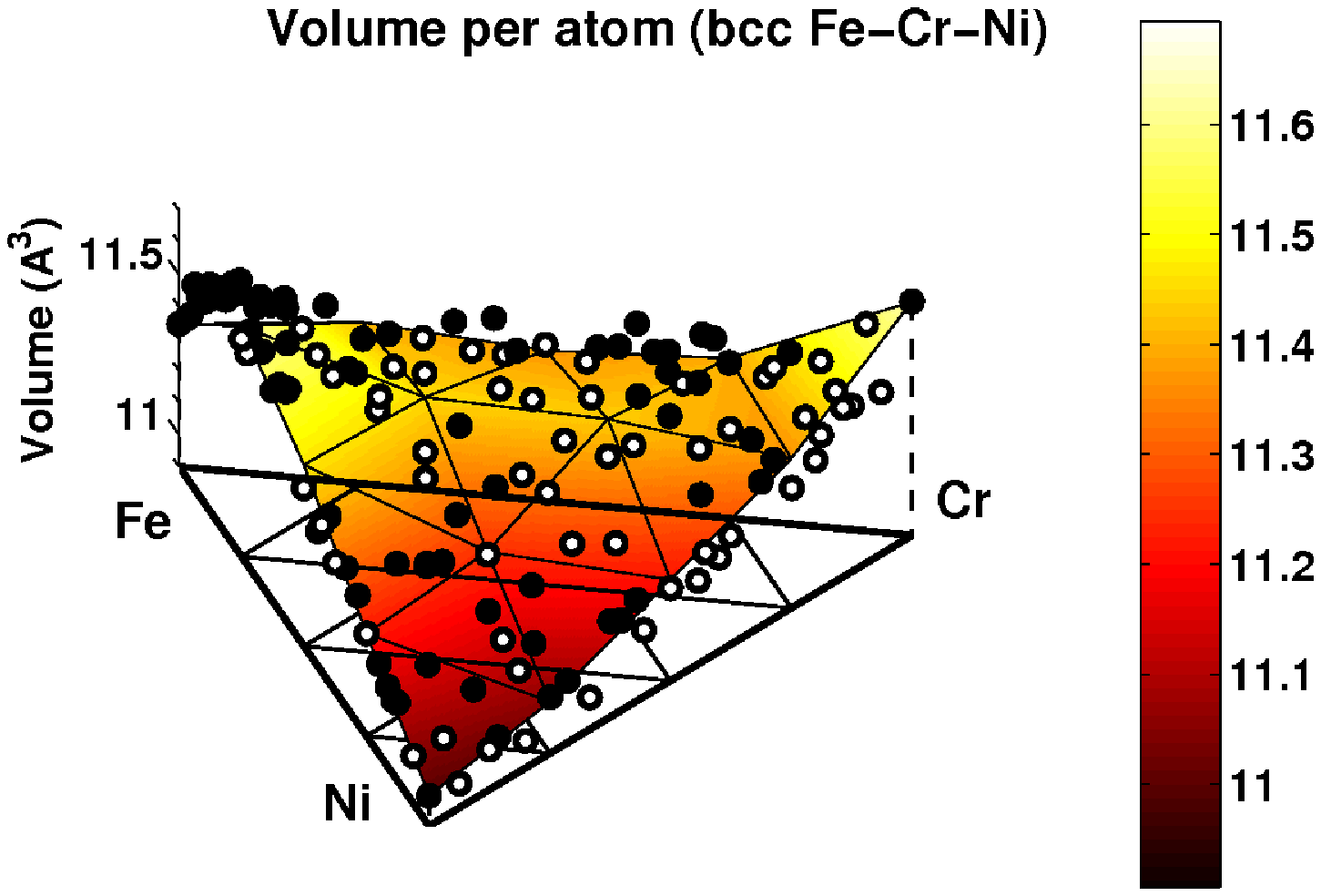}
			\end{minipage}
			\newline
			\begin{minipage}{.50\textwidth}
			  	\centering
			  	c)\includegraphics[width=.9\linewidth]{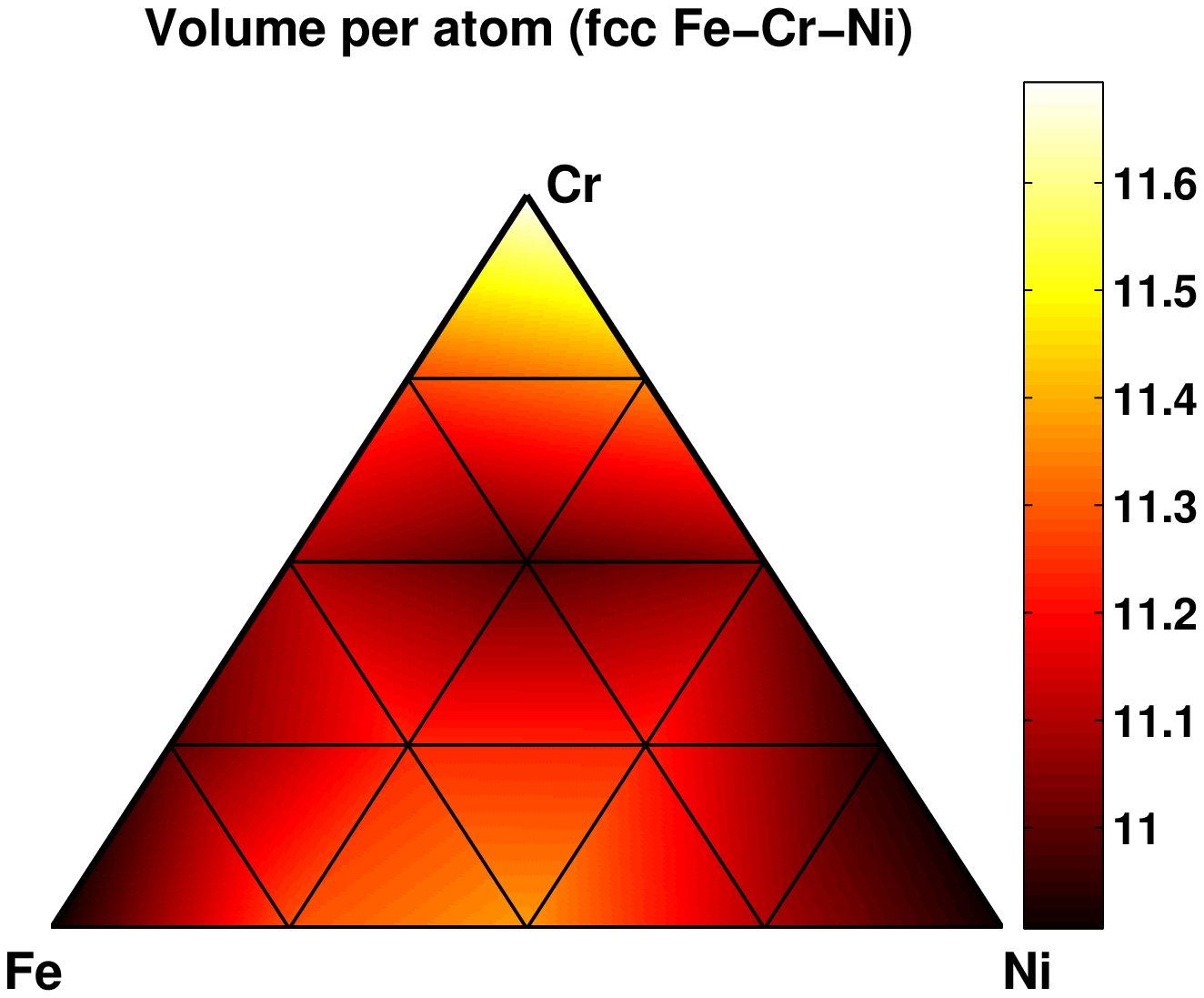}			  	
			\end{minipage}%
			\begin{minipage}{.50\textwidth}
			  	\centering
			  	d)\includegraphics[width=.9\linewidth]{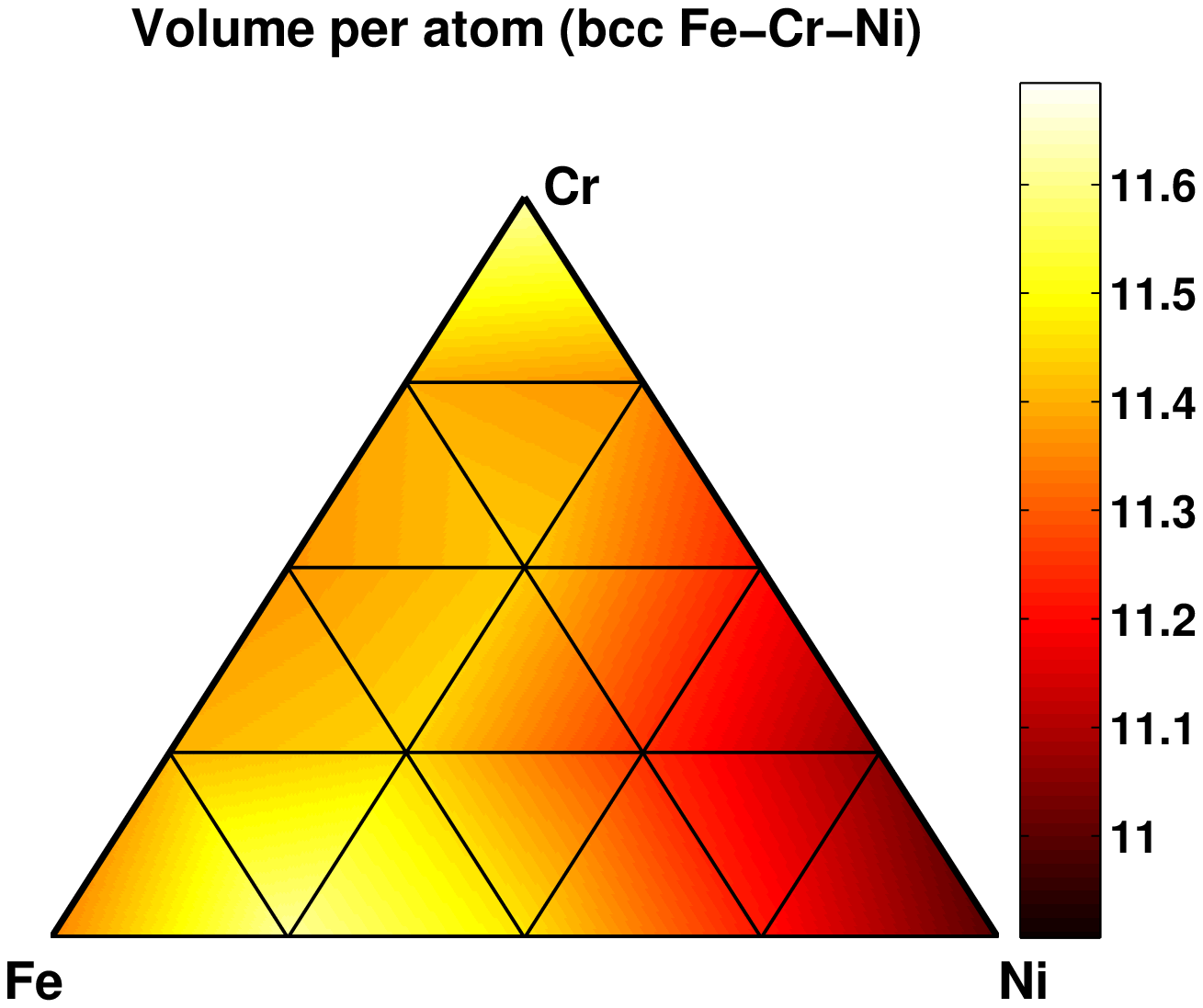}
			\end{minipage}			
			\caption{(Color online) Volumes (in $\AA^3$) of stable fcc (a, c) and bcc (b, d) ordered structures predicted by DFT calculations at 0K for various alloy compositions. Filled and open circles in (a, b) correspond to DFT data above and below the interpolated values, represented by the respective surfaces. (c, d) are the orthogonal projections of (a, b).
		}
		\label{fig:volumes_FeCrNi}
\end{figure*}

\begin{figure*}
			\centering
			\begin{minipage}{.50\textwidth}
			  	\centering
			  	a)\includegraphics[width=.9\linewidth]{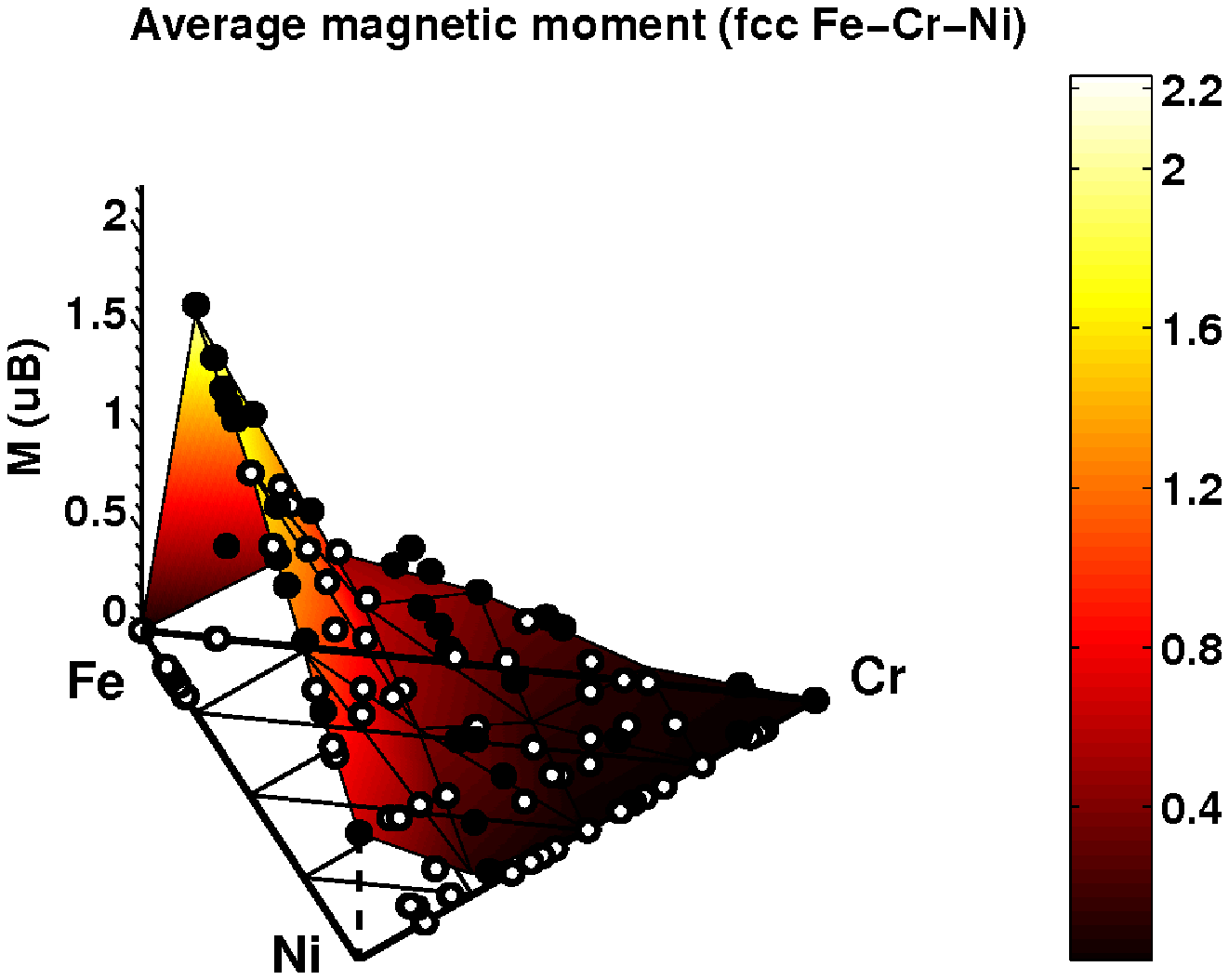}			  	
			\end{minipage}%
			\begin{minipage}{.50\textwidth}
			  	\centering
			  	b)\includegraphics[width=.9\linewidth]{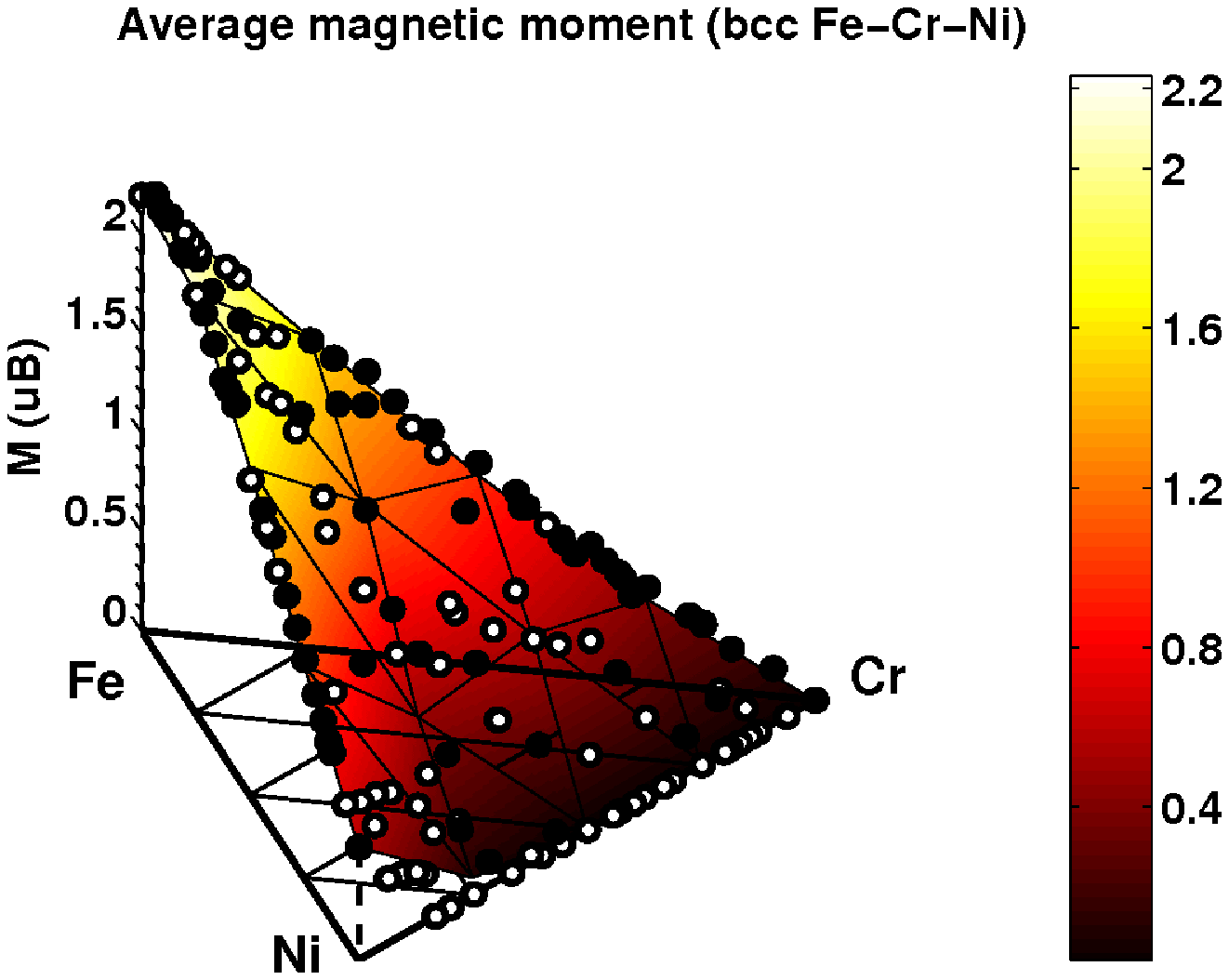}
			\end{minipage}
			\newline
			\begin{minipage}{.50\textwidth}
			  	\centering
			  	c)\includegraphics[width=.9\linewidth]{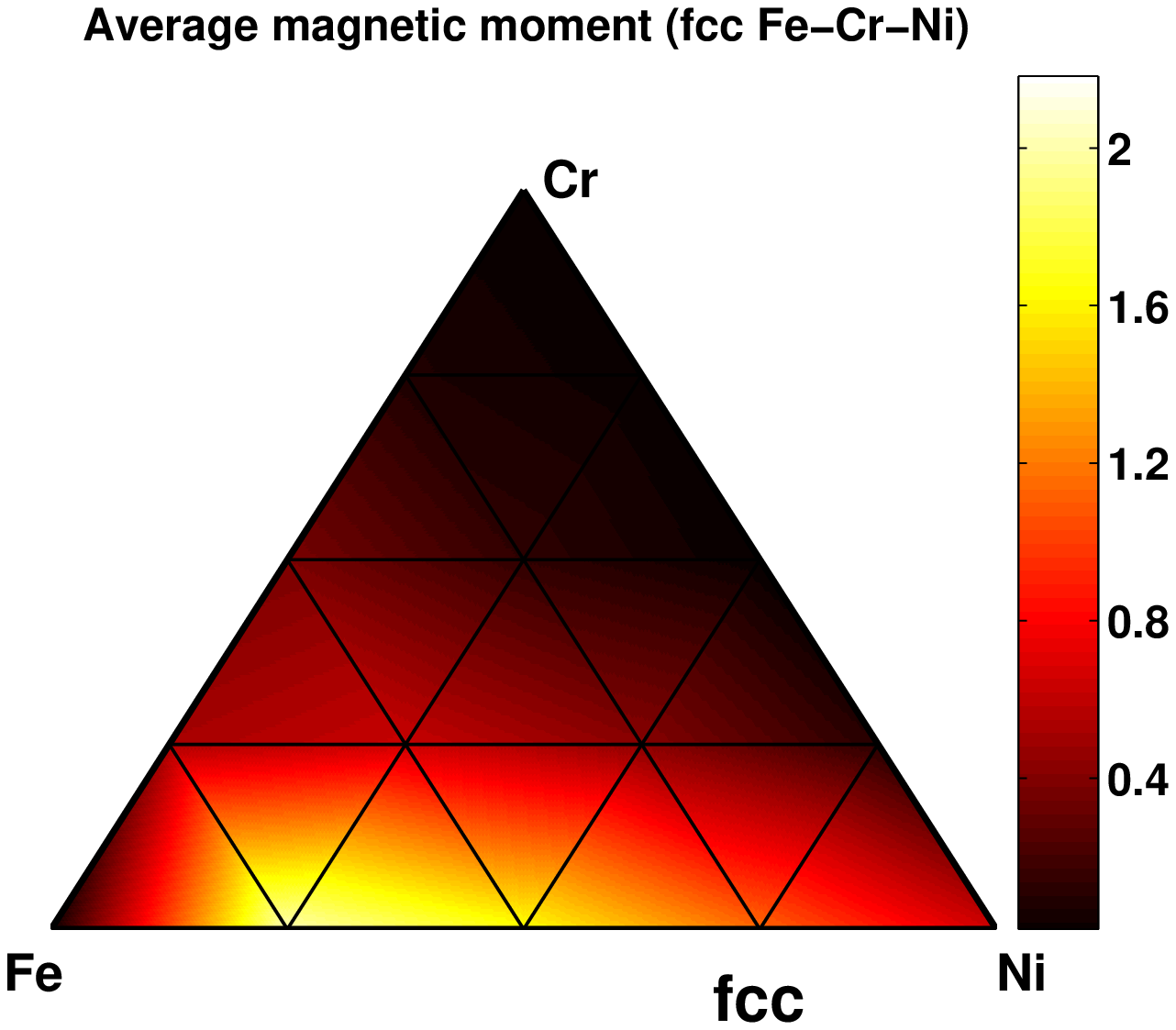}			  	
			\end{minipage}%
			\begin{minipage}{.50\textwidth}
			  	\centering
			  	d)\includegraphics[width=.9\linewidth]{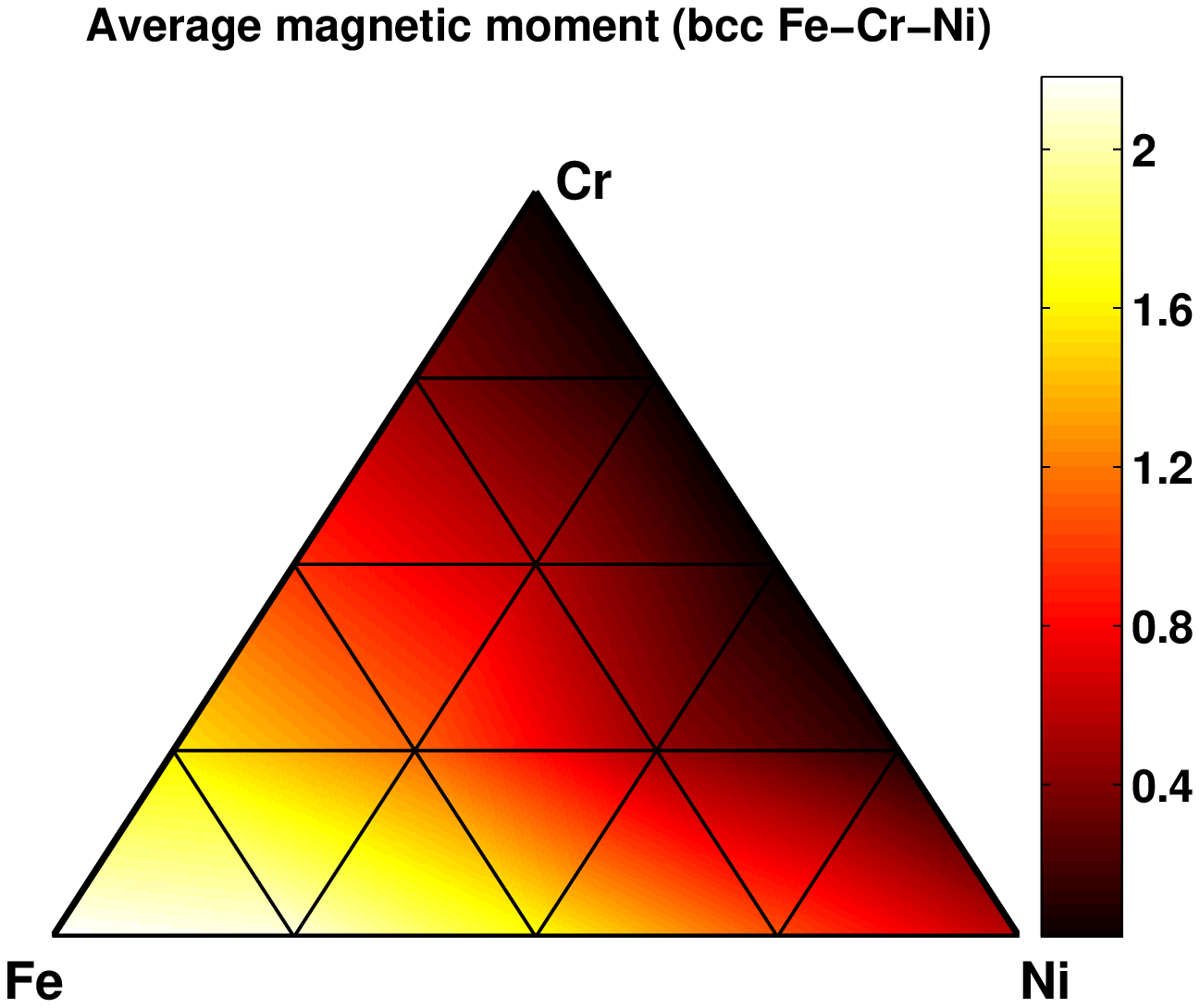}
			\end{minipage}
			\newline
			\begin{minipage}{.50\textwidth}
			  	\centering
			  	e)\includegraphics[width=.9\linewidth]{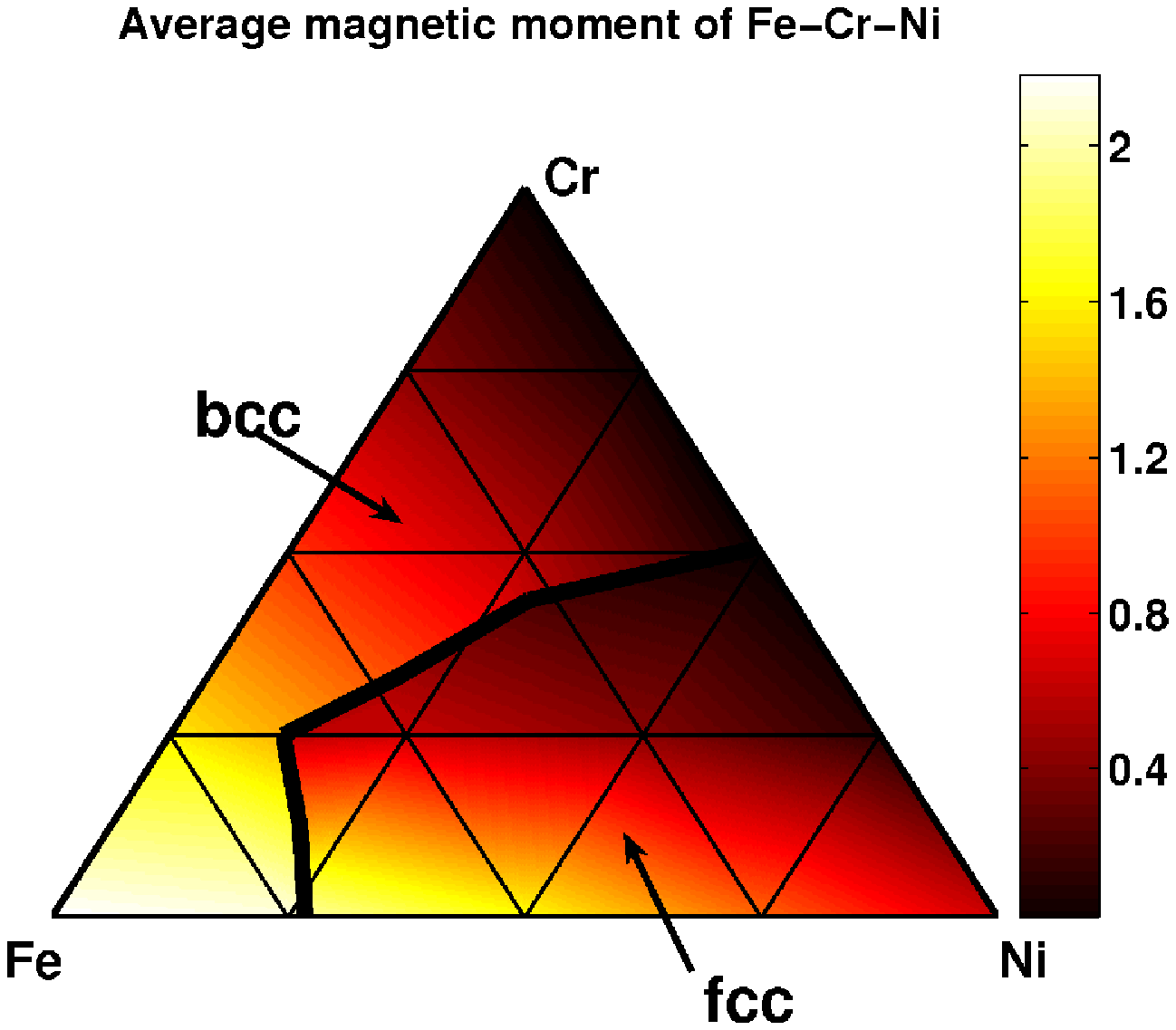}
			\end{minipage}			
			\caption{(Color online) Magnetic moment per atom (in $\mu_B$) in the most stable fcc (a, c) and bcc (b, d) ordered alloy structures predicted by DFT calculations at 0K for each alloy composition. Filled and open circles in (a, b) correspond to DFT data above and below the interpolated values represented by the respective surfaces. (c, d) are the orthogonal projections of (a, b). (e) The average atomic magnetic moment of fcc and bcc structures is discontinuous across the fcc-bcc phase transition line, shown in the figure as solid black line.
		}
		\label{fig:mag_FeCrNi}
\end{figure*}

Volumes per atom of fcc and bcc Fe-Cr-Ni ternary alloy structures computed using DFT at 0K are shown in Fig. \ref{fig:volumes_FeCrNi}. Both fcc and bcc alloy configurations exhibit the largest volume per atom in the Cr-rich corner of the diagram. Atomic volume is smallest in the Ni-rich corner. The difference between the two values is larger for fcc alloys. Atomic volumes of fcc structures exhibit a significant degree of non-linearity as functions of alloy composition. This is explained by different magnetic behaviour of fcc and bcc alloys, see Figure \ref{fig:mag_FeCrNi}(a-d), treated as a function of alloy composition. A relation between fcc-bcc phase stability and magnetic moments of the most stable structures, as well as the discontinuity in the magnitude of the average magnetic moment at the fcc-bcc phase transition line, are illustrated in Fig. \ref{fig:mag_FeCrNi}(e).

Average magnetic moments in bcc alloys are almost linear functions of Fe content. Magnetic moments are maximum for the Fe-rich alloy compositions and minimum for the anti-ferromagnetically ordered Cr-rich alloys. Fcc Fe-rich alloys do not exhibit large average magnetic moments, which are ordered anti-ferromagnetically, similarly to Fe-Ni alloys discussed in Section III.B. Alloys corresponding to the centre of the composition triangle, characterized by the approximately equal amounts of Fe, Cr and Ni, have relatively small average magnetic moments. The average magnetic moment decreases rapidly with increasing Cr content. For example, the average atomic magnetic moment in fcc (Fe$_{0.5}$Ni$_{0.5}$)$_{1-x}$Cr$_x$ alloys is 1.63, 0.97, 0.69 and 0.00 $\mu_B$ for Cr content $x$ = 0.0, 0.2, 0.33 and 0.5, respectively. These results are in agreement with experimental observations, performed at 4.2K, and showing that magnetization decreases rapidly in Fe$_{0.65}$(Cr$_x$Ni$_{1-x}$)$_{0.35}$ alloys as a function of Cr content in the interval from $x$ = 0.0 to 0.2 \cite{Rode1976}. This effect is also responsible for the observed reduction of the Curie temperature as a function of Cr content in Fe$_{0.65}$(Cr$_x$Ni$_{1-x}$)$_{0.35}$ and (Fe$_{0.5}$Ni$_{0.5}$)$_{1-x}$Cr$_x$ alloys, described in Refs. \onlinecite{Rode1976} and \onlinecite{Bansal1976}. Non-linear variation of magnetic moments as functions of alloy composition in fcc alloys results in deviations from Vegard's law. Despite the fact that Cr atoms have larger size, the volume per atom of fcc (FeNi)$_{1-x}$Cr$_x$ alloys decreases as a function of $x$, and is 11.33, 11.20, 11.09 and 10.92 $\AA^3$/atom for $x$ = 0.0, 0.2, 0.33 and 0.5, respectively.  Results for other compositions are given in Supplementary Material.

\begin{figure*}
			\centering
			\begin{minipage}{.50\textwidth}
			  	\centering
			  	a)\includegraphics[width=.85\linewidth]{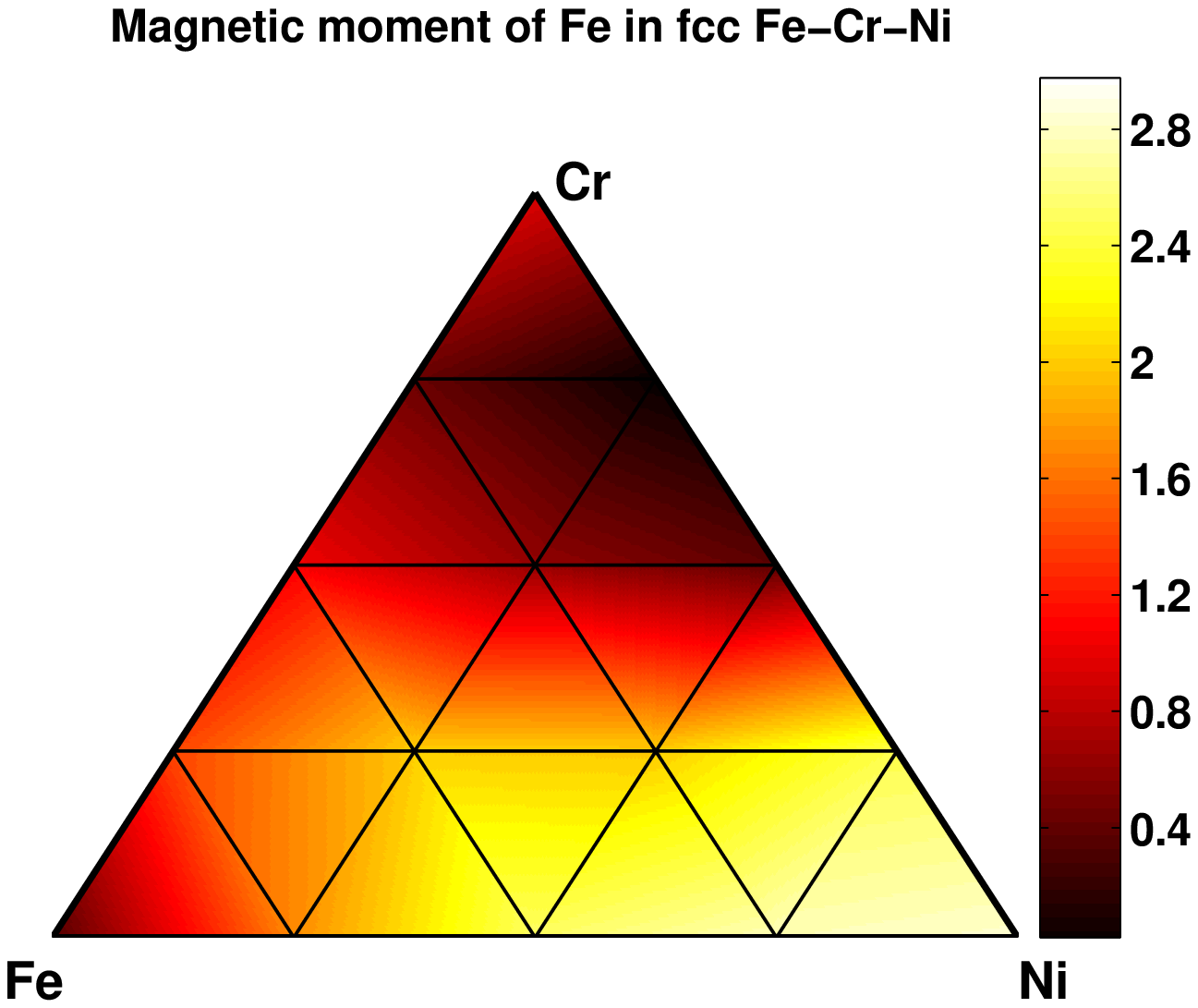}			  	
			\end{minipage}%
			\begin{minipage}{.50\textwidth}
			  	\centering
			  	b)\includegraphics[width=.85\linewidth]{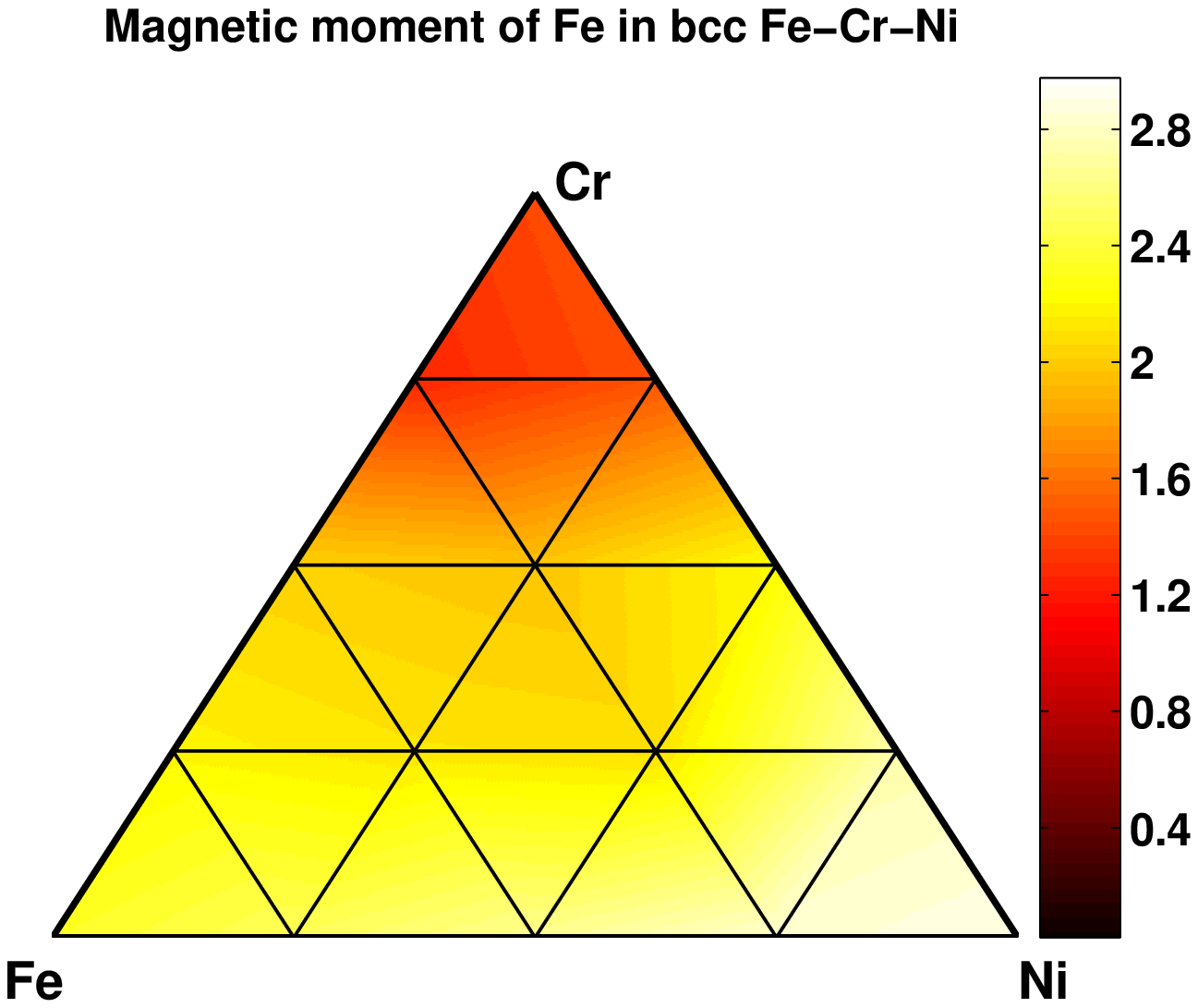}
			\end{minipage}
			\newline
			\begin{minipage}{.50\textwidth}
			  	\centering
			  	c)\includegraphics[width=.85\linewidth]{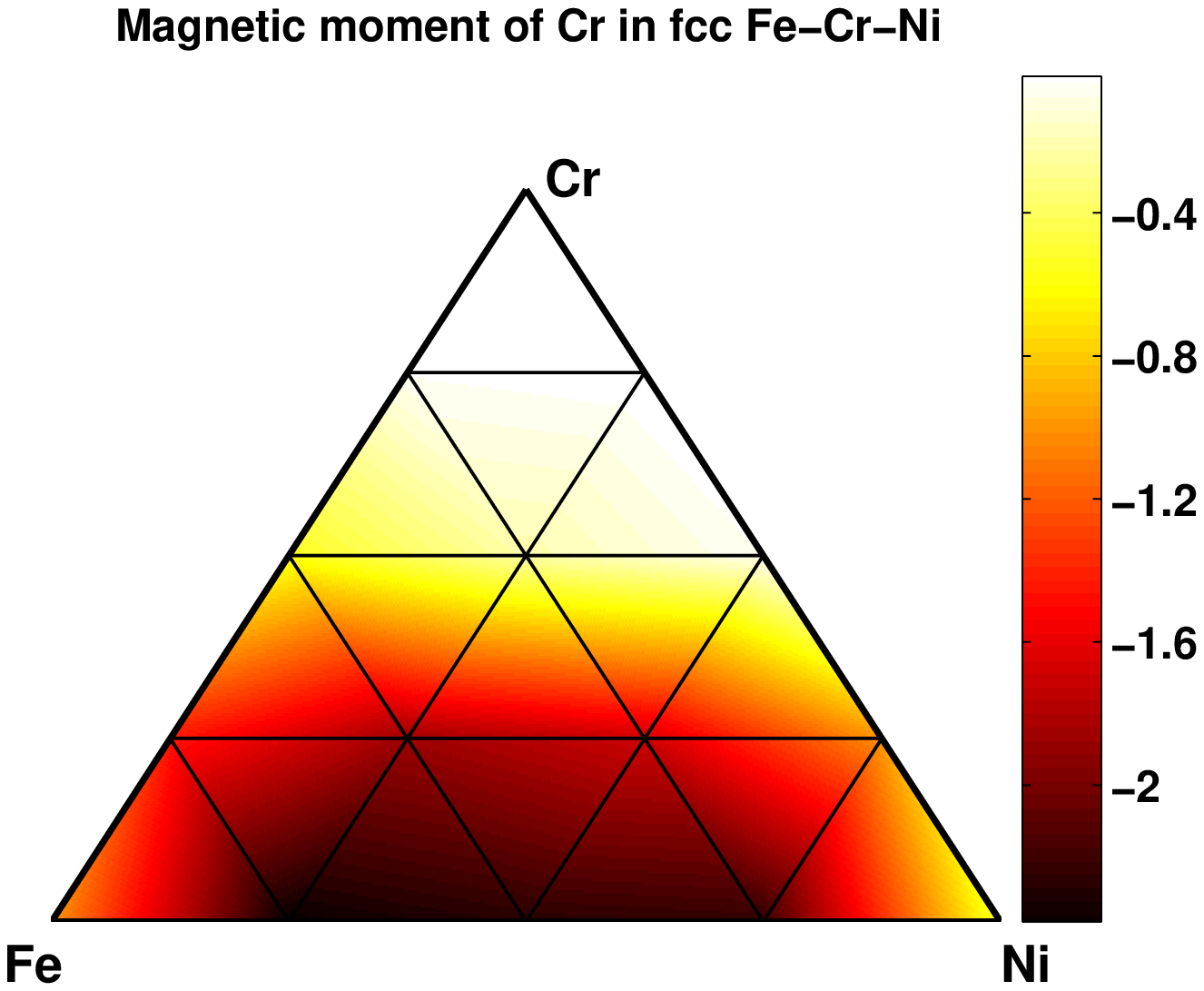}			  	
			\end{minipage}%
			\begin{minipage}{.50\textwidth}
			  	\centering
			  	d)\includegraphics[width=.85\linewidth]{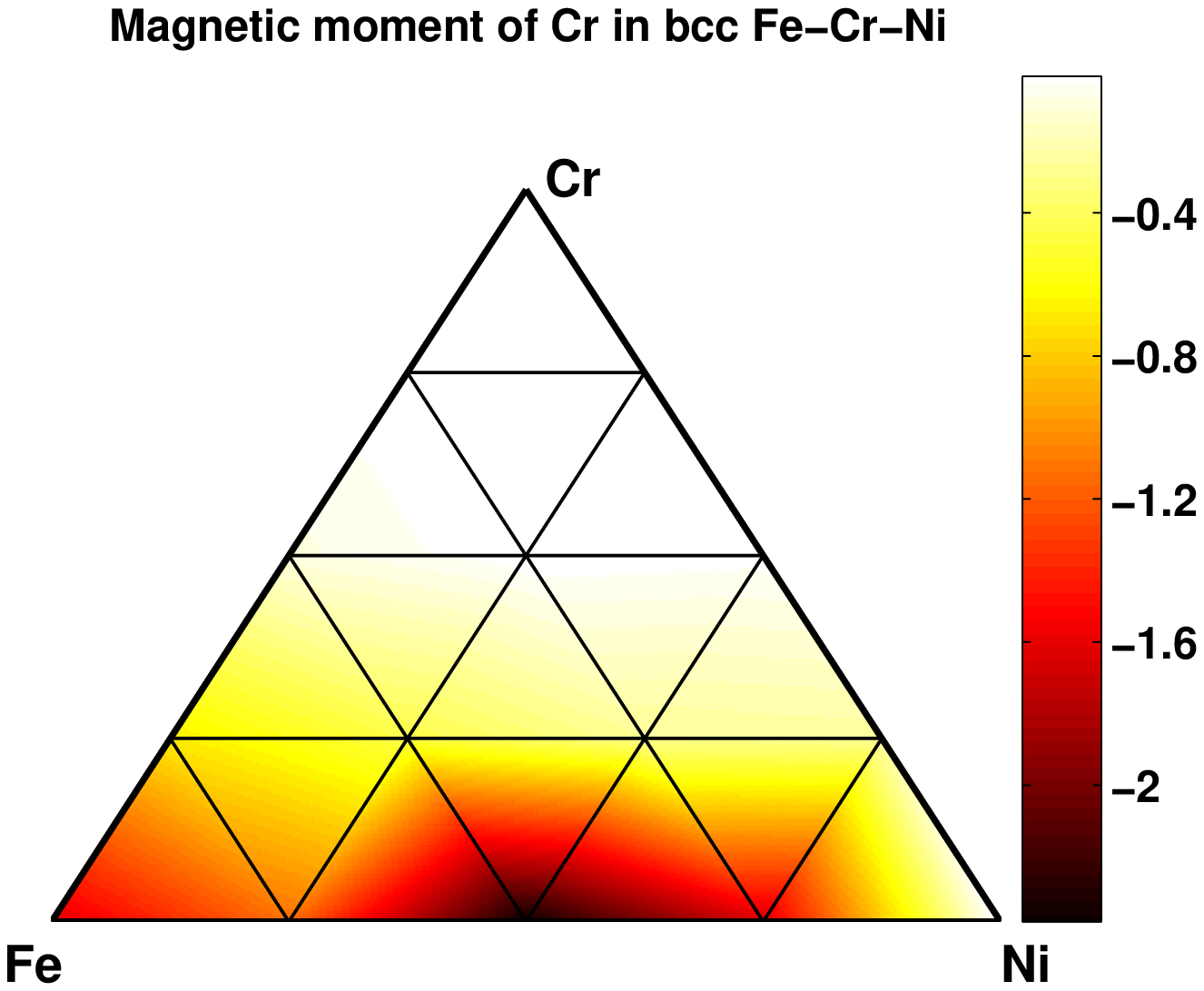}
			\end{minipage}
			\newline
			\begin{minipage}{.50\textwidth}
			  	\centering
			  	e)\includegraphics[width=.85\linewidth]{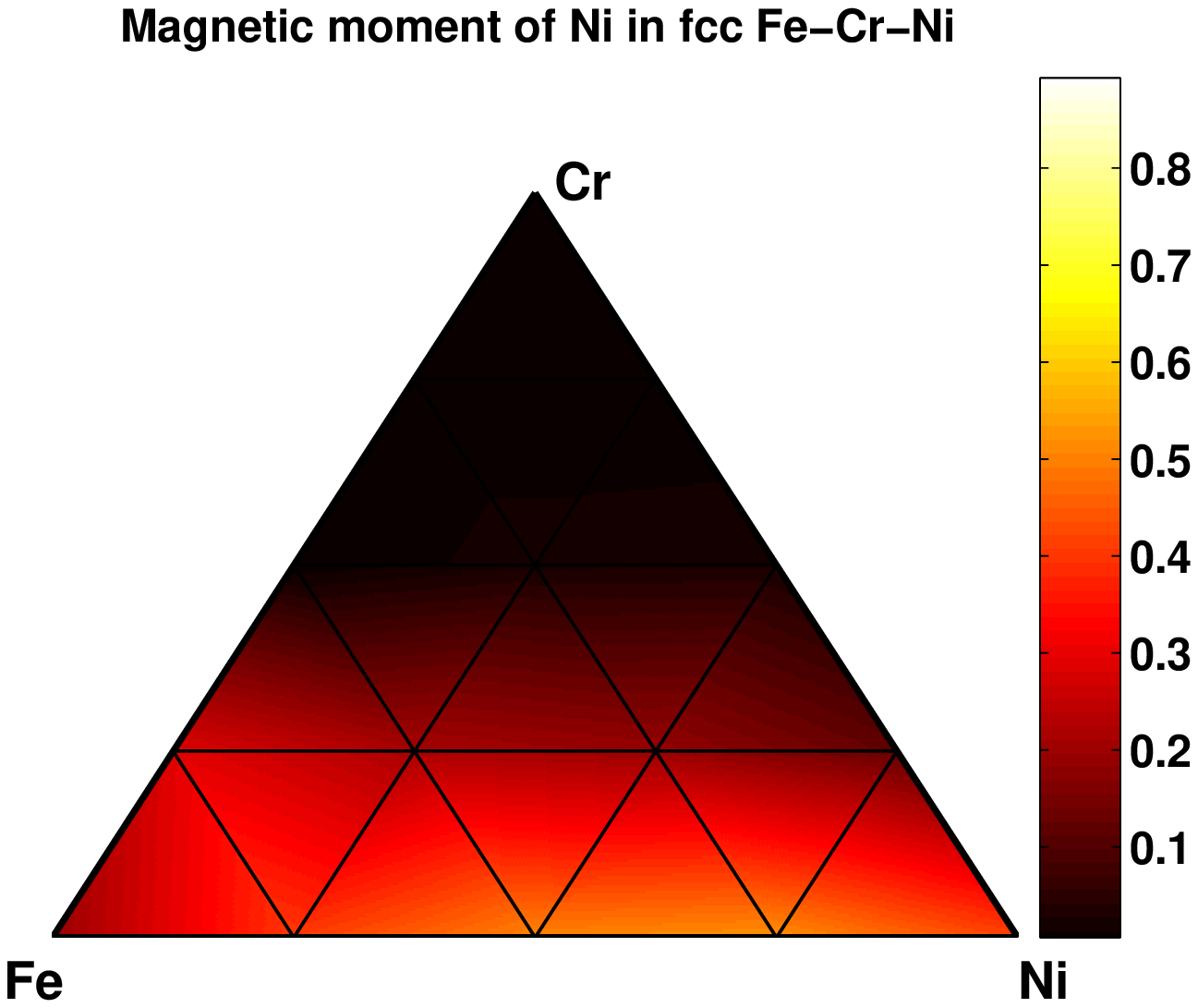}			  	
			\end{minipage}%
			\begin{minipage}{.50\textwidth}
			  	\centering
			  	f)\includegraphics[width=.85\linewidth]{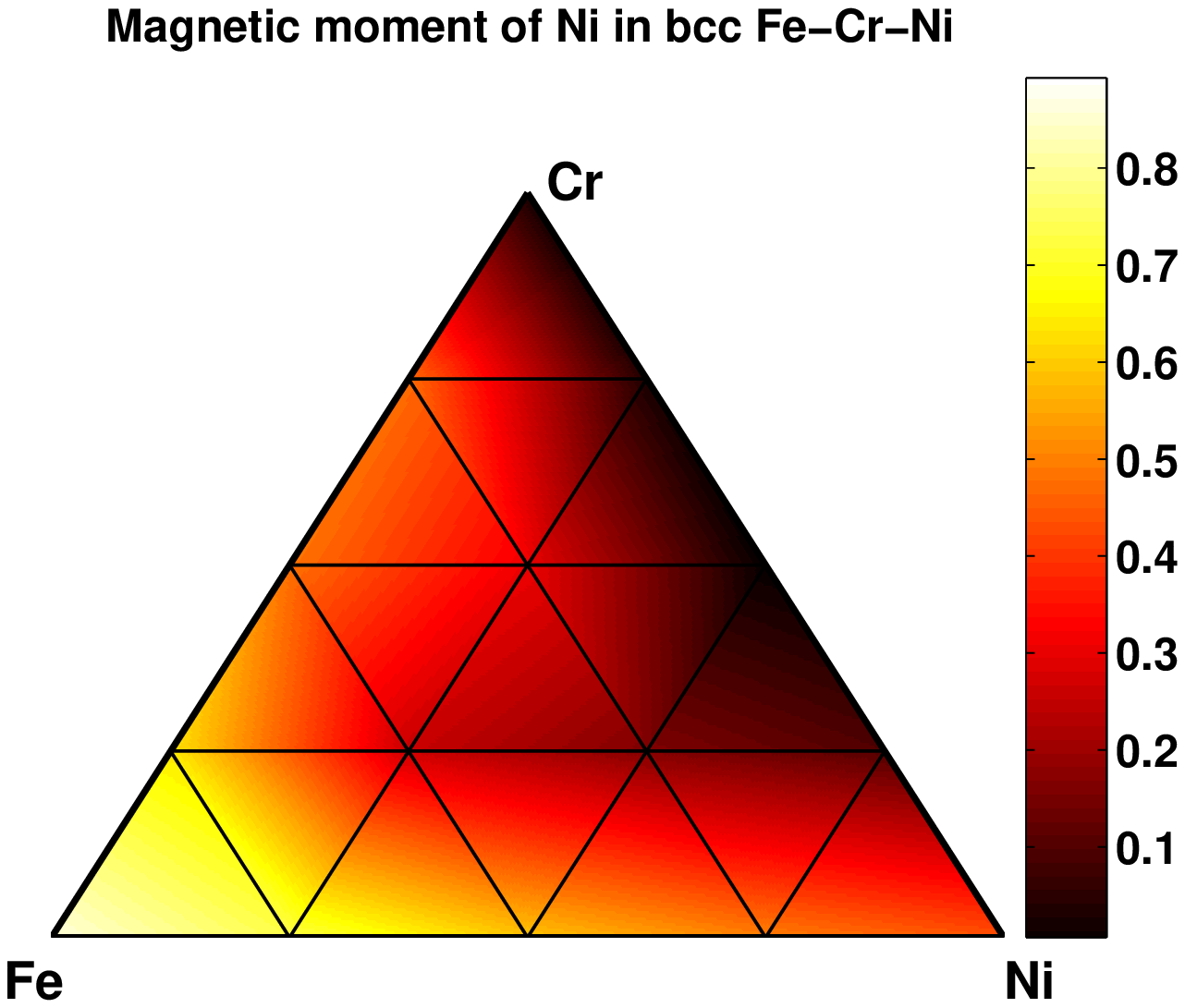}
			\end{minipage}
			\newline
			\caption{ (Color online) Magnetic moments (in $\mu_B$) of Fe (a, b), Cr (c, d) and Ni (e, f) on fcc (a, c, e) and bcc (b, d, f) lattices.
		}
		\label{fig:mag_FeCrNi_components}
\end{figure*}

Magnetic moments of each component of fcc and bcc alloys are shown in Fig. \ref{fig:mag_FeCrNi_components}. The results exhibit a rapid decrease of magnetic moments on Ni sites as functions of Cr content in fcc alloys (where magnetic moments on Ni sites in alloys containing more than 33\% Cr are close to zero). Cr atoms prefer their magnetic moments ordered anti-ferromagnetically with respect to Fe and Ni moments. Their magnitudes are larger at low Cr concentration, and even at 25 \% Cr concentration they are fairly large (-2.44 $\mu_B$ and -2.53 $\mu_B$ for Fe$_2$CrNi and FeCrNi$_2$ structures). Because of strong anti-ferromagnetic interactions between Fe and Cr atoms, structures with large magnetic moments on Cr sites also have large magnetic moments on Fe sites (2.09 $\mu_B$ and 2.31 $\mu_B$ for Fe$_2$CrNi and FeCrNi$_2$ structures). An exception from this rule is the Fe-rich corner of the diagram, where fcc structures remain anti-ferromagnetic and the mean magnetic moment as well as average magnetic moments of the constituting components are equal or close to zero.

Similarly to fcc structures, magnetic moments of Cr atoms on bcc lattice in the dilute Cr limit order anti-ferromagnetically with respect to those of Fe and Ni atoms, and their magnitudes decrease rapidly as a function of Cr content. For example, magnetic moments of Cr atoms in bcc Fe$_2$CrNi and FeCrNi$_2$ alloys are -0.19 $\mu_B$ and -0.12 $\mu_B$. In other words, they are an order of magnitude smaller than those found in fcc alloys. Furthermore, Cr-rich structures are not non-magnetic, as they are in the fcc case, but anti-ferromagnetic. Unlike Cr atoms, the magnitudes of magnetic moments on Fe and Ni sites are larger in bcc than in fcc alloys. Average moments on Fe sites are larger than 1 $\mu_B$ for most of the compositions, with the maximum value of 2.94 $\mu_B$ corresponding to FeCrNi$_{14}$ structure in the Ni-rich corner of the diagram. Average magnetic moments of Ni atoms are close to zero only in bcc Cr-rich Cr-Ni binary alloys. As the Fe content increases, the average magnetic moment of Ni atoms increases, too, reaching a maximum value of 0.86 $\mu_B$ in the Fe-rich corner, modelled by Fe$_{14}$CrNi structure.

\section{Finite temperature phase stability of Fe-Cr-Ni alloys}
\subsection{Enthalpy of formation}

The finite temperature phase stability of Fe-Cr-Ni alloys was analyzed using quasi-canonical MC simulations and ECIs derived from DFT calculations. MC simulations were performed for 63 different compositions spanning all the binary and ternary Fe-Cr-Ni alloys on a 10\% composition mesh for each of the three constituents of the alloy, and additional 12 compositions with Cr and Ni content varying from 5\% to 35\% and from 25\% to 45\%, respectively, to increase the composition mesh density in the vicinity of the fcc-bcc phase transition line.

\begin{figure*}
			\centering
			\begin{minipage}{.50\textwidth}
			  	\centering
			  	a)\includegraphics[width=.9\linewidth]{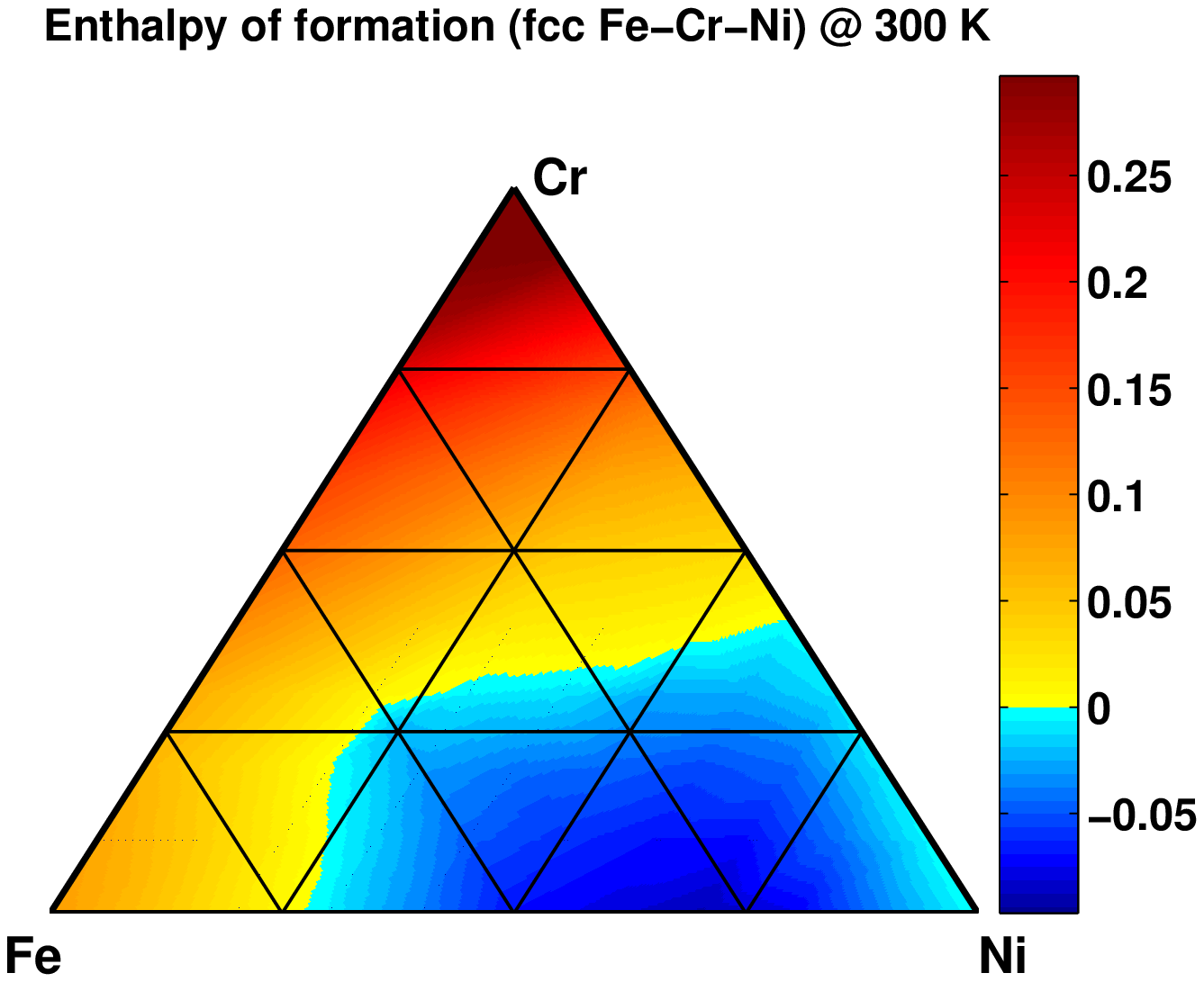}			  	
			\end{minipage}%
			\begin{minipage}{.50\textwidth}
			  	\centering
			  	b)\includegraphics[width=.9\linewidth]{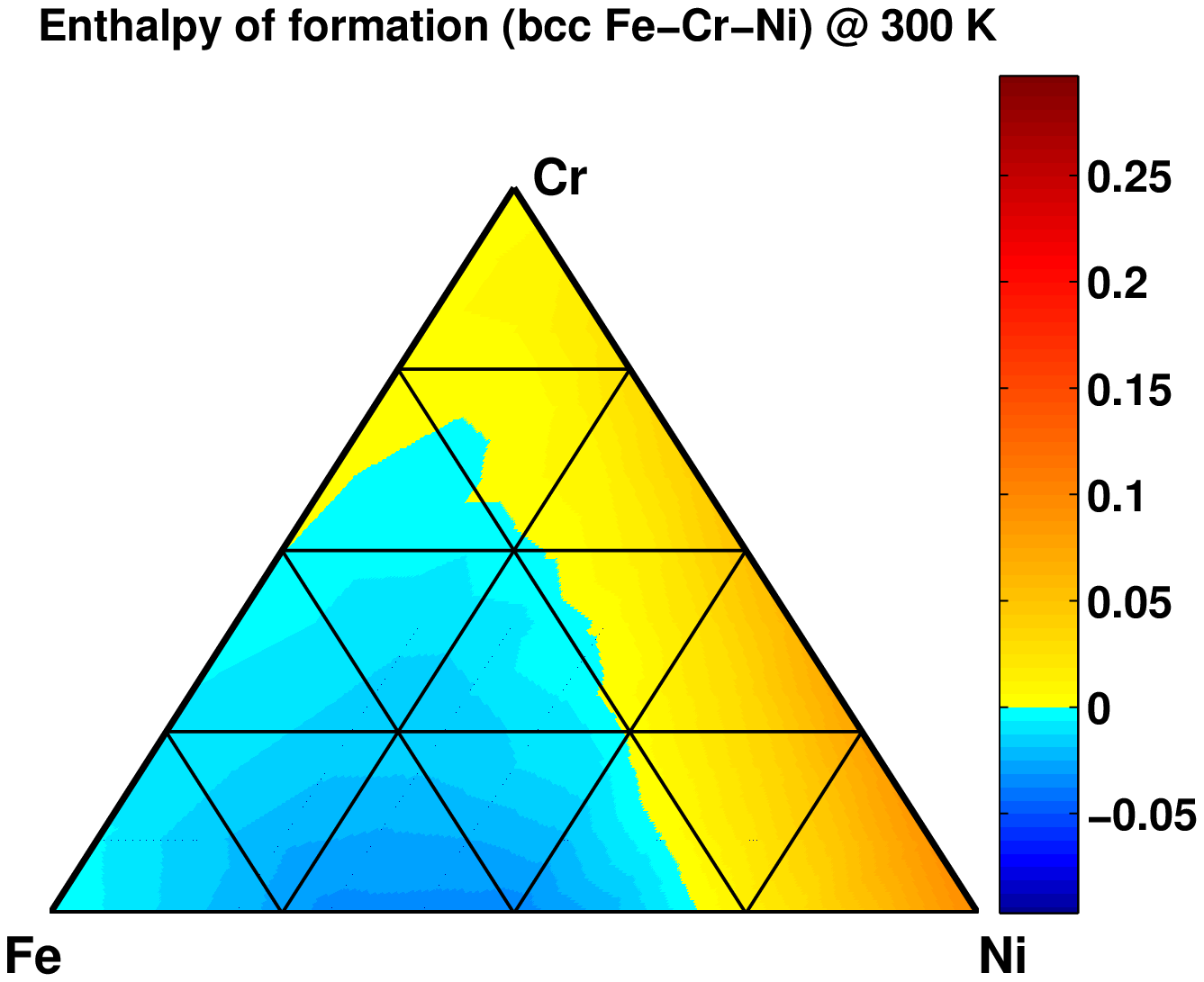}
			\end{minipage}
			\caption{
(Color online) Enthalpies of formation (in eV/atom) of fcc (a) and bcc (b) alloys calculated using MC simulations at 300K. }
		\label{fig:Enthalpies_low_temp}
		\end{figure*}

Enthalpies of formation of fcc and bcc alloys at 300 K are shown in Fig. \ref{fig:Enthalpies_low_temp}. In fcc and bcc alloys there is a large region of concentrations where enthalpies of mixing are negative, coloured blue in Fig. \ref{fig:Enthalpies_low_temp}. Negative enthalpies of formation correspond to the fact that alloys decompose into mixtures of intermetallic phases. The negative formation enthalpies of fcc Fe-Cr-Ni alloys are mainly due to the formation of fcc FeNi, FeNi$_3$, FeNi$_8$ and CrNi$_2$ binary phases and fcc Fe$_2$CrNi ternary phase. In bcc Fe-Cr-Ni alloys, a negative enthalpy of formation is primarily due to the formation of Fe-Cr $\alpha$-phase and Fe$_4$Ni$_5$ VZn-like phase, where the latter is the most stable Fe-Ni phase on bcc lattice. The fact that Fe$_4$Ni$_5$ VZn-like phase is not observed experimentally is because it is significantly less stable than the corresponding fcc phase.

\begin{figure*}
			\centering
			\begin{minipage}{.50\textwidth}
			  	\centering
			  	a)\includegraphics[width=.9\linewidth]{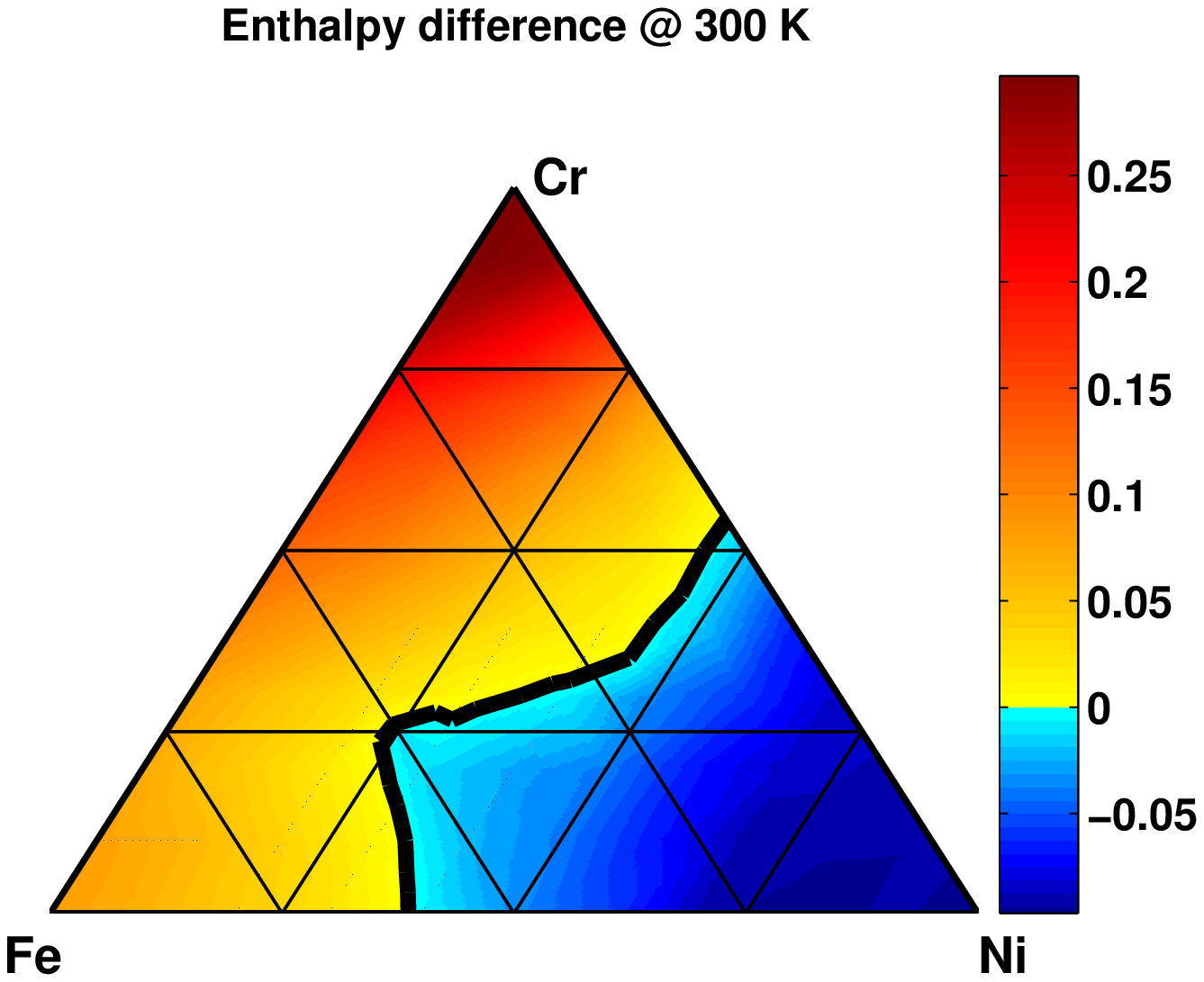}			  	
			\end{minipage}%
			\begin{minipage}{.50\textwidth}
			  	\centering
			  	b)\includegraphics[width=.9\linewidth]{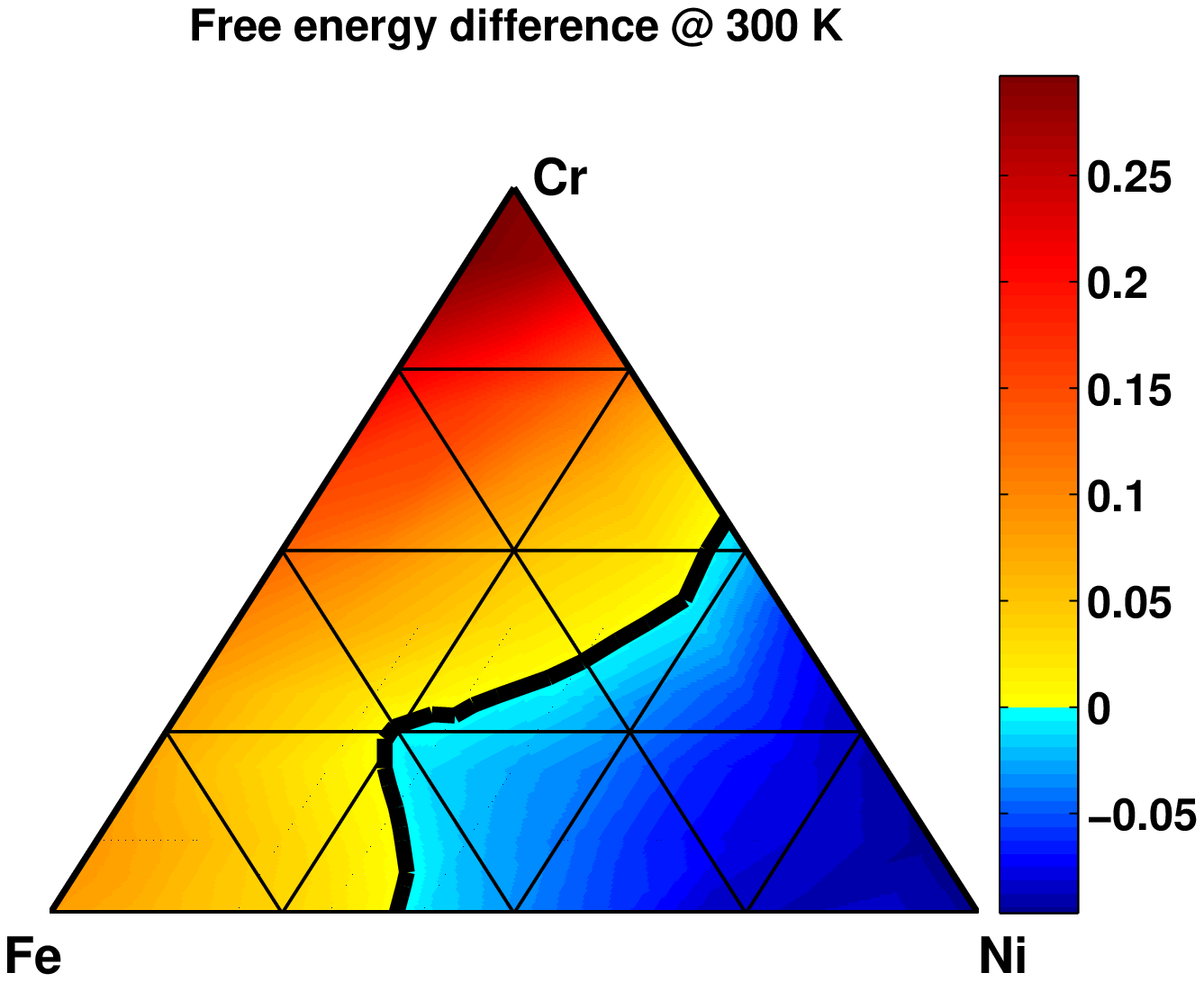}
			\end{minipage}
						\newline
			\begin{minipage}{.50\textwidth}
			  	\centering
			  	c)\includegraphics[width=.9\linewidth]{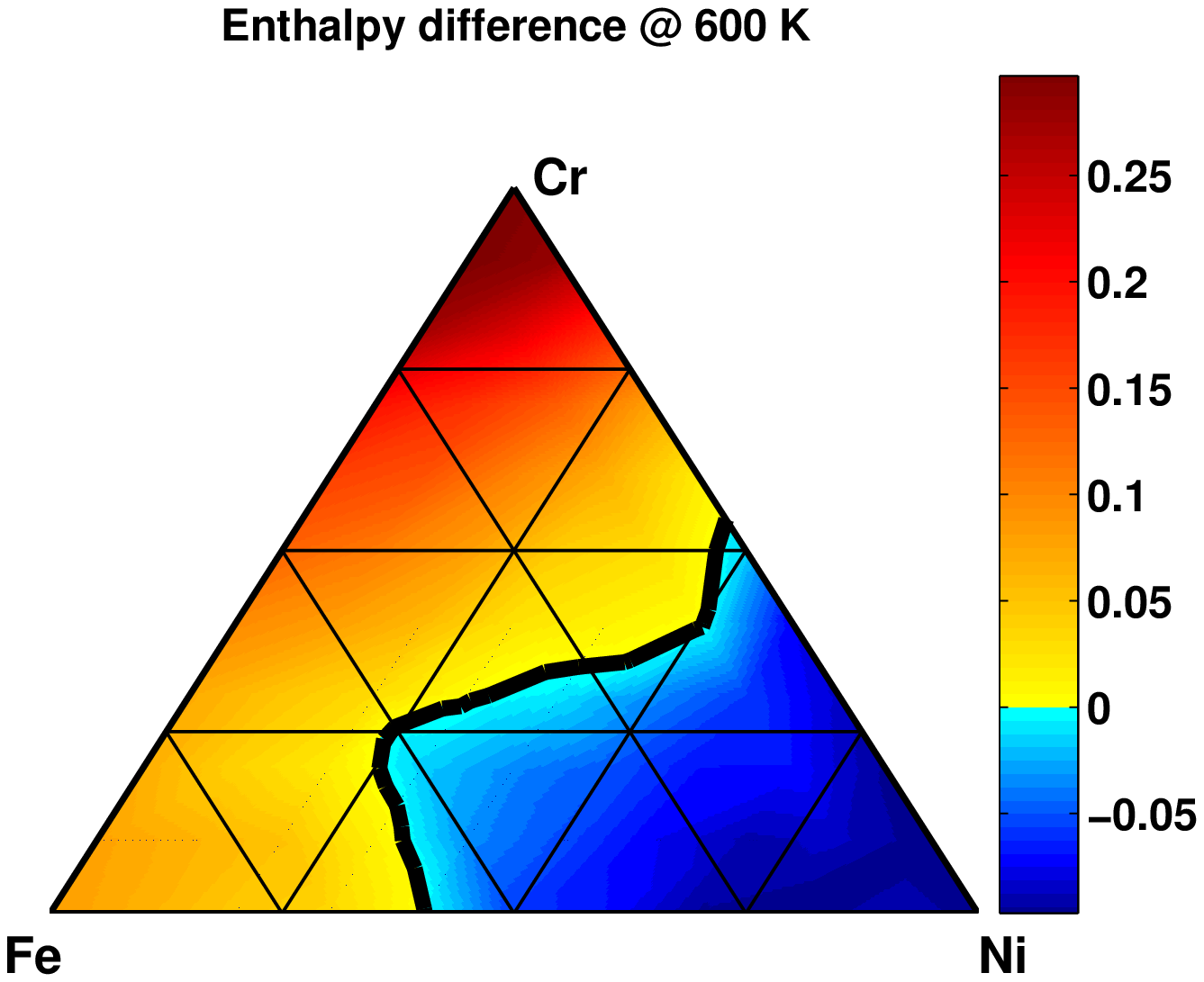}			  	
			\end{minipage}%
			\begin{minipage}{.50\textwidth}
			  	\centering
			  	d)\includegraphics[width=.9\linewidth]{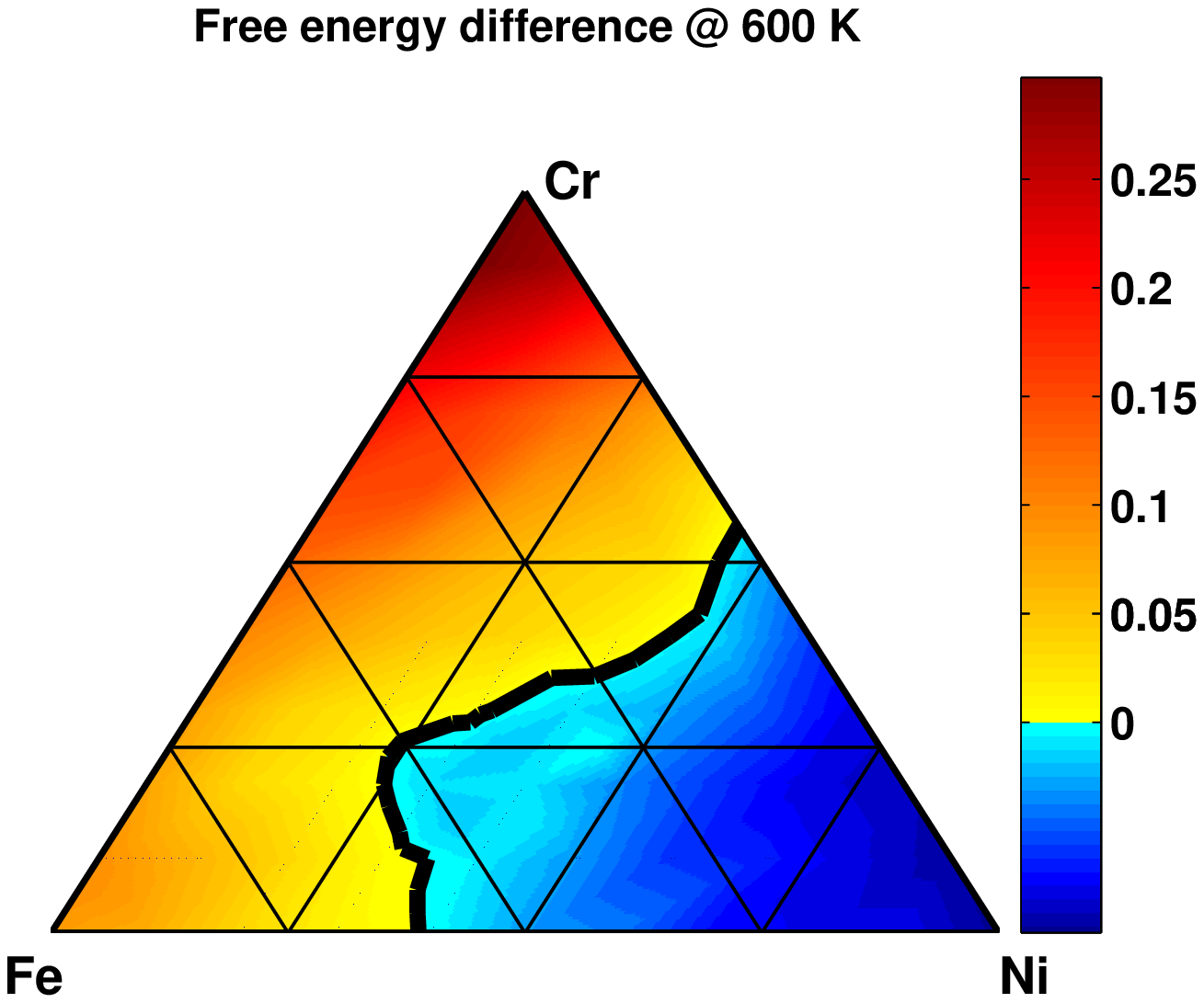}
			\end{minipage}
						\newline
			\begin{minipage}{.50\textwidth}
			  	\centering
			  	e)\includegraphics[width=.9\linewidth]{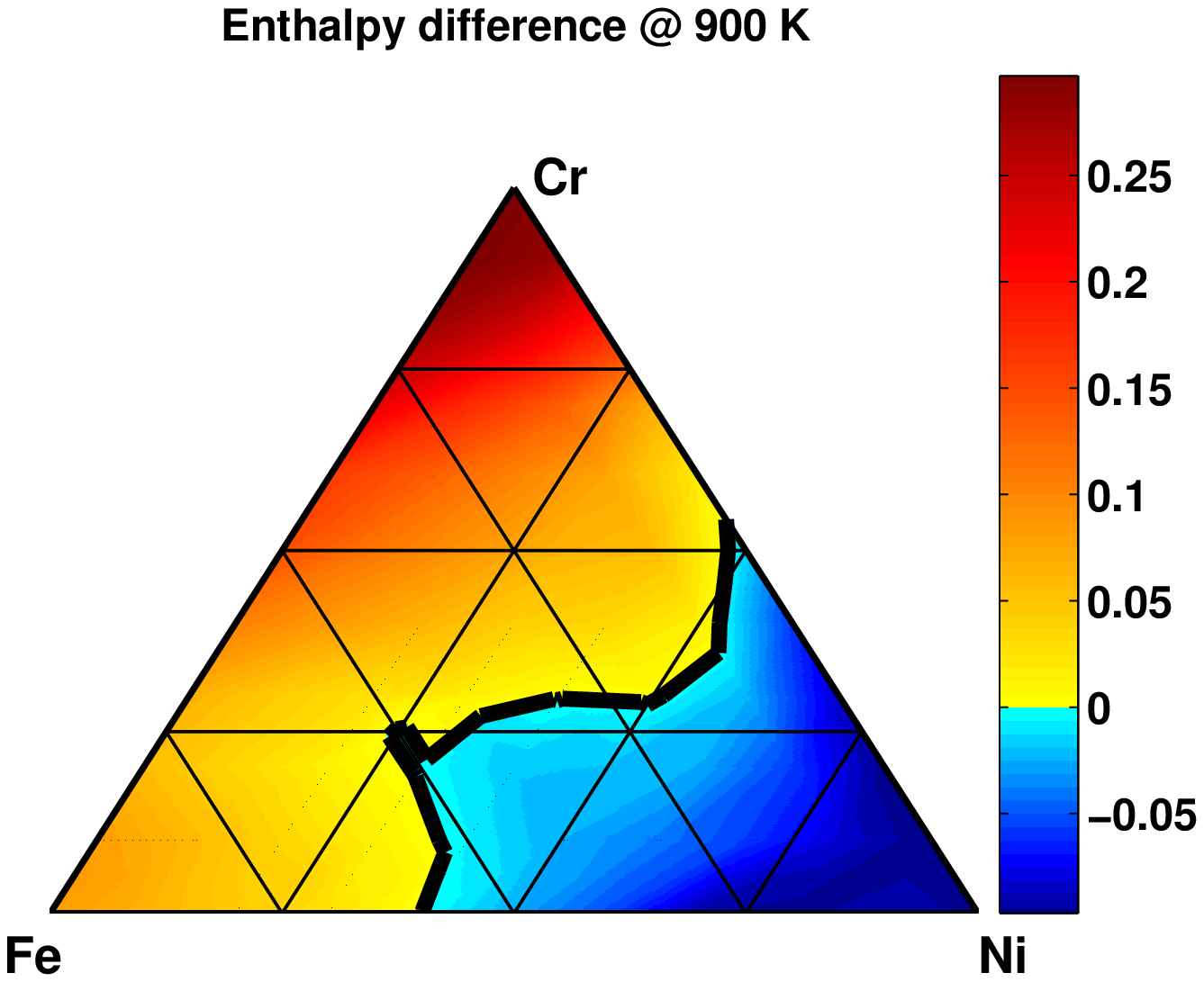}			  	
			\end{minipage}%
			\begin{minipage}{.50\textwidth}
			  	\centering
			  	f)\includegraphics[width=.9\linewidth]{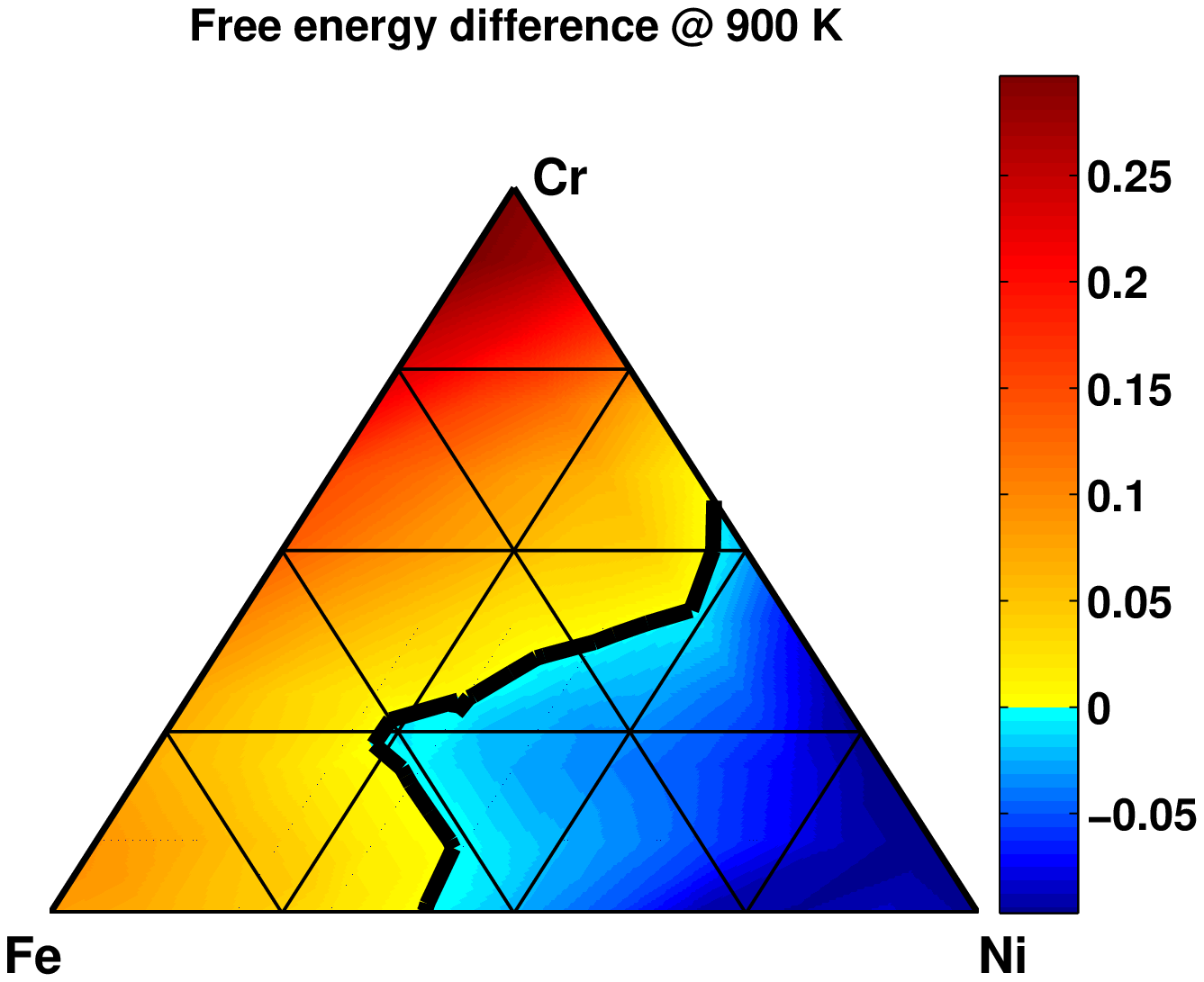}
			\end{minipage}
			\caption{
		(Color online) Difference between enthalpies of formation (a,c) and free energies of formation (b,d) of fcc and bcc alloys, in eV/atom units, predicted using MC simulations for 300 K (a,b), 600 K (c,d) and 900 K (e,f). Black solid lines separate the Ni-rich region of stability of fcc alloys and the region of stability of bcc alloys predicted using the enthalpy and free energy criteria. }
		\label{fig:Energy_diff}
\end{figure*}

Having evaluated the difference between formation enthalpies of fcc and bcc alloys at 300K, we can now separate regions of stability of fcc and bcc alloys defined by the formation enthalpy criterion, see Fig. \ref{fig:Energy_diff}(a). The fcc-bcc phase transition lines determined using the same criterion at 600 K and 900 K, see Figs. \ref{fig:Energy_diff}(c) and \ref{fig:Energy_diff}(e), show that fcc and bcc phases remain stable broadly within the same composition ranges at 300 K and 600 K, whereas at 900 K we observe that the region of stability of fcc alloys shrinks in comparison with the region of stability of bcc alloys.

Analyzing the stability of fcc and bcc alloys using their enthalpies of formation is convenient since one compares values derived from MC simulations with those computed directly by DFT for each representative alloy configuration. However, the enthalpy criterion of phase stability is valid only at relatively low temperatures. At high temperatures one should take into account the configurational entropy as well as vibrational and magnetic entropy contributions to the entropy and enthalpy of formation. In this study we do not treat vibrational entropy effects. The magnetic and configurational entropy contributions to the free energy of formation of fcc and bcc Fe-Cr-Ni alloys, and their effect on fcc-bcc phase stability, are analyzed in Section IV.D. We first discuss the enthalpies of formation, for which theoretical values can be validated by experimental data \cite{Kubaschewski1967,Dench1963}, and the magnetic contribution to the enthalpy of formation, which plays a significant part at high temperatures.

Table \ref{tab:Fe} shows that pure iron at low temperatures is stable in bcc $\alpha$-phase whereas at 1185 K \cite{Chen2001} it transforms into the fcc $\gamma$-phase, and then back into the bcc $\delta$-phase. In order to investigate the formation enthalpies of alloys at high temperature one should at least take into account the effect of thermal magnetic excitations in Fe. The following correction can then be applied to the formation enthalpies of alloys in the high temperature limit. It is proportional to the concentration of Fe and is based on results given in Fig. 2 of Ref. \onlinecite{Lavrentiev2010}:
\begin{equation}
\Delta H_{lat}^{corr}\approx c_{Fe}\left[\left(E_{lat}(T)-E_{lat}(0)\right)-\left(E_{GS}(T)-E_{bcc}(0)\right)\right]  ,
\label{eq:Correction_to_enthalpy}
\end{equation}
where $lat=fcc,bcc$, $E_{lat}(0)$ and $E_{lat}(T)$ are the energies of Fe on $lat$ at 0K and at temperature $T$ and $E_{GS}(T)$ is the temperature-dependent energy of the ground state, which is {\it either} fcc {\it or} bcc.

\begin{figure*}
			\centering
			\begin{minipage}{.50\textwidth}
			  	\centering
			  	a)\includegraphics[width=.9\linewidth]{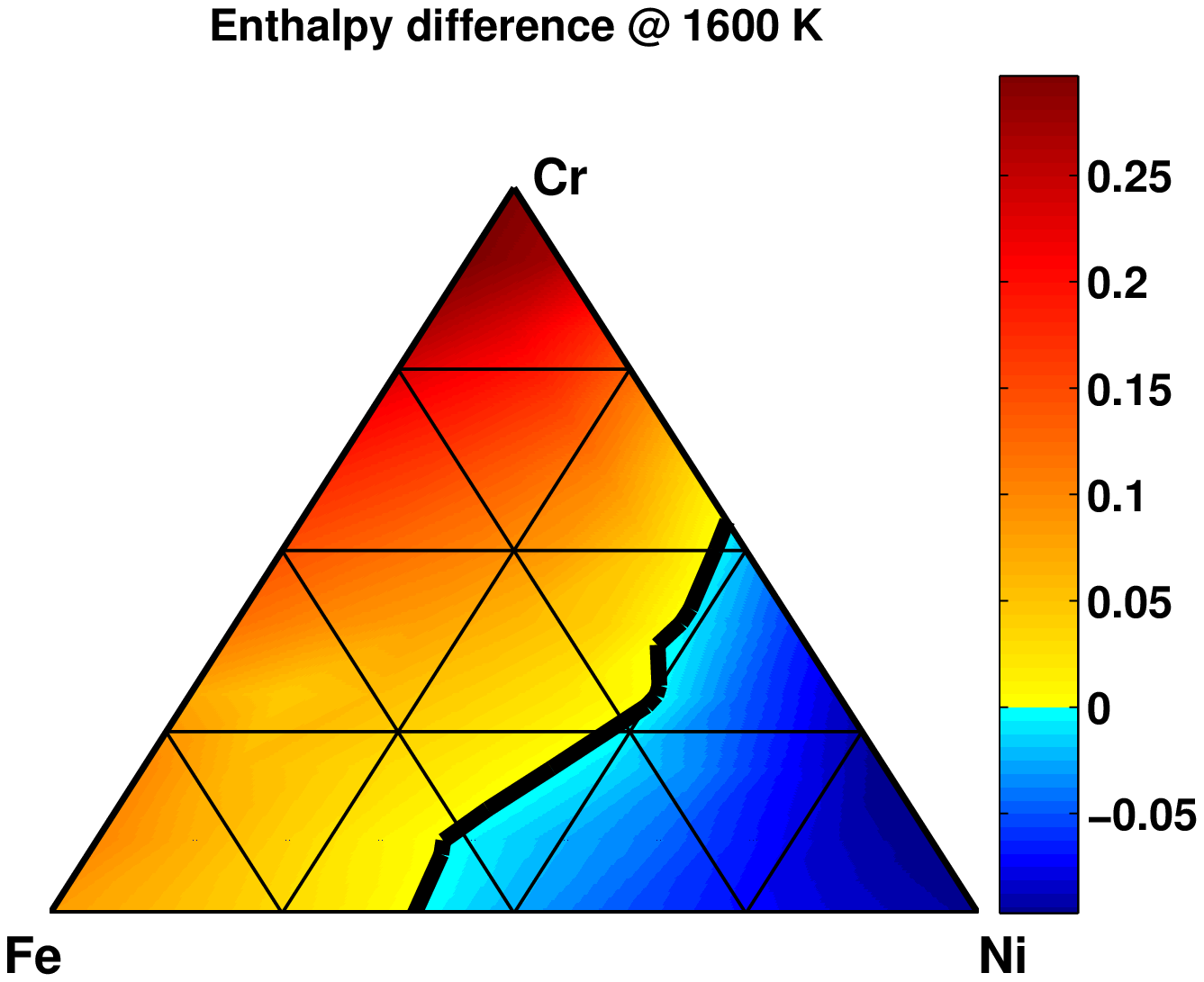}			  	
			\end{minipage}%
			\begin{minipage}{.50\textwidth}
			  	\centering
			  	b)\includegraphics[width=.9\linewidth]{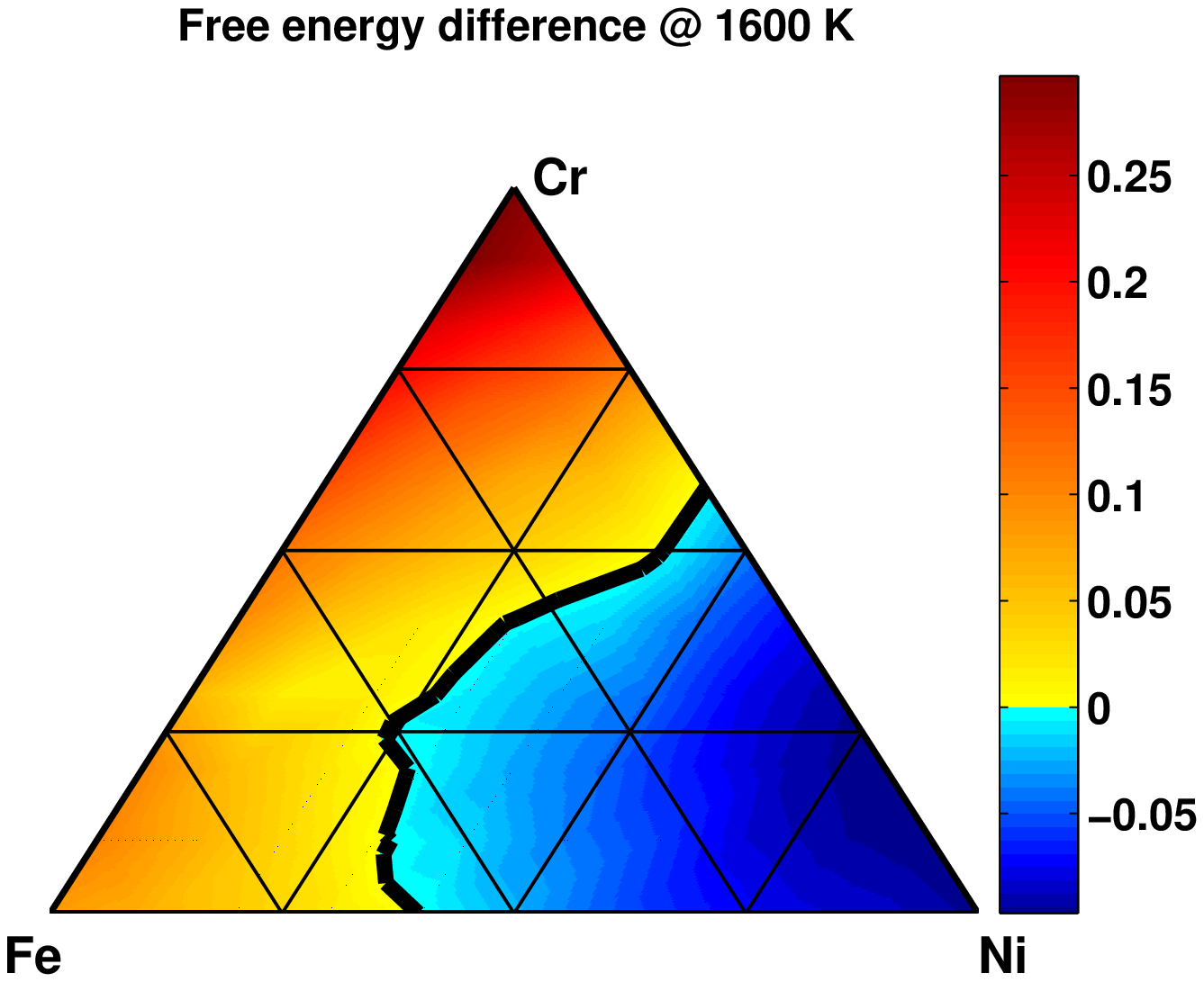}
			\end{minipage}
						\newline
			\begin{minipage}{.50\textwidth}
			  	\centering
			  	c)\includegraphics[width=.9\linewidth]{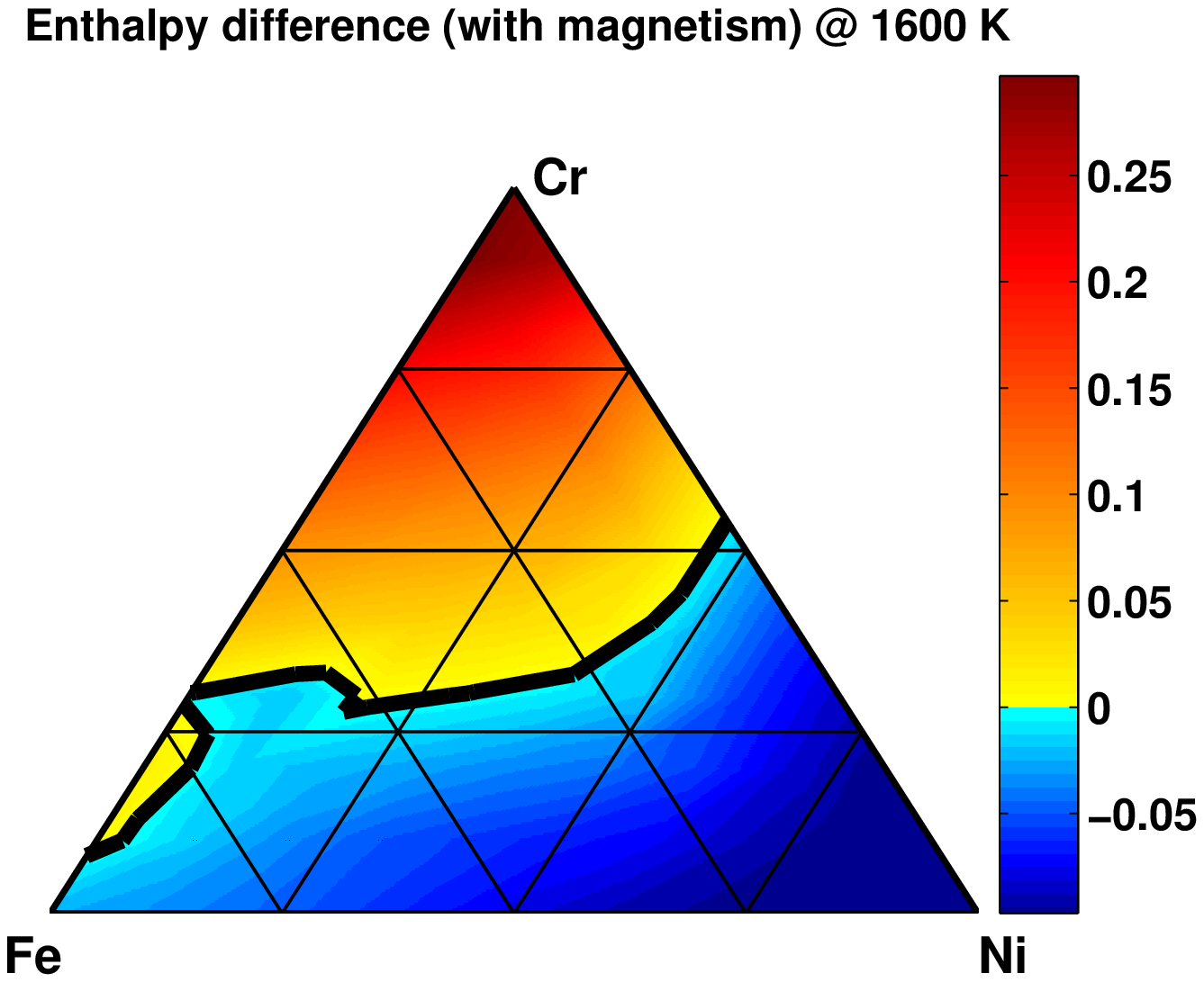}			  	
			\end{minipage}%
			\begin{minipage}{.50\textwidth}
			  	\centering
			  	d)\includegraphics[width=.9\linewidth]{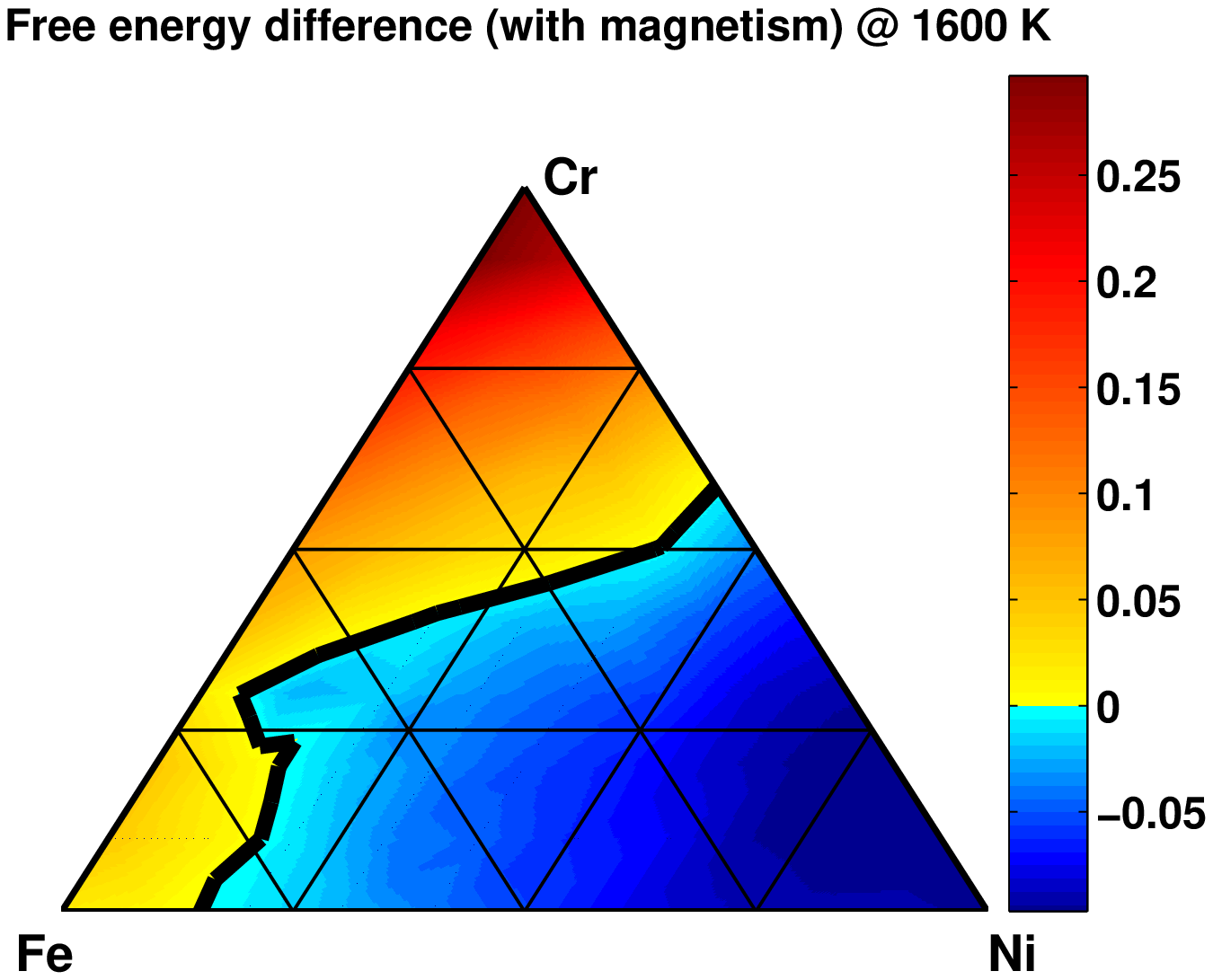}
			\end{minipage}
			\caption{
		(Color online) Difference between enthalpies of formation (a,c) and free energies of formation (b,d) of fcc and bcc alloys, in eV/atom units, calculated using MC simulations at 1600 K without (a,b) and with magnetic correction applied to the formation enthalpies (c) and free energies of formation (d). Black solid lines separate the Ni-rich region of stability of fcc alloys and the region of stability of bcc alloys predicted using the enthalpy and free energy criteria. }
		\label{fig:Energy_diff_1600K}
		\end{figure*}

\begin{figure*}
			\begin{minipage}{.50\textwidth}
			  	\centering
			  	a)\includegraphics[width=.9\linewidth]{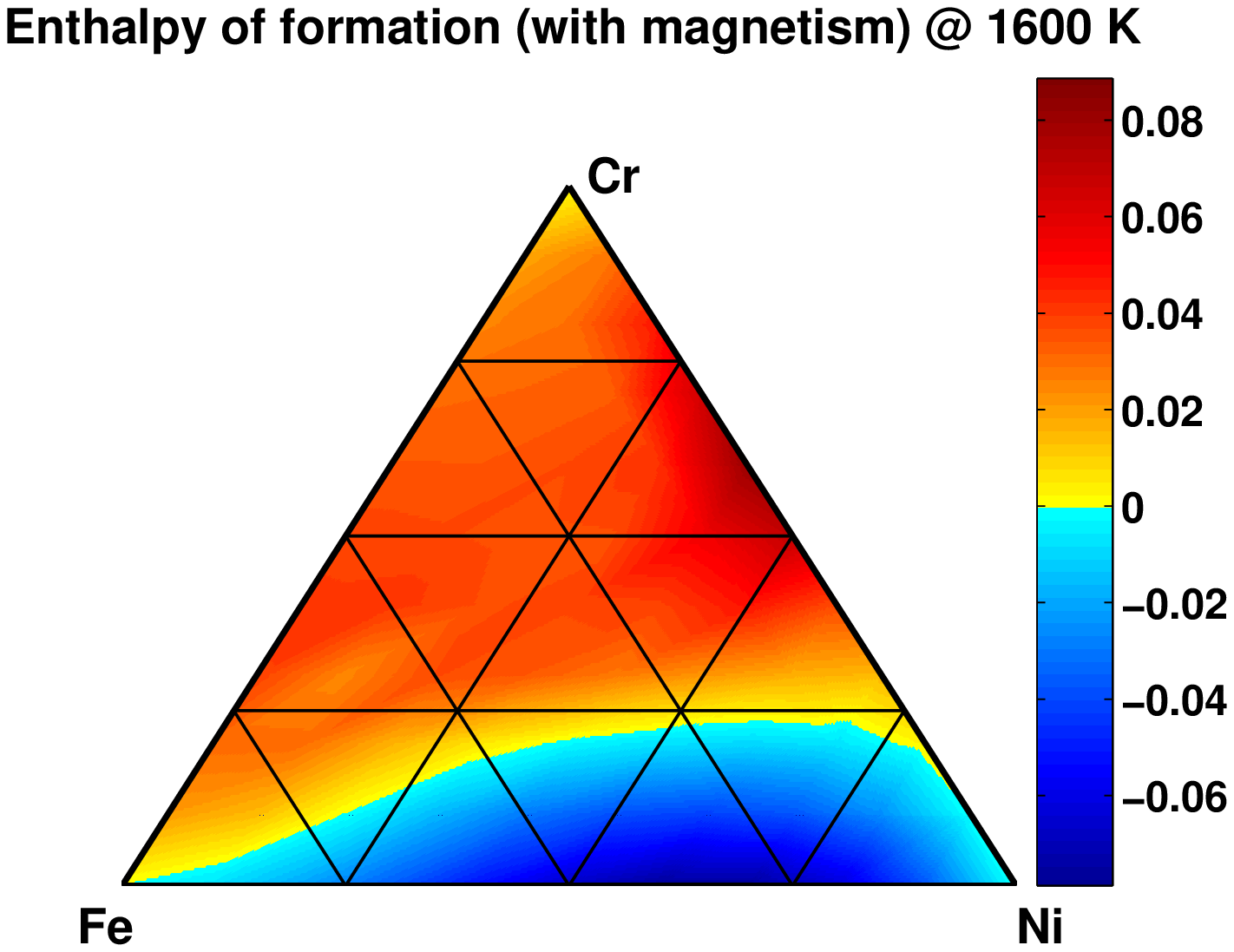}			  	
			\end{minipage}%
			\begin{minipage}{.50\textwidth}
			  	\centering
			  	b)\includegraphics[width=.9\linewidth]{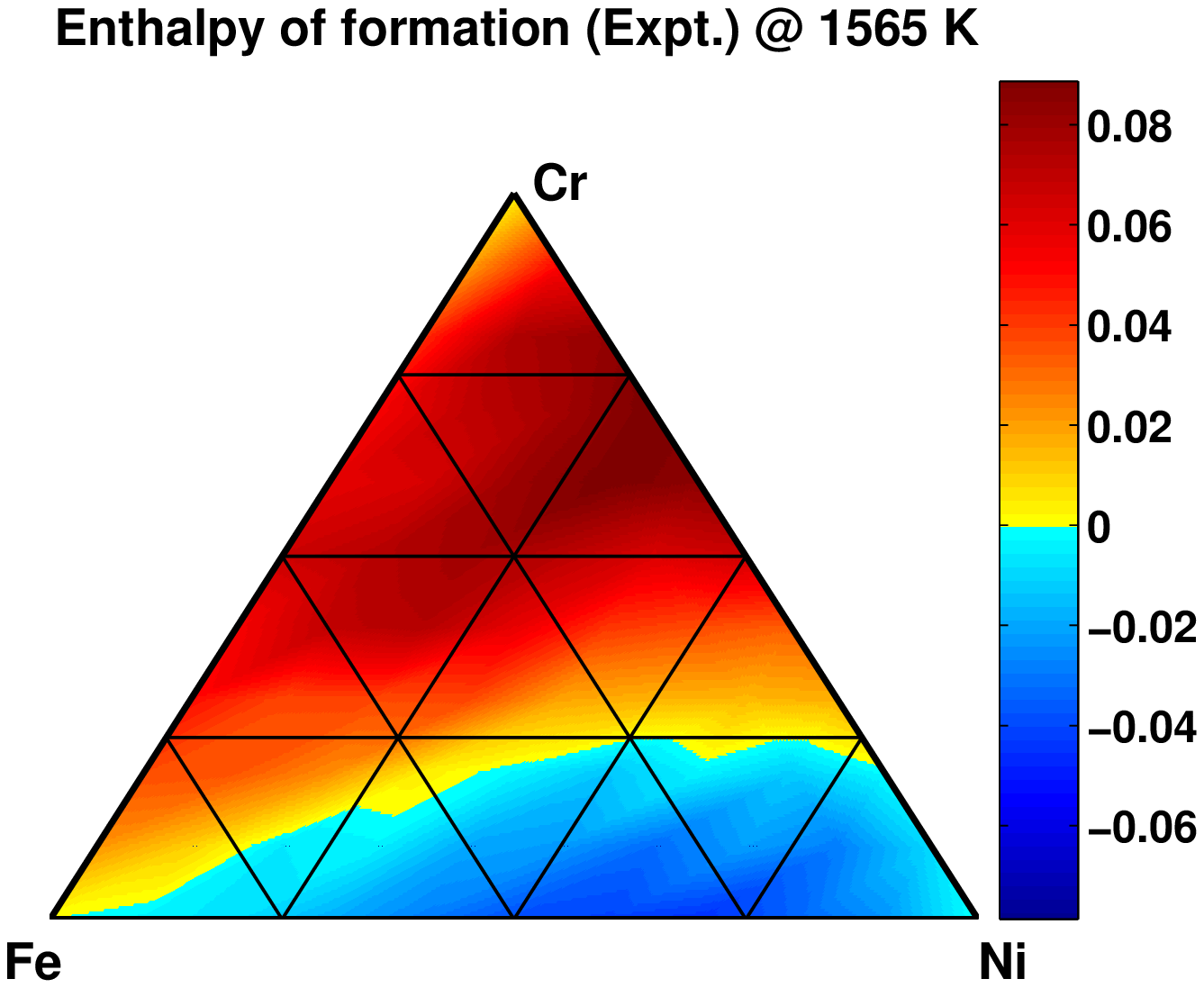}
			\end{minipage}
		\caption{
		(Color online) (a) Enthalpies of formation (in eV/atom) for the most stable crystal structures of Fe-Cr-Ni alloys computed using MC simulations at 1600K with magnetic correction applied, compared to experimental data (b) from Refs. \onlinecite{Kubaschewski1967} and \onlinecite{Dench1963}.}
		\label{fig:Enthalpy_triangle_1565K}
		\end{figure*}

The magnetic contribution to the enthalpy of formation of alloys described above is very important for predicting the position of the fcc-bcc phase transition line based on the formation enthalpy criterion. Figs. \ref{fig:Energy_diff_1600K}(a) and \ref{fig:Energy_diff_1600K}(c) show that the Ni-rich region of the composition diagram, where fcc phase has lower enthalpy of formation than bcc phase, is significantly larger and agrees better with the available experimental findings and CALPHAD simulations (see e.g. Fig. 7(a) in Ref. \onlinecite{Tomiska2004} with results at 1573 K). The enthalpies of formation of the most stable crystal structures of Fe-Cr-Ni alloys computed using MC simulations at 1600 K with magnetic correction applied, are compared to experimental data from Refs. \onlinecite{Kubaschewski1967,Dench1963} in Fig. \ref{fig:Enthalpy_triangle_1565K}, and the predictions agree with experiment very well.

The enthalpy of formation treated as a function of temperature was examined by using MC simulations more extensively for one particular composition, Fe$_{70}$Cr$_{20}$Ni$_{10}$, close to the composition of austenitic 304 and 316 steels\cite{Piochaud2014}. As shown in Fig. \ref{fig:Enthalpies_low_temp}, at 300K, 600K and 900K the Fe$_{70}$Cr$_{20}$Ni$_{10}$ alloy belongs to the Fe-rich region of stability of bcc alloys, in agreement with experimental data and CALPHAD simulations (see e.g. Fig. 6 of Ref. \onlinecite{Franke2011} referring to 500 $^{\circ}$C). At 1600K, with the above magnetic correction applied, fcc alloy is more stable than bcc alloy, and its calculated enthalpy of formation of 0.030 eV/atom obtained from MC simulations at 1600K, is close to the experimental value of 0.035 eV/atom measured at 1565 K\cite{Kubaschewski1967}.

Since austenitic stainless steels are formed by rapid cooling from approximately 1323K, we also analyze phase stability of Fe$_{70}$Cr$_{20}$Ni$_{10}$ alloy at 1300K \cite{Ferry2006,Majumdar1984}. Similarly to the 1600K case, after applying the magnetic correction, we find that fcc alloys have lower formation enthalpy than bcc alloys. The stability of various magnetic configurations of Fe$_{70}$Cr$_{20}$Ni$_{10}$ was analyzed using spin-polarized DFT calculations for the fcc structure with 256 atoms, derived from MC simulations at 1300K. As shown in Table \ref{tab:results_Fe70Cr20Ni10}, AFMSL and FM configurations are more stable than the AFMDL configuration, and the formation enthalpies of the two former ones are 0.018 eV/atom and 0.015 eV/atom higher than the value obtained from equilibrium MC simulations at 1300K. Results for the MC-generated structure are compared in Table \ref{tab:results_Fe70Cr20Ni10} also with enthalpies of formation of various magnetic configurations performed using fcc SQS with 256 atoms given in Ref. \onlinecite{Piochaud2014}. The energy of the most stable AFMSL configuration on SQS is 0.049 eV/atom higher than the energy of the most stable FM configuration realized on the MC-generated structure, and 0.064 eV/atom higher than the energy evaluated using equilibrium MC simulations at 1300K. Since the MC model simulations supplemented by magnetic correction were successfully validated against experimentally observed enthalpies of formation, as described above, we conclude that SQS-based calculations overestimate the formation enthalpy of the relevant alloy composition. Hence, even at high temperatures, configurations generated using CE combined with MC simulations describe Fe$_{70}$Cr$_{20}$Ni$_{10}$ alloy better than SQS.

\begin{table}
\caption{Enthalpies of formation of Fe$_{70}$Cr$_{20}$Ni$_{10}$ alloy derived from MC simulations at 1300K and 1600K, compared with  experimental values measured at 1565 K\cite{Kubaschewski1967}, and with DFT energies computed for SQS and MC-generated atomic structures.
        \label{tab:results_Fe70Cr20Ni10}}
\begin{ruledtabular}
    \begin{tabular}{cccc}
          & without corr. & with corr. & Expt. \\
    \hline
    \multicolumn{3}{l}{MC at 1600 K}       &  \\
    bcc   & 0.028 & 0.045 &  \\
    fcc   & 0.087 & 0.030 & 0.035 \\
    \multicolumn{3}{l}{MC at 1300 K}      &  \\
    bcc   & 0.017 & 0.029 &  \\
    fcc   & 0.051 & -0.006 &  \\
		\hline
    \multicolumn{3}{l}{DFT (SQS)$^a$}     &  \\
    fcc AFMSL & 0.115 &  &  \\
    fcc AFMDL & 0.126 &  &  \\
    fcc FM & 0.116 &  &  \\
		\hline
    \multicolumn{3}{l}{DFT (MC structure)$^b$}     &  \\
    fcc AFMSL & 0.069 &  &  \\
    fcc AFMDL & 0.105 &  &  \\
    fcc FM & 0.066 &  &  \\
    \end{tabular}%
\end{ruledtabular}
\begin{flushleft}
$^a$ SQS structure from Ref. \onlinecite{Piochaud2014}.\\
$^b$ Structure generated using MC simulations performed at 1300 K. \\
\end{flushleft}		
\end{table}

\subsection{Order-disorder transitions}

\begin{figure*}
			\centering
			\begin{minipage}{.50\textwidth}
			  	\centering
			  	a)\includegraphics[width=.85\linewidth]{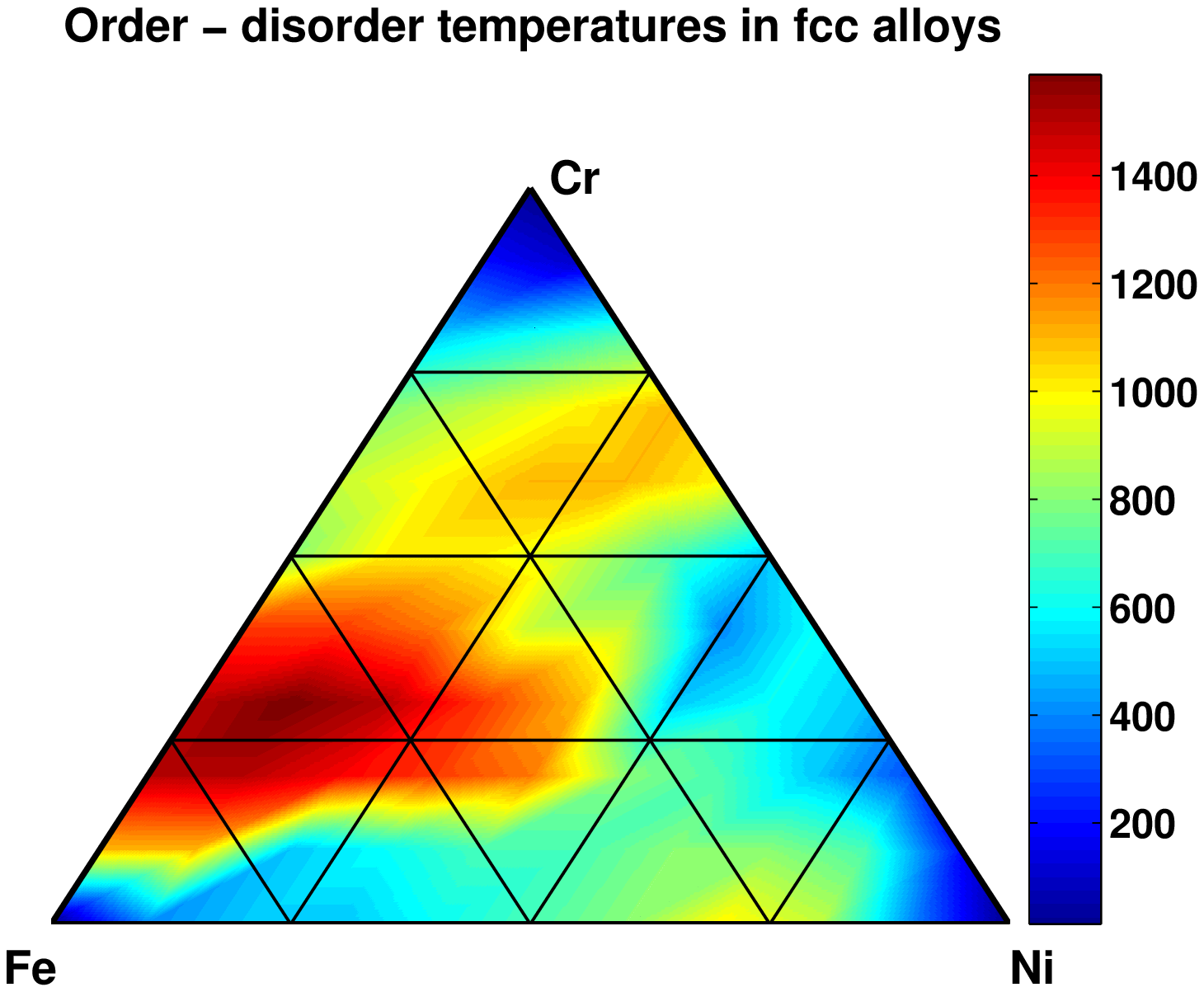}			  	
			\end{minipage}%
			\begin{minipage}{.50\textwidth}
			  	\centering
			  	b)\includegraphics[width=.85\linewidth]{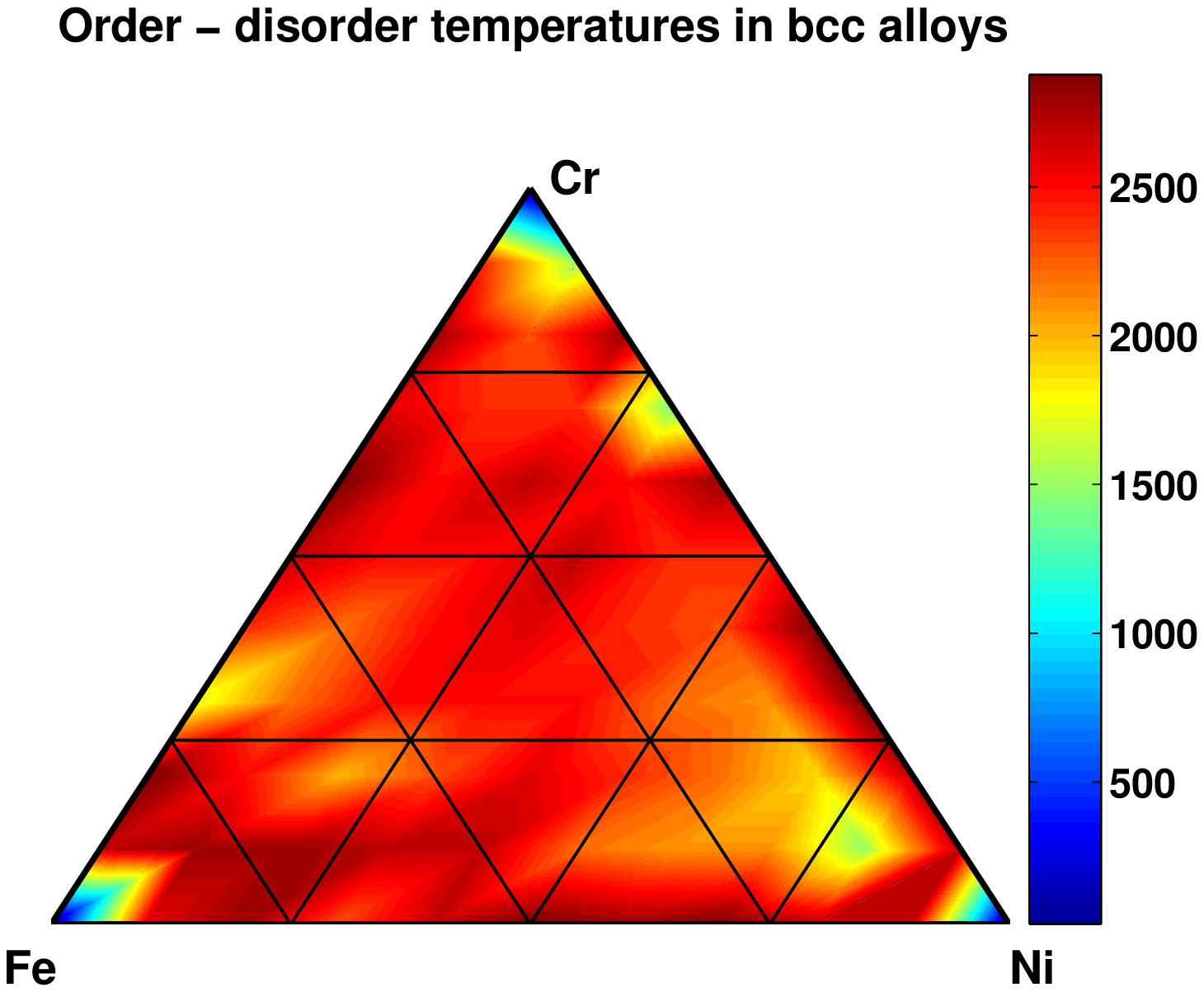}
			\end{minipage}
\caption{(Color online) Order-disorder temperatures of fcc (a) and  bcc (b) Fe-Cr-Ni alloys computed using Monte Carlo simulations. Order-disorder temperatures for pure elements are assumed to be 0 K.
        \label{fig:T_order_ternary}}
\end{figure*}

There is direct experimental evidence showing the presence of chemical order in Fe-Cr-Ni alloys. Bcc alloys at low temperatures segregate, with intermetallic Fe-Cr $\alpha$-phase representing the only known exception, whereas fcc alloys form austenitic steels exhibiting the formation of chemically ordered phases\cite{Dimitrov1986,Marwick1987,Cenedese1984,Menshikov1997}.

\begin{figure}
\includegraphics[width=\columnwidth]{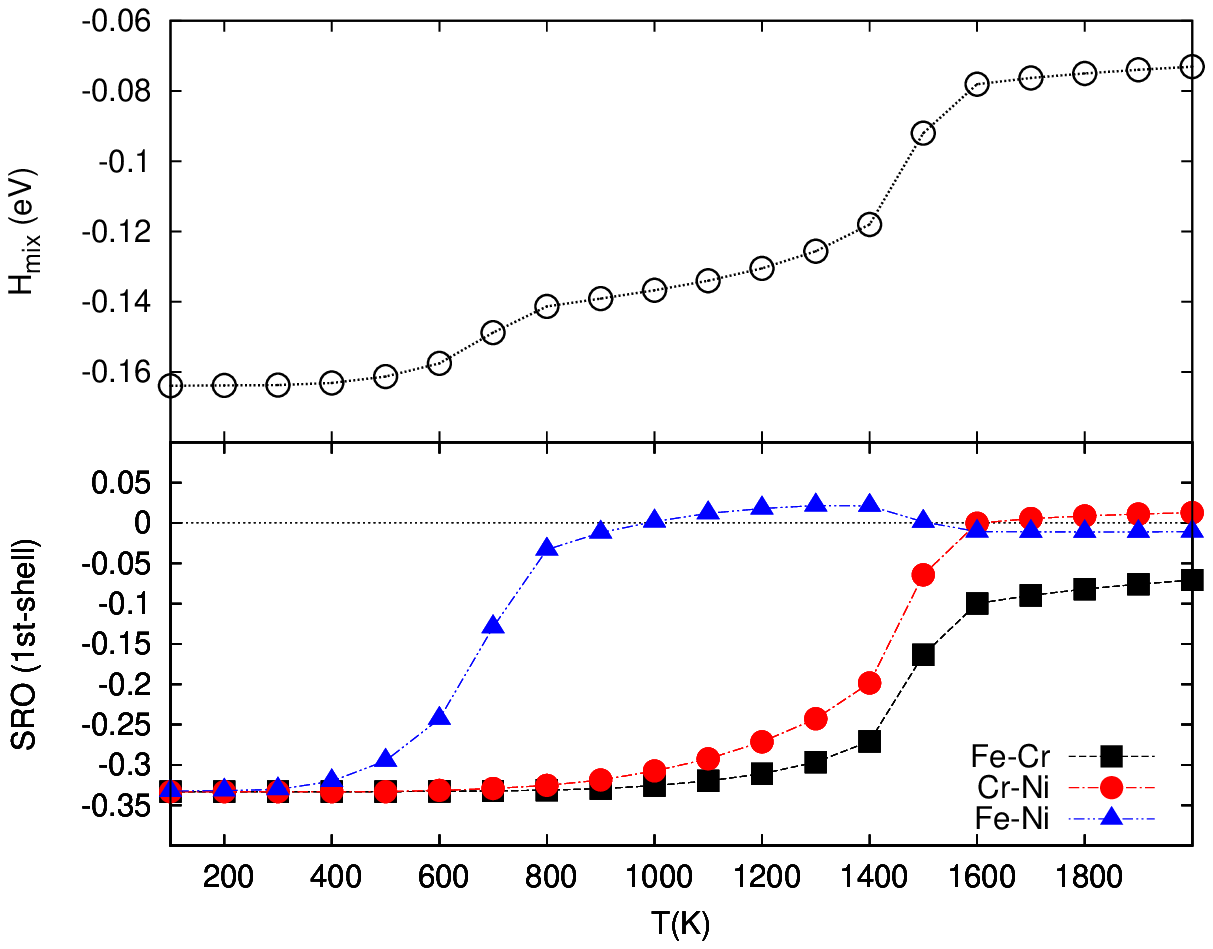}
\caption{(Color online) Enthalpies of mixing and short-range order parameters (SRO) computed as functions of temperature for Fe-Cr, Cr-Ni and Fe-Ni atomic pairs in the first coordination shell of fcc Fe$_{50}$Cr$_{25}$Ni$_{25}$ alloy.
        \label{fig:SRO_Fe2CrNi}}
\end{figure}

Order-disorder phase transition temperatures correspond to inflection points on the energy versus temperature curves. Below a phase transition temperature, a chemically ordered phase is stable. For example, the Fe$_2$CrNi intermetallic phase remains ordered below 650K, whereas chemical order between Fe and Cr, and Cr and Ni pairs of atoms in fcc Fe$_{50}$Cr$_{25}$Ni$_{25}$ alloy vanishes above 1450 K, see Fig. \ref{fig:SRO_Fe2CrNi}. Ordering temperatures computed for all the fcc ground states of Fe-Cr-Ni alloys are given in Table \ref{tab:T_ord_structures}. Fcc Fe$_3$Cr L1$_2$ phase has the highest ordering temperature of 1550K. This phase is however less stable than the bcc phase with the same composition. Ordering temperatures predicted for FeNi, FeNi$_3$ and CrNi$_2$ alloys are in reasonable agreement with experimental data. The highest order-disorder transition temperatures $T_{ord-disord}$, above which alloys can be described as disordered, were found using Monte Carlo simulations for the entire range of alloy compositions, and are shown in Fig. \ref{fig:T_order_ternary}. $T_{ord-disord}$ of fcc Fe-Cr-Ni alloys shown  in Fig. \ref{fig:T_order_ternary}(a) varies non-linearly as a function of composition, exhibiting even some local maxima. A local maximum near the FeNi$_3$ phase is in fact expected since FeNi$_3$ forms a L1$_2$ phase with relatively high ordering temperature. A local maximum around another experimentally known binary phase, CrNi$_2$ (MoPt$_2$), can also be recognized in Fig. \ref{fig:T_order_ternary}, however it is not as pronounced as in the FeNi$_3$ case. Two other maxima are less strongly pronounced. The first one corresponds to fcc alloys with Cr content between 25\% and 50\%, and is not very important for applications, since fcc alloys in this composition range are less stable than bcc alloys. More significant is the interval of compositions corresponding to Cr content from 10\% to 50\% and Ni content from 0\% to 50\%, which partially overlaps with the range of compositions of austenitic steels. Many Fe-Cr-Ni chemically ordered alloys are inside this composition range, with examples including Fe$_{64}$Cr$_{16}$Ni$_{20}$ and Fe$_{59}$Cr$_{16}$Ni$_{25}$ \cite{Dimitrov1986}, Fe$_{56}$Cr$_{21}$Ni$_{23}$ \cite{Cenedese1984}, and Fe$_{66.2}$Cr$_{17.5}$Ni$_{14.5}$Mo$_{2.8}$ \cite{Sharma1978}.

All the bcc Fe-Cr-Ni alloys are predicted to have high order-disorder temperatures, see Fig. \ref{fig:T_order_ternary}(b). The alloys exhibit short range order even at temperatures close to melting. These predicted high order-disorder temperatures are likely to be over-estimated due to the fact that our CE-based Monte Carlo simulations neglect vibrational and magnetic contributions\cite{Walle2002a,Bonny2009}.

\begin{table}
\caption{Enthalpies of mixing and order-disorder transition temperatures predicted for several intermetallic phases of Fe-Cr-Ni alloys.
        \label{tab:T_ord_structures}}
\begin{ruledtabular}
    \begin{tabular}{ccccc}
    Structure  & $\Delta H_{mix}$ (eV) & $T_{ord}$ (K) & $T_{ord}^{Expt.}$ (K) \\
    \hline
    fcc FeNi        & -0.103 & 650   & 620\cite{Massalski1990} \\
    fcc FeNi$_3$        & -0.116 & 950   & 790\cite{Massalski1990} \\
    fcc FeNi$_8$        & -0.053 & 550   &  \\
    fcc Fe$_3$Ni$_2$        & -0.082 & 550   &  \\
    fcc CrNi$_2$        & -0.155 & 750   & 863\cite{Massalski1990} \\
    fcc Cr$_2$Ni        & -0.182 & 1250  &  \\
    fcc Cr$_3$Ni        & -0.153 & 1150  &  \\
    fcc Fe$_3$Cr        & -0.103 & 1550  &  \\
    fcc FeCr$_2$        & -0.119 & 850   &  \\
    fcc FeCr$_8$        & -0.052 & 350   &  \\
    fcc Fe$_2$CrNi        & -0.164 & 650   &  \\
    \end{tabular}%
\end{ruledtabular}
\end{table}

Analysis showing how the predicted $T_{ord-disord}$ varies as a function of alloy composition confirms the experimentally observed reduction of chemical ordering in (FeNi$_3$)$_{1-x}$Cr$_x$ alloys annealed at 486 $^{\circ}$C ($=$759K) as a function of Cr content in the composition interval from $x$=0.0 to 0.17\cite{Marwick1987}. The values of $T_{ord-disord}$ obtained from MC simulations for FeNi$_3$ and (FeNi$_3$)$_{0.8}$Cr$_{0.2}$ alloys are 950K and 750K, respectively. Alloys with lower Cr content have order-disorder transition temperatures significantly higher than the annealing temperature used in the above experiments \cite{Marwick1987}. Values of $T_{ord-disord}$ in alloys with high Cr content are lower than the above annealing temperature.

\subsection{Short-range order parameters}

\begin{figure*}
			\centering
			\begin{minipage}{.50\textwidth}
			  	\centering
			  	a)\includegraphics[width=.85\linewidth]{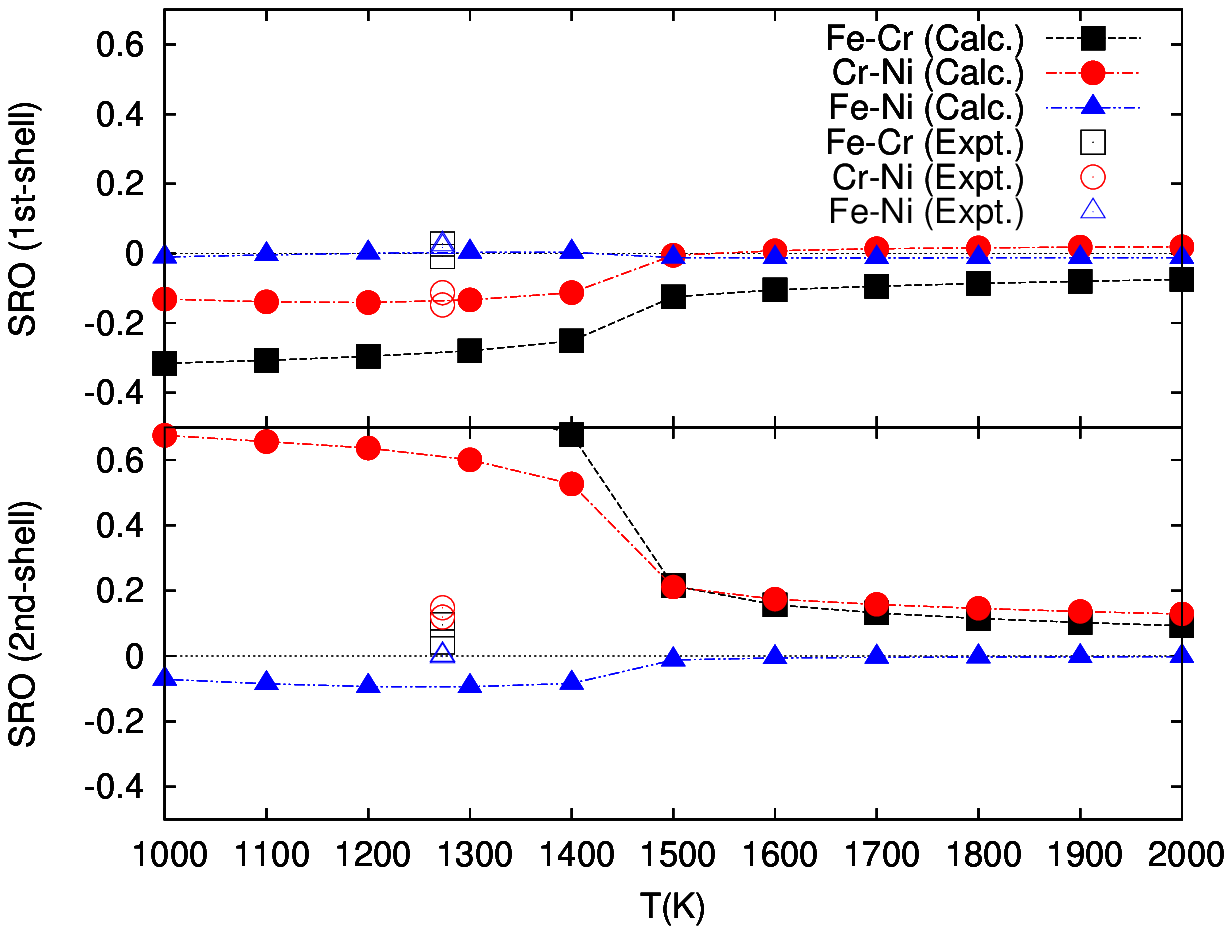}			  	
			\end{minipage}%
			\begin{minipage}{.50\textwidth}
			  	\centering
			  	b)\includegraphics[width=.85\linewidth]{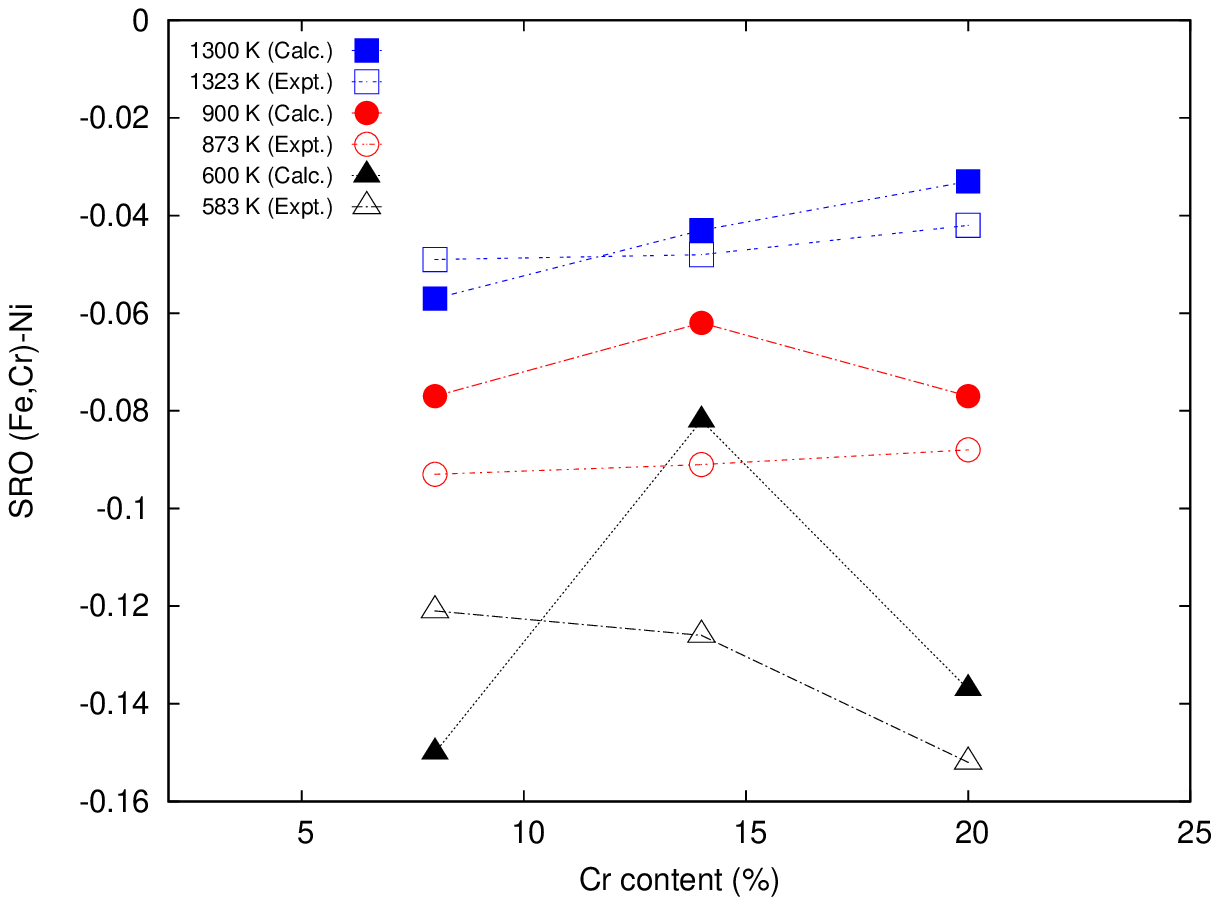}
			\end{minipage}
\caption{(Color online) (a) Short-range order parameters as functions of temperature, calculated for Fe-Cr, Cr-Ni and Fe-Ni pairs occupying two coordination shells in Fe$_{0.56}$Cr$_{0.21}$Ni$_{0.23}$ alloy, compared with experimental values from Ref. \onlinecite{Cenedese1984}; (b) 1NN SRO between Ni and average (Fe,Cr) atoms in Fe$_{42.5}$Cr$_{7.5}$Ni$_{50}$, Fe$_{38}$Cr$_{14}$Ni$_{48}$ and Fe$_{34}$Cr$_{20}$Ni$_{46}$ calculated at 600K, 900K and 1300K, and compared with experimental data taken from Ref. \onlinecite{Menshikov1997} and presented as a function of Cr content; (Fe,Cr)-Ni indicates average SRO between Ni and average (Fe,Cr) atoms obtained from Eq. \ref{eq:alpha_FeCr_Ni}.
        \label{fig:SRO_Cenedese&Menshikov}}
\end{figure*}

Chemical order in alloys is characterized by the Warren-Cowley short-range order parameters, $\alpha_1^{i-j}$ and $\alpha_2^{i-j}$, for the first (1NN) and second (2NN) nearest neighbour coordination shells. These parameters are calculated from Eq. \ref{eq:SRO_pairs} using  correlation functions deduced from MC simulations. MC simulations were performed, assuming various temperatures, for several binary and four ternary alloy compositions: Fe$_{56}$Cr$_{21}$Ni$_{23}$, Fe$_{42.5}$Cr$_{7.5}$Ni$_{50}$, Fe$_{38}$Cr$_{14}$Ni$_{48}$ and Fe$_{34}$Cr$_{20}$Ni$_{46}$, for which experimental SRO parameters were published in Refs. \onlinecite{Cenedese1984} and \onlinecite{Menshikov1997}. SRO parameters computed for binary alloys agree with experimental data, see Table \ref{tab:SRO_binary_alloys}. Comparison with experimental data for ternary Fe-Cr-Ni alloys is given in Fig. \ref{fig:SRO_Cenedese&Menshikov} and in Table \ref{tab:SRO_Cenedese&Menshikov}. MC simulations performed for fcc Fe$_{56}$Cr$_{21}$Ni$_{23}$ alloy \cite{Cenedese1984} show that it is characterized by pronounced Cr-Ni ordering, whereas at the same time there is no Fe-Ni ordering. Values of $\alpha_1^{Fe-Ni}$ and $\alpha_1^{Cr-Ni}$ are in excellent agreement with experimental observations. The calculated SRO parameter for Fe and Cr atoms is negative, in agreement with experimental observations, although the magnitude of this parameter predicted by calculations is larger. This may again be due to the fact that vibrational and magnetic contributions were neglected \cite{Walle2002a}. The effect of lattice vibrations on ordering in the bcc Fe-Cr system was noted in Refs. \onlinecite{Lavrentiev2007,Lavrentiev2010,Bonny2009}.

The effect of Cr on SRO in Fe-Cr-Ni alloys was analyzed, using MC simulations, for three compositions Fe$_{42.5}$Cr$_{7.5}$Ni$_{50}$, Fe$_{38}$Cr$_{14}$Ni$_{48}$ and Fe$_{34}$Cr$_{20}$Ni$_{46}$, which are the compositions investigated experimentally in Ref. \onlinecite{Menshikov1997}. As expected, the absolute values of SRO parameters increase with decreasing temperature for both 1NN and 2NN (see Table \ref{tab:SRO_Cenedese&Menshikov}). All the 2NN SRO parameters are positive for these three alloys. $\alpha_1^{Fe-Ni}$ and $\alpha_1^{Fe-Cr}$ are negative and their absolute values decrease as functions of Cr content. An interesting result is that the sign of $\alpha_1^{Cr-Ni}$ changes from positive for Fe$_{42.5}$Cr$_{7.5}$Ni$_{50}$ to negative for Fe$_{34}$Cr$_{20}$Ni$_{46}$ alloy. Fe$_{38}$Cr$_{14}$Ni$_{48}$ alloy with intermediate Cr content has positive $\alpha_1^{(Fe-Cr)}$ only at relatively low temperatures close to 600K. We compare our theoretical predictions with measured SRO parameters involving Ni atoms and 'average' (Fe,Cr) atoms in Fe$_{42.5}$Cr$_{7.5}$Ni$_{50}$, Fe$_{38}$Cr$_{14}$Ni$_{48}$ and Fe$_{34}$Cr$_{20}$Ni$_{46}$ alloys, quenched rapidly from 1323K, annealed at 873K and irradiated at 583K with 2.5 MeV electrons\cite{Menshikov1997}. The authors of Ref. \onlinecite{Menshikov1997} neglected ordering between Fe and Cr atoms, arguing that there was no evidence for the occurrence of stable Fe-Cr compounds at low temperatures. Their assumption was based also on experimental observations by Cenedese {\it et al.}\cite{Cenedese1984} who found that in fcc Fe$_{56}$Cr$_{21}$Ni$_{23}$ alloy only Cr and Ni atoms were ordered.

To compare results of MC simulations with experimentally measured \cite{Menshikov1997} SRO parameters involving Ni and 'average' (Fe,Cr) atoms, we treat ternary Fe-Cr-Ni alloys as a pseudo-binary alloy of composition Ni$_x$(FeCr)$_{1-x}$. We define an effective SRO parameter involving Ni and (Fe,Cr) atoms as
\begin{equation}
\alpha_n^{(Fe,Cr)-Ni} = \frac{c_{Fe}}{c_{Fe}+c_{Cr}}\alpha_n^{Fe-Ni} + \frac{c_{Cr}}{c_{Fe}+c_{Cr}}\alpha_n^{Cr-Ni}.
\label{eq:alpha_FeCr_Ni}
\end{equation}
Values of $\alpha_1^{(Fe,Cr)-Ni}$ defined in this way and calculated using MC simulations for 1300K are in excellent agreement with experimental observations for alloy samples quenched rapidly from 1323K. Despite the fact that experimental measurements for samples irradiated at 583K cannot be directly compared with MC simulations at 600K, experimental observations showing more pronounced chemical order in Fe$_{34}$Cr$_{20}$Ni$_{46}$ sample irradiated at 583K in comparison with Fe$_{38}$Cr$_{14}$Ni$_{48}$ sample are in agreement with our predictions.

\begin{figure*}
			\centering
			\begin{minipage}{.50\textwidth}
			  	\centering
			  	a)\includegraphics[width=.85\linewidth]{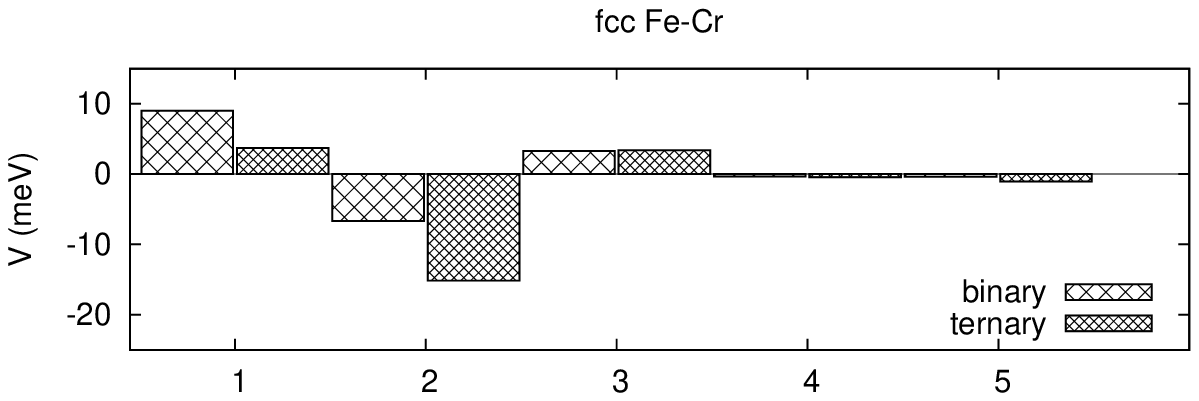}			  	
			\end{minipage}%
			\begin{minipage}{.50\textwidth}
			  	\centering
			  	b)\includegraphics[width=.85\linewidth]{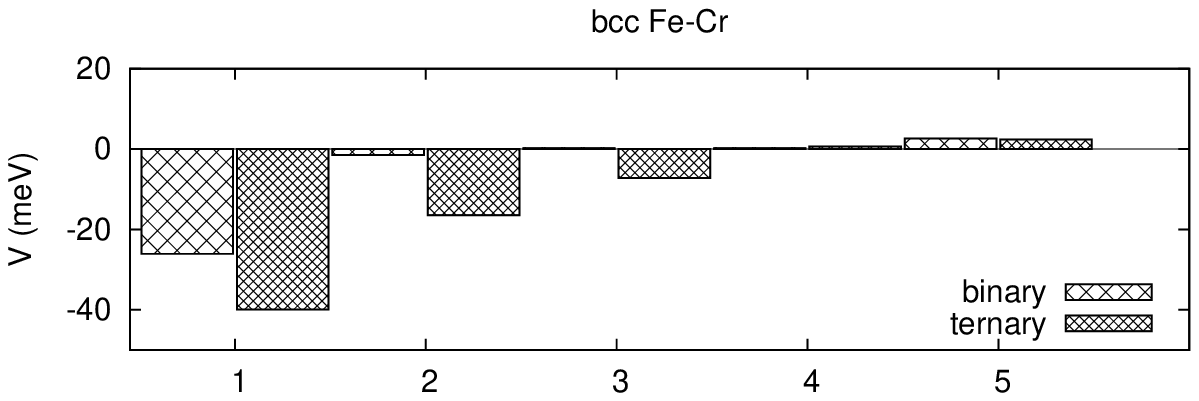}
			\end{minipage}
			\newline
			\begin{minipage}{.50\textwidth}
			  	\centering
			  	c)\includegraphics[width=.85\linewidth]{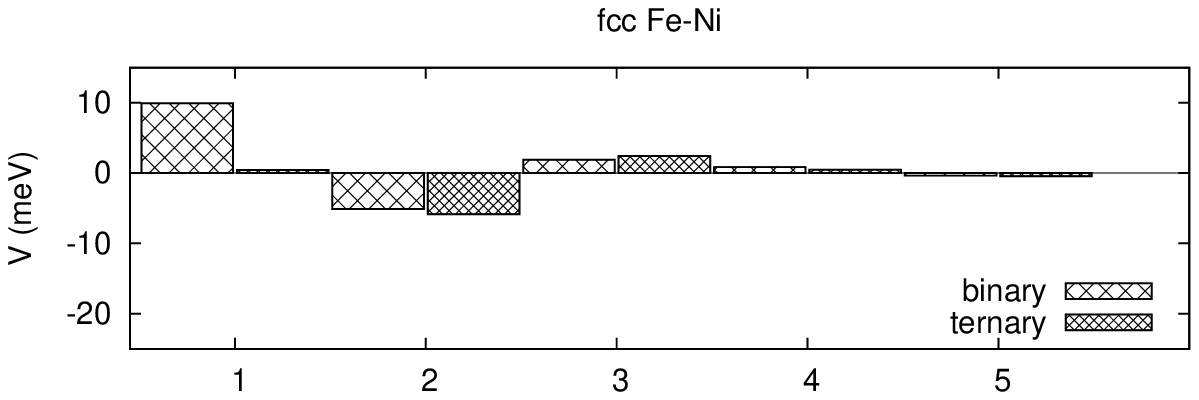}			  	
			\end{minipage}%
			\begin{minipage}{.50\textwidth}
			  	\centering
			  	d)\includegraphics[width=.85\linewidth]{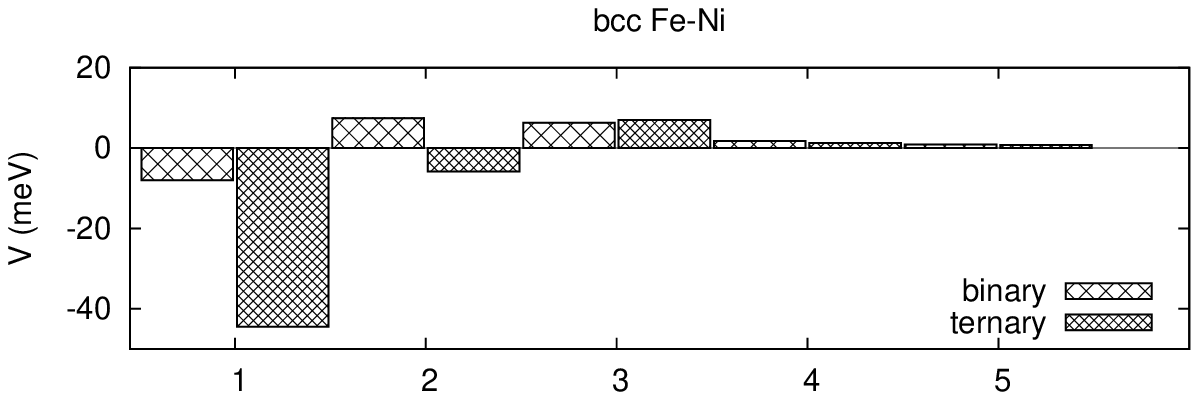}
			\end{minipage}
			\newline
			\begin{minipage}{.50\textwidth}
			  	\centering
			  	e)\includegraphics[width=.85\linewidth]{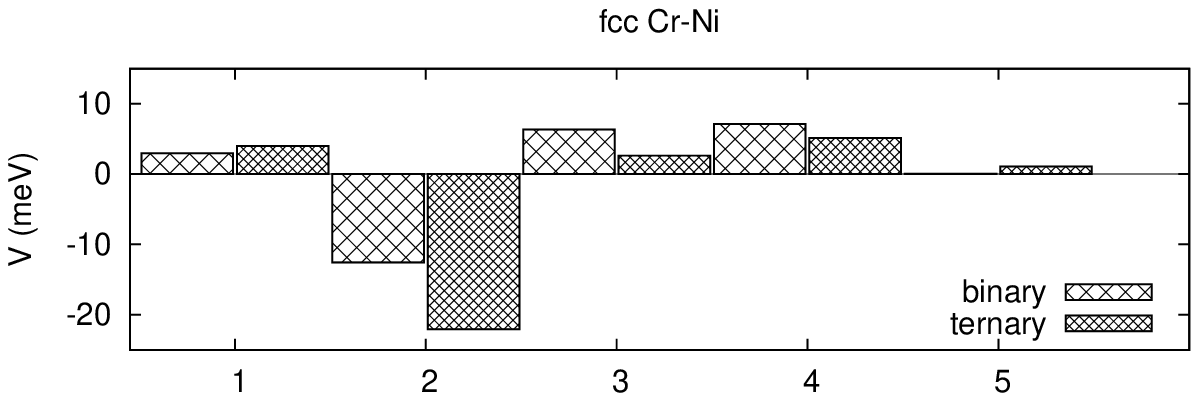}			  	
			\end{minipage}%
			\begin{minipage}{.50\textwidth}
			  	\centering
			  	f)\includegraphics[width=.85\linewidth]{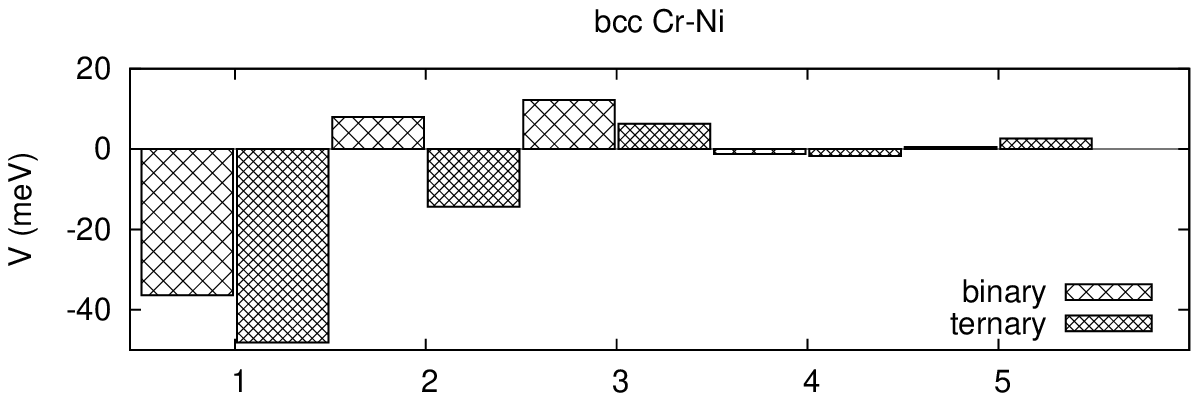}
			\end{minipage}
			\newline
			\caption{ Effective interactions between different pairs of atoms: Fe-Cr (a,b), Fe-Ni (c,d) and Cr-Ni (e,f) on fcc (a,c,e) and bcc (b,d,f) lattices in ternary Fe-Cr-Ni and binary alloys.
		}
		\label{fig:ECI_Vij}
\end{figure*}

The occurrence of chemical order in Fe-Cr-Ni alloys can be explained by analysing interactions between pairs of Fe-Cr, Fe-Ni and Cr-Ni atoms, $V_n^{ij}$. They were derived from two-body effective cluster interaction parameters for fcc and bcc Fe-Cr-Ni alloys listed in Table \ref{tab:ECI_def_ternary}. Assuming that many-body interactions are small, we find that $V_n^{ij}$ are related to $J_{2,n}^{(s)}$ and can be calculated using Eq. \ref{eq:VvsJ_Matrix}. $V_n^{Fe-Ni}$, $V_n^{Fe-Cr}$ and $V_n^{Cr-Ni}$ computed for fcc and bcc ternary alloys are compared with values derived for binary alloys in Fig. \ref{fig:ECI_Vij}. All the 1NN chemical pairwise interactions computed for bcc ternary alloys are even more negative than those computed for binary alloys. This means that repulsion between Fe-Cr, Fe-Ni and Cr-Ni atoms in ternary alloys is even stronger than in binary alloys. On the other hand, in ternary fcc alloys the 1NN chemical pairwise interaction between Fe and Ni atoms vanishes almost completely. This explains why the SRO parameter involving Fe and Ni atoms in the first nearest neighbour coordination shell measured in Ref. \onlinecite{Cenedese1984} nearly vanishes. This also explains the observed decrease of atomic ordering in FeNi$_3$ alloys following the addition of Cr \cite{Marwick1987}. Large 1NN effective pairwise Fe-Cr and Cr-Ni interactions ($V_1^{Fe-Cr}$ and $V_1^{Cr-Ni}$ are correspondingly smaller and larger in ternary alloys in comparison with binary alloys), and the relatively large 2NN effective interactions between these atoms also explain the pronounced atomic ordering in the majority of fcc Fe-Cr-Ni alloys.

The sign pattern of the first three nearest-neighbour chemical pairwise interactions in all the binary and ternary fcc alloys remains the same. The first and third nearest-neighbour (3NN) pair interactions are positive and the second nearest neighbour interaction is negative. This favours the unlike atoms occupying the 1NN and 3NN coordination shells, and the like atoms occupying the 2NN shell, see Eq.\ref{eq:CE_vs_V}. Such a pattern of signs of NN interactions favours not only intermetallic L1$_2$ and MoPt$_2$-like phases, which occur in fcc binary alloys, but also the L1$_2$-based (Cu$_2$ZnNi-like) Fe$_2$CrNi ternary phase, which is the global ground state of ternary Fe-Cr-Ni alloys.

\begin{table*}
\caption{Short-range order parameters for selected binary alloys calculated using Monte Carlo simulations at various temperatures $T$, compared with experimental data.
        \label{tab:SRO_binary_alloys}}
\begin{ruledtabular}
    \begin{tabular}{cccccccc}
          &$T$ (K) & MC (this study) & MC (Others)  & Expt.      & MC & MC \cite{Rahaman2014} & Expt. \\
\hline
    fcc alloys      & & \multicolumn{3}{c}{$\alpha_1$} & \multicolumn{3}{c}{$\alpha_2$} \\
    Fe$_{25}$Ni$_{75}$ & 1300 & -0.096 & & -0.099$^c$ & 0.097 & & 0.116$^c$ \\
    Fe$_{30}$Ni$_{70}$ & & -0.102 & & -0.088$^c$ & 0.105 & & 0.049$^c$ \\
    Fe$_{50}$Ni$_{50}$ & & -0.071 & & -0.073$^c$ & 0.082  & & 0.042$^c$ \\
    Fe$_{60}$Ni$_{40}$ & & -0.043 & & -0.058$^c$ & 0.058 & & 0.089$^c$ \\
    Fe$_{65}$Ni$_{35}$ & & -0.018 & & -0.051$^c$ & 0.049 & & 0.034$^c$ \\
    Fe$_{70}$Ni$_{30}$ & & -0.002 & & -0.033$^c$ & 0.031 & & 0.005$^c$ \\
    Fe$_{65}$Ni$_{35}$ & 1100 & -0.022 & & -0.058$^d$ & 0.076 & & 0.052$^d$ \\
    Cr$_{33}$Ni$_{67}$ & 1100 & -0.036 & -0.115$^a$ & -0.08$^e$ & 0.042 & 0.12\cite{Rahaman2014} & 0.05$^e$ \\
    Cr$_{25}$Ni$_{75}$ & 1000 & -0.047 & -0.105$^a$ & -0.07$^f$ & 0.051 & 0.10\cite{Rahaman2014} & 0.045$^f$ \\
    Cr$_{20}$Ni$_{80}$ & 800 & -0.029 & -0.125$^a$ & -0.10$^g$ & 0.057 & 0.115\cite{Rahaman2014} & 0.085$^g$ \\
\hline					
    bcc alloys & & \multicolumn{3}{c}{$\alpha_{1+2}$} &   &    &  \\
    Fe$_{95}$Cr$_{5}$ & 700 & -0.044 & -0.049$^b$ & -0.05$^h$ &    &   &  \\
    Fe$_{93.75}$Cr$_{6.25}$ & & -0.056 &  &  &    &   &  \\
    Fe$_{90}$Cr$_{10}$ & & -0.071 & -0.080$^b$ & 0.00$^h$ &   &    &  \\
    Fe$_{85}$Cr$_{15}$ & & 0.138 & 0.309$^b$ & 0.065$^h$ &   &    &  \\
    \end{tabular}%
\end{ruledtabular}
\begin{flushleft}
$^a$ Ref. \onlinecite{Rahaman2014} MC simulations. \\
$^b$ Ref. \onlinecite{Lavrentiev2007} MC simulations. \\
$^c$ Refs. \onlinecite{Gomankov1971,Menshikov1972} annealed at 1273 K. \\
$^d$ Ref. \onlinecite{Robertson1999} annealed at 1026 K. \\
$^e$ Ref. \onlinecite{Caudron1992} annealed at 1073 K. \\
$^f$ Ref. \onlinecite{Caudron1992} annealed at 993 K. \\
$^g$ Ref. \onlinecite{Schonfeld1988} annealed at 828 K. \\
$^h$ Ref. \onlinecite{Mirebeau1984} annealed at 703 K. \\
\end{flushleft}		
\end{table*}

\begin{table}
\caption{Short-range order parameters for Fe-Cr, Fe-Ni and Cr-Ni pairs in ternary alloys calculated using Monte Carlo simulations, and compared with experimental observations. (Fe,Cr)-Ni means average SRO involving Ni and average (Fe,Cr) atoms as defined by Eq. \ref{eq:alpha_FeCr_Ni}.
        \label{tab:SRO_Cenedese&Menshikov}}
\begin{ruledtabular}
    \begin{tabular}{cccccc}
          & & \multicolumn{2}{c}{$\alpha_1$} & \multicolumn{2}{c}{$\alpha_2$} \\
          & & MC   & Expt.      & MC    & Expt. \\
					\hline
    Fe$_{56}$Cr$_{21}$Ni$_{23}$ &       &       &       &       &  \\
    Fe-Ni & 1300  & 0.003 & 0.017$^a$ & -0.094 & -0.002$^a$ \\
    Fe-Cr &       & -0.280 & -0.009$^a$ & 0.781 & 0.043$^a$ \\
    Cr-Ni &       & -0.134 & -0.113$^a$ & 0.600 & 0.148$^a$ \\
    Fe$_{42.5}$Cr$_{7.5}$Ni$_{50}$ &       &       &       &       &  \\
    Fe-Ni &       & -0.069 &       & 0.073 &  \\
    Fe-Cr & 1300  & -0.080 &       & 0.087 &  \\
    Cr-Ni &       & 0.015 &       & 0.085 &  \\
    (Fe,Cr)-Ni &       & -0.057 & -0.049$^b$ & 0.075 & 0.015$^b$ \\
    Fe-Ni &       & -0.099 &       & 0.144 &  \\
    Fe-Cr & 900   & -0.158 &       & 0.281 &  \\
    Cr-Ni &       & 0.039 &       & 0.134 &  \\
    (Fe,Cr)-Ni &       & -0.077 & -0.093$^c$ & 0.142 & 0.134$^c$ \\
    Fe-Ni &       & -0.213 &       & 0.681 &  \\
    Fe-Cr & 600   & -0.403 &       & 0.975 &  \\
    Cr-Ni &       & 0.180 &       & 0.680 &  \\
    (Fe,Cr)-Ni &       & -0.150 & -0.121$^d$ & 0.681 & 0.148$^d$ \\
    Fe$_{38}$Cr$_{14}$Ni$_{48}$ &       &       &       &       &  \\
    Fe-Ni &       & -0.054 &       & 0.054 &  \\
    Fe-Cr & 1300  & -0.076 &       & 0.141 &  \\
    Cr-Ni &       & -0.014 &       & 0.115 &  \\
    (Fe,Cr)-Ni &       & -0.043$^b$ & -0.048 & 0.070 & 0.018$^b$ \\
    Fe-Ni &       & -0.076 &       & 0.116 &  \\
    Fe-Cr & 900   & -0.276 &       & 0.701 &  \\
    Cr-Ni &       & -0.023 &       & 0.382 &  \\
    (Fe,Cr)-Ni &       & -0.062 & -0.091$^c$ & 0.188 & 0.082$^c$ \\
    Fe-Ni &       & -0.149 &       & 0.648 &  \\
    Fe-Cr & 600   & -0.489 &       & 0.983 &  \\
    Cr-Ni &       & 0.097 &       & 0.814 &  \\
    (Fe,Cr)-Ni &       & -0.082 & -0.126$^d$ & 0.693 & 0.089$^d$ \\
    Fe$_{34}$Cr$_{20}$Ni$_{46}$ &       &       &       &       &  \\
    Fe-Ni &       & -0.035 &       & 0.033 &  \\
    Fe-Cr & 1300  & -0.060 &       & 0.168 &  \\
    Cr-Ni &       & -0.031 &       & 0.136 &  \\
    (Fe,Cr)-Ni &       & -0.033 & -0.042$^b$ & 0.071 & 0.017$^b$ \\
    Fe-Ni &       & -0.035 &       & 0.024 &  \\
    Fe-Cr & 900   & -0.300 &       & 0.838 &  \\
    Cr-Ni &       & -0.146 &       & 0.563 &  \\
    (Fe,Cr)-Ni &       & -0.077 & -0.088$^c$ & 0.224 & 0.086$^c$ \\
    Fe-Ni &       & -0.138 &       & 0.538 &  \\
    Fe-Cr & 600   & -0.401 &       & 0.994 &  \\
    Cr-Ni &       & -0.133 &       & 0.799 &  \\
    (Fe,Cr)-Ni &       & -0.137 & -0.152$^d$ & 0.634 & 0.098$^d$ \\
    \end{tabular}%
\begin{flushleft}
$^a$ Ref.\onlinecite{Cenedese1984} annealed at 1273 K. \\
$^b$ Ref.\onlinecite{Menshikov1997} quenched from 1323 K. \\
$^c$ Ref.\onlinecite{Menshikov1997} annealed at 873 K. \\
$^d$ Ref.\onlinecite{Menshikov1997} irradiated at 583 K with 2.5 MeV electrons. \\
\end{flushleft}		
\end{ruledtabular}
\end{table}

\subsection{Configurational entropy and free energy of formation}

While by comparing enthalpies of formations we assess the low temperature phase stability of alloys, the investigation of high temperature phase stability requires comparing formation free energies of fcc and bcc phases. Evaluating the free energy requires computing both the enthalpy of formation and the configurational entropy of the alloy. Configurational entropy is defined as

\begin{equation}
S_{conf}(T)=\int_0^T\frac{C_{conf}(T')}{T'}dT',
\label{eq:ConfEntropy}
\end{equation}
where the configurational contribution to the specific heat $C_{conf}$ is related to fluctuations of enthalpy of mixing at a given temperature \cite{Newman1999,Lavrentiev2009} through
\begin{equation}
C_{conf}(T)=\frac{\left\langle H_{mix}(T)^2\right\rangle-\left\langle H_{mix}(T)\right\rangle^2}{T^2},
\label{eq:ConfHeatCapacity}
\end{equation}
where $\left\langle H_{mix}(T)\right\rangle$ and $\left\langle H_{mix}(T)^2\right\rangle$ are the mean and mean square average enthalpies of mixing, respectively, computed by averaging over all the MC steps at the accumulation stage for a given temperature.

The accuracy of evaluation of configurational entropy depends on temperature integration step in Eq. \ref{eq:ConfEntropy} and the number of MC steps performed at the accumulation stage. Test simulations showed that choosing a sufficiently small temperature integration step is particularly significant. Calculations of configurational entropy for all the alloy compositions below were performed with 2000 MC steps per atom at the thermalization and accumulation stages and with temperature step of $\Delta T = 10$ K.

\begin{figure*}
			\centering
			\begin{minipage}{.50\textwidth}
			  	\centering
			  	a)\includegraphics[width=.85\linewidth]{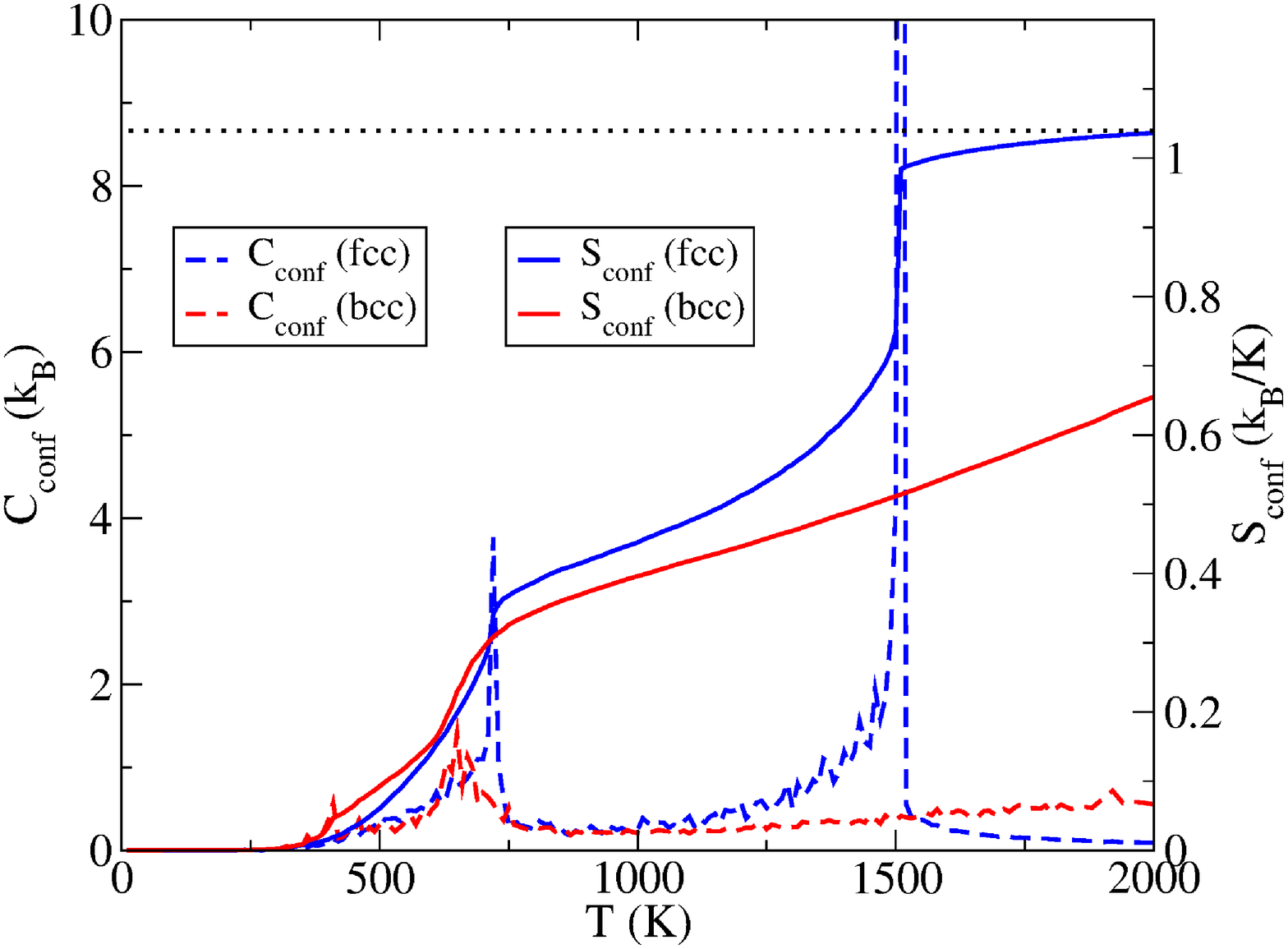}			  	
			\end{minipage}%
			\begin{minipage}{.50\textwidth}
			  	\centering
			  	b)\includegraphics[width=.85\linewidth]{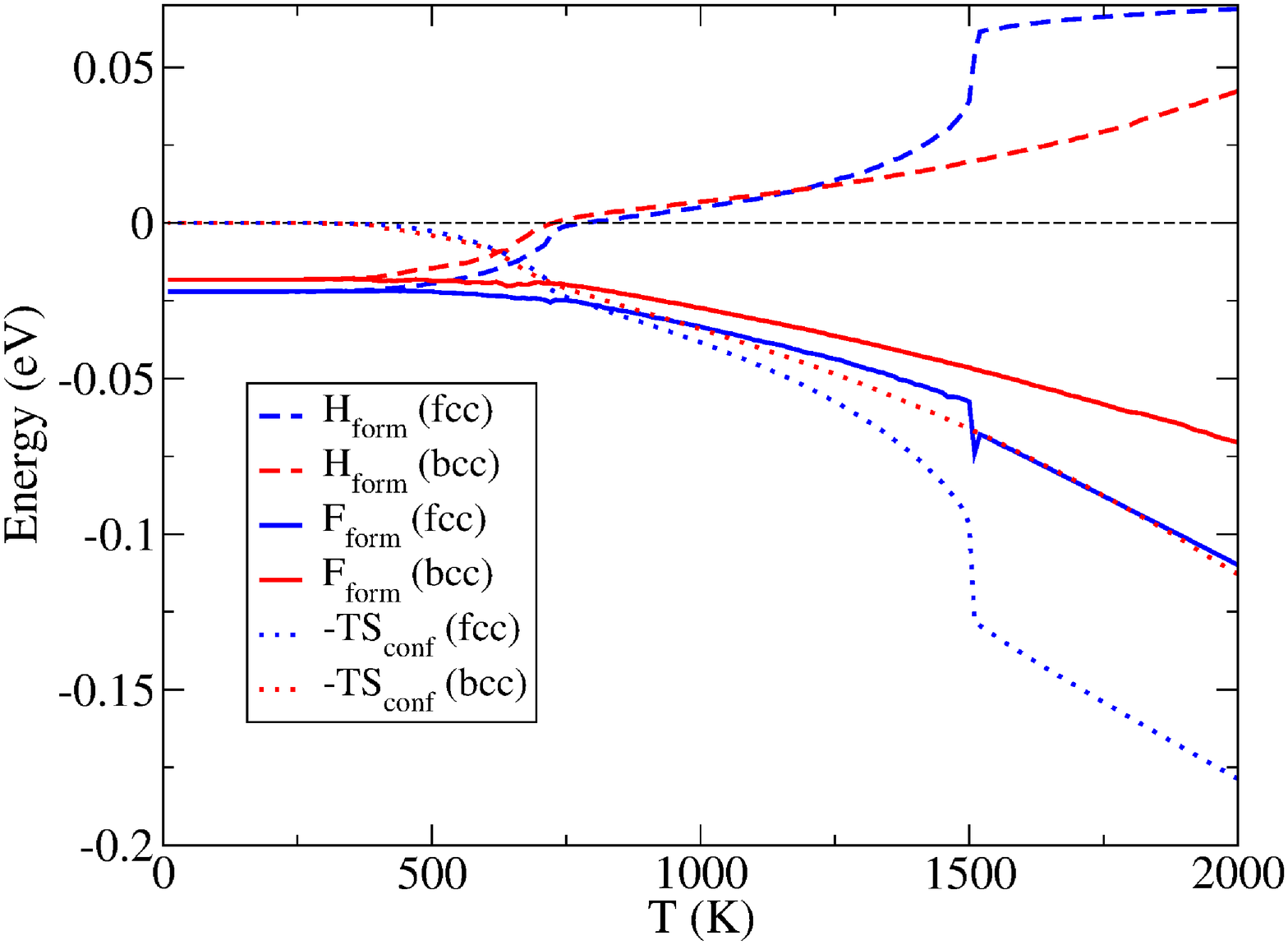}
			\end{minipage}
\caption{(Color online) (a) Configurational specific heat $C_{conf}$, configurational entropy $S_{conf}$, (b) the enthalpy of formation $H_{form}$, the product of temperature and configurational entropy $-TS_{conf}$ and the free energy of formation $F_{form}$ of fcc and bcc Fe$_2$CrNi alloys. The black dotted line is the entropy of ideal random solid solution of Fe$_2$CrNi equal to 1.04$k_B$.
        \label{fig:Entropy_FreeEn_Fe2CrNi}}
\end{figure*}

Configurational specific heat and configurational entropy of fcc and bcc Fe$_2$CrNi alloys, treated as functions of temperature, are shown in Fig. \ref{fig:Entropy_FreeEn_Fe2CrNi}(a). Configurational specific heats of both alloys exhibit sharp peaks in the vicinity of order-disorder phase transition temperatures. For example, the first peak at 700 K in the specific heat curve of fcc Fe$_2$CrNi alloy refers to the temperature of ordering of Fe and Ni atoms, whereas the second peak at 1500 K refers to the ordering temperature of Fe-Cr and Cr-Ni atoms, see Fig. \ref{fig:SRO_Fe2CrNi}. Configurational entropy of fcc Fe$_2$CrNi alloy in the high temperature limit approaches the configurational entropy of ideal random solid solution for this composition given by the formula
\begin{equation}
S_{random}(T)=-k_B\sum_ic_iln(c_i).
\label{eq:Conf_Entropy}
\end{equation}
Substituting atomic concentrations in this equation, we find that for Fe$_2$CrNi alloy the configurational entropy in the high temperature limit is equal to 1.04 $k_B$. Configurational entropy of bcc Fe$_2$CrNi alloy at 2000 K is lower. This is due to the fact that bcc alloy at 2000 K is still not fully random. For temperatures above the temperature of ordering of Fe and Ni atoms in fcc Fe$_2$CrNi alloy, the entropy of fcc alloy is always higher than that of bcc alloy. Hence, despite the fact that the enthalpy of formation of bcc alloy at high temperatures is lower than that of fcc alloy, the latter is always more stable according to the formation free energy criterion, see Fig. \ref{fig:Entropy_FreeEn_Fe2CrNi}(b).

\begin{figure*}
			\centering
			\begin{minipage}{.50\textwidth}
			  	\centering
			  	a)\includegraphics[width=.9\linewidth]{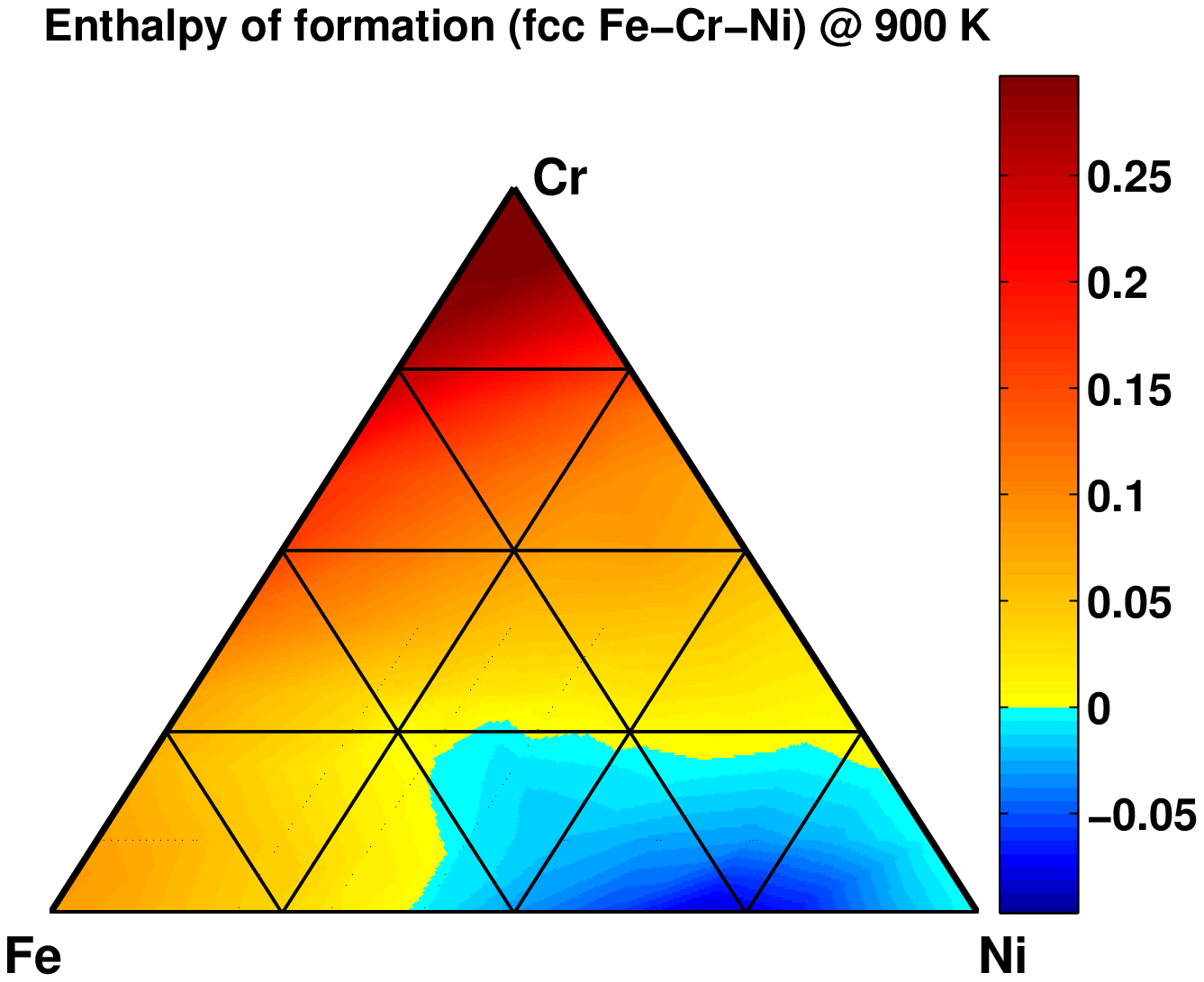}			  	
			\end{minipage}%
			\begin{minipage}{.50\textwidth}
			  	\centering
			  	b)\includegraphics[width=.9\linewidth]{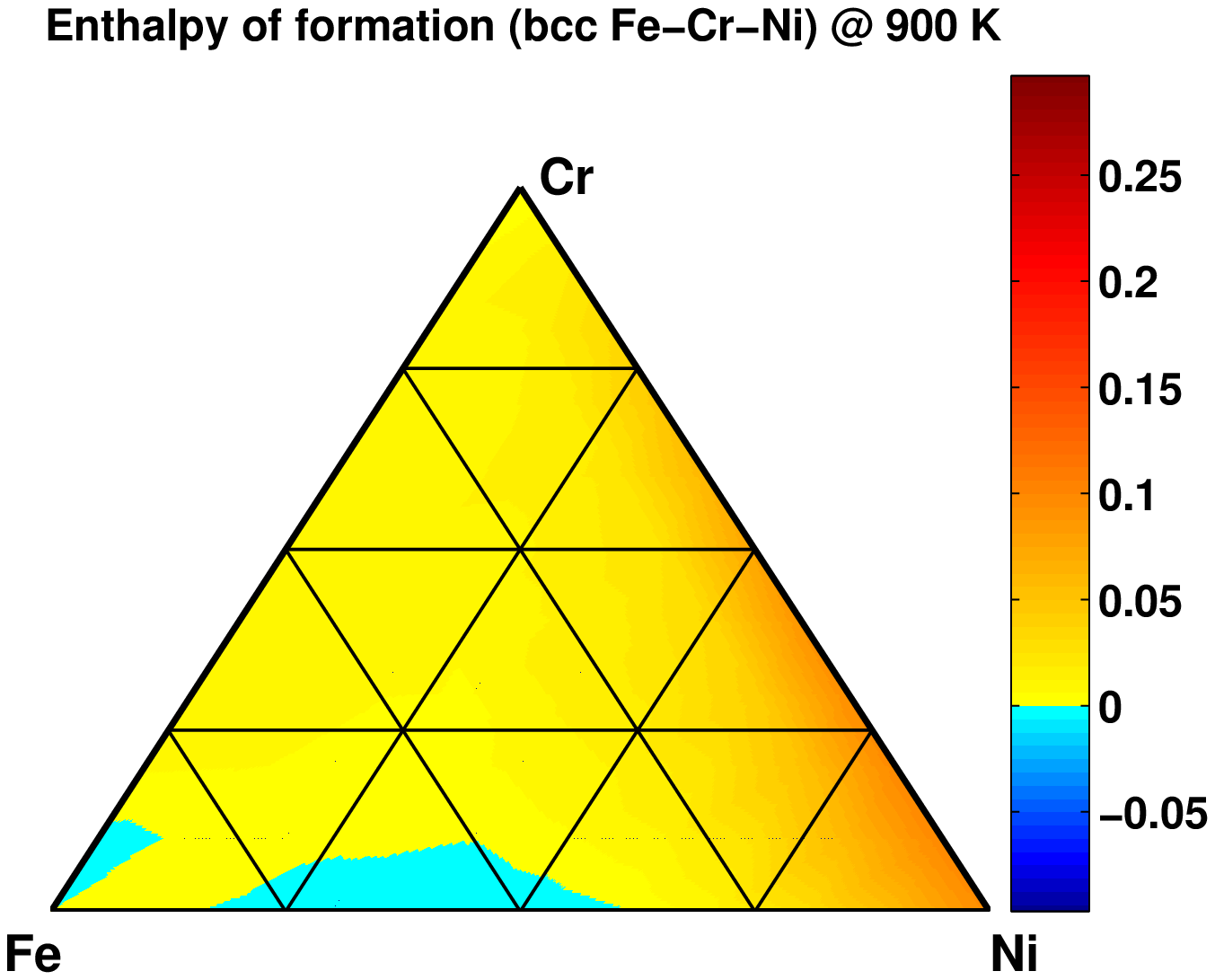}
			\end{minipage}
						\newline
			\begin{minipage}{.50\textwidth}
			  	\centering
			  	c)\includegraphics[width=.9\linewidth]{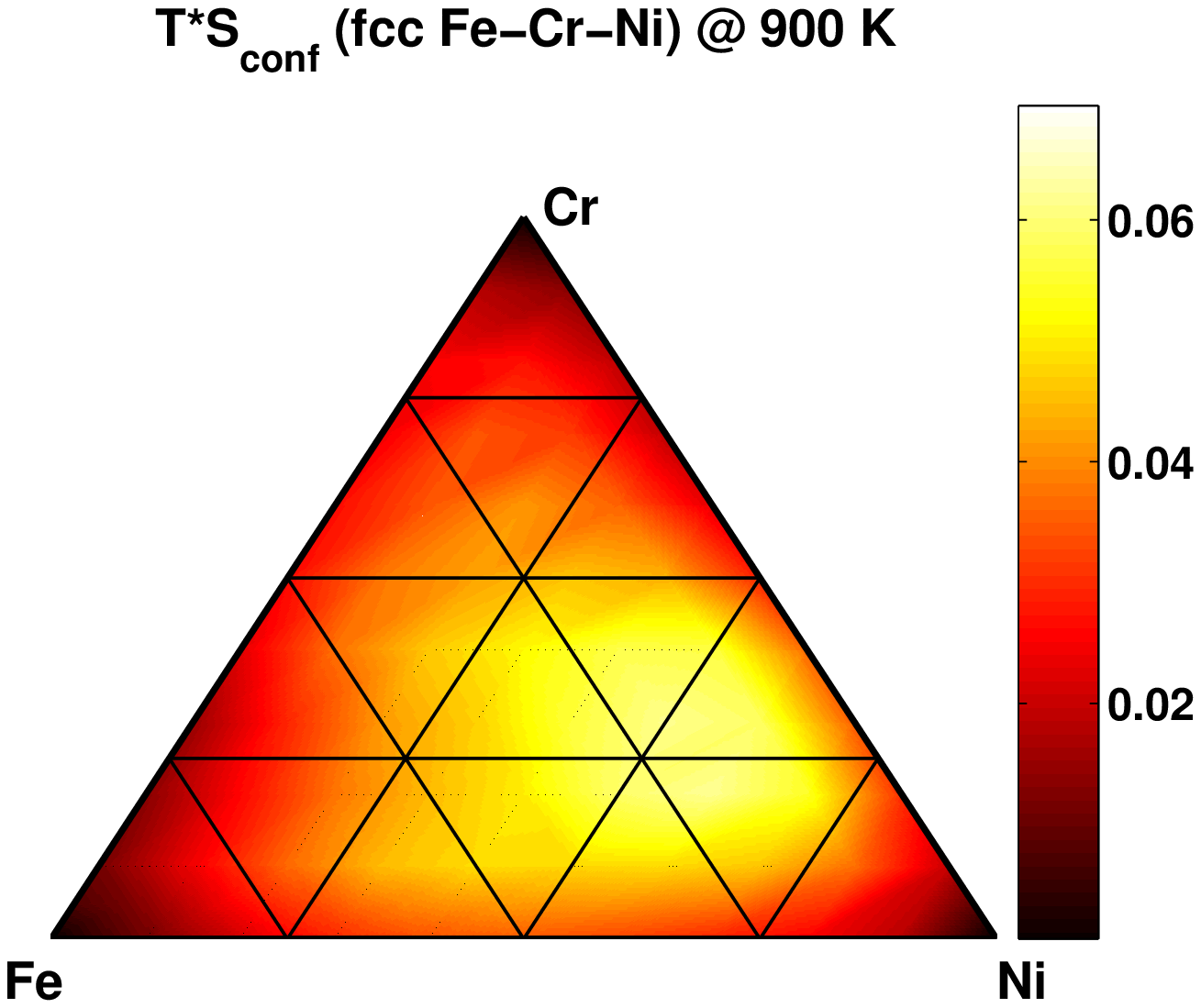}			  	
			\end{minipage}%
			\begin{minipage}{.50\textwidth}
			  	\centering
			  	d)\includegraphics[width=.9\linewidth]{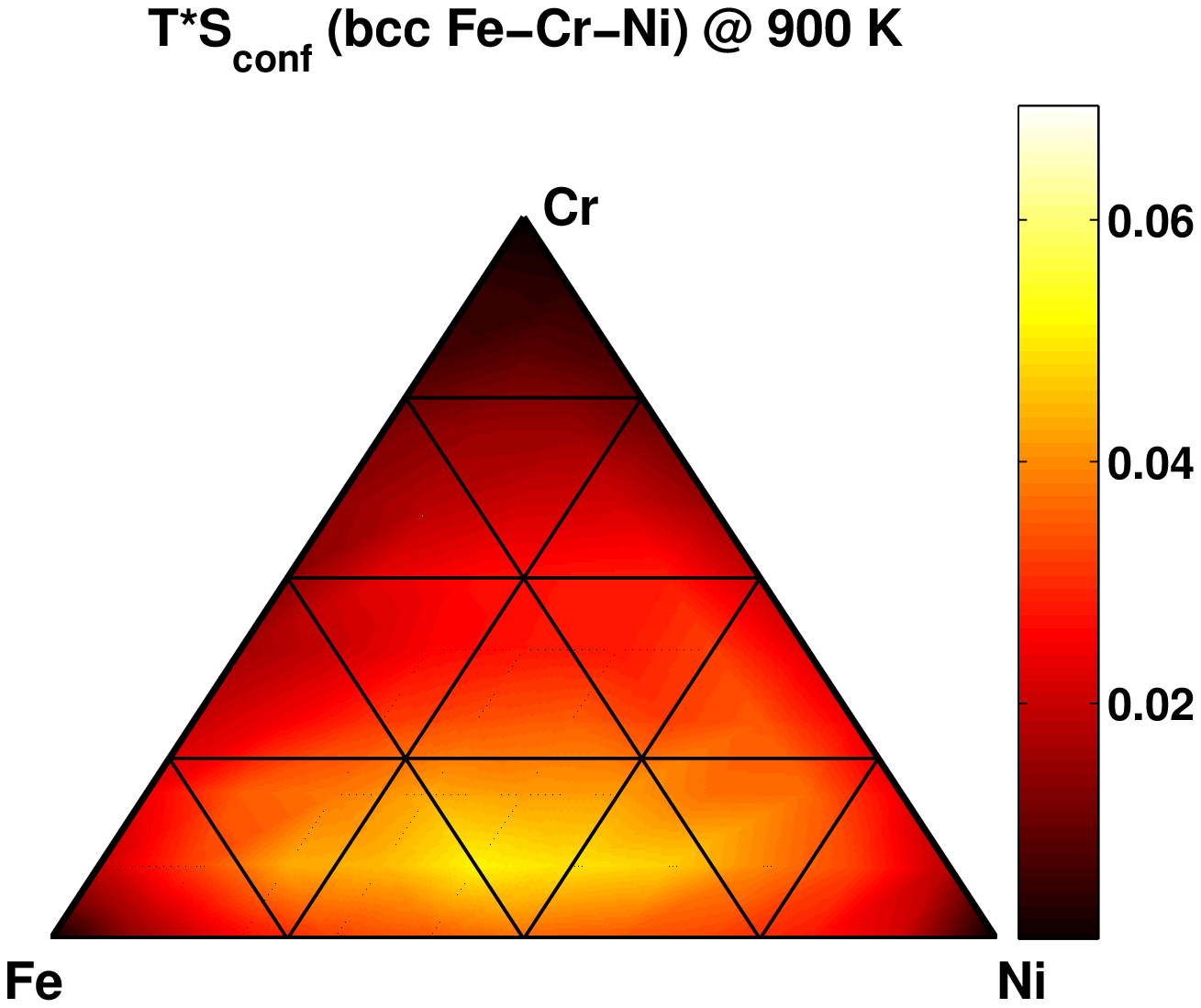}
			\end{minipage}
						\newline
			\begin{minipage}{.50\textwidth}
			  	\centering
			  	e)\includegraphics[width=.9\linewidth]{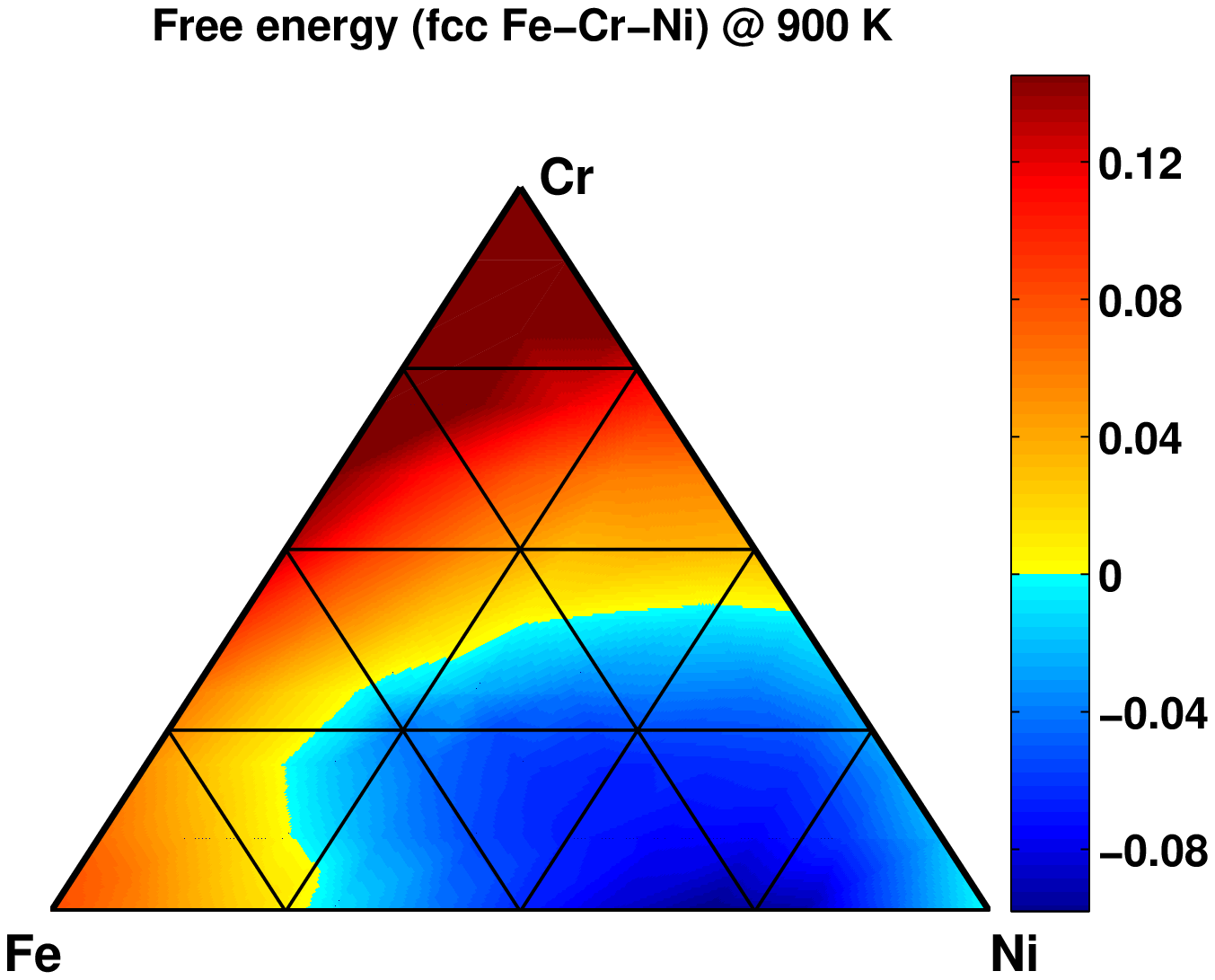}			  	
			\end{minipage}%
			\begin{minipage}{.50\textwidth}
			  	\centering
			  	f)\includegraphics[width=.9\linewidth]{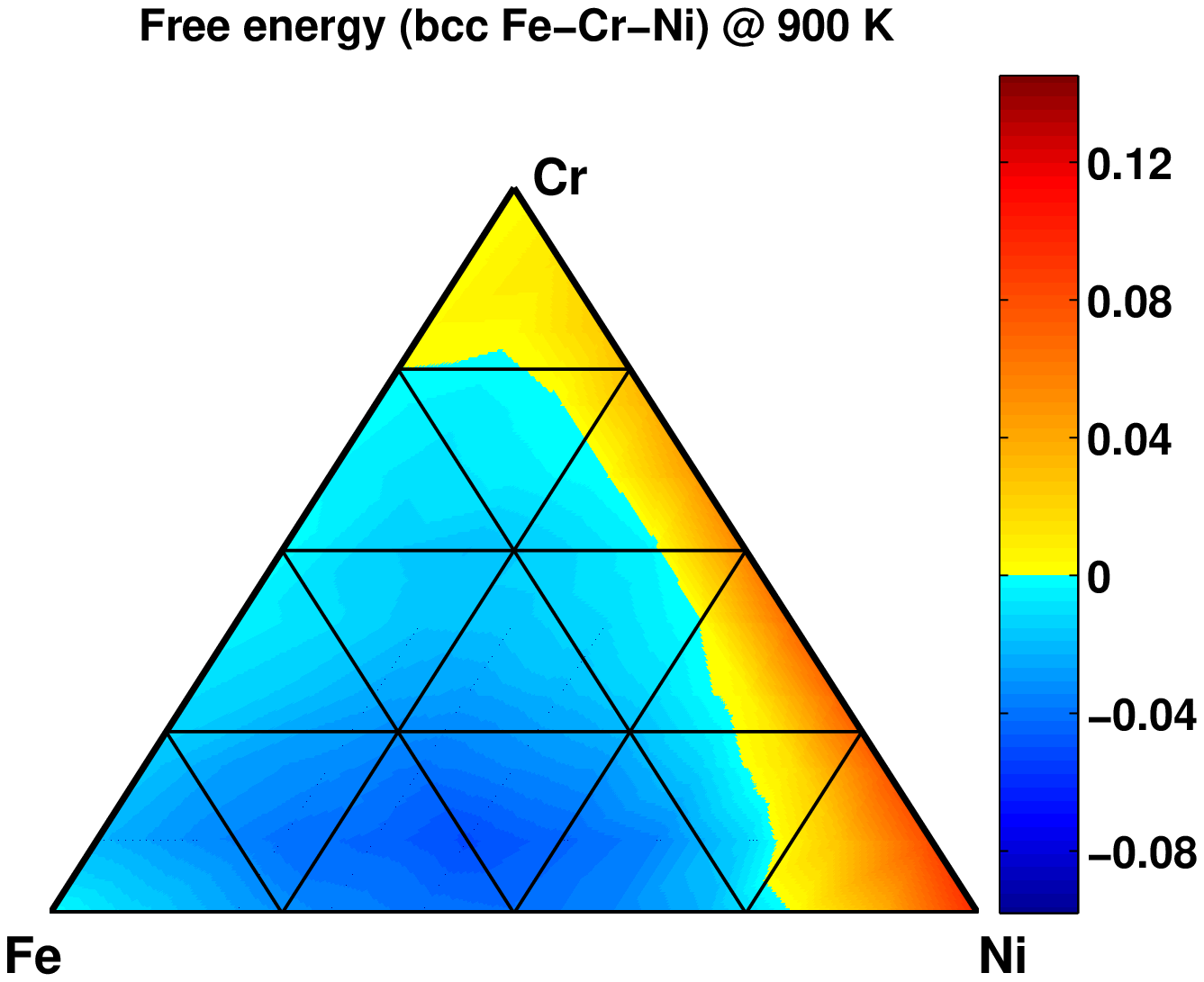}
			\end{minipage}
			\caption{
		(Color online) Enthalpies of formation (in eV/atom) (a,b), the product of temperature and configurational entropies $TS_{conf}$ (in eV/atom units) (c,d) and free energies of formation (in eV/atom units) (e,f) at 900 K computed using MC simulations for Fe-Cr-Ni alloys on fcc (a,c,e) and bcc (b,d,f) lattices. }
		\label{fig:EntropyFreeEn_900K}
\end{figure*}

Formation free energies of fcc and bcc Fe-Cr-Ni alloys at 900 K computed using MC simulations for the entire range of alloy compositions are shown in Figs. \ref{fig:EntropyFreeEn_900K}(e) and \ref{fig:EntropyFreeEn_900K}(f), together with their formation enthalpies and configurational entropies, see Figs. \ref{fig:EntropyFreeEn_900K}(a,b) and Figs. \ref{fig:EntropyFreeEn_900K}(c,d). Configurational entropy of fcc alloys is higher for most of the alloy compositions in comparison with that of bcc alloys. Hence the configurational entropy contribution to the formation free energies is more significant for fcc alloys than for bcc alloys. The region of stability of fcc alloys at 900 K defined using the free energy criterion is broader than the region of stability defined using the formation enthalpy criterion, see Figs. \ref{fig:Energy_diff}(e) and \ref{fig:Energy_diff}(f). At low temperatures the difference between fcc-bcc phase transition lines obtained using both criteria is negligible, see Figs. \ref{fig:Energy_diff}(a-d), whereas at high temperatures the role played by the configurational entropy effects is more pronounced, see Figs. \ref{fig:Energy_diff}(e-f) and \ref{fig:Energy_diff_1600K}(a-b).

\section{Finite temperature magnetic properties of Fe-Cr-Ni alloys}

\begin{figure}
\includegraphics[width=\columnwidth]{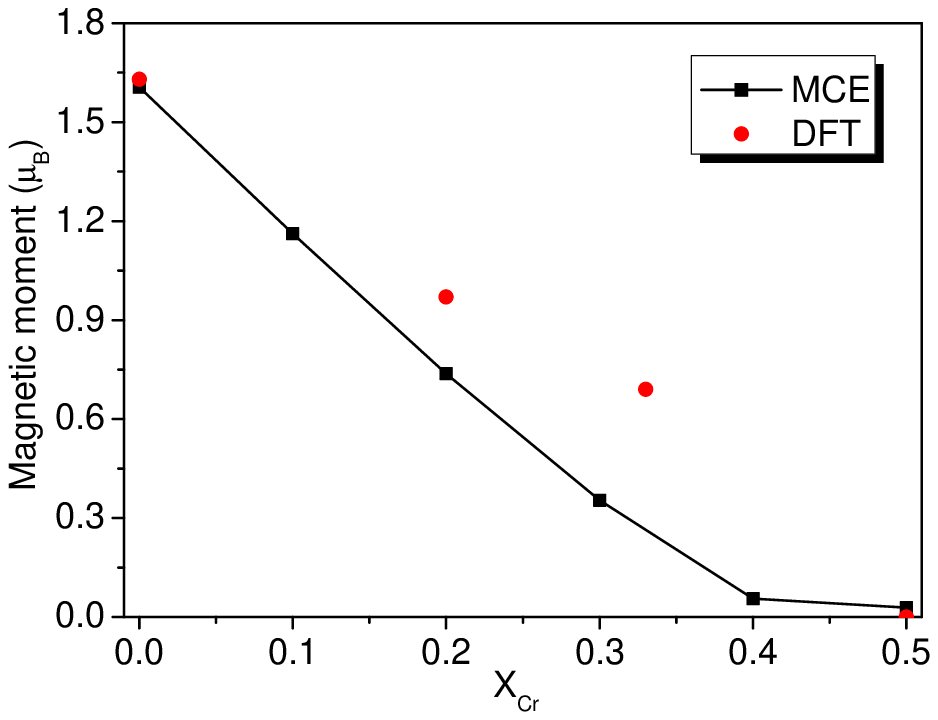}
\caption{(Color online) Total magnetic moment per atom in (Fe$_{0.5}$Ni$_{0.5}$)$_{1-x}$Cr$_{x}$ alloy as a function of chromium content $x$, predicted by MCE. DFT results (Sec. III.E) are shown by red circles.
        \label{fig:MCE_FeNi-Cr}}
\end{figure}

Using the DFT database and Magnetic Cluster Expansion, we now investigate how magnetic properties of Fe-Cr-Ni ternary alloys vary as functions of temperature. Monte Carlo MCE simulations were performed using a 16384 atom simulation cell (16$\times$16$\times$16 fcc unit cells). At each MC step, a trial random variation of the magnetic moment of a randomly chosen atom is attempted and accepted or rejected according to the Metropolis criterion. Angular and longitudinal fluctuations of magnetic moments are relatively small, resulting in the formation of non-collinear magnetic configurations. We do not consider MC moves that change the sign of magnetic moment. The gap-less spectrum of magnetic excitations (magnons) in our model is described by small tilts of magnetic moments away from their equilibrium orientations. Both the thermalization and accumulation stages include on average 40 000 MC steps per atom.
As an example of application of MCE to modelling low-temperature magnetic properties of a ternary alloy, we investigate how the total magnetic moment of a disordered (Fe$_{0.5}$Ni$_{0.5}$)$_{1-x}$Cr$_{x}$ alloy varies as a function of Cr content. We noted in Section III.E that the average magnetic moment of the alloy decreased rapidly as a function of chromium concentration. In MCE Monte Carlo simulations, ordered Fe-Ni alloy with L1$_0$ structure was chosen as the initial alloy configuration. The magnetic moment per atom in this structure was found to be 1.61 $\mu_B$, close to the DFT value of 1.63 $\mu_B$. Chromium content was then varied by replacing equal numbers of Fe and Ni atoms in their sublattices with Cr atoms, with positions of chromium atoms chosen at random. Figure \ref{fig:MCE_FeNi-Cr} shows the predicted variation of magnetic moment in the resulting alloy at low temperatures. With increasing Cr content, magnetization rapidly decreases, resulting in a completely non-magnetic system at $x_{Cr}$=0.4, in agreement with {\it ab initio} results of Section III.E, also illustrated in Figure \ref{fig:MCE_FeNi-Cr}.

\begin{figure}
\includegraphics[width=\columnwidth]{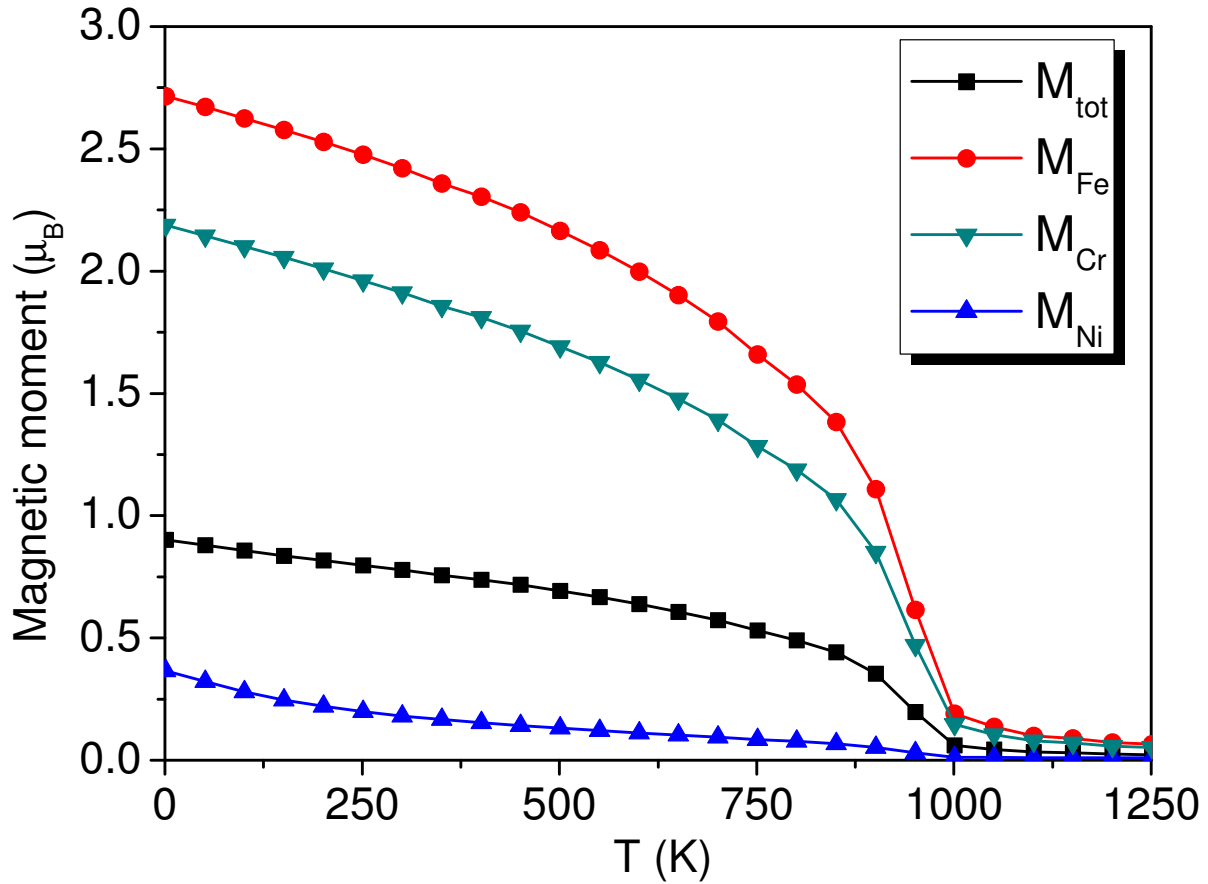}
\caption{(Color online) Temperature dependence of the total magnetic moment of ordered Fe$_2$CrNi alloy, and magnetic moments of atoms forming the alloy.
        \label{fig:MCE_Fe2CrNi}}
\end{figure}

While random alloys with composition close to Fe$_{50}$Cr$_{25}$Ni$_{25}$ are almost entirely anti-ferromagnetic, ordered alloys with the same composition have non-vanishing total magnetic moment. At low temperature, magnetic moments are collinear. The Cr moments are anti-ferromagnetically ordered with respect to Fe moments and have almost the same magnitude, while the magnetic moments of Ni atoms are smaller and ordered ferromagnetically with respect to the Fe moments. Finite temperature magnetic properties were investigated using MC simulations performed using large simulation cells. Magnetic moments of each of the three components of ordered Fe$_2$CrNi alloy, treated as functions of temperature, are shown in Figure \ref{fig:MCE_Fe2CrNi}. Their values at low temperature are in reasonable agreement with DFT, for example magnetic moments of Fe, Cr and Ni obtained from MCE simulations are 2.7, -2.2 and 0.37 $\mu_B$, whereas DFT predictions are 2.08, -2.44 and 0.15 $\mu_B$, respectively, see Table \ref{tab:Enthalpies_of_GS}. The alloy remains magnetic at fairly high temperatures close to 850-900 K. The effect is similar to that found in fcc Fe-Ni, where chemically ordered FeNi$_3$ alloy has higher Curie temperature than pure Ni. In ternary Fe$_2$CrNi alloy, the ferromagnetically ordered structure of the alloy owes its stability to strong anti-ferromagnetic interactions between (Fe, Ni) and Cr atoms.

Another important application of the Magnetic Cluster Expansion to the high-temperature properties of the Fe-Cr-Ni system is the study of the relative stability of fcc and bcc structures. Previously \cite{Lavrentiev2010} we have used the MCE in order to find the free energy difference between fcc and bcc Fe in the whole range of temperatures from 0 K to the melting point. Here, we use these results in order to estimate the magnetic correction to the free energy of alloy formation. For this correction we used formulae analogous to Eq. \ref{eq:Correction_to_enthalpy} with the free energy difference between fcc and bcc Fe taken from Fig. 2 of Ref. \onlinecite{Lavrentiev2010}. Fig. \ref{fig:Energy_diff_1600K} shows that both magnetic and configurational effects are important. However, it is also apparent that the magnetic correction to free energy difference, compare Figs. \ref{fig:Energy_diff_1600K}(b) and \ref{fig:Energy_diff_1600K}(d), is less pronounced than that applied to the enthalpy difference, see Figs. \ref{fig:Energy_diff_1600K}(a) and \ref{fig:Energy_diff_1600K}(c). This is related to the fact that even though the fcc phase of Fe is strongly stabilized at high temperatures in terms of enthalpy, the free energy of bcc Fe with magnetic effects taken into account is still lower than that of fcc Fe, see Fig. 2 of Ref. \onlinecite{Lavrentiev2010}. As discussed in Ref. \onlinecite{Lavrentiev2010}, only after the inclusion of vibrational effects can the stability of fcc phase at high temperatures be predicted correctly. It can be deduced that the vibrational contribution to free energy is also important at high temperatures  for Fe-Cr-Ni alloys. The corresponding study will be performed in our future research.

Concluding this section, we note that although the parametrization of MCE Hamiltonian involved a number of approximations, MCE predictions are in good agreement with the low temperature DFT data. In the high temperature limit, MC simulations show that interplay between chemical and magnetic degrees of freedom gives rise to the high Curie temperature of ordered Fe$_2$CrNi alloy. Further improvement in the accuracy of the MCE model is expected to provide a means for investigating temperature-dependent magnetic and configurational order over the entire range of alloy compositions.

\section{Conclusions}

We have investigated the stability of the fcc and bcc phases of ternary magnetic Fe-Cr-Ni alloys, using a combination of first-principles DFT calculations and Monte Carlo simulations, involving both conventional (CE) and magnetic (MCE) cluster expansion approaches. Detailed derivation of a general expression for the CE enthalpy of mixing for a ternary alloy system is presented, where average cluster functions are defined as products of orthogonal point functions. An explicit analytical relationship between chemical SRO and effective pair-wise interactions, involving different atomic species, is established and applied to the analysis of SRO in Fe-Cr-Ni alloys and the interpretation of experimental data. Using a DFT database of 248 fcc and 246 bcc structures, we assessed fcc and bcc phase stability of this ternary alloy system. Effective cluster interaction parameters for fcc and bcc binaries and ternaries have been derived and cross-validated against DFT data. Strong deviations from Vegard's law for atomic volumes treated as functions of alloy composition stem from magnetic interactions. The predicted average total and local magnetic moments treated as functions of Ni concentration in the ground-state bcc and fcc structures of Fe-Ni alloys are in good agreement with experimental data. Calculations have not only helped identify ground-state structures of the three binary alloys, but also predicted the fcc-like Fe$_{2}$CrNi ternary compound as the most stable ground-state ternary intermetallic system with negative enthalpy of formation of -0.164 eV/atom and the lowest order-disoder transition temperature of 650 K. Both DFT and MCE simulations show that the phase stability of the Fe$_{2}$CrNi structure is primarily determined by strong anti-ferromagnetic interactions between Fe and Ni atoms with Cr atoms.

Analysis of the relative phase stability of ternary fcc and bcc phases at various temperatures in terms of formation enthalpies and formation free energies shows that configurational entropy plays an important part at high temperature in stabilizing fcc alloys with respect to bcc alloys. Preliminary incorporation of magnetic entropy for free energy differences in the Fe-rich corner shows that non-collinear magnetic effects are important at high temperatures. Excellent agreement between calculations and experimental data on enthalpies of formation at 1600 K also shows that magnetic contribution plays a significant part, correcting the deficiencies of conventional CE treatment of Fe-Cr-Ni alloys. We have calculated the Warren-Cowley short range order parameters at various temperatures and found good agreement with experimental data on binary and ternary alloys. Particular attention has been devoted to Fe$_{56}$Cr$_{21}$Ni$_{23}$, Fe$_{38}$Cr$_{14}$Ni$_{48}$ and Fe$_{34}$Cr$_{20}$Ni$_{46}$ alloys close to the centre of the composition triangle, to rationalize how SRO varies in Fe-Cr, Fe-Ni and Ni-Cr binary alloys at various temperatures. The fact that SRO decreases significantly for Fe-Ni pairs as a function of Cr concentration agrees with experimental observations. This important aspect of alloy thermodynamics is also related to the fact that interaction between Cr and both Fe and Ni is strongly anti-ferromagnetic, explaining large negative values of SRO predicted for Fe-Cr and Ni-Cr atomic pairs.

Our study shows that it is now possible to treat thermodynamics of Fe-Cr-Ni starting from first principles, taking into account both chemical and magnetic interactions in this traditionally important but complex ternary magnetic alloy. By comparing MC configurations generated using effective cluster interactions with those created by the SQS method, we are able to demonstrate that the former are energetically more stable for all the magnetic structures considered here, as illustrated by the case of Fe$_{70}$Cr$_{20}$Ni$_{10}$ alloy. This provides vital information about the reference structures required for modelling point defects in ternary alloys, where defect properties exhibit high sensitivity not only to the average alloy composition but also to the local chemical and magnetic environment of a defect site \cite{Nguyen-Manh2012,Muzyk2011,Muzyk2013}.

\begin{acknowledgments}
This work was funded by the Accelerated Metallurgy Project, which is co-funded by the European Commission in the 7th Framework Programme (Contract NMP4-LA-2011-263206), by the European Space Agency and by the individual partner organizations. This work was also part-funded by the RCUK Energy Programme (Grant Number EP/I501045) and by the European Union's Horizon 2020 research and innovation programme under grant agreement number 633053. To obtain further information on the data and models underlying this paper please contact PublicationsManager@ccfe.ac.uk. The views and opinions expressed herein do not necessarily reflect those of the European Commission. The authors would like to thank Charlotte Becquart, Maria Ganchenkova and George Smith for stimulating and helpful discussions. DNM would like to acknowledge the Juelich supercomputer centre for the provision of High-Performances Computer for Fusion (HPC-FF) facilities as well as the International Fusion Energy Research Centre (IFERC) for the provision of a supercomputer (Helios) at the Computational Simulation Centre (CSC) in Rokkasho (Japan).
\end{acknowledgments}

\bibliography{FeCrNi_references_v1}

\begin{thebibliography}{104}%
\makeatletter
\providecommand \@ifxundefined [1]{%
 \@ifx{#1\undefined}
}%
\providecommand \@ifnum [1]{%
 \ifnum #1\expandafter \@firstoftwo
 \else \expandafter \@secondoftwo
 \fi
}%
\providecommand \@ifx [1]{%
 \ifx #1\expandafter \@firstoftwo
 \else \expandafter \@secondoftwo
 \fi
}%
\providecommand \natexlab [1]{#1}%
\providecommand \enquote  [1]{``#1''}%
\providecommand \bibnamefont  [1]{#1}%
\providecommand \bibfnamefont [1]{#1}%
\providecommand \citenamefont [1]{#1}%
\providecommand \href@noop [0]{\@secondoftwo}%
\providecommand \href [0]{\begingroup \@sanitize@url \@href}%
\providecommand \@href[1]{\@@startlink{#1}\@@href}%
\providecommand \@@href[1]{\endgroup#1\@@endlink}%
\providecommand \@sanitize@url [0]{\catcode `\\12\catcode `\$12\catcode
  `\&12\catcode `\#12\catcode `\^12\catcode `\_12\catcode `\%12\relax}%
\providecommand \@@startlink[1]{}%
\providecommand \@@endlink[0]{}%
\providecommand \url  [0]{\begingroup\@sanitize@url \@url }%
\providecommand \@url [1]{\endgroup\@href {#1}{\urlprefix }}%
\providecommand \urlprefix  [0]{URL }%
\providecommand \Eprint [0]{\href }%
\providecommand \doibase [0]{http://dx.doi.org/}%
\providecommand \selectlanguage [0]{\@gobble}%
\providecommand \bibinfo  [0]{\@secondoftwo}%
\providecommand \bibfield  [0]{\@secondoftwo}%
\providecommand \translation [1]{[#1]}%
\providecommand \BibitemOpen [0]{}%
\providecommand \bibitemStop [0]{}%
\providecommand \bibitemNoStop [0]{.\EOS\space}%
\providecommand \EOS [0]{\spacefactor3000\relax}%
\providecommand \BibitemShut  [1]{\csname bibitem#1\endcsname}%
\let\auto@bib@innerbib\@empty
\bibitem [{\citenamefont {Guillaume}(1897)}]{Guillaume1897}%
  \BibitemOpen
  \bibfield  {author} {\bibinfo {author} {\bibfnamefont {C.~E.}\ \bibnamefont
  {Guillaume}},\ }\href@noop {} {\bibfield  {journal} {\bibinfo  {journal} {C.
  R. Acad. Sci.}\ }\textbf {\bibinfo {volume} {125}},\ \bibinfo {pages} {235}
  (\bibinfo {year} {1897})}\BibitemShut {NoStop}%
\bibitem [{\citenamefont {Arnold}\ and\ \citenamefont
  {Elmen}(1923)}]{Arnold1923}%
  \BibitemOpen
  \bibfield  {author} {\bibinfo {author} {\bibfnamefont {H.}~\bibnamefont
  {Arnold}}\ and\ \bibinfo {author} {\bibfnamefont {G.~W.}\ \bibnamefont
  {Elmen}},\ }\href@noop {} {\bibfield  {journal} {\bibinfo  {journal} {J.
  Franklin Inst.}\ }\textbf {\bibinfo {volume} {195}},\ \bibinfo {pages} {621}
  (\bibinfo {year} {1923})}\BibitemShut {NoStop}%
\bibitem [{\citenamefont {Klueh}\ and\ \citenamefont {Harries}(2001)}]{Klueh}%
  \BibitemOpen
  \bibfield  {author} {\bibinfo {author} {\bibfnamefont {R.~L.}\ \bibnamefont
  {Klueh}}\ and\ \bibinfo {author} {\bibfnamefont {D.~R.}\ \bibnamefont
  {Harries}},\ }\href@noop {} {\emph {\bibinfo {title} {{High-Chromium Ferritic
  and Martensitic Steels for Nuclear Applications}}}}\ (\bibinfo  {publisher}
  {American Society for Testing of Materials (ASTM)},\ \bibinfo {address}
  {USA},\ \bibinfo {year} {2001})\BibitemShut {NoStop}%
\bibitem [{\citenamefont {Toyama}\ \emph {et~al.}(2012)\citenamefont {Toyama},
  \citenamefont {Nozawa}, \citenamefont {{van Renterghem}}, \citenamefont
  {Matsukawa}, \citenamefont {Hatakeyama}, \citenamefont {Nagai}, \citenamefont
  {{Al Mazouzi}},\ and\ \citenamefont {{van Dyck}}}]{Toyama2012}%
  \BibitemOpen
  \bibfield  {author} {\bibinfo {author} {\bibfnamefont {T.}~\bibnamefont
  {Toyama}}, \bibinfo {author} {\bibfnamefont {Y.}~\bibnamefont {Nozawa}},
  \bibinfo {author} {\bibfnamefont {W.}~\bibnamefont {{van Renterghem}}},
  \bibinfo {author} {\bibfnamefont {Y.}~\bibnamefont {Matsukawa}}, \bibinfo
  {author} {\bibfnamefont {M.}~\bibnamefont {Hatakeyama}}, \bibinfo {author}
  {\bibfnamefont {Y.}~\bibnamefont {Nagai}}, \bibinfo {author} {\bibfnamefont
  {A.}~\bibnamefont {{Al Mazouzi}}}, \ and\ \bibinfo {author} {\bibfnamefont
  {S.}~\bibnamefont {{van Dyck}}},\ }\href {\doibase
  10.1016/j.jnucmat.2011.11.072} {\bibfield  {journal} {\bibinfo  {journal} {J.
  Nucl. Mater.}\ }\textbf {\bibinfo {volume} {425}},\ \bibinfo {pages} {71}
  (\bibinfo {year} {2012})}\BibitemShut {NoStop}%
\bibitem [{\citenamefont {Rowcliffe}\ \emph {et~al.}(2009)\citenamefont
  {Rowcliffe}, \citenamefont {Mansur}, \citenamefont {Hoelzer},\ and\
  \citenamefont {Nanstad}}]{Rowcliffe2009}%
  \BibitemOpen
  \bibfield  {author} {\bibinfo {author} {\bibfnamefont {A.~F.}\ \bibnamefont
  {Rowcliffe}}, \bibinfo {author} {\bibfnamefont {L.~K.}\ \bibnamefont
  {Mansur}}, \bibinfo {author} {\bibfnamefont {D.~T.}\ \bibnamefont {Hoelzer}},
  \ and\ \bibinfo {author} {\bibfnamefont {R.~K.}\ \bibnamefont {Nanstad}},\
  }\href {\doibase 10.1016/j.jnucmat.2009.03.023} {\bibfield  {journal}
  {\bibinfo  {journal} {J. Nucl. Mater.}\ }\textbf {\bibinfo {volume} {392}},\
  \bibinfo {pages} {341} (\bibinfo {year} {2009})}\BibitemShut {NoStop}%
\bibitem [{\citenamefont {Stork}\ \emph {et~al.}(2014)\citenamefont {Stork},
  \citenamefont {Agostini}, \citenamefont {Boutard}, \citenamefont
  {Buckthorpe}, \citenamefont {Diegele}, \citenamefont {Dudarev}, \citenamefont
  {English}, \citenamefont {Federici}, \citenamefont {Gilbert}, \citenamefont
  {Gonzalez}, \citenamefont {Ibarra}, \citenamefont {Linsmeier}, \citenamefont
  {Puma}, \citenamefont {Marbach}, \citenamefont {Packer}, \citenamefont {Raj},
  \citenamefont {Rieth}, \citenamefont {Tran}, \citenamefont {Ward},\ and\
  \citenamefont {Zinkle}}]{Stork2014}%
  \BibitemOpen
  \bibfield  {author} {\bibinfo {author} {\bibfnamefont {D.}~\bibnamefont
  {Stork}}, \bibinfo {author} {\bibfnamefont {P.}~\bibnamefont {Agostini}},
  \bibinfo {author} {\bibfnamefont {J.-L.}\ \bibnamefont {Boutard}}, \bibinfo
  {author} {\bibfnamefont {D.}~\bibnamefont {Buckthorpe}}, \bibinfo {author}
  {\bibfnamefont {E.}~\bibnamefont {Diegele}}, \bibinfo {author} {\bibfnamefont
  {S.~L.}\ \bibnamefont {Dudarev}}, \bibinfo {author} {\bibfnamefont
  {C.}~\bibnamefont {English}}, \bibinfo {author} {\bibfnamefont
  {G.}~\bibnamefont {Federici}}, \bibinfo {author} {\bibfnamefont {M.~R.}\
  \bibnamefont {Gilbert}}, \bibinfo {author} {\bibfnamefont {S.}~\bibnamefont
  {Gonzalez}}, \bibinfo {author} {\bibfnamefont {A.}~\bibnamefont {Ibarra}},
  \bibinfo {author} {\bibfnamefont {C.}~\bibnamefont {Linsmeier}}, \bibinfo
  {author} {\bibfnamefont {A.~L.}\ \bibnamefont {Puma}}, \bibinfo {author}
  {\bibfnamefont {G.}~\bibnamefont {Marbach}}, \bibinfo {author} {\bibfnamefont
  {L.~W.}\ \bibnamefont {Packer}}, \bibinfo {author} {\bibfnamefont
  {B.}~\bibnamefont {Raj}}, \bibinfo {author} {\bibfnamefont {M.}~\bibnamefont
  {Rieth}}, \bibinfo {author} {\bibfnamefont {M.~Q.}\ \bibnamefont {Tran}},
  \bibinfo {author} {\bibfnamefont {D.~J.}\ \bibnamefont {Ward}}, \ and\
  \bibinfo {author} {\bibfnamefont {S.~J.}\ \bibnamefont {Zinkle}},\ }\href
  {\doibase 10.1016/j.fusengdes.2013.11.007} {\bibfield  {journal} {\bibinfo
  {journal} {Fusion Eng. Des.}\ }\textbf {\bibinfo {volume} {89}},\ \bibinfo
  {pages} {1586} (\bibinfo {year} {2014})}\BibitemShut {NoStop}%
\bibitem [{\citenamefont {Boutard}\ \emph {et~al.}(2008)\citenamefont
  {Boutard}, \citenamefont {Alamo}, \citenamefont {R},\ and\ \citenamefont
  {M}}]{Boutard2008}%
  \BibitemOpen
  \bibfield  {author} {\bibinfo {author} {\bibfnamefont {J.~L.}\ \bibnamefont
  {Boutard}}, \bibinfo {author} {\bibfnamefont {A.}~\bibnamefont {Alamo}},
  \bibinfo {author} {\bibfnamefont {L.}~\bibnamefont {R}}, \ and\ \bibinfo
  {author} {\bibfnamefont {R.}~\bibnamefont {M}},\ }\href@noop {} {\bibfield
  {journal} {\bibinfo  {journal} {C. R. Physique}\ }\textbf {\bibinfo {volume}
  {9}},\ \bibinfo {pages} {287} (\bibinfo {year} {2008})}\BibitemShut {NoStop}%
\bibitem [{\citenamefont {Satoh}\ \emph {et~al.}(2007)\citenamefont {Satoh},
  \citenamefont {Abe}, \citenamefont {Matsui},\ and\ \citenamefont
  {Yamagata}}]{Satoh2007}%
  \BibitemOpen
  \bibfield  {author} {\bibinfo {author} {\bibfnamefont {Y.}~\bibnamefont
  {Satoh}}, \bibinfo {author} {\bibfnamefont {S.}~\bibnamefont {Abe}}, \bibinfo
  {author} {\bibfnamefont {H.}~\bibnamefont {Matsui}}, \ and\ \bibinfo {author}
  {\bibfnamefont {I.}~\bibnamefont {Yamagata}},\ }\href {\doibase
  10.1016/j.jnucmat.2007.03.057} {\bibfield  {journal} {\bibinfo  {journal} {J.
  Nucl. Mater.}\ }\textbf {\bibinfo {volume} {367-370}},\ \bibinfo {pages}
  {972} (\bibinfo {year} {2007})}\BibitemShut {NoStop}%
\bibitem [{\citenamefont {Ferry}(2006)}]{Ferry2006}%
  \BibitemOpen
  \bibfield  {author} {\bibinfo {author} {\bibfnamefont {M.}~\bibnamefont
  {Ferry}},\ }\href@noop {} {\emph {\bibinfo {title} {{Direct Strip Casting of
  Metals and Alloys}}}}\ (\bibinfo  {publisher} {Woodhead Publishing Limited},\
  \bibinfo {address} {Cambridge},\ \bibinfo {year} {2006})\BibitemShut
  {NoStop}%
\bibitem [{\citenamefont {Rees}\ \emph {et~al.}(1949)\citenamefont {Rees},
  \citenamefont {Burns},\ and\ \citenamefont {Cook}}]{Rees1949}%
  \BibitemOpen
  \bibfield  {author} {\bibinfo {author} {\bibfnamefont {W.~P.}\ \bibnamefont
  {Rees}}, \bibinfo {author} {\bibfnamefont {B.~D.}\ \bibnamefont {Burns}}, \
  and\ \bibinfo {author} {\bibfnamefont {A.~J.}\ \bibnamefont {Cook}},\
  }\href@noop {} {\bibfield  {journal} {\bibinfo  {journal} {J. Iron Steel
  Inst.}\ }\textbf {\bibinfo {volume} {162}},\ \bibinfo {pages} {325} (\bibinfo
  {year} {1949})}\BibitemShut {NoStop}%
\bibitem [{\citenamefont {Hattersley}\ and\ \citenamefont
  {Hume-Rothery}(1966)}]{Hattersley1966}%
  \BibitemOpen
  \bibfield  {author} {\bibinfo {author} {\bibfnamefont {B.}~\bibnamefont
  {Hattersley}}\ and\ \bibinfo {author} {\bibfnamefont {W.}~\bibnamefont
  {Hume-Rothery}},\ }\href@noop {} {\bibfield  {journal} {\bibinfo  {journal}
  {J. Iron Steel Inst.}\ }\textbf {\bibinfo {volume} {204}},\ \bibinfo {pages}
  {683} (\bibinfo {year} {1966})}\BibitemShut {NoStop}%
\bibitem [{\citenamefont {Cook}\ and\ \citenamefont {Brown}(1952)}]{Cook1952}%
  \BibitemOpen
  \bibfield  {author} {\bibinfo {author} {\bibfnamefont {A.~J.}\ \bibnamefont
  {Cook}}\ and\ \bibinfo {author} {\bibfnamefont {B.~R.}\ \bibnamefont
  {Brown}},\ }\href@noop {} {\bibfield  {journal} {\bibinfo  {journal} {J. Iron
  Steel Inst.}\ }\textbf {\bibinfo {volume} {171}},\ \bibinfo {pages} {345}
  (\bibinfo {year} {1952})}\BibitemShut {NoStop}%
\bibitem [{\citenamefont {K\"ormann}\ \emph {et~al.}(2014)\citenamefont
  {K\"ormann}, \citenamefont {Breidi}, \citenamefont {Dudarev}, \citenamefont
  {Dupin}, \citenamefont {Ghosh}, \citenamefont {Hickel}, \citenamefont
  {Korzhavyi}, \citenamefont {Mu\~noz},\ and\ \citenamefont
  {Ohnuma}}]{Koermann2014}%
  \BibitemOpen
  \bibfield  {author} {\bibinfo {author} {\bibfnamefont {F.}~\bibnamefont
  {K\"ormann}}, \bibinfo {author} {\bibfnamefont {A.~A.~H.}\ \bibnamefont
  {Breidi}}, \bibinfo {author} {\bibfnamefont {S.~L.}\ \bibnamefont {Dudarev}},
  \bibinfo {author} {\bibfnamefont {N.}~\bibnamefont {Dupin}}, \bibinfo
  {author} {\bibfnamefont {G.}~\bibnamefont {Ghosh}}, \bibinfo {author}
  {\bibfnamefont {T.}~\bibnamefont {Hickel}}, \bibinfo {author} {\bibfnamefont
  {P.}~\bibnamefont {Korzhavyi}}, \bibinfo {author} {\bibfnamefont {J.~A.}\
  \bibnamefont {Mu\~noz}}, \ and\ \bibinfo {author} {\bibfnamefont
  {I.}~\bibnamefont {Ohnuma}},\ }\href@noop {} {\bibfield  {journal} {\bibinfo
  {journal} {Phys. Stat. Sol. B}\ }\textbf {\bibinfo {volume} {251}},\ \bibinfo
  {pages} {53} (\bibinfo {year} {2014})}\BibitemShut {NoStop}%
\bibitem [{\citenamefont {Cacciamani}\ \emph {et~al.}(2010)\citenamefont
  {Cacciamani}, \citenamefont {Dinsdale}, \citenamefont {Palumbo},\ and\
  \citenamefont {Pasturel}}]{Cacciamani2010}%
  \BibitemOpen
  \bibfield  {author} {\bibinfo {author} {\bibfnamefont {G.}~\bibnamefont
  {Cacciamani}}, \bibinfo {author} {\bibfnamefont {A.}~\bibnamefont
  {Dinsdale}}, \bibinfo {author} {\bibfnamefont {M.}~\bibnamefont {Palumbo}}, \
  and\ \bibinfo {author} {\bibfnamefont {A.}~\bibnamefont {Pasturel}},\ }\href
  {\doibase 10.1016/j.intermet.2010.02.026} {\bibfield  {journal} {\bibinfo
  {journal} {Intermetallics}\ }\textbf {\bibinfo {volume} {18}},\ \bibinfo
  {pages} {1148} (\bibinfo {year} {2010})}\BibitemShut {NoStop}%
\bibitem [{\citenamefont {Franke}\ and\ \citenamefont
  {Seifert}(2011)}]{Franke2011}%
  \BibitemOpen
  \bibfield  {author} {\bibinfo {author} {\bibfnamefont {P.}~\bibnamefont
  {Franke}}\ and\ \bibinfo {author} {\bibfnamefont {H.~J.}\ \bibnamefont
  {Seifert}},\ }\href {\doibase 10.1016/j.calphad.2010.10.006} {\bibfield
  {journal} {\bibinfo  {journal} {Calphad}\ }\textbf {\bibinfo {volume} {35}},\
  \bibinfo {pages} {148} (\bibinfo {year} {2011})}\BibitemShut {NoStop}%
\bibitem [{\citenamefont {Klaver}\ \emph {et~al.}(2012)\citenamefont {Klaver},
  \citenamefont {Hepburn},\ and\ \citenamefont {Ackland}}]{Klaver2012}%
  \BibitemOpen
  \bibfield  {author} {\bibinfo {author} {\bibfnamefont {T.~P.~C.}\
  \bibnamefont {Klaver}}, \bibinfo {author} {\bibfnamefont {D.~J.}\
  \bibnamefont {Hepburn}}, \ and\ \bibinfo {author} {\bibfnamefont {G.~J.}\
  \bibnamefont {Ackland}},\ }\href {\doibase 10.1103/PhysRevB.85.174111}
  {\bibfield  {journal} {\bibinfo  {journal} {Phys. Rev. B}\ }\textbf {\bibinfo
  {volume} {85}},\ \bibinfo {pages} {174111} (\bibinfo {year}
  {2012})}\BibitemShut {NoStop}%
\bibitem [{\citenamefont {Hepburn}\ \emph {et~al.}(2013)\citenamefont
  {Hepburn}, \citenamefont {Ferguson}, \citenamefont {Gardner},\ and\
  \citenamefont {Ackland}}]{Hepburn2013}%
  \BibitemOpen
  \bibfield  {author} {\bibinfo {author} {\bibfnamefont {D.~J.}\ \bibnamefont
  {Hepburn}}, \bibinfo {author} {\bibfnamefont {D.}~\bibnamefont {Ferguson}},
  \bibinfo {author} {\bibfnamefont {S.}~\bibnamefont {Gardner}}, \ and\
  \bibinfo {author} {\bibfnamefont {G.~J.}\ \bibnamefont {Ackland}},\ }\href
  {\doibase 10.1103/PhysRevB.88.024115} {\bibfield  {journal} {\bibinfo
  {journal} {Phys. Rev. B}\ }\textbf {\bibinfo {volume} {88}},\ \bibinfo
  {pages} {024115} (\bibinfo {year} {2013})}\BibitemShut {NoStop}%
\bibitem [{\citenamefont {Vitos}\ \emph {et~al.}(2002)\citenamefont {Vitos},
  \citenamefont {Korzhavyi},\ and\ \citenamefont {Johansson}}]{Vitos2002}%
  \BibitemOpen
  \bibfield  {author} {\bibinfo {author} {\bibfnamefont {L.}~\bibnamefont
  {Vitos}}, \bibinfo {author} {\bibfnamefont {P.~A.}\ \bibnamefont
  {Korzhavyi}}, \ and\ \bibinfo {author} {\bibfnamefont {B.}~\bibnamefont
  {Johansson}},\ }\href {\doibase 10.1103/PhysRevLett.88.155501} {\bibfield
  {journal} {\bibinfo  {journal} {Phys. Rev. Lett.}\ }\textbf {\bibinfo
  {volume} {88}},\ \bibinfo {pages} {155501} (\bibinfo {year}
  {2002})}\BibitemShut {NoStop}%
\bibitem [{\citenamefont {Vitos}\ \emph {et~al.}(2006)\citenamefont {Vitos},
  \citenamefont {Korzhavyi},\ and\ \citenamefont {Johansson}}]{Vitos2006}%
  \BibitemOpen
  \bibfield  {author} {\bibinfo {author} {\bibfnamefont {L.}~\bibnamefont
  {Vitos}}, \bibinfo {author} {\bibfnamefont {P.~A.}\ \bibnamefont
  {Korzhavyi}}, \ and\ \bibinfo {author} {\bibfnamefont {B.}~\bibnamefont
  {Johansson}},\ }\href {\doibase 10.1103/PhysRevLett.96.117210} {\bibfield
  {journal} {\bibinfo  {journal} {Phys. Rev. Lett.}\ }\textbf {\bibinfo
  {volume} {96}},\ \bibinfo {pages} {117210} (\bibinfo {year}
  {2006})}\BibitemShut {NoStop}%
\bibitem [{\citenamefont {Delczeg}\ \emph {et~al.}(2012)\citenamefont
  {Delczeg}, \citenamefont {Johansson},\ and\ \citenamefont
  {Vitos}}]{Delczeg2012}%
  \BibitemOpen
  \bibfield  {author} {\bibinfo {author} {\bibfnamefont {L.}~\bibnamefont
  {Delczeg}}, \bibinfo {author} {\bibfnamefont {B.}~\bibnamefont {Johansson}},
  \ and\ \bibinfo {author} {\bibfnamefont {L.}~\bibnamefont {Vitos}},\ }\href
  {\doibase 10.1103/PhysRevB.85.174101} {\bibfield  {journal} {\bibinfo
  {journal} {Phys. Rev. B}\ }\textbf {\bibinfo {volume} {85}},\ \bibinfo
  {pages} {174101} (\bibinfo {year} {2012})}\BibitemShut {NoStop}%
\bibitem [{\citenamefont {Piochaud}\ \emph {et~al.}(2014)\citenamefont
  {Piochaud}, \citenamefont {Klaver}, \citenamefont {Adjanor}, \citenamefont
  {Olsson}, \citenamefont {Domain},\ and\ \citenamefont
  {Becquart}}]{Piochaud2014}%
  \BibitemOpen
  \bibfield  {author} {\bibinfo {author} {\bibfnamefont {J.~B.}\ \bibnamefont
  {Piochaud}}, \bibinfo {author} {\bibfnamefont {T.~P.~C.}\ \bibnamefont
  {Klaver}}, \bibinfo {author} {\bibfnamefont {G.}~\bibnamefont {Adjanor}},
  \bibinfo {author} {\bibfnamefont {P.}~\bibnamefont {Olsson}}, \bibinfo
  {author} {\bibfnamefont {C.}~\bibnamefont {Domain}}, \ and\ \bibinfo {author}
  {\bibfnamefont {C.~S.}\ \bibnamefont {Becquart}},\ }\href {\doibase
  10.1103/PhysRevB.89.024101} {\bibfield  {journal} {\bibinfo  {journal} {Phys.
  Rev. B}\ }\textbf {\bibinfo {volume} {89}},\ \bibinfo {pages} {024101}
  (\bibinfo {year} {2014})}\BibitemShut {NoStop}%
\bibitem [{\citenamefont {Zunger}\ \emph {et~al.}(1990)\citenamefont {Zunger},
  \citenamefont {Wei}, \citenamefont {Ferreira},\ and\ \citenamefont
  {Bernard}}]{Zunger1990}%
  \BibitemOpen
  \bibfield  {author} {\bibinfo {author} {\bibfnamefont {A.}~\bibnamefont
  {Zunger}}, \bibinfo {author} {\bibfnamefont {S.~H.}\ \bibnamefont {Wei}},
  \bibinfo {author} {\bibfnamefont {L.~G.}\ \bibnamefont {Ferreira}}, \ and\
  \bibinfo {author} {\bibfnamefont {J.~E.}\ \bibnamefont {Bernard}},\
  }\href@noop {} {\bibfield  {journal} {\bibinfo  {journal} {Phys. Rev. Lett.}\
  }\textbf {\bibinfo {volume} {65}},\ \bibinfo {pages} {353} (\bibinfo {year}
  {1990})}\BibitemShut {NoStop}%
\bibitem [{\citenamefont {Dimitrov}\ and\ \citenamefont
  {Dimitrov}(1986)}]{Dimitrov1986}%
  \BibitemOpen
  \bibfield  {author} {\bibinfo {author} {\bibfnamefont {O.}~\bibnamefont
  {Dimitrov}}\ and\ \bibinfo {author} {\bibfnamefont {C.}~\bibnamefont
  {Dimitrov}},\ }\href@noop {} {\bibfield  {journal} {\bibinfo  {journal} {J.
  Phys. F: Met. Phys.}\ }\textbf {\bibinfo {volume} {16}},\ \bibinfo {pages}
  {969} (\bibinfo {year} {1986})}\BibitemShut {NoStop}%
\bibitem [{\citenamefont {Cenedese}\ \emph {et~al.}(1984)\citenamefont
  {Cenedese}, \citenamefont {Bley},\ and\ \citenamefont
  {Lefebvre}}]{Cenedese1984}%
  \BibitemOpen
  \bibfield  {author} {\bibinfo {author} {\bibfnamefont {P.}~\bibnamefont
  {Cenedese}}, \bibinfo {author} {\bibfnamefont {F.}~\bibnamefont {Bley}}, \
  and\ \bibinfo {author} {\bibfnamefont {S.}~\bibnamefont {Lefebvre}},\
  }\href@noop {} {\bibfield  {journal} {\bibinfo  {journal} {Acta Cryst.}\
  }\textbf {\bibinfo {volume} {A40}},\ \bibinfo {pages} {228} (\bibinfo {year}
  {1984})}\BibitemShut {NoStop}%
\bibitem [{\citenamefont {Menshikov}\ \emph {et~al.}(1997)\citenamefont
  {Menshikov}, \citenamefont {Dimitrov},\ and\ \citenamefont
  {Teplykh}}]{Menshikov1997}%
  \BibitemOpen
  \bibfield  {author} {\bibinfo {author} {\bibfnamefont {A.~Z.}\ \bibnamefont
  {Menshikov}}, \bibinfo {author} {\bibfnamefont {C.}~\bibnamefont {Dimitrov}},
  \ and\ \bibinfo {author} {\bibfnamefont {A.~E.}\ \bibnamefont {Teplykh}},\
  }\href@noop {} {\bibfield  {journal} {\bibinfo  {journal} {J. Phys. III
  France}\ }\textbf {\bibinfo {volume} {7}},\ \bibinfo {pages} {1899} (\bibinfo
  {year} {1997})}\BibitemShut {NoStop}%
\bibitem [{\citenamefont {Marwick}\ \emph {et~al.}(1987)\citenamefont
  {Marwick}, \citenamefont {Piller},\ and\ \citenamefont
  {Cranshaw}}]{Marwick1987}%
  \BibitemOpen
  \bibfield  {author} {\bibinfo {author} {\bibfnamefont {A.~D.}\ \bibnamefont
  {Marwick}}, \bibinfo {author} {\bibfnamefont {R.~C.}\ \bibnamefont {Piller}},
  \ and\ \bibinfo {author} {\bibfnamefont {T.~E.}\ \bibnamefont {Cranshaw}},\
  }\href@noop {} {\bibfield  {journal} {\bibinfo  {journal} {J. Phys. F: Met.
  Phys.}\ }\textbf {\bibinfo {volume} {17}},\ \bibinfo {pages} {37} (\bibinfo
  {year} {1987})}\BibitemShut {NoStop}%
\bibitem [{\citenamefont {Sanchez}\ \emph {et~al.}(1984)\citenamefont
  {Sanchez}, \citenamefont {Ducastelle},\ and\ \citenamefont
  {Gratias}}]{Sanchez1984}%
  \BibitemOpen
  \bibfield  {author} {\bibinfo {author} {\bibfnamefont {J.~M.}\ \bibnamefont
  {Sanchez}}, \bibinfo {author} {\bibfnamefont {F.}~\bibnamefont {Ducastelle}},
  \ and\ \bibinfo {author} {\bibfnamefont {D.}~\bibnamefont {Gratias}},\
  }\href@noop {} {\bibfield  {journal} {\bibinfo  {journal} {Physica A}\
  }\textbf {\bibinfo {volume} {128}},\ \bibinfo {pages} {334} (\bibinfo {year}
  {1984})}\BibitemShut {NoStop}%
\bibitem [{\citenamefont {Connolly}\ and\ \citenamefont
  {Williams}(1983)}]{Connolly1983}%
  \BibitemOpen
  \bibfield  {author} {\bibinfo {author} {\bibfnamefont {J.~W.~D.}\
  \bibnamefont {Connolly}}\ and\ \bibinfo {author} {\bibfnamefont {A.~R.}\
  \bibnamefont {Williams}},\ }\href@noop {} {\bibfield  {journal} {\bibinfo
  {journal} {Phys. Rev. B}\ }\textbf {\bibinfo {volume} {27}},\ \bibinfo
  {pages} {5169} (\bibinfo {year} {1983})}\BibitemShut {NoStop}%
\bibitem [{\citenamefont {Ruban}\ and\ \citenamefont
  {Abrikosov}(2008)}]{Ruban2008}%
  \BibitemOpen
  \bibfield  {author} {\bibinfo {author} {\bibfnamefont {A.~V.}\ \bibnamefont
  {Ruban}}\ and\ \bibinfo {author} {\bibfnamefont {I.~A.}\ \bibnamefont
  {Abrikosov}},\ }\href {\doibase 10.1088/0034-4885/71/4/046501} {\bibfield
  {journal} {\bibinfo  {journal} {Rep. Prog. Phys.}\ }\textbf {\bibinfo
  {volume} {71}},\ \bibinfo {pages} {046501} (\bibinfo {year}
  {2008})}\BibitemShut {NoStop}%
\bibitem [{\citenamefont {Klaver}\ \emph {et~al.}(2006)\citenamefont {Klaver},
  \citenamefont {Drautz},\ and\ \citenamefont {Finnis}}]{Klaver2006}%
  \BibitemOpen
  \bibfield  {author} {\bibinfo {author} {\bibfnamefont {T.~P.~C.}\
  \bibnamefont {Klaver}}, \bibinfo {author} {\bibfnamefont {R.}~\bibnamefont
  {Drautz}}, \ and\ \bibinfo {author} {\bibfnamefont {M.~W.}\ \bibnamefont
  {Finnis}},\ }\href {\doibase 10.1103/PhysRevB.74.094435} {\bibfield
  {journal} {\bibinfo  {journal} {Phys. Rev. B}\ }\textbf {\bibinfo {volume}
  {74}},\ \bibinfo {pages} {094435} (\bibinfo {year} {2006})}\BibitemShut
  {NoStop}%
\bibitem [{\citenamefont {Nguyen-Manh}\ \emph {et~al.}(2007)\citenamefont
  {Nguyen-Manh}, \citenamefont {Lavrentiev},\ and\ \citenamefont
  {Dudarev}}]{Nguyen-Manh2007}%
  \BibitemOpen
  \bibfield  {author} {\bibinfo {author} {\bibfnamefont {D.}~\bibnamefont
  {Nguyen-Manh}}, \bibinfo {author} {\bibfnamefont {M.~Y.}\ \bibnamefont
  {Lavrentiev}}, \ and\ \bibinfo {author} {\bibfnamefont {S.~L.}\ \bibnamefont
  {Dudarev}},\ }\href {\doibase 10.1007/s10820-007-9079-4} {\bibfield
  {journal} {\bibinfo  {journal} {J. Comput.-Aided Mater. Des.}\ }\textbf
  {\bibinfo {volume} {14}},\ \bibinfo {pages} {159} (\bibinfo {year}
  {2007})}\BibitemShut {NoStop}%
\bibitem [{\citenamefont {Nguyen-Manh}\ \emph {et~al.}(2012)\citenamefont
  {Nguyen-Manh}, \citenamefont {Lavrentiev}, \citenamefont {Muzyk},\ and\
  \citenamefont {Dudarev}}]{Nguyen-Manh2012}%
  \BibitemOpen
  \bibfield  {author} {\bibinfo {author} {\bibfnamefont {D.}~\bibnamefont
  {Nguyen-Manh}}, \bibinfo {author} {\bibfnamefont {M.~Y.}\ \bibnamefont
  {Lavrentiev}}, \bibinfo {author} {\bibfnamefont {M.}~\bibnamefont {Muzyk}}, \
  and\ \bibinfo {author} {\bibfnamefont {S.~L.}\ \bibnamefont {Dudarev}},\
  }\href {\doibase 10.1007/s10853-012-6657-y} {\bibfield  {journal} {\bibinfo
  {journal} {Journal of Materials Science}\ }\textbf {\bibinfo {volume} {47}},\
  \bibinfo {pages} {7385} (\bibinfo {year} {2012})}\BibitemShut {NoStop}%
\bibitem [{\citenamefont {Lavrentiev}\ \emph {et~al.}(2007)\citenamefont
  {Lavrentiev}, \citenamefont {Drautz}, \citenamefont {Nguyen-Manh},
  \citenamefont {Klaver},\ and\ \citenamefont {Dudarev}}]{Lavrentiev2007}%
  \BibitemOpen
  \bibfield  {author} {\bibinfo {author} {\bibfnamefont {M.~Y.}\ \bibnamefont
  {Lavrentiev}}, \bibinfo {author} {\bibfnamefont {R.}~\bibnamefont {Drautz}},
  \bibinfo {author} {\bibfnamefont {D.}~\bibnamefont {Nguyen-Manh}}, \bibinfo
  {author} {\bibfnamefont {T.~P.~C.}\ \bibnamefont {Klaver}}, \ and\ \bibinfo
  {author} {\bibfnamefont {S.~L.}\ \bibnamefont {Dudarev}},\ }\href {\doibase
  10.1103/PhysRevB.75.014208} {\bibfield  {journal} {\bibinfo  {journal} {Phys.
  Rev. B}\ }\textbf {\bibinfo {volume} {75}},\ \bibinfo {pages} {014208}
  (\bibinfo {year} {2007})}\BibitemShut {NoStop}%
\bibitem [{\citenamefont {Barabash}\ \emph {et~al.}(2009)\citenamefont
  {Barabash}, \citenamefont {Chepulskii}, \citenamefont {Blum},\ and\
  \citenamefont {Zunger}}]{Barabash2009}%
  \BibitemOpen
  \bibfield  {author} {\bibinfo {author} {\bibfnamefont {S.~V.}\ \bibnamefont
  {Barabash}}, \bibinfo {author} {\bibfnamefont {R.~V.}\ \bibnamefont
  {Chepulskii}}, \bibinfo {author} {\bibfnamefont {V.}~\bibnamefont {Blum}}, \
  and\ \bibinfo {author} {\bibfnamefont {A.}~\bibnamefont {Zunger}},\ }\href
  {\doibase 10.1103/PhysRevB.80.220201} {\bibfield  {journal} {\bibinfo
  {journal} {Phys. Rev. B}\ }\textbf {\bibinfo {volume} {80}},\ \bibinfo
  {pages} {220201} (\bibinfo {year} {2009})}\BibitemShut {NoStop}%
\bibitem [{\citenamefont {Ekholm}\ \emph {et~al.}(2010)\citenamefont {Ekholm},
  \citenamefont {Zapolsky}, \citenamefont {Ruban}, \citenamefont {Vernyhora},
  \citenamefont {Ledue},\ and\ \citenamefont {Abrikosov}}]{Ekholm2010}%
  \BibitemOpen
  \bibfield  {author} {\bibinfo {author} {\bibfnamefont {M.}~\bibnamefont
  {Ekholm}}, \bibinfo {author} {\bibfnamefont {H.}~\bibnamefont {Zapolsky}},
  \bibinfo {author} {\bibfnamefont {A.~V.}\ \bibnamefont {Ruban}}, \bibinfo
  {author} {\bibfnamefont {I.}~\bibnamefont {Vernyhora}}, \bibinfo {author}
  {\bibfnamefont {D.}~\bibnamefont {Ledue}}, \ and\ \bibinfo {author}
  {\bibfnamefont {I.~A.}\ \bibnamefont {Abrikosov}},\ }\href {\doibase
  10.1103/PhysRevLett.105.167208} {\bibfield  {journal} {\bibinfo  {journal}
  {Phys. Rev. Let.}\ }\textbf {\bibinfo {volume} {105}},\ \bibinfo {pages}
  {167208} (\bibinfo {year} {2010})}\BibitemShut {NoStop}%
\bibitem [{\citenamefont {Rahaman}\ \emph {et~al.}(2014)\citenamefont
  {Rahaman}, \citenamefont {Johansson},\ and\ \citenamefont
  {Ruban}}]{Rahaman2014}%
  \BibitemOpen
  \bibfield  {author} {\bibinfo {author} {\bibfnamefont {M.}~\bibnamefont
  {Rahaman}}, \bibinfo {author} {\bibfnamefont {B.}~\bibnamefont {Johansson}},
  \ and\ \bibinfo {author} {\bibfnamefont {A.~V.}\ \bibnamefont {Ruban}},\
  }\href {\doibase 10.1103/PhysRevB.89.064103} {\bibfield  {journal} {\bibinfo
  {journal} {Phys. Rev. B}\ }\textbf {\bibinfo {volume} {89}},\ \bibinfo
  {pages} {064103} (\bibinfo {year} {2014})}\BibitemShut {NoStop}%
\bibitem [{\citenamefont {Majumdar}\ and\ \citenamefont
  {Blanckenhagen}(1984)}]{Majumdar1984}%
  \BibitemOpen
  \bibfield  {author} {\bibinfo {author} {\bibfnamefont {A.~K.}\ \bibnamefont
  {Majumdar}}\ and\ \bibinfo {author} {\bibfnamefont {P.~v.}\ \bibnamefont
  {Blanckenhagen}},\ }\href@noop {} {\bibfield  {journal} {\bibinfo  {journal}
  {Phys. Rev. B}\ }\textbf {\bibinfo {volume} {29}},\ \bibinfo {pages} {4079}
  (\bibinfo {year} {1984})}\BibitemShut {NoStop}%
\bibitem [{\citenamefont {van~de Walle}(2009)}]{Walle2009}%
  \BibitemOpen
  \bibfield  {author} {\bibinfo {author} {\bibfnamefont {A.}~\bibnamefont
  {van~de Walle}},\ }\href {\doibase 10.1016/j.calphad.2008.12.005} {\bibfield
  {journal} {\bibinfo  {journal} {Calphad}\ }\textbf {\bibinfo {volume} {33}},\
  \bibinfo {pages} {266} (\bibinfo {year} {2009})}\BibitemShut {NoStop}%
\bibitem [{\citenamefont {Sandberg}\ \emph {et~al.}(2007)\citenamefont
  {Sandberg}, \citenamefont {Slabanja},\ and\ \citenamefont
  {Holmestad}}]{Sandberg2007}%
  \BibitemOpen
  \bibfield  {author} {\bibinfo {author} {\bibfnamefont {N.}~\bibnamefont
  {Sandberg}}, \bibinfo {author} {\bibfnamefont {M.}~\bibnamefont {Slabanja}},
  \ and\ \bibinfo {author} {\bibfnamefont {R.}~\bibnamefont {Holmestad}},\
  }\href {\doibase 10.1016/j.commatsci.2007.01.001} {\bibfield  {journal}
  {\bibinfo  {journal} {Comp. Mater. Sci.}\ }\textbf {\bibinfo {volume} {40}},\
  \bibinfo {pages} {309} (\bibinfo {year} {2007})}\BibitemShut {NoStop}%
\bibitem [{\citenamefont {Wolverton}\ and\ \citenamefont
  {de~Fontaine}(1994)}]{Wolverton1994}%
  \BibitemOpen
  \bibfield  {author} {\bibinfo {author} {\bibfnamefont {C.}~\bibnamefont
  {Wolverton}}\ and\ \bibinfo {author} {\bibfnamefont {D.}~\bibnamefont
  {de~Fontaine}},\ }\href@noop {} {\bibfield  {journal} {\bibinfo  {journal}
  {Phys. Rev. B}\ }\textbf {\bibinfo {volume} {49}},\ \bibinfo {pages} {8627}
  (\bibinfo {year} {1994})}\BibitemShut {NoStop}%
\bibitem [{\citenamefont {Ducastelle}(1991)}]{Ducastelle1991}%
  \BibitemOpen
  \bibfield  {author} {\bibinfo {author} {\bibfnamefont {F.}~\bibnamefont
  {Ducastelle}},\ }\href@noop {} {\emph {\bibinfo {title} {{Order and phase
  stability in alloys}}}}\ (\bibinfo  {publisher} {North-Holland},\ \bibinfo
  {address} {Amsterdam},\ \bibinfo {year} {1991})\BibitemShut {NoStop}%
\bibitem [{\citenamefont {Asta}\ \emph {et~al.}(1991)\citenamefont {Asta},
  \citenamefont {Wolverton}, \citenamefont {de~Fontaine},\ and\ \citenamefont
  {Dreyss\'{e}}}]{Asta1991}%
  \BibitemOpen
  \bibfield  {author} {\bibinfo {author} {\bibfnamefont {M.}~\bibnamefont
  {Asta}}, \bibinfo {author} {\bibfnamefont {C.}~\bibnamefont {Wolverton}},
  \bibinfo {author} {\bibfnamefont {D.}~\bibnamefont {de~Fontaine}}, \ and\
  \bibinfo {author} {\bibfnamefont {H.}~\bibnamefont {Dreyss\'{e}}},\
  }\href@noop {} {\bibfield  {journal} {\bibinfo  {journal} {Phys. Rev. B}\
  }\textbf {\bibinfo {volume} {44}},\ \bibinfo {pages} {4907} (\bibinfo {year}
  {1991})}\BibitemShut {NoStop}%
\bibitem [{\citenamefont {de~Fontaine}(1971)}]{DeFontaine1971}%
  \BibitemOpen
  \bibfield  {author} {\bibinfo {author} {\bibfnamefont {D.}~\bibnamefont
  {de~Fontaine}},\ }\href@noop {} {\bibfield  {journal} {\bibinfo  {journal}
  {J. Appl. Cryst.}\ }\textbf {\bibinfo {volume} {4}},\ \bibinfo {pages} {15}
  (\bibinfo {year} {1971})}\BibitemShut {NoStop}%
\bibitem [{\citenamefont {van~de Walle}\ \emph {et~al.}(2002)\citenamefont
  {van~de Walle}, \citenamefont {Asta},\ and\ \citenamefont
  {Ceder}}]{Walle2002}%
  \BibitemOpen
  \bibfield  {author} {\bibinfo {author} {\bibfnamefont {A.}~\bibnamefont
  {van~de Walle}}, \bibinfo {author} {\bibfnamefont {M.}~\bibnamefont {Asta}},
  \ and\ \bibinfo {author} {\bibfnamefont {G.}~\bibnamefont {Ceder}},\
  }\href@noop {} {\bibfield  {journal} {\bibinfo  {journal} {Calphad}\ }\textbf
  {\bibinfo {volume} {26}},\ \bibinfo {pages} {539} (\bibinfo {year}
  {2002})}\BibitemShut {NoStop}%
\bibitem [{\citenamefont {Lavrentiev}\ \emph {et~al.}(2010)\citenamefont
  {Lavrentiev}, \citenamefont {Nguyen-Manh},\ and\ \citenamefont
  {Dudarev}}]{Lavrentiev2010}%
  \BibitemOpen
  \bibfield  {author} {\bibinfo {author} {\bibfnamefont {M.~Y.}\ \bibnamefont
  {Lavrentiev}}, \bibinfo {author} {\bibfnamefont {D.}~\bibnamefont
  {Nguyen-Manh}}, \ and\ \bibinfo {author} {\bibfnamefont {S.~L.}\ \bibnamefont
  {Dudarev}},\ }\href {\doibase 10.1103/PhysRevB.81.184202} {\bibfield
  {journal} {\bibinfo  {journal} {Phys. Rev. B}\ }\textbf {\bibinfo {volume}
  {81}},\ \bibinfo {pages} {184202} (\bibinfo {year} {2010})}\BibitemShut
  {NoStop}%
\bibitem [{\citenamefont {Lavrentiev}\ \emph
  {et~al.}(2011{\natexlab{a}})\citenamefont {Lavrentiev}, \citenamefont
  {Soulairol}, \citenamefont {Fu}, \citenamefont {Nguyen-Manh},\ and\
  \citenamefont {Dudarev}}]{Lavrentiev2011a}%
  \BibitemOpen
  \bibfield  {author} {\bibinfo {author} {\bibfnamefont {M.~Y.}\ \bibnamefont
  {Lavrentiev}}, \bibinfo {author} {\bibfnamefont {R.}~\bibnamefont
  {Soulairol}}, \bibinfo {author} {\bibfnamefont {C.-C.}\ \bibnamefont {Fu}},
  \bibinfo {author} {\bibfnamefont {D.}~\bibnamefont {Nguyen-Manh}}, \ and\
  \bibinfo {author} {\bibfnamefont {S.~L.}\ \bibnamefont {Dudarev}},\
  }\href@noop {} {\bibfield  {journal} {\bibinfo  {journal} {Phys. Rev. B}\
  }\textbf {\bibinfo {volume} {84}},\ \bibinfo {pages} {144203} (\bibinfo
  {year} {2011}{\natexlab{a}})}\BibitemShut {NoStop}%
\bibitem [{\citenamefont {Lavrentiev}\ \emph {et~al.}(2014)\citenamefont
  {Lavrentiev}, \citenamefont {Wr\'obel}, \citenamefont {Nguyen-Manh},\ and\
  \citenamefont {Dudarev}}]{Lavrentiev2014}%
  \BibitemOpen
  \bibfield  {author} {\bibinfo {author} {\bibfnamefont {M.~Y.}\ \bibnamefont
  {Lavrentiev}}, \bibinfo {author} {\bibfnamefont {J.~S.}\ \bibnamefont
  {Wr\'obel}}, \bibinfo {author} {\bibfnamefont {D.}~\bibnamefont
  {Nguyen-Manh}}, \ and\ \bibinfo {author} {\bibfnamefont {S.~L.}\ \bibnamefont
  {Dudarev}},\ }\href {\doibase 10.1039/c4cp01366b} {\bibfield  {journal}
  {\bibinfo  {journal} {Phys. Chem. Chem. Phys.}\ }\textbf {\bibinfo {volume}
  {16}},\ \bibinfo {pages} {16049} (\bibinfo {year} {2014})}\BibitemShut
  {NoStop}%
\bibitem [{\citenamefont {Lavrentiev}\ \emph {et~al.}(2009)\citenamefont
  {Lavrentiev}, \citenamefont {Nguyen-Manh},\ and\ \citenamefont
  {Dudarev}}]{Lavrentiev2009}%
  \BibitemOpen
  \bibfield  {author} {\bibinfo {author} {\bibfnamefont {M.~Y.}\ \bibnamefont
  {Lavrentiev}}, \bibinfo {author} {\bibfnamefont {D.}~\bibnamefont
  {Nguyen-Manh}}, \ and\ \bibinfo {author} {\bibfnamefont {S.~L.}\ \bibnamefont
  {Dudarev}},\ }\href {\doibase 10.1016/j.jnucmat.2008.12.052} {\bibfield
  {journal} {\bibinfo  {journal} {J. Nucl. Mater.}\ }\textbf {\bibinfo {volume}
  {386-388}},\ \bibinfo {pages} {22} (\bibinfo {year} {2009})}\BibitemShut
  {NoStop}%
\bibitem [{\citenamefont {Lavrentiev}\ \emph
  {et~al.}(2011{\natexlab{b}})\citenamefont {Lavrentiev}, \citenamefont
  {Nguyen-Manh},\ and\ \citenamefont {Dudarev}}]{Lavrentiev2011}%
  \BibitemOpen
  \bibfield  {author} {\bibinfo {author} {\bibfnamefont {M.~Y.}\ \bibnamefont
  {Lavrentiev}}, \bibinfo {author} {\bibfnamefont {D.}~\bibnamefont
  {Nguyen-Manh}}, \ and\ \bibinfo {author} {\bibfnamefont {S.~L.}\ \bibnamefont
  {Dudarev}},\ }\href {\doibase 10.4028/www.scientific.net/SSP.172-174.1002}
  {\bibfield  {journal} {\bibinfo  {journal} {Solid State Phenom.}\ }\textbf
  {\bibinfo {volume} {172-174}},\ \bibinfo {pages} {1002} (\bibinfo {year}
  {2011}{\natexlab{b}})}\BibitemShut {NoStop}%
\bibitem [{\citenamefont {Hortamani}\ \emph {et~al.}(2009)\citenamefont
  {Hortamani}, \citenamefont {Sandratskii}, \citenamefont {Kratzer},\ and\
  \citenamefont {Mertig}}]{Hortamani2009}%
  \BibitemOpen
  \bibfield  {author} {\bibinfo {author} {\bibfnamefont {M.}~\bibnamefont
  {Hortamani}}, \bibinfo {author} {\bibfnamefont {L.}~\bibnamefont
  {Sandratskii}}, \bibinfo {author} {\bibfnamefont {P.}~\bibnamefont
  {Kratzer}}, \ and\ \bibinfo {author} {\bibfnamefont {I.}~\bibnamefont
  {Mertig}},\ }\href@noop {} {\bibfield  {journal} {\bibinfo  {journal} {New J.
  Phys.}\ }\textbf {\bibinfo {volume} {11}},\ \bibinfo {pages} {125009}
  (\bibinfo {year} {2009})}\BibitemShut {NoStop}%
\bibitem [{\citenamefont {Liot}\ and\ \citenamefont
  {Abrikosov}(2009)}]{Liot2009}%
  \BibitemOpen
  \bibfield  {author} {\bibinfo {author} {\bibfnamefont {F.}~\bibnamefont
  {Liot}}\ and\ \bibinfo {author} {\bibfnamefont {I.~A.}\ \bibnamefont
  {Abrikosov}},\ }\href@noop {} {\bibfield  {journal} {\bibinfo  {journal}
  {Phys. Rev. B}\ }\textbf {\bibinfo {volume} {79}},\ \bibinfo {pages} {014202}
  (\bibinfo {year} {2009})}\BibitemShut {NoStop}%
\bibitem [{\citenamefont {Kresse}\ and\ \citenamefont
  {Furthm\"{u}ller}(1996{\natexlab{a}})}]{Kresse1996}%
  \BibitemOpen
  \bibfield  {author} {\bibinfo {author} {\bibfnamefont {G.}~\bibnamefont
  {Kresse}}\ and\ \bibinfo {author} {\bibfnamefont {J.}~\bibnamefont
  {Furthm\"{u}ller}},\ }\href@noop {} {\bibfield  {journal} {\bibinfo
  {journal} {Comp. Mater. Sci.}\ }\textbf {\bibinfo {volume} {6}},\ \bibinfo
  {pages} {15} (\bibinfo {year} {1996}{\natexlab{a}})}\BibitemShut {NoStop}%
\bibitem [{\citenamefont {Kresse}\ and\ \citenamefont
  {Furthm\"{u}ller}(1996{\natexlab{b}})}]{Kresse1996a}%
  \BibitemOpen
  \bibfield  {author} {\bibinfo {author} {\bibfnamefont {G.}~\bibnamefont
  {Kresse}}\ and\ \bibinfo {author} {\bibfnamefont {J.}~\bibnamefont
  {Furthm\"{u}ller}},\ }\href {\doibase 10.1103/PhysRevB.54.11169} {\bibfield
  {journal} {\bibinfo  {journal} {Phys. Rev. B}\ }\textbf {\bibinfo {volume}
  {54}},\ \bibinfo {pages} {11169} (\bibinfo {year}
  {1996}{\natexlab{b}})}\BibitemShut {NoStop}%
\bibitem [{\citenamefont {Perdew}\ \emph {et~al.}(1996)\citenamefont {Perdew},
  \citenamefont {Burke},\ and\ \citenamefont {Ernzerhof}}]{Perdew1996}%
  \BibitemOpen
  \bibfield  {author} {\bibinfo {author} {\bibfnamefont {J.~P.}\ \bibnamefont
  {Perdew}}, \bibinfo {author} {\bibfnamefont {K.}~\bibnamefont {Burke}}, \
  and\ \bibinfo {author} {\bibfnamefont {M.}~\bibnamefont {Ernzerhof}},\ }\href
  {\doibase 10.1103/PhysRevLett.77.3865} {\bibfield  {journal} {\bibinfo
  {journal} {Phys. Rev. Lett.}\ }\textbf {\bibinfo {volume} {77}},\ \bibinfo
  {pages} {3865} (\bibinfo {year} {1996})}\BibitemShut {NoStop}%
\bibitem [{\citenamefont {Monkhorst}\ and\ \citenamefont
  {Pack}(1976)}]{Monkhorst1976}%
  \BibitemOpen
  \bibfield  {author} {\bibinfo {author} {\bibfnamefont {H.~J.}\ \bibnamefont
  {Monkhorst}}\ and\ \bibinfo {author} {\bibfnamefont {J.~D.}\ \bibnamefont
  {Pack}},\ }\href@noop {} {\bibfield  {journal} {\bibinfo  {journal} {Phys.
  Rev. B}\ }\textbf {\bibinfo {volume} {13}},\ \bibinfo {pages} {5188}
  (\bibinfo {year} {1976})}\BibitemShut {NoStop}%
\bibitem [{\citenamefont {Barabash}\ \emph {et~al.}(2006)\citenamefont
  {Barabash}, \citenamefont {Blum}, \citenamefont {M\"{u}ller},\ and\
  \citenamefont {Zunger}}]{Barabash2006}%
  \BibitemOpen
  \bibfield  {author} {\bibinfo {author} {\bibfnamefont {S.~V.}\ \bibnamefont
  {Barabash}}, \bibinfo {author} {\bibfnamefont {V.}~\bibnamefont {Blum}},
  \bibinfo {author} {\bibfnamefont {S.}~\bibnamefont {M\"{u}ller}}, \ and\
  \bibinfo {author} {\bibfnamefont {A.}~\bibnamefont {Zunger}},\ }\href
  {\doibase 10.1103/PhysRevB.74.035108} {\bibfield  {journal} {\bibinfo
  {journal} {Phys. Rev. B}\ }\textbf {\bibinfo {volume} {74}},\ \bibinfo
  {pages} {035108} (\bibinfo {year} {2006})}\BibitemShut {NoStop}%
\bibitem [{\citenamefont {Ceder}\ \emph {et~al.}(1994)\citenamefont {Ceder},
  \citenamefont {Garbulsky}, \citenamefont {Avis},\ and\ \citenamefont
  {Fukuda}}]{Garbulsky1994}%
  \BibitemOpen
  \bibfield  {author} {\bibinfo {author} {\bibfnamefont {G.}~\bibnamefont
  {Ceder}}, \bibinfo {author} {\bibfnamefont {G.~D.}\ \bibnamefont
  {Garbulsky}}, \bibinfo {author} {\bibfnamefont {D.}~\bibnamefont {Avis}}, \
  and\ \bibinfo {author} {\bibfnamefont {K.}~\bibnamefont {Fukuda}},\
  }\href@noop {} {\bibfield  {journal} {\bibinfo  {journal} {Phys. Rev. B}\
  }\textbf {\bibinfo {volume} {49}},\ \bibinfo {pages} {1} (\bibinfo {year}
  {1994})}\BibitemShut {NoStop}%
\bibitem [{\citenamefont {Acet}\ \emph {et~al.}(1994)\citenamefont {Acet},
  \citenamefont {Z\"ahres}, \citenamefont {Wassermann},\ and\ \citenamefont
  {Pepperhoff}}]{Acet1994}%
  \BibitemOpen
  \bibfield  {author} {\bibinfo {author} {\bibfnamefont {M.}~\bibnamefont
  {Acet}}, \bibinfo {author} {\bibfnamefont {H.}~\bibnamefont {Z\"ahres}},
  \bibinfo {author} {\bibfnamefont {E.~F.}\ \bibnamefont {Wassermann}}, \ and\
  \bibinfo {author} {\bibfnamefont {W.}~\bibnamefont {Pepperhoff}},\
  }\href@noop {} {\bibfield  {journal} {\bibinfo  {journal} {Phys. Rev. B}\
  }\textbf {\bibinfo {volume} {49}},\ \bibinfo {pages} {6012} (\bibinfo {year}
  {1994})}\BibitemShut {NoStop}%
\bibitem [{\citenamefont {Crangle}\ and\ \citenamefont
  {Hallam}(1963)}]{Crangle1963}%
  \BibitemOpen
  \bibfield  {author} {\bibinfo {author} {\bibfnamefont {J.}~\bibnamefont
  {Crangle}}\ and\ \bibinfo {author} {\bibfnamefont {G.~C.}\ \bibnamefont
  {Hallam}},\ }\href@noop {} {\bibfield  {journal} {\bibinfo  {journal} {Proc.
  Roy. Soc. A}\ }\textbf {\bibinfo {volume} {272}},\ \bibinfo {pages} {119}
  (\bibinfo {year} {1963})}\BibitemShut {NoStop}%
\bibitem [{\citenamefont {Abrahams}\ \emph {et~al.}(1962)\citenamefont
  {Abrahams}, \citenamefont {Guttman},\ and\ \citenamefont
  {Kasper}}]{Abrahams1962}%
  \BibitemOpen
  \bibfield  {author} {\bibinfo {author} {\bibfnamefont {S.~C.}\ \bibnamefont
  {Abrahams}}, \bibinfo {author} {\bibfnamefont {L.}~\bibnamefont {Guttman}}, \
  and\ \bibinfo {author} {\bibfnamefont {J.~S.}\ \bibnamefont {Kasper}},\
  }\href@noop {} {\bibfield  {journal} {\bibinfo  {journal} {Phys. Rev.}\
  }\textbf {\bibinfo {volume} {127}},\ \bibinfo {pages} {2052} (\bibinfo {year}
  {1962})}\BibitemShut {NoStop}%
\bibitem [{\citenamefont {Kittel}(1971)}]{Kittel1971}%
  \BibitemOpen
  \bibfield  {author} {\bibinfo {author} {\bibfnamefont {C.}~\bibnamefont
  {Kittel}},\ }\href@noop {} {\emph {\bibinfo {title} {{Introduction to Solid
  State Physics}}}}\ (\bibinfo  {publisher} {Wiley},\ \bibinfo {address} {New
  York},\ \bibinfo {year} {1971})\BibitemShut {NoStop}%
\bibitem [{\citenamefont {Herper}\ \emph {et~al.}(1999)\citenamefont {Herper},
  \citenamefont {Hoffmann},\ and\ \citenamefont {Entel}}]{Herper1999}%
  \BibitemOpen
  \bibfield  {author} {\bibinfo {author} {\bibfnamefont {H.~C.}\ \bibnamefont
  {Herper}}, \bibinfo {author} {\bibfnamefont {E.}~\bibnamefont {Hoffmann}}, \
  and\ \bibinfo {author} {\bibfnamefont {P.}~\bibnamefont {Entel}},\ }\href
  {\doibase 10.1103/PhysRevB.60.3839} {\bibfield  {journal} {\bibinfo
  {journal} {Phys. Rev. B}\ }\textbf {\bibinfo {volume} {60}},\ \bibinfo
  {pages} {3839} (\bibinfo {year} {1999})}\BibitemShut {NoStop}%
\bibitem [{\citenamefont {Moruzzi}\ \emph {et~al.}(1989)\citenamefont
  {Moruzzi}, \citenamefont {Marcus},\ and\ \citenamefont
  {K\"ubler}}]{Moruzzi1989}%
  \BibitemOpen
  \bibfield  {author} {\bibinfo {author} {\bibfnamefont {V.~L.}\ \bibnamefont
  {Moruzzi}}, \bibinfo {author} {\bibfnamefont {P.~M.}\ \bibnamefont {Marcus}},
  \ and\ \bibinfo {author} {\bibfnamefont {J.}~\bibnamefont {K\"ubler}},\
  }\href@noop {} {\bibfield  {journal} {\bibinfo  {journal} {Phys. Rev. B}\
  }\textbf {\bibinfo {volume} {39}},\ \bibinfo {pages} {6957} (\bibinfo {year}
  {1989})}\BibitemShut {NoStop}%
\bibitem [{\citenamefont {Wijn}(1991)}]{Landolt}%
  \BibitemOpen
  \bibinfo {editor} {\bibfnamefont {H.~P.~J.}\ \bibnamefont {Wijn}},\ ed.,\
  \href {\doibase 10.1007/978-3-642-58218-9} {\emph {\bibinfo {title}
  {{Magnetic Properties of Metals}}}},\ Landolt-B\"orstein. Numerical Data and
  Functional Relationships in Science and Technology. Volume 19\ (\bibinfo
  {publisher} {Springer Berlin Heidelberg},\ \bibinfo {address} {Berlin,
  Heidelberg},\ \bibinfo {year} {1991})\BibitemShut {NoStop}%
\bibitem [{\citenamefont {Chamberod}\ \emph {et~al.}(1979)\citenamefont
  {Chamberod}, \citenamefont {Laugier},\ and\ \citenamefont
  {Penisson}}]{Chamberod1979}%
  \BibitemOpen
  \bibfield  {author} {\bibinfo {author} {\bibfnamefont {A.}~\bibnamefont
  {Chamberod}}, \bibinfo {author} {\bibfnamefont {J.}~\bibnamefont {Laugier}},
  \ and\ \bibinfo {author} {\bibfnamefont {J.~M.}\ \bibnamefont {Penisson}},\
  }\href@noop {} {\bibfield  {journal} {\bibinfo  {journal} {J. Magn. Magn.
  Mater.}\ }\textbf {\bibinfo {volume} {10}},\ \bibinfo {pages} {139} (\bibinfo
  {year} {1979})}\BibitemShut {NoStop}%
\bibitem [{\citenamefont {Mohri}\ and\ \citenamefont {Chen}(2004)}]{Mohri2004}%
  \BibitemOpen
  \bibfield  {author} {\bibinfo {author} {\bibfnamefont {T.}~\bibnamefont
  {Mohri}}\ and\ \bibinfo {author} {\bibfnamefont {Y.}~\bibnamefont {Chen}},\
  }\href@noop {} {\bibfield  {journal} {\bibinfo  {journal} {J. Alloys Compd.}\
  }\textbf {\bibinfo {volume} {383}},\ \bibinfo {pages} {23} (\bibinfo {year}
  {2004})}\BibitemShut {NoStop}%
\bibitem [{\citenamefont {Tucker}(2008)}]{Tucker2008}%
  \BibitemOpen
  \bibfield  {author} {\bibinfo {author} {\bibfnamefont {J.~D.}\ \bibnamefont
  {Tucker}},\ }\href@noop {} {\emph {\bibinfo {title} {{Ab initio - based
  modelling of radiation effects in the Ni-Fe-Cr system}}}}\ (\bibinfo
  {publisher} {University of Wisconsin-Madison},\ \bibinfo {address}
  {Wisconsin},\ \bibinfo {year} {2008})\BibitemShut {NoStop}%
\bibitem [{\citenamefont {Abrikosov}\ \emph {et~al.}(2007)\citenamefont
  {Abrikosov}, \citenamefont {Kissavos}, \citenamefont {Liot}, \citenamefont
  {Alling}, \citenamefont {Simak}, \citenamefont {Peil},\ and\ \citenamefont
  {Ruban}}]{Abrikosov2007}%
  \BibitemOpen
  \bibfield  {author} {\bibinfo {author} {\bibfnamefont {I.~A.}\ \bibnamefont
  {Abrikosov}}, \bibinfo {author} {\bibfnamefont {A.~E.}\ \bibnamefont
  {Kissavos}}, \bibinfo {author} {\bibfnamefont {F.}~\bibnamefont {Liot}},
  \bibinfo {author} {\bibfnamefont {B.}~\bibnamefont {Alling}}, \bibinfo
  {author} {\bibfnamefont {S.~I.}\ \bibnamefont {Simak}}, \bibinfo {author}
  {\bibfnamefont {O.}~\bibnamefont {Peil}}, \ and\ \bibinfo {author}
  {\bibfnamefont {A.~V.}\ \bibnamefont {Ruban}},\ }\href {\doibase
  10.1103/PhysRevB.76.014434} {\bibfield  {journal} {\bibinfo  {journal} {Phys.
  Rev. B}\ }\textbf {\bibinfo {volume} {76}},\ \bibinfo {pages} {014434}
  (\bibinfo {year} {2007})}\BibitemShut {NoStop}%
\bibitem [{\citenamefont {Crisan}\ \emph {et~al.}(2002)\citenamefont {Crisan},
  \citenamefont {Entel}, \citenamefont {Ebert}, \citenamefont {Akai},
  \citenamefont {Johnson},\ and\ \citenamefont {Staunton}}]{Crisan2002}%
  \BibitemOpen
  \bibfield  {author} {\bibinfo {author} {\bibfnamefont {V.}~\bibnamefont
  {Crisan}}, \bibinfo {author} {\bibfnamefont {P.}~\bibnamefont {Entel}},
  \bibinfo {author} {\bibfnamefont {H.}~\bibnamefont {Ebert}}, \bibinfo
  {author} {\bibfnamefont {H.}~\bibnamefont {Akai}}, \bibinfo {author}
  {\bibfnamefont {D.~D.}\ \bibnamefont {Johnson}}, \ and\ \bibinfo {author}
  {\bibfnamefont {J.~B.}\ \bibnamefont {Staunton}},\ }\href {\doibase
  10.1103/PhysRevB.66.014416} {\bibfield  {journal} {\bibinfo  {journal} {Phys.
  Rev. B}\ }\textbf {\bibinfo {volume} {66}},\ \bibinfo {pages} {014416}
  (\bibinfo {year} {2002})}\BibitemShut {NoStop}%
\bibitem [{\citenamefont {Ruban}\ \emph {et~al.}(2005)\citenamefont {Ruban},
  \citenamefont {Katsnelson}, \citenamefont {Olovsson}, \citenamefont {Simak},\
  and\ \citenamefont {Abrikosov}}]{Ruban2005}%
  \BibitemOpen
  \bibfield  {author} {\bibinfo {author} {\bibfnamefont {A.~V.}\ \bibnamefont
  {Ruban}}, \bibinfo {author} {\bibfnamefont {M.~I.}\ \bibnamefont
  {Katsnelson}}, \bibinfo {author} {\bibfnamefont {W.}~\bibnamefont
  {Olovsson}}, \bibinfo {author} {\bibfnamefont {S.~I.}\ \bibnamefont {Simak}},
  \ and\ \bibinfo {author} {\bibfnamefont {I.~A.}\ \bibnamefont {Abrikosov}},\
  }\href {\doibase 10.1103/PhysRevB.71.054402} {\bibfield  {journal} {\bibinfo
  {journal} {Phys. Rev. B}\ }\textbf {\bibinfo {volume} {71}},\ \bibinfo
  {pages} {054402} (\bibinfo {year} {2005})}\BibitemShut {NoStop}%
\bibitem [{\citenamefont {Ruban}\ \emph {et~al.}(2007)\citenamefont {Ruban},
  \citenamefont {Khmelevskyi}, \citenamefont {Mohn},\ and\ \citenamefont
  {Johansson}}]{Ruban2007}%
  \BibitemOpen
  \bibfield  {author} {\bibinfo {author} {\bibfnamefont {A.~V.}\ \bibnamefont
  {Ruban}}, \bibinfo {author} {\bibfnamefont {S.}~\bibnamefont {Khmelevskyi}},
  \bibinfo {author} {\bibfnamefont {P.}~\bibnamefont {Mohn}}, \ and\ \bibinfo
  {author} {\bibfnamefont {B.}~\bibnamefont {Johansson}},\ }\href {\doibase
  10.1103/PhysRevB.75.054402} {\bibfield  {journal} {\bibinfo  {journal} {Phys.
  Rev. B}\ }\textbf {\bibinfo {volume} {76}},\ \bibinfo {pages} {014420}
  (\bibinfo {year} {2007})}\BibitemShut {NoStop}%
\bibitem [{\citenamefont {Massalski}\ \emph {et~al.}(1990)\citenamefont
  {Massalski}, \citenamefont {Okamoto}, \citenamefont {Subramanion},\ and\
  \citenamefont {Kacprzak}}]{Massalski1990}%
  \BibitemOpen
  \bibinfo {editor} {\bibfnamefont {T.~B.}\ \bibnamefont {Massalski}}, \bibinfo
  {editor} {\bibfnamefont {H.}~\bibnamefont {Okamoto}}, \bibinfo {editor}
  {\bibfnamefont {P.~K.}\ \bibnamefont {Subramanion}}, \ and\ \bibinfo {editor}
  {\bibfnamefont {L.}~\bibnamefont {Kacprzak}},\ eds.,\ \href@noop {} {\emph
  {\bibinfo {title} {{Binary Alloy Phase Diagrams}}}},\ \bibinfo {edition}
  {2nd}\ ed.\ (\bibinfo  {publisher} {American Society for Metals},\ \bibinfo
  {address} {Metals Park, OH},\ \bibinfo {year} {1990})\BibitemShut {NoStop}%
\bibitem [{\citenamefont {Reuter}\ \emph {et~al.}(1989)\citenamefont {Reuter},
  \citenamefont {Williams},\ and\ \citenamefont {Goldstein}}]{Reuter1989}%
  \BibitemOpen
  \bibfield  {author} {\bibinfo {author} {\bibfnamefont {K.~B.}\ \bibnamefont
  {Reuter}}, \bibinfo {author} {\bibfnamefont {D.~B.}\ \bibnamefont
  {Williams}}, \ and\ \bibinfo {author} {\bibfnamefont {J.~I.}\ \bibnamefont
  {Goldstein}},\ }\href@noop {} {\bibfield  {journal} {\bibinfo  {journal}
  {Metall. Trans. A}\ }\textbf {\bibinfo {volume} {20}},\ \bibinfo {pages}
  {719} (\bibinfo {year} {1989})}\BibitemShut {NoStop}%
\bibitem [{\citenamefont {Mohri}\ \emph {et~al.}(2009)\citenamefont {Mohri},
  \citenamefont {Chen},\ and\ \citenamefont {Jufuku}}]{Mohri2009}%
  \BibitemOpen
  \bibfield  {author} {\bibinfo {author} {\bibfnamefont {T.}~\bibnamefont
  {Mohri}}, \bibinfo {author} {\bibfnamefont {Y.}~\bibnamefont {Chen}}, \ and\
  \bibinfo {author} {\bibfnamefont {Y.}~\bibnamefont {Jufuku}},\ }\href@noop {}
  {\bibfield  {journal} {\bibinfo  {journal} {Calphad}\ }\textbf {\bibinfo
  {volume} {33}},\ \bibinfo {pages} {244} (\bibinfo {year} {2009})}\BibitemShut
  {NoStop}%
\bibitem [{\citenamefont {Lu}\ \emph {et~al.}(1991)\citenamefont {Lu},
  \citenamefont {Wei}, \citenamefont {Zunger}, \citenamefont {Frota-Pessoa},\
  and\ \citenamefont {Ferreira}}]{Lu1991}%
  \BibitemOpen
  \bibfield  {author} {\bibinfo {author} {\bibfnamefont {Z.~W.}\ \bibnamefont
  {Lu}}, \bibinfo {author} {\bibfnamefont {S.~H.}\ \bibnamefont {Wei}},
  \bibinfo {author} {\bibfnamefont {A.}~\bibnamefont {Zunger}}, \bibinfo
  {author} {\bibfnamefont {S.}~\bibnamefont {Frota-Pessoa}}, \ and\ \bibinfo
  {author} {\bibfnamefont {L.~G.}\ \bibnamefont {Ferreira}},\ }\href@noop {}
  {\bibfield  {journal} {\bibinfo  {journal} {Phys. Rev. B}\ }\textbf {\bibinfo
  {volume} {44}},\ \bibinfo {pages} {512} (\bibinfo {year} {1991})}\BibitemShut
  {NoStop}%
\bibitem [{\citenamefont {Entel}\ \emph {et~al.}(1993)\citenamefont {Entel},
  \citenamefont {Hoffmann}, \citenamefont {Mohn}, \citenamefont {Schwarz},\
  and\ \citenamefont {Moruzzi}}]{Entel1993}%
  \BibitemOpen
  \bibfield  {author} {\bibinfo {author} {\bibfnamefont {P.}~\bibnamefont
  {Entel}}, \bibinfo {author} {\bibfnamefont {E.}~\bibnamefont {Hoffmann}},
  \bibinfo {author} {\bibfnamefont {P.}~\bibnamefont {Mohn}}, \bibinfo {author}
  {\bibfnamefont {K.}~\bibnamefont {Schwarz}}, \ and\ \bibinfo {author}
  {\bibfnamefont {V.~L.}\ \bibnamefont {Moruzzi}},\ }\href@noop {} {\bibfield
  {journal} {\bibinfo  {journal} {Phys. Rev. B}\ }\textbf {\bibinfo {volume}
  {47}},\ \bibinfo {pages} {8706} (\bibinfo {year} {1993})}\BibitemShut
  {NoStop}%
\bibitem [{\citenamefont {Aldred}(1976)}]{Aldred1976}%
  \BibitemOpen
  \bibfield  {author} {\bibinfo {author} {\bibfnamefont {A.~T.}\ \bibnamefont
  {Aldred}},\ }\href@noop {} {\bibfield  {journal} {\bibinfo  {journal} {Phys.
  Rev. B}\ }\textbf {\bibinfo {volume} {14}},\ \bibinfo {pages} {219} (\bibinfo
  {year} {1976})}\BibitemShut {NoStop}%
\bibitem [{\citenamefont {Aldred}\ \emph {et~al.}(1976)\citenamefont {Aldred},
  \citenamefont {Rainford}, \citenamefont {Kouvel},\ and\ \citenamefont
  {Hicks}}]{Aldred1976a}%
  \BibitemOpen
  \bibfield  {author} {\bibinfo {author} {\bibfnamefont {A.~T.}\ \bibnamefont
  {Aldred}}, \bibinfo {author} {\bibfnamefont {B.~D.}\ \bibnamefont
  {Rainford}}, \bibinfo {author} {\bibfnamefont {J.~S.}\ \bibnamefont
  {Kouvel}}, \ and\ \bibinfo {author} {\bibfnamefont {T.~J.}\ \bibnamefont
  {Hicks}},\ }\href@noop {} {\bibfield  {journal} {\bibinfo  {journal} {Phys.
  Rev. B}\ }\textbf {\bibinfo {volume} {14}},\ \bibinfo {pages} {228} (\bibinfo
  {year} {1976})}\BibitemShut {NoStop}%
\bibitem [{\citenamefont {Kajzar}\ and\ \citenamefont
  {Parette}(1980)}]{Kajzar1980}%
  \BibitemOpen
  \bibfield  {author} {\bibinfo {author} {\bibfnamefont {F.}~\bibnamefont
  {Kajzar}}\ and\ \bibinfo {author} {\bibfnamefont {G.}~\bibnamefont
  {Parette}},\ }\href@noop {} {\bibfield  {journal} {\bibinfo  {journal} {Phys.
  Rev. B}\ }\textbf {\bibinfo {volume} {22}},\ \bibinfo {pages} {5471}
  (\bibinfo {year} {1980})}\BibitemShut {NoStop}%
\bibitem [{\citenamefont {Nguyen-Manh}\ \emph {et~al.}(2008)\citenamefont
  {Nguyen-Manh}, \citenamefont {Lavrentiev},\ and\ \citenamefont
  {Dudarev}}]{Nguyen-Manh2008}%
  \BibitemOpen
  \bibfield  {author} {\bibinfo {author} {\bibfnamefont {D.}~\bibnamefont
  {Nguyen-Manh}}, \bibinfo {author} {\bibfnamefont {M.~Y.}\ \bibnamefont
  {Lavrentiev}}, \ and\ \bibinfo {author} {\bibfnamefont {S.~L.}\ \bibnamefont
  {Dudarev}},\ }\href@noop {} {\bibfield  {journal} {\bibinfo  {journal} {C. R.
  Physique}\ }\textbf {\bibinfo {volume} {9}},\ \bibinfo {pages} {379}
  (\bibinfo {year} {2008})}\BibitemShut {NoStop}%
\bibitem [{\citenamefont {Nguyen-Manh}\ and\ \citenamefont
  {Dudarev}(2009)}]{Nguyen-Manh2009}%
  \BibitemOpen
  \bibfield  {author} {\bibinfo {author} {\bibfnamefont {D.}~\bibnamefont
  {Nguyen-Manh}}\ and\ \bibinfo {author} {\bibfnamefont {S.~L.}\ \bibnamefont
  {Dudarev}},\ }\href@noop {} {\bibfield  {journal} {\bibinfo  {journal} {Phys.
  Rev. B}\ }\textbf {\bibinfo {volume} {80}},\ \bibinfo {pages} {104440}
  (\bibinfo {year} {2009})}\BibitemShut {NoStop}%
\bibitem [{\citenamefont {Olsson}\ \emph {et~al.}(2003)\citenamefont {Olsson},
  \citenamefont {Abrikosov}, \citenamefont {Vitos},\ and\ \citenamefont
  {Wallenius}}]{Olsson2003}%
  \BibitemOpen
  \bibfield  {author} {\bibinfo {author} {\bibfnamefont {P.}~\bibnamefont
  {Olsson}}, \bibinfo {author} {\bibfnamefont {I.~A.}\ \bibnamefont
  {Abrikosov}}, \bibinfo {author} {\bibfnamefont {L.}~\bibnamefont {Vitos}}, \
  and\ \bibinfo {author} {\bibfnamefont {J.}~\bibnamefont {Wallenius}},\ }\href
  {\doibase 10.1016/S0022-3115(03)00207-1} {\bibfield  {journal} {\bibinfo
  {journal} {J. Nucl. Mater.}\ }\textbf {\bibinfo {volume} {321}},\ \bibinfo
  {pages} {84} (\bibinfo {year} {2003})}\BibitemShut {NoStop}%
\bibitem [{\citenamefont {Olsson}\ \emph {et~al.}(2006)\citenamefont {Olsson},
  \citenamefont {Abrikosov},\ and\ \citenamefont {Wallenius}}]{Olsson2006}%
  \BibitemOpen
  \bibfield  {author} {\bibinfo {author} {\bibfnamefont {P.}~\bibnamefont
  {Olsson}}, \bibinfo {author} {\bibfnamefont {I.~A.}\ \bibnamefont
  {Abrikosov}}, \ and\ \bibinfo {author} {\bibfnamefont {J.}~\bibnamefont
  {Wallenius}},\ }\href {\doibase 10.1103/PhysRevB.73.104416} {\bibfield
  {journal} {\bibinfo  {journal} {Phys. Rev. B}\ }\textbf {\bibinfo {volume}
  {73}},\ \bibinfo {pages} {104416} (\bibinfo {year} {2006})}\BibitemShut
  {NoStop}%
\bibitem [{\citenamefont {Zhang}\ \emph {et~al.}(2009)\citenamefont {Zhang},
  \citenamefont {Johansson},\ and\ \citenamefont {Vitos}}]{Zhang2009}%
  \BibitemOpen
  \bibfield  {author} {\bibinfo {author} {\bibfnamefont {H.}~\bibnamefont
  {Zhang}}, \bibinfo {author} {\bibfnamefont {B.}~\bibnamefont {Johansson}}, \
  and\ \bibinfo {author} {\bibfnamefont {L.}~\bibnamefont {Vitos}},\ }\href
  {\doibase 10.1103/PhysRevB.79.224201} {\bibfield  {journal} {\bibinfo
  {journal} {Phys. Rev. B}\ }\textbf {\bibinfo {volume} {79}},\ \bibinfo
  {pages} {224201} (\bibinfo {year} {2009})}\BibitemShut {NoStop}%
\bibitem [{\citenamefont {Erhart}\ \emph {et~al.}(2008)\citenamefont {Erhart},
  \citenamefont {Sadigh},\ and\ \citenamefont {Caro}}]{Erhart2008}%
  \BibitemOpen
  \bibfield  {author} {\bibinfo {author} {\bibfnamefont {P.}~\bibnamefont
  {Erhart}}, \bibinfo {author} {\bibfnamefont {B.}~\bibnamefont {Sadigh}}, \
  and\ \bibinfo {author} {\bibfnamefont {A.}~\bibnamefont {Caro}},\ }\href@noop
  {} {\bibfield  {journal} {\bibinfo  {journal} {Appl. Phys. Lett.}\ }\textbf
  {\bibinfo {volume} {92}},\ \bibinfo {pages} {141904} (\bibinfo {year}
  {2008})}\BibitemShut {NoStop}%
\bibitem [{\citenamefont {Mirebeau}\ \emph {et~al.}(1984)\citenamefont
  {Mirebeau}, \citenamefont {Hennion},\ and\ \citenamefont
  {Parette}}]{Mirebeau1984}%
  \BibitemOpen
  \bibfield  {author} {\bibinfo {author} {\bibfnamefont {I.}~\bibnamefont
  {Mirebeau}}, \bibinfo {author} {\bibfnamefont {M.}~\bibnamefont {Hennion}}, \
  and\ \bibinfo {author} {\bibfnamefont {G.}~\bibnamefont {Parette}},\
  }\href@noop {} {\bibfield  {journal} {\bibinfo  {journal} {Phys. Rev. Lett.}\
  }\textbf {\bibinfo {volume} {53}},\ \bibinfo {pages} {687} (\bibinfo {year}
  {1984})}\BibitemShut {NoStop}%
\bibitem [{\citenamefont {Pearson}(1958)}]{Pearson1958}%
  \BibitemOpen
  \bibfield  {author} {\bibinfo {author} {\bibfnamefont {W.~B.}\ \bibnamefont
  {Pearson}},\ }\href@noop {} {\emph {\bibinfo {title} {{A Handbook of Lattice
  Spacings and Structures of Metals and Alloys}}}}\ (\bibinfo  {publisher}
  {Pergamon Press},\ \bibinfo {address} {London},\ \bibinfo {year}
  {1958})\BibitemShut {NoStop}%
\bibitem [{\citenamefont {Rode}\ \emph {et~al.}(1976)\citenamefont {Rode},
  \citenamefont {Deryabin},\ and\ \citenamefont {Damashke}}]{Rode1976}%
  \BibitemOpen
  \bibfield  {author} {\bibinfo {author} {\bibfnamefont {V.}~\bibnamefont
  {Rode}}, \bibinfo {author} {\bibfnamefont {A.}~\bibnamefont {Deryabin}}, \
  and\ \bibinfo {author} {\bibfnamefont {G.}~\bibnamefont {Damashke}},\ }\href
  {\doibase 10.1109/TMAG.1976.1059034} {\bibfield  {journal} {\bibinfo
  {journal} {IEEE Trans. Magn.}\ }\textbf {\bibinfo {volume} {12}},\ \bibinfo
  {pages} {404} (\bibinfo {year} {1976})}\BibitemShut {NoStop}%
\bibitem [{\citenamefont {Bansal}\ and\ \citenamefont
  {Chandra}(1976)}]{Bansal1976}%
  \BibitemOpen
  \bibfield  {author} {\bibinfo {author} {\bibfnamefont {C.}~\bibnamefont
  {Bansal}}\ and\ \bibinfo {author} {\bibfnamefont {G.}~\bibnamefont
  {Chandra}},\ }\href {\doibase 10.1016/0038-1098(76)90445-2} {\bibfield
  {journal} {\bibinfo  {journal} {Solid State Comm.}\ }\textbf {\bibinfo
  {volume} {19}},\ \bibinfo {pages} {107} (\bibinfo {year} {1976})}\BibitemShut
  {NoStop}%
\bibitem [{\citenamefont {Kubaschewski}\ and\ \citenamefont
  {Stuart}(1967)}]{Kubaschewski1967}%
  \BibitemOpen
  \bibfield  {author} {\bibinfo {author} {\bibfnamefont {O.}~\bibnamefont
  {Kubaschewski}}\ and\ \bibinfo {author} {\bibfnamefont {L.~E.~H.}\
  \bibnamefont {Stuart}},\ }\href@noop {} {\bibfield  {journal} {\bibinfo
  {journal} {J. Chem. Eng. Data}\ }\textbf {\bibinfo {volume} {12}},\ \bibinfo
  {pages} {418} (\bibinfo {year} {1967})}\BibitemShut {NoStop}%
\bibitem [{\citenamefont {Dench}(1963)}]{Dench1963}%
  \BibitemOpen
  \bibfield  {author} {\bibinfo {author} {\bibfnamefont {W.~A.}\ \bibnamefont
  {Dench}},\ }\href@noop {} {\bibfield  {journal} {\bibinfo  {journal} {Trans.
  Faraday Soc.}\ }\textbf {\bibinfo {volume} {59}},\ \bibinfo {pages} {1279}
  (\bibinfo {year} {1963})}\BibitemShut {NoStop}%
\bibitem [{\citenamefont {Chen}\ and\ \citenamefont
  {Sundman}(2001)}]{Chen2001}%
  \BibitemOpen
  \bibfield  {author} {\bibinfo {author} {\bibfnamefont {Q.}~\bibnamefont
  {Chen}}\ and\ \bibinfo {author} {\bibfnamefont {B.}~\bibnamefont {Sundman}},\
  }\href@noop {} {\bibfield  {journal} {\bibinfo  {journal} {J. Phase Equil.}\
  }\textbf {\bibinfo {volume} {22}},\ \bibinfo {pages} {631} (\bibinfo {year}
  {2001})}\BibitemShut {NoStop}%
\bibitem [{\citenamefont {Tomiska}(2004)}]{Tomiska2004}%
  \BibitemOpen
  \bibfield  {author} {\bibinfo {author} {\bibfnamefont {J.}~\bibnamefont
  {Tomiska}},\ }\href {\doibase 10.1016/j.jallcom.2004.02.027} {\bibfield
  {journal} {\bibinfo  {journal} {J. Alloys Compd.}\ }\textbf {\bibinfo
  {volume} {379}},\ \bibinfo {pages} {176} (\bibinfo {year}
  {2004})}\BibitemShut {NoStop}%
\bibitem [{\citenamefont {Sharma}\ \emph {et~al.}(1978)\citenamefont {Sharma},
  \citenamefont {Sonnenberg}, \citenamefont {Antesberger},\ and\ \citenamefont
  {Kesternich}}]{Sharma1978}%
  \BibitemOpen
  \bibfield  {author} {\bibinfo {author} {\bibfnamefont {B.~D.}\ \bibnamefont
  {Sharma}}, \bibinfo {author} {\bibfnamefont {K.}~\bibnamefont {Sonnenberg}},
  \bibinfo {author} {\bibfnamefont {G.}~\bibnamefont {Antesberger}}, \ and\
  \bibinfo {author} {\bibfnamefont {W.}~\bibnamefont {Kesternich}},\
  }\href@noop {} {\bibfield  {journal} {\bibinfo  {journal} {Phil. Mag. A}\
  }\textbf {\bibinfo {volume} {37}},\ \bibinfo {pages} {777} (\bibinfo {year}
  {1978})}\BibitemShut {NoStop}%
\bibitem [{\citenamefont {van~de Walle}\ and\ \citenamefont
  {Ceder}(2002)}]{Walle2002a}%
  \BibitemOpen
  \bibfield  {author} {\bibinfo {author} {\bibfnamefont {A.}~\bibnamefont
  {van~de Walle}}\ and\ \bibinfo {author} {\bibfnamefont {G.}~\bibnamefont
  {Ceder}},\ }\href@noop {} {\bibfield  {journal} {\bibinfo  {journal} {Rev.
  Mod. Phys.}\ }\textbf {\bibinfo {volume} {74}},\ \bibinfo {pages} {11}
  (\bibinfo {year} {2002})}\BibitemShut {NoStop}%
\bibitem [{\citenamefont {Bonny}\ \emph {et~al.}(2009)\citenamefont {Bonny},
  \citenamefont {Pasianot}, \citenamefont {Malerba}, \citenamefont {Caro},
  \citenamefont {Olsson},\ and\ \citenamefont {Lavrentiev}}]{Bonny2009}%
  \BibitemOpen
  \bibfield  {author} {\bibinfo {author} {\bibfnamefont {G.}~\bibnamefont
  {Bonny}}, \bibinfo {author} {\bibfnamefont {R.~C.}\ \bibnamefont {Pasianot}},
  \bibinfo {author} {\bibfnamefont {L.}~\bibnamefont {Malerba}}, \bibinfo
  {author} {\bibfnamefont {A.}~\bibnamefont {Caro}}, \bibinfo {author}
  {\bibfnamefont {P.}~\bibnamefont {Olsson}}, \ and\ \bibinfo {author}
  {\bibfnamefont {M.~Y.}\ \bibnamefont {Lavrentiev}},\ }\href {\doibase
  10.1016/j.jnucmat.2008.12.001} {\bibfield  {journal} {\bibinfo  {journal} {J.
  Nucl. Mater.}\ }\textbf {\bibinfo {volume} {385}},\ \bibinfo {pages} {268}
  (\bibinfo {year} {2009})}\BibitemShut {NoStop}%
\bibitem [{\citenamefont {Gomankov}\ \emph {et~al.}(1971)\citenamefont
  {Gomankov}, \citenamefont {Puzei}, \citenamefont {Sigaev}, \citenamefont
  {Kozis},\ and\ \citenamefont {Maltsev}}]{Gomankov1971}%
  \BibitemOpen
  \bibfield  {author} {\bibinfo {author} {\bibfnamefont {V.~I.}\ \bibnamefont
  {Gomankov}}, \bibinfo {author} {\bibfnamefont {I.~M.}\ \bibnamefont {Puzei}},
  \bibinfo {author} {\bibfnamefont {V.~N.}\ \bibnamefont {Sigaev}}, \bibinfo
  {author} {\bibfnamefont {E.~V.}\ \bibnamefont {Kozis}}, \ and\ \bibinfo
  {author} {\bibfnamefont {E.~I.}\ \bibnamefont {Maltsev}},\ }\href@noop {}
  {\bibfield  {journal} {\bibinfo  {journal} {Pis'ma Zh. Eksp. Teor. Fiz.}\
  }\textbf {\bibinfo {volume} {13}},\ \bibinfo {pages} {600} (\bibinfo {year}
  {1971})}\BibitemShut {NoStop}%
\bibitem [{\citenamefont {Menshikov}\ \emph {et~al.}(1972)\citenamefont
  {Menshikov}, \citenamefont {Arkhipov}, \citenamefont {Zakharov},\ and\
  \citenamefont {Sidorov}}]{Menshikov1972}%
  \BibitemOpen
  \bibfield  {author} {\bibinfo {author} {\bibfnamefont {A.~Z.}\ \bibnamefont
  {Menshikov}}, \bibinfo {author} {\bibfnamefont {V.~Y.}\ \bibnamefont
  {Arkhipov}}, \bibinfo {author} {\bibfnamefont {A.~I.}\ \bibnamefont
  {Zakharov}}, \ and\ \bibinfo {author} {\bibfnamefont {S.~K.}\ \bibnamefont
  {Sidorov}},\ }\href@noop {} {\bibfield  {journal} {\bibinfo  {journal} {Fiz.
  Met. Metalloved.}\ }\textbf {\bibinfo {volume} {34}},\ \bibinfo {pages} {309}
  (\bibinfo {year} {1972})}\BibitemShut {NoStop}%
\bibitem [{\citenamefont {Robertson}\ \emph {et~al.}(1999)\citenamefont
  {Robertson}, \citenamefont {Ice}, \citenamefont {Sparks}, \citenamefont
  {Jiang}, \citenamefont {Zschack}, \citenamefont {Bley}, \citenamefont
  {Lefebvre},\ and\ \citenamefont {Bessiere}}]{Robertson1999}%
  \BibitemOpen
  \bibfield  {author} {\bibinfo {author} {\bibfnamefont {J.~L.}\ \bibnamefont
  {Robertson}}, \bibinfo {author} {\bibfnamefont {G.~E.}\ \bibnamefont {Ice}},
  \bibinfo {author} {\bibfnamefont {C.~J.}\ \bibnamefont {Sparks}}, \bibinfo
  {author} {\bibfnamefont {X.}~\bibnamefont {Jiang}}, \bibinfo {author}
  {\bibfnamefont {P.}~\bibnamefont {Zschack}}, \bibinfo {author} {\bibfnamefont
  {F.}~\bibnamefont {Bley}}, \bibinfo {author} {\bibfnamefont {S.}~\bibnamefont
  {Lefebvre}}, \ and\ \bibinfo {author} {\bibfnamefont {M.}~\bibnamefont
  {Bessiere}},\ }\href@noop {} {\bibfield  {journal} {\bibinfo  {journal}
  {Phys. Rev. Lett.}\ }\textbf {\bibinfo {volume} {82}},\ \bibinfo {pages}
  {2911} (\bibinfo {year} {1999})}\BibitemShut {NoStop}%
\bibitem [{\citenamefont {Caudron}\ \emph {et~al.}(1992)\citenamefont
  {Caudron}, \citenamefont {Sarfati}, \citenamefont {Barrachin}, \citenamefont
  {Finel}, \citenamefont {Ducastelle},\ and\ \citenamefont
  {Solal}}]{Caudron1992}%
  \BibitemOpen
  \bibfield  {author} {\bibinfo {author} {\bibfnamefont {R.}~\bibnamefont
  {Caudron}}, \bibinfo {author} {\bibfnamefont {M.}~\bibnamefont {Sarfati}},
  \bibinfo {author} {\bibfnamefont {M.}~\bibnamefont {Barrachin}}, \bibinfo
  {author} {\bibfnamefont {A.}~\bibnamefont {Finel}}, \bibinfo {author}
  {\bibfnamefont {F.}~\bibnamefont {Ducastelle}}, \ and\ \bibinfo {author}
  {\bibfnamefont {F.}~\bibnamefont {Solal}},\ }\href@noop {} {\bibfield
  {journal} {\bibinfo  {journal} {J. Phys. I France}\ }\textbf {\bibinfo
  {volume} {2}},\ \bibinfo {pages} {1145} (\bibinfo {year} {1992})}\BibitemShut
  {NoStop}%
\bibitem [{\citenamefont {Sch\"{o}nfeld}\ \emph {et~al.}(1988)\citenamefont
  {Sch\"{o}nfeld}, \citenamefont {Reinhard}, \citenamefont {Kostorz},\ and\
  \citenamefont {B\"{u}hrer}}]{Schonfeld1988}%
  \BibitemOpen
  \bibfield  {author} {\bibinfo {author} {\bibfnamefont {B.}~\bibnamefont
  {Sch\"{o}nfeld}}, \bibinfo {author} {\bibfnamefont {L.}~\bibnamefont
  {Reinhard}}, \bibinfo {author} {\bibfnamefont {G.}~\bibnamefont {Kostorz}}, \
  and\ \bibinfo {author} {\bibfnamefont {W.~B.}\ \bibnamefont {B\"{u}hrer}},\
  }\href@noop {} {\bibfield  {journal} {\bibinfo  {journal} {Phys. Stat. Sol.
  B}\ }\textbf {\bibinfo {volume} {148}},\ \bibinfo {pages} {457} (\bibinfo
  {year} {1988})}\BibitemShut {NoStop}%
\bibitem [{\citenamefont {Newman}\ and\ \citenamefont {T}(1999)}]{Newman1999}%
  \BibitemOpen
  \bibfield  {author} {\bibinfo {author} {\bibfnamefont {M.~E.~J.}\
  \bibnamefont {Newman}}\ and\ \bibinfo {author} {\bibfnamefont {B.~G.}\
  \bibnamefont {T}},\ }\href@noop {} {\emph {\bibinfo {title} {{Monte Carlo
  Methods in Statistical Physics}}}}\ (\bibinfo  {publisher} {Clarendon
  Press},\ \bibinfo {address} {Oxford},\ \bibinfo {year} {1999})\BibitemShut
  {NoStop}%
\bibitem [{\citenamefont {Muzyk}\ \emph {et~al.}(2011)\citenamefont {Muzyk},
  \citenamefont {Nguyen-Manh}, \citenamefont {Kurzyd\l{}owski}, \citenamefont
  {Baluc},\ and\ \citenamefont {Dudarev}}]{Muzyk2011}%
  \BibitemOpen
  \bibfield  {author} {\bibinfo {author} {\bibfnamefont {M.}~\bibnamefont
  {Muzyk}}, \bibinfo {author} {\bibfnamefont {D.}~\bibnamefont {Nguyen-Manh}},
  \bibinfo {author} {\bibfnamefont {K.~J.}\ \bibnamefont {Kurzyd\l{}owski}},
  \bibinfo {author} {\bibfnamefont {N.~L.}\ \bibnamefont {Baluc}}, \ and\
  \bibinfo {author} {\bibfnamefont {S.~L.}\ \bibnamefont {Dudarev}},\ }\href
  {\doibase 10.1103/PhysRevB.84.104115} {\bibfield  {journal} {\bibinfo
  {journal} {Phys. Rev. B}\ }\textbf {\bibinfo {volume} {84}},\ \bibinfo
  {pages} {104115} (\bibinfo {year} {2011})}\BibitemShut {NoStop}%
\bibitem [{\citenamefont {Muzyk}\ \emph {et~al.}(2013)\citenamefont {Muzyk},
  \citenamefont {Nguyen-Manh}, \citenamefont {Wr\'{o}bel}, \citenamefont
  {Kurzyd\l{}owski}, \citenamefont {Baluc},\ and\ \citenamefont
  {Dudarev}}]{Muzyk2013}%
  \BibitemOpen
  \bibfield  {author} {\bibinfo {author} {\bibfnamefont {M.}~\bibnamefont
  {Muzyk}}, \bibinfo {author} {\bibfnamefont {D.}~\bibnamefont {Nguyen-Manh}},
  \bibinfo {author} {\bibfnamefont {J.}~\bibnamefont {Wr\'{o}bel}}, \bibinfo
  {author} {\bibfnamefont {K.~J.}\ \bibnamefont {Kurzyd\l{}owski}}, \bibinfo
  {author} {\bibfnamefont {N.~L.}\ \bibnamefont {Baluc}}, \ and\ \bibinfo
  {author} {\bibfnamefont {S.~L.}\ \bibnamefont {Dudarev}},\ }\href {\doibase
  10.1016/j.jnucmat.2012.10.025} {\bibfield  {journal} {\bibinfo  {journal} {J.
  Nucl. Mater.}\ }\textbf {\bibinfo {volume} {442}},\ \bibinfo {pages} {S680}
  (\bibinfo {year} {2013})}\BibitemShut {NoStop}%
\end{thebibliography}%

\newpage
\appendix

\section{Averaged cluster functions for triple clusters}

Similarly to Eq. \ref{eq:pair_cluster_funct}, the average cluster functions for triple clusters ($n$-th nearest neighbours) are defined as

\begin{eqnarray}
\left\langle \Gamma_{3,n}^{(ijk)}\right\rangle&=&\langle\gamma_i,\gamma_j,\gamma_k\rangle
\nonumber \\
&=&\sum_p\sum_q\sum_ry_n^{pqr}\gamma_i(\sigma_p)\gamma_j(\sigma_q)\gamma_k(\sigma_r),
\label{eq:triple_cluster_funct}
\end{eqnarray}

where $y_n^{pqr}$ is the probability of finding $p$, $q$ and $r$ atoms in the $n$-th nearest neighbour coordination shell. In particular, we write
\begin{eqnarray}
\left\langle \Gamma_{3,n}^{(111)}\right\rangle&=&\frac{1}{8}\left(-8y_n^{AAA}+12y_n^{AAB}+12y_n^{AAC}-6y_n^{ABB}\right.
\nonumber \\
&-&\left.6y_n^{ABC}-6y_n^{ACC}+y_n^{BBB}\right. \nonumber \\
&+&\left.3y_n^{BBC}+3y_n^{BCC}+y_n^{CCC}\right) \nonumber \\
\left\langle \Gamma_{3,n}^{(112)}\right\rangle&=&\frac{\sqrt{3}}{8}\left(-4y_n^{AAB}+4y_n^{AAC}+4y_n^{ABB}-4y_n^{ACC}\right.
\nonumber \\
&-&\left.y_n^{BBB}-y_n^{BBC}+y_n^{BCC}+y_n^{CCC}\right) \nonumber \\
\left\langle \Gamma_{3,n}^{(122)}\right\rangle&=&\frac{3}{8}\left(-2y_n^{ABB}+2y_n^{ABC}-2y_n^{ACC}+y_n^{BBB}\right.
\nonumber \\
&-&\left.y_n^{BBC}-y_n^{BCC}+y_n^{CCC}\right) \nonumber \\
\left\langle \Gamma_{3,n}^{(222)}\right\rangle&=&\frac{3\sqrt{3}}{8}\left(-y_n^{BBB}+3y_n^{BBC}\right.
\nonumber \\
&-&\left.3y_n^{BCC}+y_n^{CCC}\right)
\label{eq:Phi_triple}
\end{eqnarray}

Rewriting Eq. \ref{eq:CE_expanded_1} using the average point, pair and triple correlation functions from Eqs. \ref{eq:Gamma_point_clust}, \ref{eq:Gamma_pairs} and \ref{eq:Phi_triple}, the configurational enthalpy of mixing of ternary alloys can now be expressed as a function of concentrations, $c_i$, and the average pair and 3-body probabilities, $y_n^{ij}$ and  $y_n^{ijk}$ as
\begin{widetext}

\begin{eqnarray}
\Delta H_{CE}(\vec{\sigma}) &=& J_1^{(0)}+J_1^{(1)}\left(1-3c_A\right) + J_1^{(2)}\frac{\sqrt{3}}{2}\left(c_C-c_B\right) \nonumber \\
&+&\sum_{n}^{pairs} \left[\frac{1}{4}m_{2,n}^{(11)}J_{2,n}^{(11)}\left(1+3y_n^{AA}-6y_n^{AB}-6y_n^{AC}\right)  \right. \nonumber \\
&+& \frac{\sqrt{3}}{4}m_{2,n}^{(12)}J_{2,n}^{(12)}\left(-y_n^{BB}+y_n^{CC}+2y_n^{AB}-2y_n^{AC}\right) + \left.\frac{3}{4}m_{2,n}^{(22)}J_{2,n}^{(22)}\left(y_n^{BB}+y_n^{CC}-2y_n^{BC}\right) \right] \nonumber \\
&+& \sum_{n}^{triples} \left[ \frac{1}{8}m_{3,n}^{(111)}J_{3,n}^{(111)}\left(-8y_n^{AAA}+12y_n^{AAB}+12y_n^{AAC}
\right.\right. \nonumber \\
&-& \left. 6y_n^{ABB}-6y_n^{ABC}-6y_n^{ACC}+y_n^{BBB}+3y_n^{BBC}+3y_n^{BCC}+y_n^{CCC}\right) \nonumber \\
&+& \frac{\sqrt{3}}{8}\left(m_{3,n}^{(112)}J_{3,n}^{(112)}+m_{3,n}^{(121)}J_{3,n}^{(121)}+m_{3,n}^{(211)}J_{3,n}^{(211)}\right)  \nonumber \\
&\cdot& \left(-4y_n^{AAB}+4y_n^{AAC} + 4y_n^{ABB}-4y_n^{ACC}-y_n^{BBB}-y_n^{BBC}+y_n^{BCC}+y_n^{CCC}\right) \nonumber \\
&+& \frac{3}{8}\left(m_{3,n}^{(122)}J_{3,n}^{(122)}+m_{3,n}^{(212)}J_{3,n}^{(212)}+m_{3,n}^{(221)}J_{3,n}^{(221)}\right) \nonumber \\
&\cdot & \left(-2y_n^{ABB}+2y_n^{ABC}-2y_n^{ACC}+y_n^{BBB}-y_n^{BBC}-y_n^{BCC}+y_n^{CCC}\right) \nonumber \\
&+& \left. \frac{3\sqrt{3}}{8}m_{3,n}^{(222)}J_{3,n}^{(222)}\left(-y_n^{BBB}+3y_n^{BBC}-3y_n^{BCC}+y_n^{CCC} \right)
 \right] \nonumber \\
&+& \sum_{n}^{multibody} \ldots
\label{eq:CE_expanded_Suppl}
\end{eqnarray}
\end{widetext}

\section{Input ternary structures for bcc alloys}

Input ternary CE structures for bcc alloys are constructed from binary structures \cite{Nguyen-Manh2007}, by replacing atoms A or B at non-equivalent atomic positions by atoms C. \\
a) A$_3$B$_{13}$ - based on 13sc222 \\
Space group: $P4/mmm$ (no. 123) \\
Wyckoff positions: \\
A$_1$ 1$c$ ($\frac{1}{2},\frac{1}{2},0$) \\
A$_2$ 1$d$ ($\frac{1}{2},\frac{1}{2},\frac{1}{2}$) \\
A$_3$ 1$a$ (0,0,0) \\
B$_1$ 2$f$ ($0,\frac{1}{2},0$) \\
B$_2$ 8$r$ ($\frac{3}{4},\frac{3}{4},\frac{3}{4}$) \\
B$_3$ 2$e$ ($0,\frac{1}{2},\frac{1}{2}$) \\
B$_4$ 1$b$ ($0,0,\frac{1}{2}$) \\
b) A$_5$B$_{11}$ - based on 11sc222 \\
Space group: $P4/mmm$ (no. 123) \\
Wyckoff positions: \\
A$_1$ 1$a$ (0,0,0) \\
A$_2$ 2$f$ ($0,\frac{1}{2},0$) \\
A$_3$ 2$e$ ($0,\frac{1}{2},\frac{1}{2}$) \\
B$_1$ 8$r$ ($\frac{3}{4},\frac{3}{4},\frac{3}{4}$) \\
B$_2$ 1$b$ ($0,0,\frac{1}{2}$) \\
B$_3$ 1$c$ ($\frac{1}{2},\frac{1}{2},0$) \\
B$_4$ 1$d$ ($\frac{1}{2},\frac{1}{2},\frac{1}{2}$) \\
c) A$_3$B$_5$ - based on PdTi \\
Space group: $P4/mmm$ (no. 123) \\
Wyckoff positions: \\
A$_1$ 2$h$ ($\frac{1}{2},\frac{1}{2},\frac{1}{8}$) \\
A$_2$ 1$b$ ($0,0,\frac{1}{2}$) \\
B$_1$ 1$a$ ($0,0,0$) \\
B$_2$ 2$g$ ($0,0,\frac{1}{4}$) \\
B$_3$ 2$h$ ($\frac{1}{2},\frac{1}{2},\frac{3}{8}$) \\
d) A$_3$B$_5$ - based on tP8-L53-1 \\
Space group: $P4/mmm$ (no. 123) \\
Wyckoff positions: \\
A$_1$ 2$h$ ($\frac{1}{2},\frac{1}{2},\frac{7}{8}$) \\
A$_2$ 1$a$ ($0,0,0$) \\
B$_1$ 2$g$ ($0,0,\frac{1}{4}$) \\
B$_2$ 2$h$ ($\frac{1}{2},\frac{1}{2},\frac{3}{8}$) \\
B$_3$ 1$b$ ($0,0,\frac{1}{2}$) \\
e) A$_3$B$_4$ - based on tI14-L34-2 \\
Space group: $I4/mmm$ (no. 139) \\
Wyckoff positions: \\
A$_1$ 4$e$ ($0,0,-\frac{1}{14}$) \\
A$_2$ 2$b$ ($0,0,\frac{1}{2}$) \\
B$_1$ 4$e$ ($0,0,\frac{9}{14}$) \\
B$_2$ 4$e$ ($0,0,\frac{3}{14}$) \\
f) A$_7$B$_9$ - based on 9sc222 \\
Space group: $Pm-3m$ (no. 221) \\
Wyckoff positions: \\
A$_1$ 1$a$ ($0,0,0$) \\
A$_2$ 3$d$ ($\frac{1}{2},0,0$) \\
A$_3$ 3$c$ ($0,\frac{1}{2},\frac{1}{2}$) \\
B$_1$ 8$g$ ($\frac{1}{4},\frac{1}{4},\frac{1}{4}$) \\
B$_2$ 1$b$ ($\frac{1}{2},\frac{1}{2},\frac{1}{2}$) \\
g) A$_4$B$_4$ - based on tP8-L44-1 \\
Space group: $P4/nmm$ (no. 129) \\
Wyckoff positions: \\
A$_1$ 2$c$ ($\frac{1}{4},\frac{1}{4},\frac{1}{4}$) \\
A$_2$ 2$c$ ($\frac{1}{4},\frac{1}{4},\frac{1}{2}$) \\
B$_1$ 2$c$ ($\frac{1}{4},\frac{1}{4},\frac{3}{4}$) \\
B$_2$ 2$c$ ($\frac{1}{4},\frac{1}{4},0$) \\
h) A$_5$B$_4$ - based on VZn \\
Space group: $I4/mmm$ (no. 139) \\
Wyckoff positions: \\
A$_1$ 2$a$ (0,0,0) \\
A$_2$ 8$h$ ($\frac{1}{3},\frac{1}{3},0$) \\
B$_1$ 8$i$ ($\frac{2}{3},0,0$) \\
i) A$_4$B$_3$ - based on tI14-L34-1 \\
Space group: $I4/mmm$ (no. 139) \\
Wyckoff positions: \\
A$_1$ 4$e$ (0,0,$\frac{9}{14}$) \\
A$_2$ 4$e$ (0,0,$-\frac{1}{14}$) \\
B$_1$ 4$e$ (0,0,$\frac{3}{14}$) \\
B$_2$ 2$b$ (0,0,$\frac{1}{2}$) \\
j) A$_3$B$_2$ - based on tI10-L32-1 \\
Space group: $I4/mmm$ (no. 139) \\
Wyckoff positions: \\
A$_1$ 4$e$ (0,0,-$\frac{1}{10}$) \\
A$_2$ 2$b$ (0,0,$\frac{1}{2}$) \\
B$_1$ 4$e$ (0,0,$\frac{7}{10}$) \\

\section{Effective cluster interactions for binary alloys}

Table \ref{tab:ECI_binary} contains a complete set of effective cluster interactions for fcc and bcc Fe-Ni, Fe-Cr, and Cr-Ni binary alloys.

\begin{table*}
  \centering
  \caption{Number of points, $|\omega|$, labels $n$, multiplicities, $m_{|\omega|,n}$, and effective cluster interactions, $J_{|\omega|,n}$ (in meV), for fcc and bcc binary alloys: Fe-Cr, Fe-Ni and Cr-Ni. }
\begin{ruledtabular}
    \begin{tabular}{cccccccccc}
   $|\omega|$ & $n$ & \multicolumn{2}{c}{$m_{|\omega|,n}$}  & \multicolumn{6}{c}{$J_{|\omega|,n}$} \\
       &    & $fcc$   & $bcc$   & $fcc$ Fe-Cr   & $bcc$ Fe-Cr   & $fcc$ Fe-Ni   & $bcc$ Fe-Ni   & $fcc$ Cr-Ni   & $bcc$ Cr-Ni \\
    \hline
    1     & 0     & 1     & 1     & -64.130 & 85.814 & -61.528 & -49.833 & -121.455 & 67.238 \\
    1     & 1     & 1     & 1     & -40.636 & 50.986 & -52.464 & -45.967 & 8.527 & -84.613 \\
    2     & 1     & 6     & 4     & 9.034 & -26.088 & 9.926 & -8.198 & 2.982 & -36.395 \\
    2     & 2     & 3     & 3     & -6.667 & -1.530 & -5.087 & 7.622 & -12.547 & 7.956 \\
    2     & 3     & 12    & 6     & 3.308 & 0.219 & 1.901 & 6.252 & 6.350 & 12.229 \\
    2     & 4     & 6     & 12    & -0.332 & 0.219 & 0.851 & 1.724 & 7.129 & -1.245 \\
    2     & 5     & 12    & 4     & -0.386 & 2.601 & -0.347 & 0.731 & 0.048 & 0.510 \\
    3     & 1     & 8     & 12    & -4.548 & 0.721 & 3.310 & 2.927 & 4.165 & 5.918 \\
    3     & 2     & 12    & 12    & 3.547 & -1.976 & 0.572 & 1.870 & 3.926 & 1.599 \\
    3     & 3     & 24    &       & 0.681 &       & 0.853 &       & -2.319 &  \\
    4     & 1     & 2     & 6     & 0.557 & 1.278 & -2.667 & 1.026 & 3.770 & -0.191 \\
    4     & 2     & 12    &       & 0.181 &       & -0.007 &       & 0.945 &  \\
    5     & 1     & 6     & 12    & 1.693 & -2.907 & -0.982 & -0.622 & -5.171 & -0.542 \\

    \end{tabular}%
		\end{ruledtabular}
  \label{tab:ECI_binary}%
\end{table*}%

\end{document}